\documentclass[useAMS,usenatbib,usegraphicx]{mn2e}
 
\usepackage[T1]{fontenc}
\usepackage{pslatex,color,tabularx,hyperref}
\usepackage{amsmath,amsfonts,amssymb,amsxtra,eufrak,url}

\def\idm#1{{\mbox{\scriptsize #1}}}
\def\v#1{{\pmb #1}}
\def\vec#1{{\pmb #1}}
\newcommand{\ab}{a_{\idm{b}}}
\newcommand{\ac}{a_{\idm{c}}}
\newcommand{\ad}{a_{\idm{d}}}
\newcommand{\aee}{a_{\idm{e}}}
\newcommand{\af}{a_{\idm{f}}}
\newcommand{\eb}{e_{\idm{b}}}
\newcommand{\ec}{e_{\idm{c}}}
\newcommand{\ed}{e_{\idm{d}}}
\newcommand{\ee}{e_{\idm{e}}}

\newcommand{\mb}{m_{\idm{b}}}
\newcommand{\mc}{m_{\idm{c}}}
\newcommand{\md}{m_{\idm{d}}}
\newcommand{\me}{m_{\idm{e}}}
\newcommand{\tobs}{t_{\idm{obs}}}
\newcommand{\au}{\mbox{au}}
\newcommand{\msun}{\mbox{M}_{\odot}}

\newcommand{\mJ}{\mbox{m}_{\idm{Jup}}}

\newcommand{\Mmean}{\mathcal{M}}

\newcommand{\cchi}{\sqrt{\chi_{\nu}^2}}

\newcommand{\critLL}{\theta_{\idm{1:2:4:8}}}

\def\mechanic{{\tt Mechanic}}
\def\moa{{MCOA}}
\def\spitzer{{\sc Spitzer}}
\def\hershel{{\sc Herschel}}
\def\kepler{{\sc Kepler}}

\usepackage{color}
\definecolor{myred}{rgb}{0.9,0.1,0.1}
\definecolor{myblue}{rgb}{0.2,0.0,0.7}
\definecolor{mybrown}{rgb}{0.9,0.3,0.2}

\newcommand\corr[1]{{\color{mybrown} #1}}
\renewcommand\corr[1]{{\color{black} #1}} 

\newcommand\hide[1]{}

\title[Multiple mean motion resonances in the HR 8799 planetary system]
{Multiple mean motion resonances in the HR 8799 planetary system}
\author[Go\'zdziewski \& Migaszewski]%
{Krzysztof Go\'zdziewski$^{1}$\thanks{E-mail: k.gozdziewski@astri.umk.pl}
\& Cezary Migaszewski$^{1}$\thanks{E-mail: c.migaszewski@astri.umk.pl}\\
$^1$ Centre for Astronomy, 
Faculty of Physics, Astronomy and Informatics,
Nicolaus Copernicus University, 
Grudziadzka 5, 87-100 Toru\'n, Poland
}
\begin{document}
%
\date{Accepted .... Received ...; in original form ...}
\pagerange{\pageref{firstpage}--\pageref{lastpage}} \pubyear{2012}
\maketitle
\label{firstpage}

\begin{abstract}
HR~8799 is a nearby star hosting at least four $\sim 10$\,$\mJ$ planets in
wide orbits up to $\sim 70$~au, detected through the direct,
high-contrast infrared imaging.  Large companions and debris disks reported interior
to $\sim 10$~au, and exterior to $\sim 100$~au indicate massive
protoplanetary disc in the past.  The dynamical state of the HR 8799 system
is not yet fully resolved, due to limited astrometric data covering tiny
orbital arcs.  We construct a new, orbital model of the HR~8799 system, assuming rapid migration
of the planets after their formation in wider orbits.  We found that the
HR~8799 planets are likely involved in double Laplace resonance, 1e:2d:4c:8b~MMR. 
Quasi-circular planetary orbits are coplanar with the stellar equator and inclined by
$\sim 25^{\circ}$ to the sky plane.  This best-fit orbital configuration matches astrometry, debris
disk models, and mass estimates from cooling models.  The multiple MMR is
stable for the age of the star $\sim 160$~Myr, for at least 1~Gyr unless
significant perturbations to the $N$-body dynamics are present.  We predict
four configurations with the fifth hypothetical innermost planet HR~8799f in
$\sim 9.7$~au, or $\sim 7.5$~au orbit, extending the MMR chain to triple
Laplace resonance 1f:2e:4d:8c:16b~MMR or to the 1f:3e:6d:12c:24b~MMR,
respectively.  Our findings may establish strong boundary conditions for the
system formation and its early history.
\end{abstract}

%
\section{Introduction}
\label{sec:introduction}
%
Numerous ground based and space surveys of extrasolar planets brought thousands of new detections since the pioneering radial velocity (RV) observations \citep{Walker2012}. Besides the RV method,  the most successful observational techniques are transits, micro-lensing, eclipse timing and direct imaging \citep{Perryman2011,Bhattacharjee2013}. The observations revealed a few hundreds of confirmed and well characterized extrasolar planets, exhibiting a rich diversity of orbital architectures, masses, densities, radii, as well as spectral types and evolutionary stages of single and binary parent stars \citep{Howard2013}. Most of the extrasolar planets have been detected within a few astronomical units (au) of their host stars. Only the direct imaging brought a handful detections of massive planets beyond 10~au distance which roughly compares to the orbit of Saturn in the Solar system.  This natural but extremely demanding observational technique  is called the Holy Grail of exoplanet searching \citep{Bhattacharjee2013}. The direct imaging may provide information on masses, radii, chemical composition, atmospheres and orbital architecture \citep{Oppenheimer2013} resulting in a complete characterization of extrasolar planetary systems. The main limitations are the angular resolution and contrast requirements, reaching more than 20 stellar magnitudes.  Therefore the prime targets are only massive, young and still self-radiating giant planets or brown dwarfs, \corr{in orbits} beyond a few au of their host stars.  A recent survey \citep{Wahhaj2013} of 57 debris disk stars \corr{shows} at 95\% confidence that $<13$\% of these stars have a $>5~\mJ$ planet beyond 80~au, and $<21$\% of debris disk stars have a $>3~\mJ$ planet outside of 40~au.  The HR~8799 \citep{Marois2008,Marois2010} belongs to a rare sample of stars hosting planets discovered by the direct imaging \citep{Konopacky2011}. The HR 8799 system remains truly exceptional as the only {\em multiple} and dynamically compact  configuration of four giant planets in $\sim 7-10$\,$\mJ$ mass range. 

Since the discovery, the HR~8799 system receives an enormous attention. Tens
of papers and proceedings are devoted to the age, companion masses, the
orbital architecture of this system and its stability, \corr{debris disks},
long term-evolution and formation \cite[e.g.][]{Matthews2014,Marleau2014,
Oppenheimer2013,Baines2012,Esposito2013,Currie2012,Sudol2012,Soummer2011,Bergfors2011,Currie2011,Marshall2010,MoroMartin2010,Fabrycky2010,Gozdziewski2009,Su2009,Reidemeister2009},
to mention just a~handful of these works. However, two questions seem still opened:
what is the dynamical state of the HR~8799 system, and how this system has
formed \citep{Marois2010}. 

None of simple, analytic criteria of stability apply to the HR~8799 system.  The early dynamical analyses  \citep{Fabrycky2010,Gozdziewski2009,Reidemeister2009} of the three-planet systems announced in the discovery paper \citep{Marois2008} revealed that apparently circular, wide ($\sim 100$~au) orbits are separated by less than 3--4~mutual Hill radii.  Such configuration must self-destruct statistically in  100,000~years time-scale \citep{Chambers1996,Chatterjee2008} unless a protecting mechanism is present. Such a mechanism maintaining the stability for the star lifetime can be the mean motion resonance (MMR). Indeed, three outer planets hosted by the the HR~8799 are most likely involved in stable Laplace 1d:2c:4b~MMR \citep{Fabrycky2010,Gozdziewski2009,Reidemeister2009,Soummer2011,Marshall2010}, similarly to innermost moons of Jupiter (Io, Europa and Callisto). (This is further confirmed in this paper). The orbits are low-eccentric, coplanar and inclined by $\sim 20^{\circ}$--$30^{\circ}$ to the sky plane. Companion masses are estimated in 5-7\,$\mJ$ for planet~b and in 7-10\,$\mJ$ for planets~c and~d, in accord with evolution theories and cooling rates of sub-stellar objects \citep{Baraffe2003}. This \corr{early understanding} of the HR~8799 system with three outer planets is roughly consistent across the literature.  

However, after the discovery of the fourth planet \citep{Marois2010}, there is no unique nor certain orbital model that predicts long-term stable dynamical evolution anymore
\citep{Marois2010,Currie2011,Konopacky2011,Sudol2012,Currie2012,Esposito2013}. This problem might be expected.  The initial condition of the four-planet system involves almost \corr{30~free parameters (osculating orbital elements and masses)}. The observational time window of $\sim 15$~years is very narrow as compared to the orbital periods between $\sim 50$~years and $\sim 500$~years, i.e., only $\sim 30\%$ to $\sim 3\%$ of the innermost and outermost orbital arcs, respectively.  A small number of $\sim 60$ observations with significant uncertainties {($\sim 1\%$)} results in the ratio of measurements per the degree of freedom close to~4. In contrast, to well characterize planetary orbits by the RV technique, this ratio should be 6--7, provided that data cover roughly 1--2 longest orbital periods. Any determination of the HR~8799 orbits by the common kinematic (Keplerian) or $N$-body (Newtonian) approaches is badly constrained. \corr{Actually, it is not clear}, whether this young system is dynamically stable at all, and \corr{could} be disrupted due to strong mutual \corr{interactions and dynamical chaos} present \citep[e.g.,][]{Gozdziewski2009,Sudol2012,Esposito2013}.

The second unsolved problem \corr{regards close proximity} of two inner planets to the parent star, interior to $\sim 25$~au orbit.  The present planet formation theories cannot explain creation of all four planets {\em in situ} by one mechanism~\citep{Marois2010}. Because the spectral studies reveal similar sizes, chemical compounds and age of the planets~\citep{Marois2010}, also the birth conditions of all planets likely have been similar. [See however the new results by \cite{Oppenheimer2013}]. Therefore, if all planets formed relatively far from the star, through gravitational fragmentation and/or the core accretion [see a discussion in \citep[][]{Currie2011}], they {likely have been moved} to their present orbits from wider orbits. This might be possible via the planet-planet scattering \citep{Chatterjee2008} or the planetary migration. The planet-planet scattering might explain the marginal stability of the system reported in many papers. However, the
\spitzer{}  and \hershel{} observations \citep{Chen2006,Su2009,Matthews2014} detected two coplanar, \corr{massive debris disks} and an extended spherical halo around HR~8799. The inner warm disk reminds the asteroid belt, while the outer cold disk is similar to the Kuiper Belt in the Solar system. The presence of at least four giant planets and extended debris features are suggestive for a particularly massive protoplanetary disk. The presence of such a large disk might support and indicate the fragmentation formation and migration scenario. The migration could be rapid in such a presumably massive disk, in accord with estimates of the $100\,\au$ migration time-scale as short as $10^4$~yrs \citep{Baruteu2011}. 

In this paper we report a possible solution of both these problems, although focusing mainly on the orbital properties the HR~8799 system and its global dynamics. We derived a long-term stable $N$-body model of the HR~8799 system in accord with astrometric observations and companion mass estimates published to date. This self-consistent model relies on three basic assumptions, which actually reflect the results in the extensive literature:  i) the current HR~8788 planetary system emerged due to joint migration of four (or even more) massive planets that have been formed in wide orbits, ii) the system is coplanar or almost coplanar,
iii) the system is long-term stable and the stability is maintained by the mean motion resonances.  These assumptions lead us to construct a new optimization algorithm \corr{for} finding such configurations, which are strictly stable and fully consistent with \corr{astrometry and astrophysical} mass constraints. 

Our approach differs from common methods of modeling planetary systems by a crucial aspect. We assume that the orbital elements are {\em not free nor independent} parameters of the data model. Instead, the orbital elements are constrained by the dynamical evolution governed by planetary migration, hence the orbits are instantly coupled. This component of the optimization  might be thought as a generalization of the self-consistent $N$-body fitting \citep{Laughlin2001} \corr{constrained by the dynamical stability \cite[e.g.][]{Gozdziewski2008,Gozdziewski2009} which imply a complex discretization of the parameter space}.
The method makes use of a heuristic model of migration \citep{Moore2013} and theoretical estimates of the masses varied in prescribed, yet reasonable ranges \corr{derived on the grounds of recent cooling models}, see the very recent paper
by \cite{Marleau2014}. 

This paper is structured as follows. After Introduction, we present a short review of the recent literature devoted mostly to the orbital models of the HR~8799 system.  Section~\ref{sec:method} presents our new approach of the optimization of astrometric data through constraining it by the planetary migration. We call this method  {the Migration Constrained Optimization Algorithm
(\moa{} from hereafter)}.  Section~\ref{sec:architectures} regards the results and details of orbital architectures of the HR~8799 system derived with the
\moa{}. We re-analyse older data in \citep{Marois2008} and \corr{early models} including three outer planets b, c, d~as well as the most recent and complete literature dataset and the best-fitting four-planet model.
A single-epoch characterization of three- and four-planet systems is discussed. We also consider ephemeris of the fifth, hypothetical planet interior to the innermost planet~e. Section \ref{sec:stability} regards \corr{important aspects of the stability analysis}. Conclusions are given in Section~\ref{sec:conclusions}.  \corr{At the end, we provide a} compilation of the observational data and our ephemeris of the four- and five-planet models (Tabs.~\ref{tab:ephemeris1}--\ref{tab:ephemeris5}) discussed in Sect.~\ref{sec:architectures}.
 
%
\section{Literature models of the HR~8799~system}
\label{sec:literature}
Shortly after the discovery, \cite{Fabrycky2010} found that the stability of the three-planet system may be protected by the three-body Laplace 1d:2c:4b~MMR.  The system locked in this MMR could survive even if the planets have masses as large as $\sim 20$\,$\mJ$.  Other solutions with large eccentricities and large mutual inclinations also were found.   \cite{Gozdziewski2009} found very narrow stable zones in the phase space of the system and a few extreme solutions, like the 1d:1c~MMR between two inner planets.   \cite{Reidemeister2009} concluded that all companions in the discovery paper \citep{Marois2008} may be stable in the mass ranges of (5, 7, 7)\,$\mJ$, (7, 10, 10)$\mJ$ to (11, 13, 13)$\mJ$ provided that the Laplace resonance is present.  They also found that the inclination of the orbital plane must be larger than $\sim 20^{\circ}$.   \cite{Soummer2011} extended the observational window by ten years from the analysis of the HST images of the HR~8799. They confirmed that stable resonances (1d:1c~MMR, 1d:2c~MMR, or 1d:2c:4b~MMR) found in the previous papers are still compatible with the extended astrometric data.  They found stable Laplace resonance with low-eccentric orbit of planet~d $e_{\idm{d}}\sim 0.1$ and moderate inclination of the system $\sim 28^{\circ}$ assuming coplanar configuration with circular orbits of the outermost planets.  After releasing these constraints they could limit the inclination to ($27.3^{\circ}-31.4^{\circ}$) range and $e_{\idm{d}}<0.46$.  \cite{Currie2012} confirmed these results recently by estimating the inclination of HR~8799~d as $I_{\idm{d}}>25^{\circ}$ and finding all eccentricities smaller than $0.18$--$0.3$, with a strong indication of not face-on orbits.   \cite{Bergfors2011} observed the system with NACO and VLT.  They found that planet HR 8799~d is inclined with respect to the line of sight, suggesting that its orbit is slightly eccentric or non-coplanar with the outer planets and debris disk.

The very recent works devoted to the dynamical analysis of the four-planet system including HR~8799~e bring sometimes mutually contradicting conclusions.  \cite{Marois2010} revisited the stability analysis in \citep{Fabrycky2010}. By varying parameters of planet~e in the four-planet system with single 1d:2c~MMR or double 1d:2c:4b~MMR they found a few~solutions surviving 160~Myr in samples of 100,000 trial models, regarding masses in the range of 30~Myr (5,7,7,7)\,$\mJ$ and 60~Myr (7,10,10,10)\,$\mJ$ for planets b, c, d and~e, respectively. Semi-major axis of planet~e has been changed between 12.5~au and 14.5~au. Stability analysis in \citep{Marois2010} suggest a younger age and lower planet masses. \cite{Konopacky2011} determined eccentricity limits to less than 0.4~for all planets. They found that the addition of the fourth planet makes it very difficult to find any stable configuration for masses greater than 7\,$\mJ$, also suggesting lower masses in a younger system ($\sim 30$~Myr). \cite{Currie2011} combined the results from planet evolution models and the stability analysis to limit the masses of planets HR~8799b, c, d, e to the ranges of 6-7$\mJ$, 7-10\,$\mJ$, 7-10\,$\mJ$, and 7-10\,$\mJ$, respectively. \cite{Esposito2013} show that planet~e can not form the 1e:2d:4c~MMR if its orbit is circular and coplanar with planets~d and~c, while such orbits are allowed for the 2e:5d~MMR. They found significant stable regions for masses in the $5\,\mJ$~range, below the current estimates based on the stellar age of 30~Myr and astrophysical models of cooling sub-stellar objects. \cite{Sudol2012} found the system marginally stable, surviving less than $\sim 5$~Myr at inclinations in the range of $\sim ~10^{\circ}$ and $\sim 31$~Myr at larger inclinations $\sim 30^{\circ}$.  The most stable systems also favour planet~e closer to the star than is observed. They conclude that the planetary masses must be less than 7-10-10-10\,$\mJ$ and the system is young. Planets~b and~c could be in eccentric or mutually inclined orbits with respect to planet~d.

\cite{Baines2012} estimate HR~8799 mass $\sim 1.51\,\msun$ and two ages $\sim 30$~Myr and $\sim 90$~Myr, depending on the evolutionary track contracting toward the zero-age main sequence (ZAMS) or expanding from it. These estimates of the HR~8799~age suggest that the companions are indeed  planets. The very recent work of \cite{Oppenheimer2013} brings different spectra of the planets suggesting a greater diversity of these objects than previously found. The \spitzer{} and \hershel{} infrared spectra were used to resolve two coplanar debris belts \citep{Chen2006,Reidemeister2009,Su2009,Hughes2011,Patience2011,Matthews2014} divided by the radial gap between $\sim 15$--$90$~au, and an extended dust halo surrounding the whole system. \cite{Patience2011} measured the first spatially resolved map of the HR~8799 disk at 350~$\mu$m and detected an arc of emission with a bright clump at a distance consistent  with simulations of dust trapped in a 1b:2~MMR with the outermost planet. This result is suggestive for the planets migrated to their current locations and that the eccentricity is low, if the dust is trapped in a resonance \citep{Patience2011}. \cite{Su2009} and \cite{Hinkley2011} report a dust-free hole interior to $\sim 6$~au in the inner warm belt. Assuming an age of 30~Myr and adopting the \cite{Baraffe2003} evolutionary models, \cite{Hinkley2011} determined upper limit of a companion mass of 80, 60, and 11\,$\mJ$ at projected orbital separations of 0.8~au, 1~au, and 3-10~au, respectively, ruling out a brown dwarf or a small star between $0.8$--$10$~au. 

%
\section{Optimization constrained by migration}
\label{sec:method}

{\em A stable} low-order MMR in mutually interacting planetary system is a dynamical state forcing only certain, somehow discrete configurations of the planets. This might be understood and visualized as narrow islands of stable motions in the orbital and physical parameter space \citep[e.g.,][]{Gozdziewski2009,Gozdziewski2008}. A particular multiple MMR determines orbital periods (and semi-major axes); stability constraints permit only certain ranges of eccentricities; the relative orbital phases are limited by a critical argument of the MMR.  Therefore, the best-fitting orbital elements are {\em not free nor independent} parameters of the data model. The orbital elements must be constrained by the dynamical evolution governed by planetary migration and orbits are instantly coupled.  Planets migrate as a whole dissipative dynamical system that synchronizes itself to a certain state (MMR) which is a kind of an equilibrium in the phase space. We then search for those ``equilibria'' which fit the observations at some epoch.  A crucial aspect of a practical realization of this idea is that  a heuristic rather than fully realistic model of the planetary migration \citep{Moore2013} is required to establish a chain of multiple, low-order MMRs. The coupled migration component serves as a kind of implicit constraint of the optimization process. The hard part of this task is that we not know a'priori which initial parameters (masses, initial orbits, migration rates) lead to the best-fitting or acceptable configuration. 

%
\subsection{A heuristic model of planetary migration}
%
Mechanisms of MMRs formation are widely studied as the result of the planetary migration \citep[see a review by][and references therein]{Papaloizou2006}.  Planetary migration is \corr{a sophisticated physical process depending subtly} on many parameters.  Here, we use a simplistic, heuristic $2$-dimensional model of the migration in \citep{Moore2013}.  In this model, the migration is driven by the  drag force in the form of:
\begin{equation}
\vec{F} = -\frac{\vec{v}}{2\,\tau_a} - \frac{\vec{v} - \vec{v}_c}{\tau_e},
\label{eq:moore}
\end{equation}
where $\vec{v}$ is an astrocentric velocity of a planet, $\vec{v}_c$ is the Keplerian velocity of this planet in a circular orbit at a given radius, $\tau_a$ and $\tau_e$ are the migration and circularization rates of orbits, respectively. After \cite{Moore2013}, we assume that $\tau_e = K \tau_a$, where $K$ is a constant between~$1$ and $100$. We assume that $\tau_a$ is a function of  astrocentric distance $r$ of the planet and time, i.e.,
\begin{equation}
\frac{1}{\tau_a} = \frac{1}{\tau_1} + \frac{1}{\tau_2}, \quad \tau_1 = \tau_{1,0} \, r^{\alpha_1} \, \exp(t/T_1), \quad \tau_2 = \tau_{2,0} \, r^{\alpha_2} \, \exp(t/T_2),
\label{eq:ratios}
\end{equation}
where $\alpha_i$, $\tau_i$ and $T_i$ ($i=1,2$) are constant factors in wide ranges, i.e., $\tau_{i,0} \in [10^6, 10^8]\,$yr, $T_i \in [10^6, 10^8]\,$yr, $\alpha_1 \in [-2.0, -0.1]$, $\alpha_2 \in [0.1, 2.0]$. These ranges are sufficient to encompass different types of migration, both convergent and divergent. The first term, with $\alpha_1<0$, leads to the convergent migration, while the second term with $\alpha_2>0$ accounts for the divergent migration. For $\alpha_1 < -1.0$, the period ratios of pairs of subsequent planets $P_{i+1}/P_i$ (where $P_{i+1} > P_i$) decrease in time, while for $\alpha_1 > -1.0$ the period ratios  increase.   

The migration resulting from such wide ranges of the parameters may occur on very different time-scales. For instance, for the convergent term, $\alpha_1 = -1.3$ and $\tau_{1,0} = 10^8\,$yr, $\tau_1 \approx 0.4\,$Myr ar $r=70$~au, which is consistent with the time-scale implied by the type~II migration \citep[e.g.,][]{Kratter2010}. 

To establish the orbital architecture of a studied system, there is no need to resolve all details of physical processes forcing the migration. We found that the heuristic model is sufficient for the optimization. Although permitted ranges of the migration parameters may seem unreasonably wide, we do not narrow these ranges to avoid too strict assumptions on the time-scales, for instance, on their dependencies on $r$.

\begin{figure*}
\centerline{
\hbox{
{\includegraphics[width=0.49\textwidth]{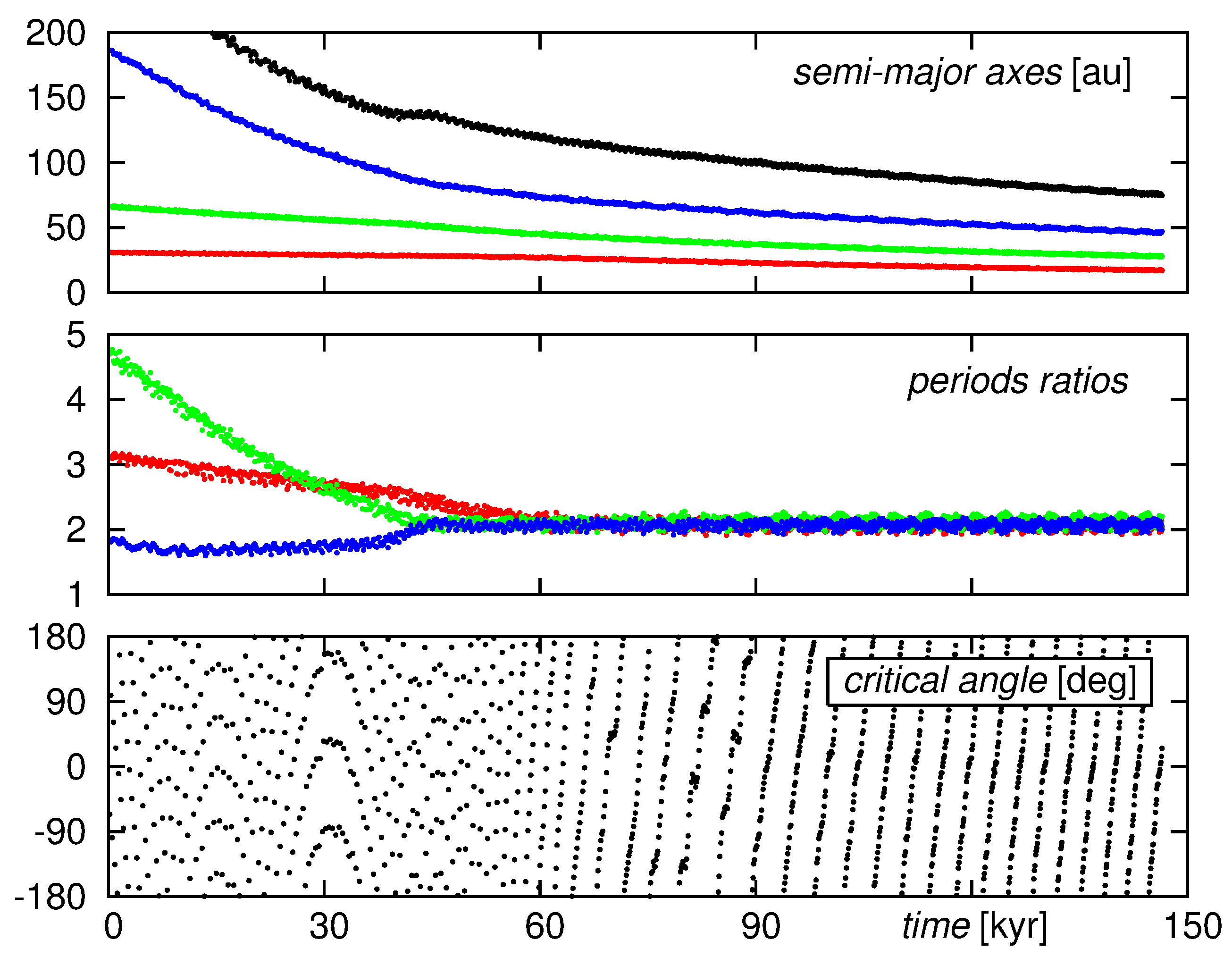}}
{\includegraphics[width=0.49\textwidth]{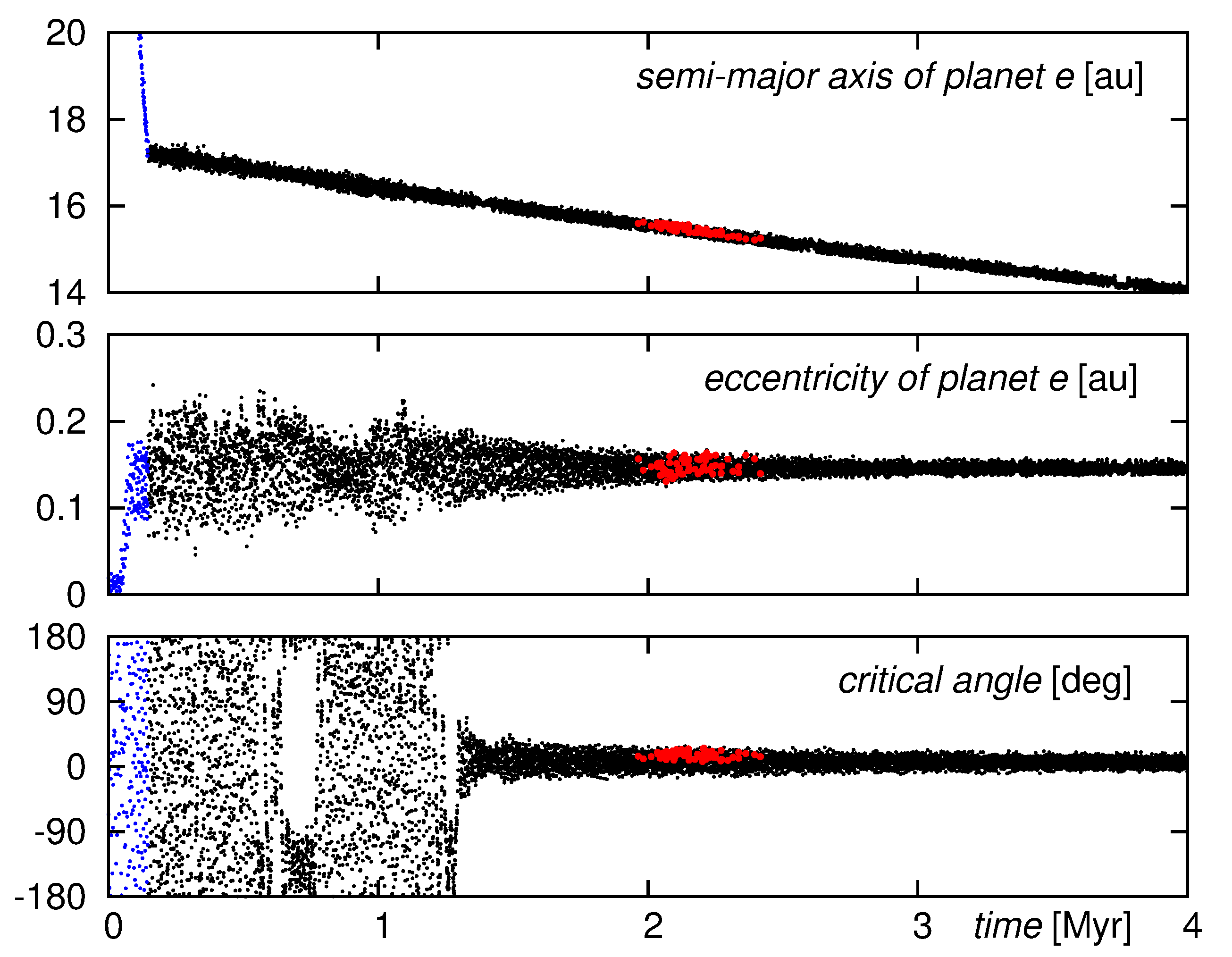}}
}
}
\caption{
Temporal evolution of a four-planet system and trapping the system into 1e:2d:4c:8b~MMR. {\em Left panel:} Evolution of the semi-major axes $a_i$ (i=b, c, d, e), periods ratios and the critical angle of the double Laplace MMR in a fast migration regime. {\em Right panel:} Evolution of $a_{\idm{e}}$, $e_{\idm{e}}$ and the critical angle
\corr{$\theta_{\idm{1:2:4:8}}$} in a slow migration regime. Red symbols mark orbits well fitting
the observations (formal $\cchi < 2$). 
} 
\label{fig:fig1}
\end{figure*}

\begin{figure*}
\centerline{
\hbox{
{\includegraphics[width=0.49\textwidth]{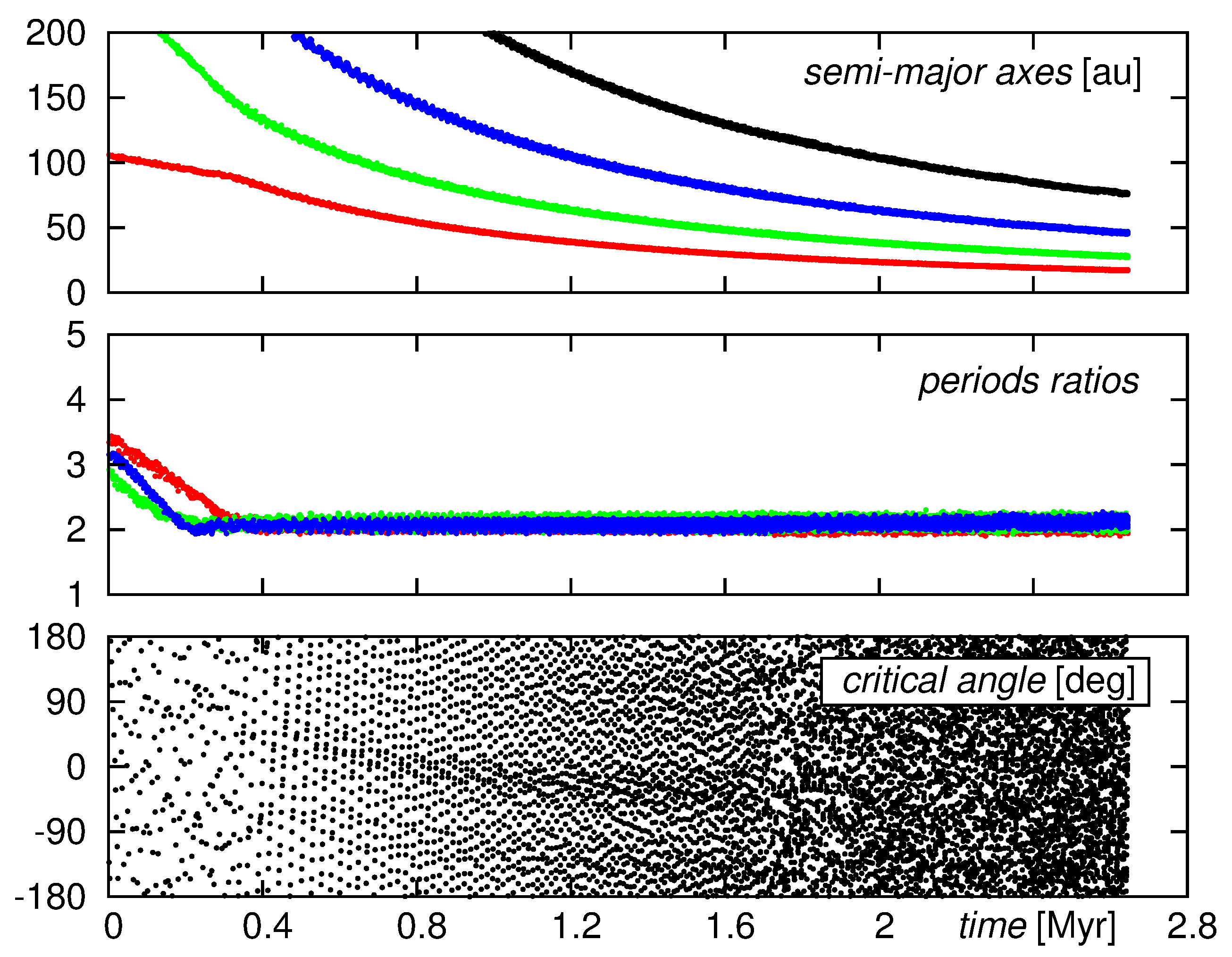}}
{\includegraphics[width=0.49\textwidth]{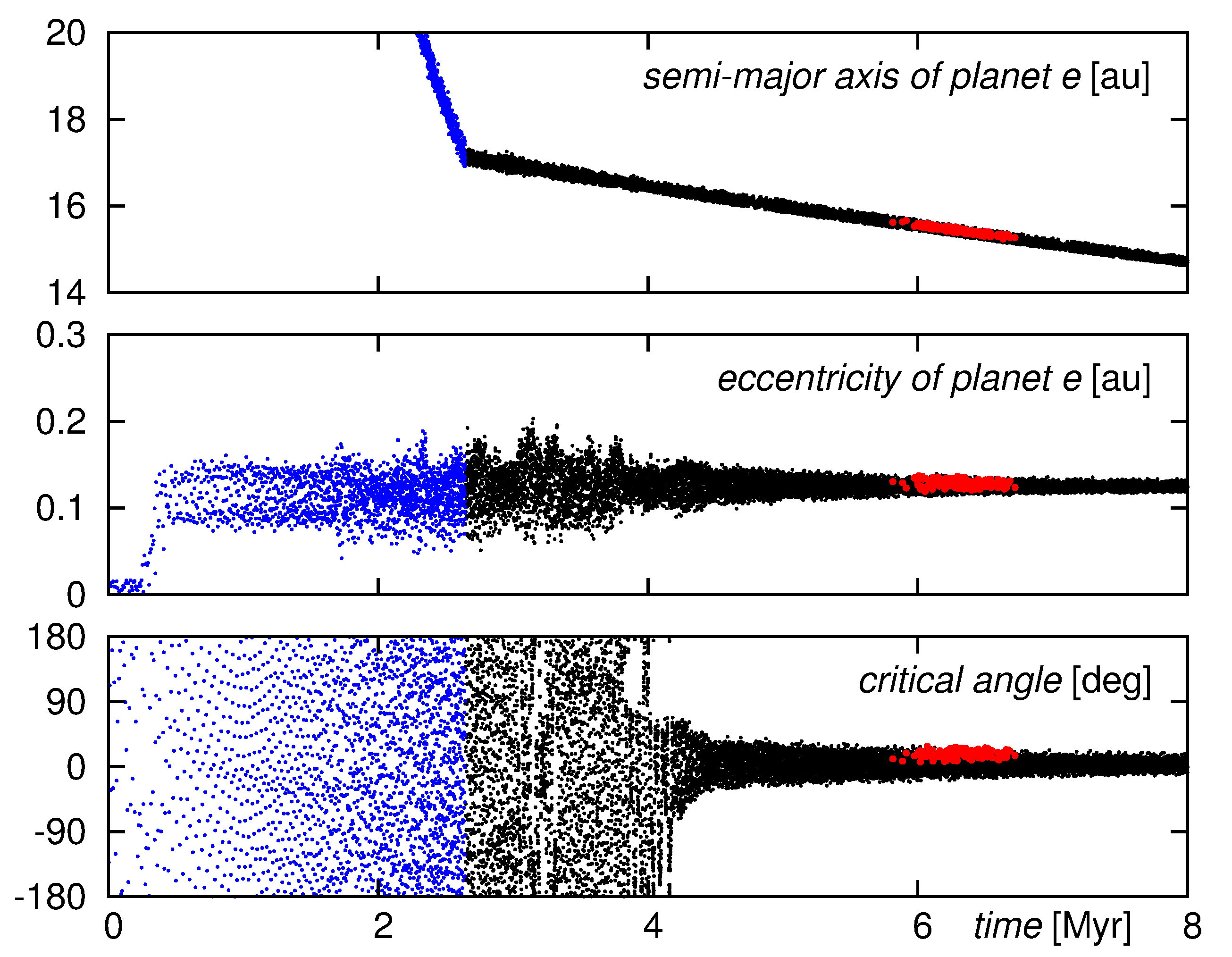}}
}
}
\caption{
Same as on Fig.~\ref{fig:fig1} but for different initial orbits and evolution parameters (initial conditions).
}
\label{fig:fig2}
\end{figure*}

\begin{figure*}
\centerline{
\hbox{
{\includegraphics[width=0.49\textwidth]{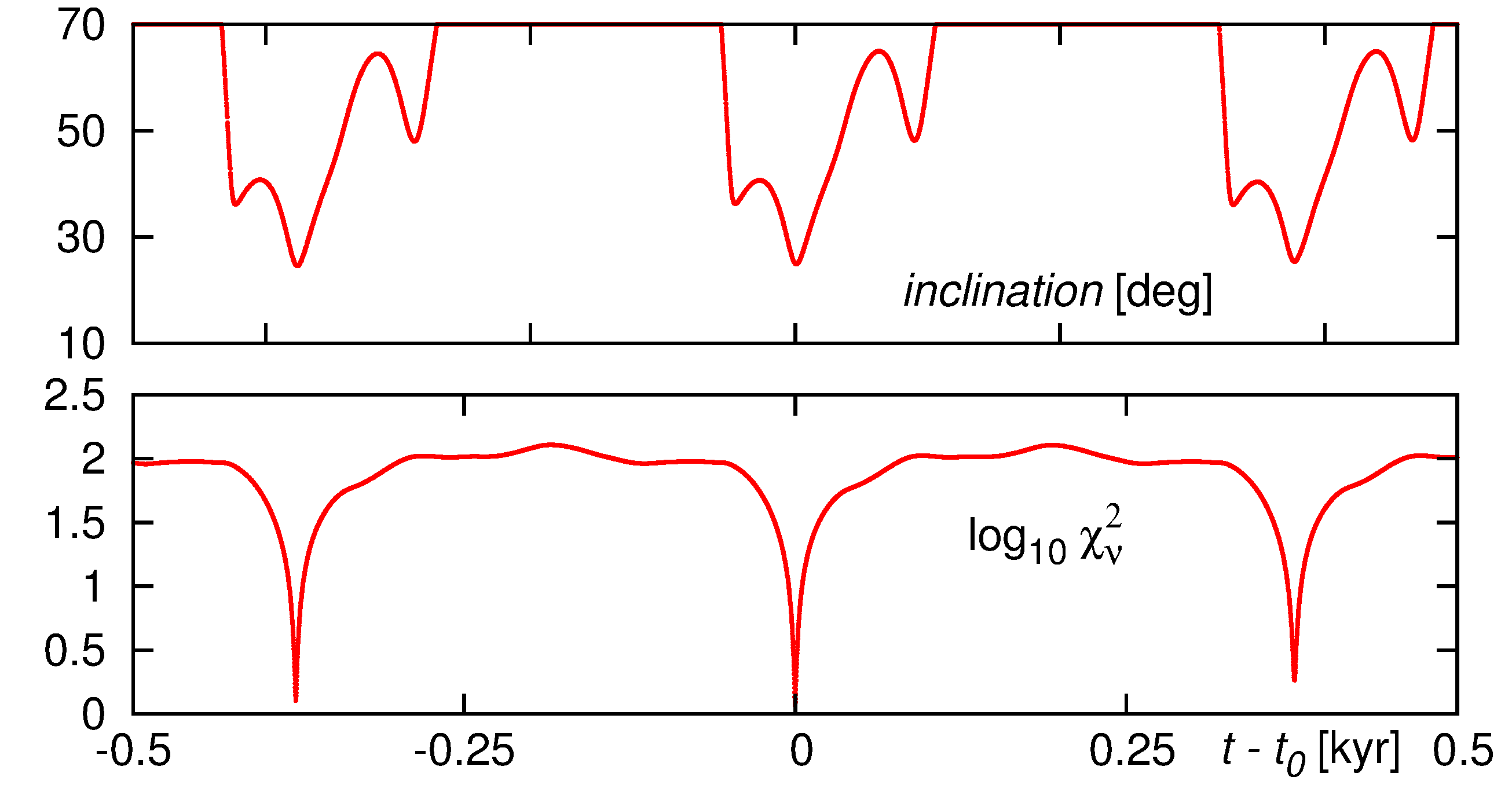}}
{\includegraphics[width=0.49\textwidth]{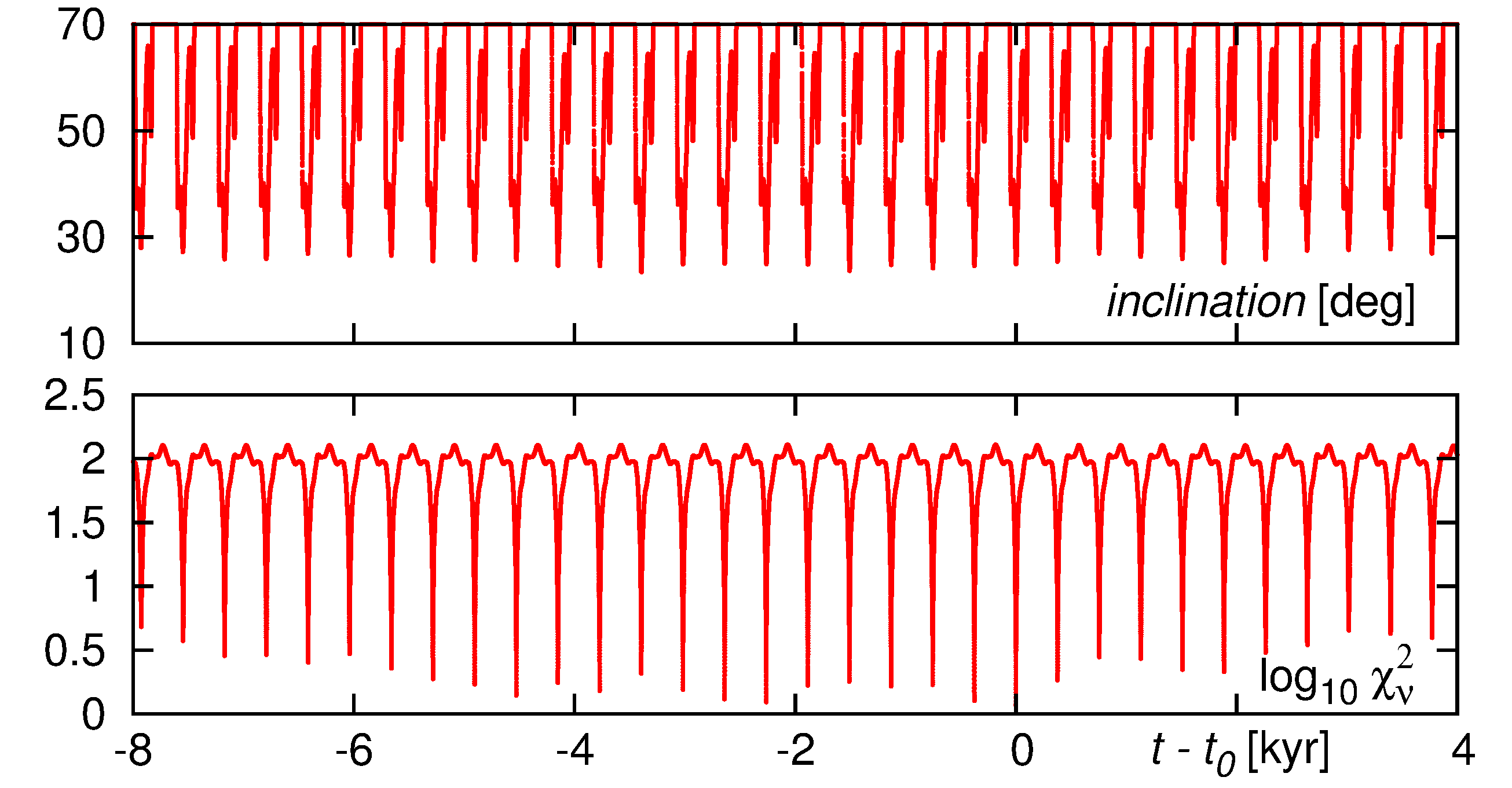}}
}
}
\caption{
Illustration of the optimization algorithm constrained by planetary migration. Evolution of the best-fitting inclination ({\em top}) and $\cchi$ ({\em bottom}) in time around the best-fit solution.
}
\label{fig:fig3}
\end{figure*}
%

%
\subsection{Optimization and migration algorithm in the real world}
%

\subsubsection{Parametrisation and initial conditions}

To initiate a single migration track of the HR~8977 planetary system, we choose at random initial, circular orbits (semi-major axes), planetary masses and the migration parameters.  An initial semi-major axis of an $n$-th planet increases exponentially in accord with $a_n = a_1 \exp{[\beta \left( n - 1 \right)]}$, where $\beta \in [0.2, 1.8]$.  Orbital longitudes of the planets are chosen from the $[0, 360^{\circ}]$ range.  Masses of the planets are constrained through astrophysical cooling models.  For instance, \cite{Marois2008} found $\me = \left( 9 \pm 4 \right)\,\mJ$, $\md = \left( 10 \pm 3 \right)\,\mJ$, $\mc = \left( 10 \pm 3 \right)\,\mJ$, $\mb = 7^{+3}_{-2}\,\mJ$. The mass estimates depend on the star age, which is  a matter of ongoing debate. Ages among 30~Myr, 60~Myr,
90~Myr, 160~Myr and even 1~Gyr are quoted \citep{Baines2012,Oppenheimer2013}. Because our dynamical model provide independent mass constraints, we considered a relatively smooth mass ranges, roughly between [2,13]~$\mJ$ for all planets, up to $\sim 20$\,$\mJ$ in a few experiments.  We integrate the $N$-body equations of motion with the dissipative term (\ref{eq:moore}) added to the right-hand sides, until the migrating system becomes so compact that the orbits cannot be consistent with the observations anymore.

In this way, a given initial condition determines an evolutionary track of the system.  The problem is to find the best-fitting parameters $(\tobs, I, \Omega, \omega_{\idm{rot}})$, where $t_{\idm{obs}}$ is an arbitrary time interval counted from the beginning of the migration and $(\Omega, I, \omega_{\idm{rot}})$ are \corr{3--1--3} Euler angles.  The best-fitting epoch $\tobs$ is a particular moment during the migration that corresponds to correct -- possibly optimal -- sizes of the orbits and relative orbital phases of the planets at these orbits.  Three \corr{relevant Euler angles $(\Omega, I, \omega_{\idm{rot}})$} are required to describe the orientation of an arbitrary orbital frame w.r.t the sky plane (the observer's frame).  This orientation is parameterised naturally through classic 3-1-3 sequence of rotations: $I$ is the inclination of the orbital plane, $\Omega$ is the longitude of ascending node, and $\omega_{\idm{rot}}$ is an angle measured in the orbital plane.
We search for systems evolving in such a manner that being observed in a certain epoch from a certain direction, they look like the current HR~8799 system.  Such particular epoch and orientation of the orbital frame together with the migration model and its parameters determine the current orbital architecture of the system.  Obviously, the migration model, Eqs.\ref{eq:moore},\ref{eq:ratios}, is simplistic and very similar final configurations may be obtained when starting from different initial states.  However, we stress that we do not aim to answer the question {\em How the HR~8799 looked like at the early stages of its evolution?}. We focus entirely on its observed, geometrical architecture and present astrometry.

Searching for the best-fitting solutions in a whole possible range of $\tobs$ would be unreasonably time and CPU consuming. The actual optimization of the measurements (searching for the minimum of $\cchi$) begins when the migrating system looks roughly similar to the HR~8799. This happens when  the semi-major axes of four/five-planet system are confined to the following ranges:   $\aee \in [12, 17]\,\au$, $\ad \in [23, 29]\,\au$, $\ac \in [37, 48]\,\au$, $\ab \in [62, 75]\,\au$, and $\af \in [5, 12]$\,au, respectively. (Note that we also consider five-planet systems with a hypothetical, yet undetected planet HR~8799~f, see Sect.~\ref{sec:architectures}).

To illustrate these ideas,  the left-hand panel of Fig.~\ref{fig:fig1} shows an evolution of an example configuration. The top panel is for the evolution of semi-major axes. The system migrates inward in relatively short time-scale of the order of $10^5$~years. The time axis of $(a_i, t)-$ graphs {ends at a moment at which $a_i$ for all planets reach the upper limits} of semi-major axes quoted in the previous paragraph. 

The middle left-hand panel in Fig.~\ref{fig:fig1} shows the evolution of the periods ratios of pairs of subsequent planets. The red color is for $P_{\idm{d}}/P_{\idm{e}}$, green color marks $P_{\idm{c}}/P_{\idm{d}}$, while blue color is for $P_{\idm{b}}/P_{\idm{c}}$. Initial periods ratios are $\sim 3$, $\sim 5$ and $\sim 2$ for subsequent pairs of the planets. After a few of $10^4$~years, all subsequent pairs of planets are already locked in $2:1$~MMRs. \corr{For each pair of subsequent planets, one of the two-planet critical resonant arguments 
\[
\theta_{\idm{1:2}} = \lambda_1 - 2\,\lambda_2 + \omega_i, \quad i=1,2,
\]
where $\lambda_{1,2}$ and $\omega_{1,2}$ are the mean longitudes and pericenter arguments}, 
is already librating around $0^{\circ}$. However, the system evolves chaotically and the four-body resonant argument circulates in the whole permitted range of $[-180^{\circ}, +180^{\circ}]$. The four-body critical resonant argument is a generalization of the Laplace resonant argument, i.e.,
$\theta_{\idm{1:2:4}} \equiv \lambda_1 - 3\,\lambda_2 + 2\,\lambda_3$, where the mean longitudes $\lambda_i$, $i=1,2,3$, are ordered from the innermost to the outermost planet. For the three-planet Laplace 1d:2c:4b~MMR there is only one combination of the longitudes forming the critical resonant argument. For a four-planet 1e:2d:4c:8b~MMR  (actually, this is the most likely orbital configuration of the HR~8799 system found in this paper) there are more critical arguments possible. We chose the following critical argument:
\[
\theta_{\idm{1:2:4:8}} = \lambda_1 - 2\,\lambda_2 - \lambda_3 + 2\,\lambda_4
\equiv \lambda_{\idm{e}} - 2\,\lambda_{\idm{d}} - \lambda_{\idm{c}} + 2\,\lambda_{\idm{b}},
\] 
which librates around $0^{\circ}$ in the four-planet configurations of the HR~8799 system. The bottom left-hand panel of Fig.~\ref{fig:fig1} shows the time evolution of the critical resonance $\theta_{\idm{1:2:4:8}}$ for some time of typical migration run.

\subsubsection{Fine tuning of the migration rates}

\corr{The orbital evolution of a planetary system occurs in three different time-scales}. The shortest time scale is related to the orbital periods. The intermediate time-scale is related to the long-term conservative evolution of the system (rotations of periastrons, secular and resonant modulations of the eccentricities). The longest time-scale corresponds to the migration.   

After the system reaches a reasonably accurate configuration, the migration coefficients are changed to slow down the migration substantially.  This is being done mostly for technical reasons, to avoid overlooking a proper configuration possibly well with the observations.  If the resonant system is observed at certain epoch, its osculating Keplerian elements (semi-major axes, eccentricities, arguments of pericenters, mean longitudes) take some particular values.  A configuration which fits the observations consists of appropriate shapes of the orbits, their orientations as well as orbital phases.  Due to the MMR lock implying quasi-periodic conjunctions of the planets, such particular quasi-periodic, {\em relative} {orbital setup repeats} in space.  This repetition may take quite a long time interval.  In such a case, too fast migration during the repetition interval may tight the orbits too much and we can skip a correct, optimal configuration.  At some instant the planets may be phased correctly by the MMR lock but their orbits are still {\em too extended} to fit the observations.  However, after the next repetition period, appropriate orbital phases might appear when the migrating system is already {\em too compact} to fit the data.  A slower migration increases a chance of fixing proper orbital phases simultaneously with other orbital elements of the planets.

A slow enough migration is crucial because it helps to lock the system into the multiple-body MMR.  But the migration cannot be also {\em too slow} or arbitrarily slow.  At the initial stages of migration, the system is strongly chaotic (see Fig.~\ref{fig:fig1}) until the four-planet MMR lock appears and the whole system may be ''missed'' at all due to a self-disruption.  The resonance locking should then occur in a time-scale shorter than a characteristic instability time-scale.  An optimal migration rate was chosen after a number of numerical experiments.  We must underline that the multiple-body MMRs trapping does not occur always and ``easily''. The \moa{} requires much care and a fine tuning in this respect.

\corr{
The MCOA may be further simplified and optimized for CPU resources. Once the orbital  phases and eccentricities are constrained through a resonance, the migration causes the semi-major axes decay. Because the $N$-body dynamics are essentially scale-free, the scale of the system may be a free (fifth) geometrical parameter of the model. We did not explore this approach in this paper, but likely that makes the MCOA run more quickly and find better fits to the observations. (Credits of this improvement go to the Reviewer).
}

\subsubsection{Example evolutionary tracks of migrating systems}

These thoughts and the results of accompanying experiments are illustrated in Fig.~\ref{fig:fig1}. The right-hand panels of this Figure present the orbital evolution during a slow migration. Subsequent panels, from the top to the bottom  are for graphs of $\aee(t)$, $\ee(t)$ and
$\theta_{\idm{1:2:4:8}}(t)$, respectively. The migration rate for planet~e is less than
$\sim1\,\au/$Myr. After  $\sim 1\,$Myr, the systems locks into exact 1e:2d:4c:8b~MMR. Amplitudes of eccentricities $\ee(t)$ and critical angle
$\theta_{\idm{1:2:4:8}}$ steadily decrease. After $\sim 2$~Myr the simulated configuration looks like the actually observed HR~8799 system. Middle parts of the elements graphs are marked in red indicating reasonably good solutions that provide $\cchi < 2$. 

This simulation concerns the actual evolution of the best-fitting four-planet model IVa, which is found rigorously stable (see Sect.~\ref{sec:architectures}). However, even very similar configurations can be obtained from quite different initial systems. Indeed, the next Figure~\ref{fig:fig2} shows the results for a solution with marginally worse $\cchi$. In this case the initial orbits are much more extended. A desired trapping into the 1e:2d:4c:8b~MMR takes place $\sim 2\,$Myr before the system migrates into the observed configuration. 

Figure~\ref{fig:fig3} presents 1-dim scans of temporarily best-fitting inclination $I$ and an instant $\cchi$ as functions of $\tobs$ {during the stage of slow migration}. These data are found for the best-fitting configuration with $\cchi \approx 1.15$ and $I \approx 25^{\circ}$. After each period of $\sim 400\,$yr, which is actually the orbital period of planet~b, $\cchi$ possesses minima of varying depth. Although the planets have orbital phases close to best-fitting values, other osculating elements might be still not optimal, hence we should examine a time sequence of such local minima of $\cchi$ to find the best model.  If the orbital phases are distinct from their observed values, the temporal best-fitting inclination as well as the two remaining orientation angles do not take {any reasonable values}.  

\subsubsection{Non-standard features and limitations of the \moa{}}

It might be surprising that the \moa{} is supposed to constrain much more parameters than the number of observations. We underline here that only four~parameters of the migration model are free: three Euler angles describing orientation of the orbital frame w.r.t. the observer frame and the osculating epoch. All remaining orbital elements are self-constrained by the common migration and trapping the planets in a multiple MMR. 
 
By the design and assumptions, our approach can be useful for predicting positions of planets in multiple systems when only a few observational epochs are available (see Sec.~\ref{sec:architectures}). The \moa{} might also help to verify if  putative candidates are bounded to the star. In principle, the minimal number of required parameters (4) is equal to the number of data provided by two~astrometric measurements of two companions. This may be provided by single-epoch detection of two massive, spatially close planets or sub-stellar objects which are presumably involved in a low-order MMR (Sect.~\ref{sec:architectures}). 
 
The \moa{} will almost certainly fail if planetary masses are small, and non-resonant configurations are possible. However, we consider here a particular class of planetary systems whose parameters are biased through natural limits of the direct imaging. The \moa{} may successfully model such systems by making use of their essential physical and orbital characteristics.

A complete modeling of four- or five-planet HR~8799 configurations involves the optimization and stability analysis, and requires significant computational resources. For this work, we needed roughly $\sim$100~CPU cores for 2-3 weeks, to sample densely the space of the initial parameters and evolutionary tracks, and to gather possibly large statistics of the best-fitting models. \corr{The maximal Lyapunov characteristic exponent (MLCE) criterion } \citep{Cincotta2000,Cincotta2003} and direct numerical integrations were used to verify stable models and to reconstruct the MMRs structures (see Sect.~\ref{sec:stability} for details).  This step of dynamical post-analysis of the best-fitting models is also CPU intensive. We estimate its cost for $\sim 1000$~CPU cores for 2-3 weeks. 

%
\section{Architectures of the HR 8799 system}
\label{sec:architectures}
%
To recover likely and dynamically stable models of the HR~8799 system, we consider a few orbital architectures as well as five distinct sets of observations (see Tabs.~\ref{tab:data1}--\ref{tab:data4}). Dataset D1 comprises of all astrometric measurements published till August, 2013.  This dataset {\em diminished} by observations of planet~e is called dataset~D5. We analyze also dataset D2 that consists of astrometric measurements published in the discovery paper \citep{Marois2008}. Datasets D3 and D4 mimic single-image detection of three and four planets. These sets consist of only three and four most accurate single epoch measurements, selected from the discovery paper \citep{Marois2008} (dataset D3, epoch 2008.61) and from a more recent work by \cite{Currie2011} (dataset D4, epoch 2009.77).

We optimized a few combinations of orbital architectures labelled with Roman numbers III (three planets), IV (four planets) and V (five planets) with datasets D1--D5. In all optimization models, we assumed that masses of the planets are bounded  to $\sim[2, 13]$~$\mJ$ range, roughly within limits determined in \citep{Marois2008,Marois2010}, \corr{yet in a~few experiments we tested also masses up to $\sim 20~\mJ$ limit}. The masses were randomly varied within these bounds in each single run. The mass of the parent star was fixed to 1.56 Solar mass \citep{Marois2010}, very close to the most recent estimate \citep{Baines2012}.    The optimization algorithm was run at least $\sim 10^4$--$10^5$ times for each dataset-model combination. Each single run was initiated from random initial orbits and migration parameters (\ref{eq:moore}) which were set in wide, yet reasonable and carefully tested ranges. As the result, we gather a set of initial conditions representing the best-fitting orbital parameters. We consider all solutions within $3\sigma$ and $6\sigma$ confidence interval of a given, best-fitting {\em stable} model.

Finally, we investigated the orbital stability of all $6\sigma$ models  through direct, long-term $N$-body integrations up to the upper limit of the system age (160~Myr); yet we did a few experiments for 500~Myr and 1~Gyr intervals (Sect.~\ref{sec:stability}). The results of the long-term integrations are interpreted through the event time $T_{\idm{E}}$ -- an interval of crossing orbits or an ejection of a planet from the system. We also reconstructed the local structure of the phase space (resonances widths) by computing the maximal Lyapunov exponent expressed through the fast indicator MEGNO \citep{Cincotta2003}. The dynamical maps in selected orbital parameter planes were computed in resolutions up to $720\times360$.  All details of the stability experiments and regarding a time-calibration of the fast indicator are given in Section~\ref{sec:stability}. 
%
\subsection{The nominal model IVa: four planets, dataset D1}
\begin{figure*}
\centerline{
\hbox{
{\includegraphics[width=0.72\textwidth]{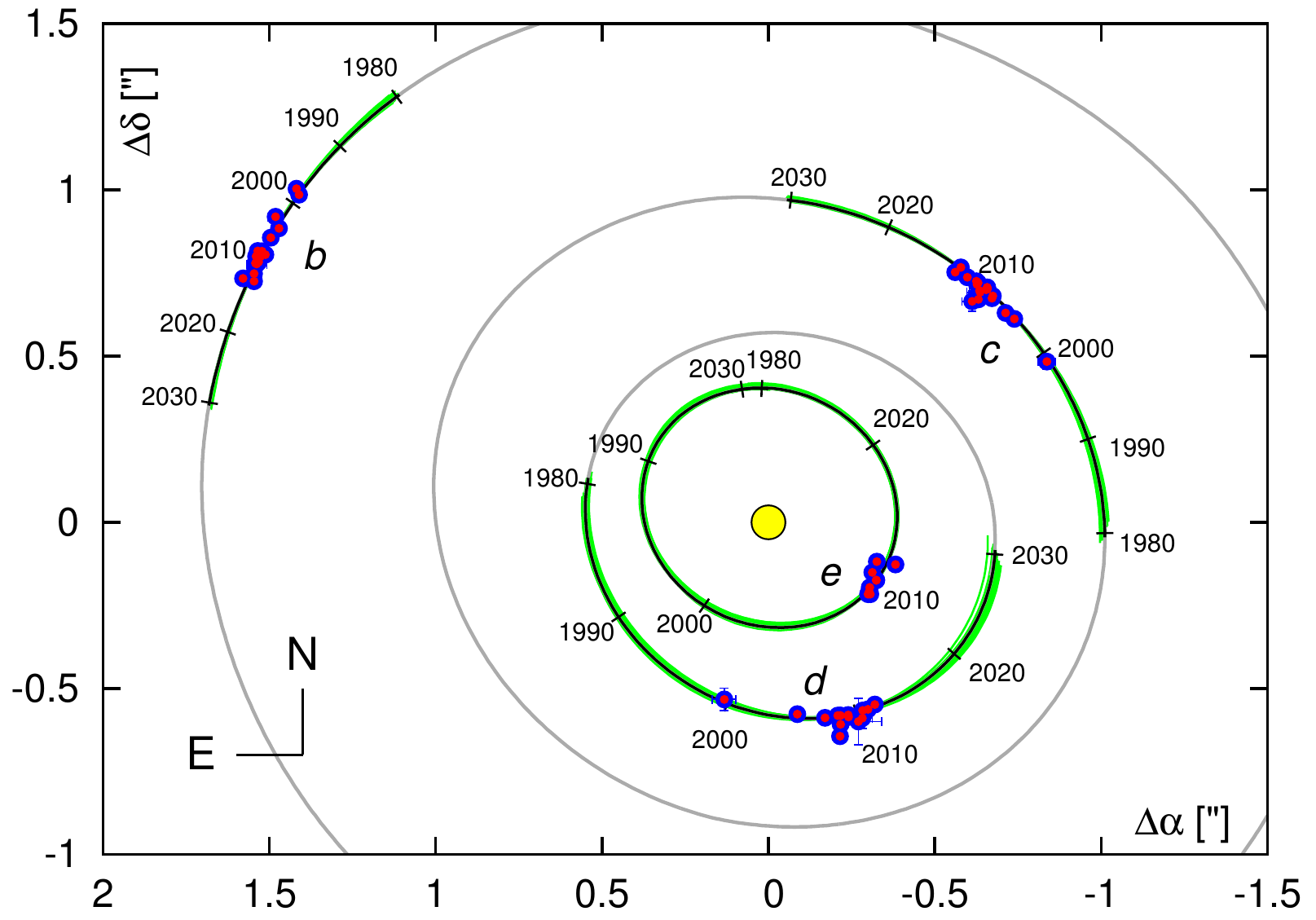}}
}
}
\caption{
Relative astrometric positions of the planets (red filled circles), 
orbital arcs for the best-fitting model~IVa (black curves), and stable solutions
within the $3\sigma$ confidence level of the best-fitting model  (green curves). 
}
\label{fig:fig4}
\end{figure*}

\begin{figure}
\centerline{
\vbox{
\hbox{\includegraphics[width=0.49\textwidth]{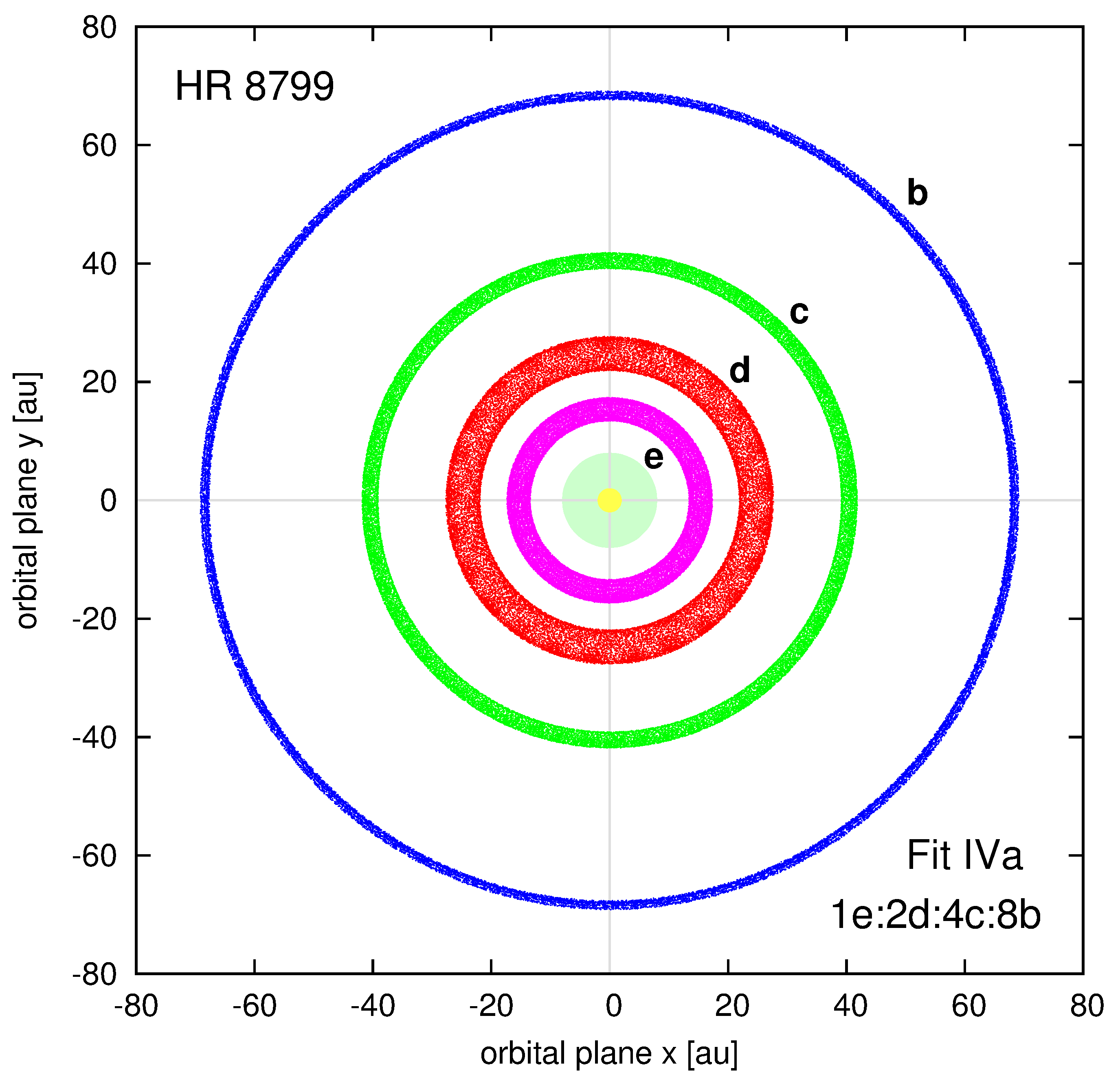}}
}
}
\caption{
Orbital architecture of the HR 8799 system in accord with the best fitting four-planet model IVa (Table~1). Temporal positions of HR~8799 planets are shown for 160~Myr in the astrocentric coordinate frame, coplanar with the orbits. The inner shaded circle in the center has an approximate radius of the last stable orbit of a mass-less particle perturbed by the giant planets.  This might correspond to the inner warm disk investigated in a number of papers
\protect\citep[e.g.,][]{MoroMartin2010,Hinkley2011,Su2009,Reidemeister2009}.}
\label{fig:fig5}
\end{figure}

%
Model IVa combines four-planet, coplanar system and dataset D1 comprising of all available observations in the literature. The osculating astrocentric Keplerian elements together with their uncertainties are given in Table~\ref{tab:table1}. This model provides $\cchi \approx 1.15$. We did not find any other four-planet configuration consistent equally well with the observations.   Figure~\ref{fig:fig4} shows the best-fitting orbits over-plotted onto the sky plane together with the observations. The green curves show very small deviations between all stable orbits in the derived statistics within $3\sigma$ confidence level of the best-fitting model. Table~\ref{tab:ephemeris1} displays the $(E,N)$ ephemeris between epochs~$1995.0$ and~$2020.0$. Temporal evolution of all orbits for 160~Myr is shown in Fig.~\ref{fig:fig5}.   This best-fitting, stable and unique orbital configuration consistent with all astrometric data corresponds to {\em exact} first order MMR, double Laplace resonance 1e:2d:4c:8b~MMR. The ratios of the orbital periods for subsequent pairs of planets are very close to $2$. This might be the first report of such {\em long-term stable} MMR in the literature, although this type of solution was already investigated \citep{Esposito2013}.  

\begin{table*}
\caption{
Orbital osculating elements of the best-fitting solution IVa at the epoch $1998.83$. The stellar mass $m_0 = 1.56\,\msun$. {Note that due to the geometry the $\Omega$ angle may take two values that differ by $180^{\circ}$, however the {\em pericenter longitude} $\varpi$ is preserved after the rotation of the nodal line by $180^{\circ}$}.
}  
\label{tab:table1}
\begin{tabular}{l c c c c c c c}
\hline
& $m\,[\mJ]$ & $a\,$[au] & e & $I\,$[deg] & $\Omega\,$[deg] & $\varpi\,$[deg] & $\Mmean\,$[deg] \\
\hline
HR 8799 e & $9 \pm 2$ & $15.4 \pm 0.2$ & $0.13 \pm 0.03$ & & & $176 \pm 6$ & $326 \pm 5$\\
HR 8799 d & $9 \pm 3$ & $25.4 \pm 0.3$ & $0.12 \pm 0.02$ & $25 \pm 3$ & $64 \pm 3$ & $91 \pm 3$ & $58 \pm 3$\\
HR 8799 c & $9 \pm 3$ & $39.4 \pm 0.3$ & $0.05 \pm 0.02$ & & ($244\pm 3$) & $151 \pm 6$ & $148 \pm 6$\\
HR 8799 b & $7 \pm 2$ & $69.1 \pm 0.2$ & $0.020 \pm 0.003$ & & & $95 \pm 10$ & $321 \pm 10$\\
\hline
\end{tabular}
\end{table*}

The inclination of coplanar orbits to the sky plane is well constrained $I \approx 25^{\circ} \pm 3^{\circ}$. Statistical analysis of the rotational speed of A5 stars imply the inclination $\sim 23~$degrees of the HR~8799 equator to the sky plane \citep{Royer2007,Kaye1998,Wright2011}. 
\corr{In a very recent study of the {{\sc Herschel}} far-infrared and submilimeter observations of the outer debris disk, \cite{Matthews2014} measured
its inclination to $26^{\circ}\pm 3^{\circ}$  from face on and position angle of $64^{\circ}$ $E$ of $N$, closely matching parameters of our model IVa. 
These results, derived independently on astrometry and on different data and observations, are very suggestive for the planetary system perfectly coplanar with the stellar equator and remnants of the protoplanetary disk.} 

An extensive stability study reveals that the best-fitting four-planet configuration is strictly quasi-periodic, with the maximal Lyapunov exponent equal to~0. The direct numerical integrations do not show any sign of instability up to 1~Gyrs (for details see Sect.~\ref{sec:stability}). The system is locked deeply in a four-body  Laplace MMR. The critical argument $\critLL \equiv \lambda_{\idm{e}} - 2\,\lambda_{\idm{d}} - \lambda_{\idm{c}} + 2\,\lambda_{\idm{b}}$
librates around $0^{\circ}$ with a semi-amplitude less than $20^{\circ}$. The resonance width across the semi-major axes is between $2$~au for planet~b and $\sim 0.3$~au for the innermost planet~e.

\begin{figure*}
\centerline{
\hbox{
\includegraphics[width=0.48\textwidth]{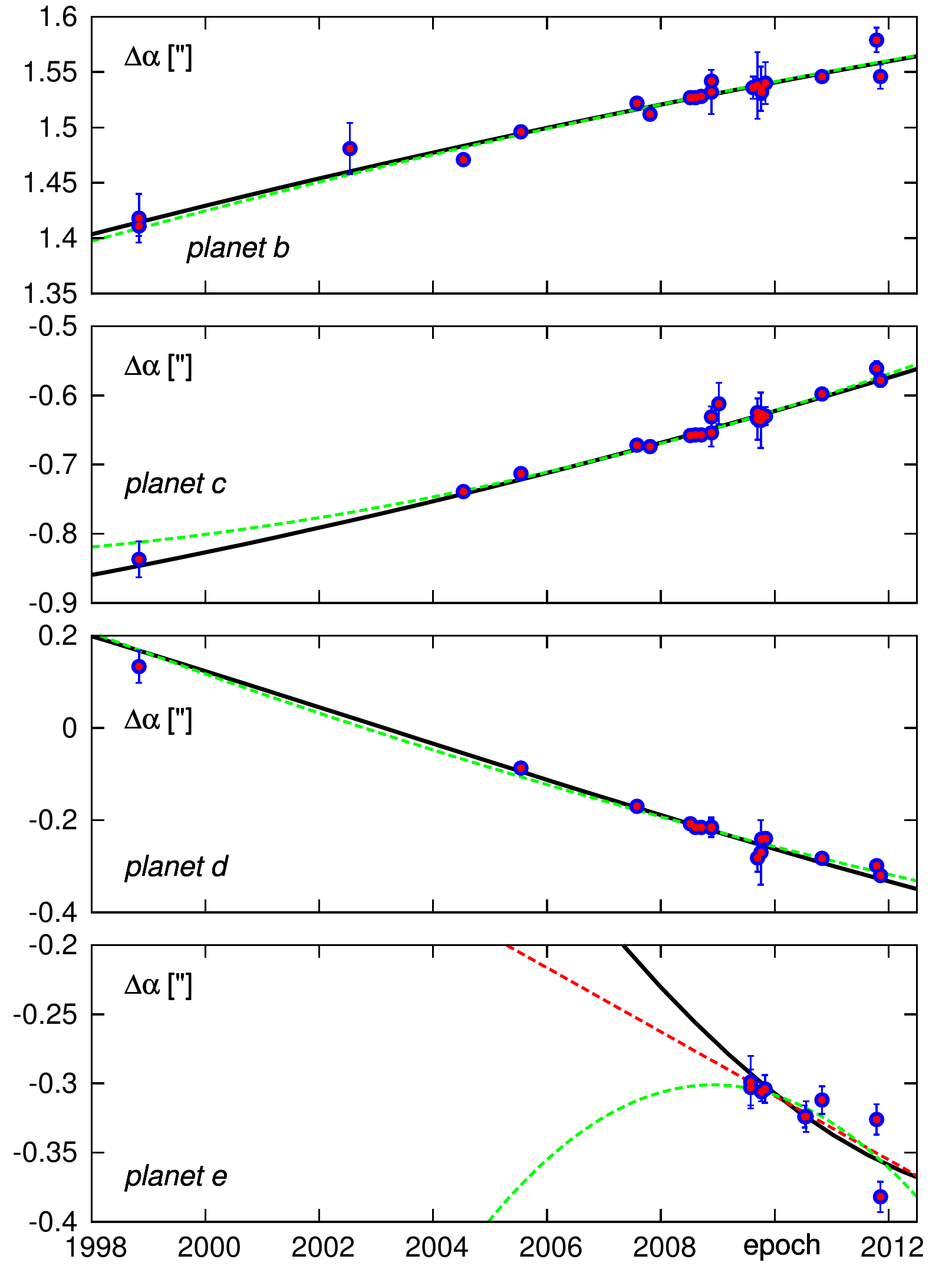}
\includegraphics[width=0.48\textwidth]{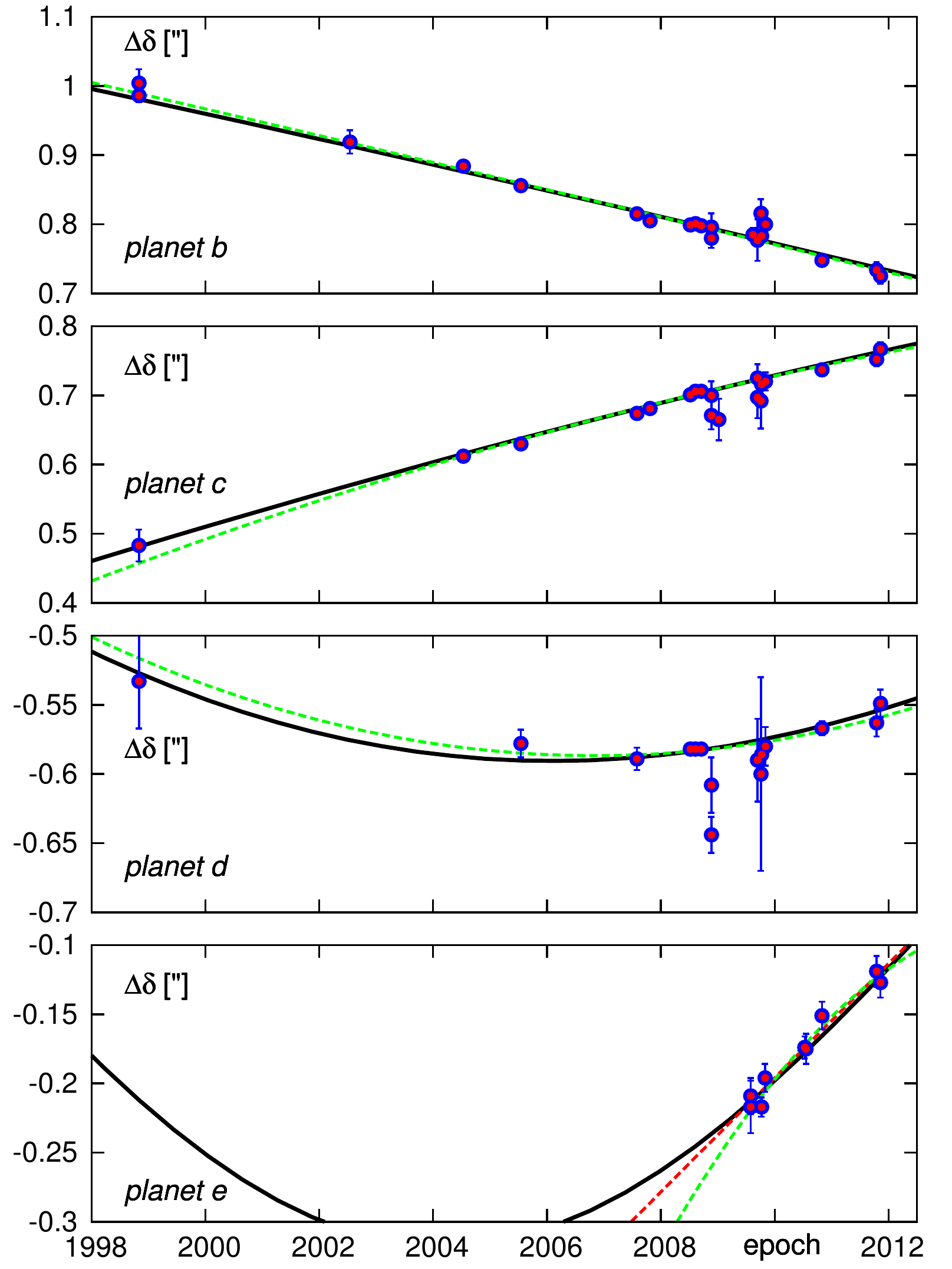}
}
}
\caption{
Astrocentric, astrometric sky-plane coordinates $x \equiv \Delta\alpha$ [arc sec] ({\em left panel}) and $y \equiv \Delta\delta$ [arc sec] ({\em right panel}) for subsequent planets  (red filled circles) and orbital arcs for the best-fitting model IVa (black curves). Dashed curves are for best-fitting quadratic and linear Gibbs series expansion to the measurements (dataset D1), respectively.
}
\label{fig:fig6}
\end{figure*}

Figure~\ref{fig:fig6} shows the $(E, N)$ sky coordinates determined by the dynamically derived best-fitting solution~IVa (black solid curves), as $x(t) \equiv \Delta\alpha(t)$ and $y(t) \equiv \Delta\delta(t)$, respectively, versus  observation epoch~t over-plotted with data points.  Green dashed curves are for the best-fitting quadratic functions of $x(t)$ and $y(t)$, respectively.  The number of measurements seems too small to constrain the parabolic model for planet~HR8799~e. The dispersion of data is large, hence we also computed  the best-fitting linear model (red dashed lines).

Best-fitting parameters of the linear and parabolic models of the sky coordinates express Keplerian (kinematic) approximations of the astrocentric radius $\v{r}(t)$ and orbital velocity $\v{v}(t)$ of each planet. These parameters are the first and second order terms of the Gibbs  series $(f,g,\dot f, \dot g)$ in the well known formulae
\[
  \v{r}(t) = f \v{r}_0 + g \v{v}_0, \quad 
  \v{v}(t) = \dot f \v{r}_0 + {\dot g} \v{v}_0,
\]
where $\v{r}_0 \equiv \v{r}(t_0)$, $\v{v}_0 \equiv \v{v}(t_0$) is the initial condition at epoch $t_0$. The linear model was used in previous works to bound the space of permitted orbital elements \cite[e.g.][]{Fabrycky2010,Gozdziewski2009,Reidemeister2009}. Due to narrow observational window this space is huge and the orbits are kinematically unconstrained. A comparison of the dynamical best-fitting solution IVa found in this work with the linear and parabolic approximations of the data (Fig.~\ref{fig:fig6}) shows that this {\em rigorously stable} initial condition  closely matches the kinematic (or geometric) description of the measurements.  Moreover, a close inspection of panels in Fig.~{\ref{fig:fig6}} for~planets b, c, and~d reveals that the dynamic model fits the measurements {\em even better} than the first terms of the Gibbs functions, particularly at the earliest HST images. This is suggestive for \corr{noticeable deviations of the $N$-body orbits from their first and second order  Keplerian approximations}.

Moreover, the model orbit of planet~e (shown in bottom panel of Fig.~\ref{fig:fig6}) seems to pass in between three apparently accurate observations. This may indicate underestimated errors of the recent astrometric measurements in \citep{Esposito2013}. These data points are likely responsible for orbital fits implying a large eccentricity of planet~e or a significant non-coplanarity \cite[][]{Esposito2013}. Both the linear and quadratic approximations of the Gibbs series are ambiguous that leads to badly constrained initial condition for planet~e
if kinematic or even plain dynamical models are applied.

\begin{figure*}
\centerline{ 
\vbox{
\hbox{
\includegraphics[width=0.49\textwidth]{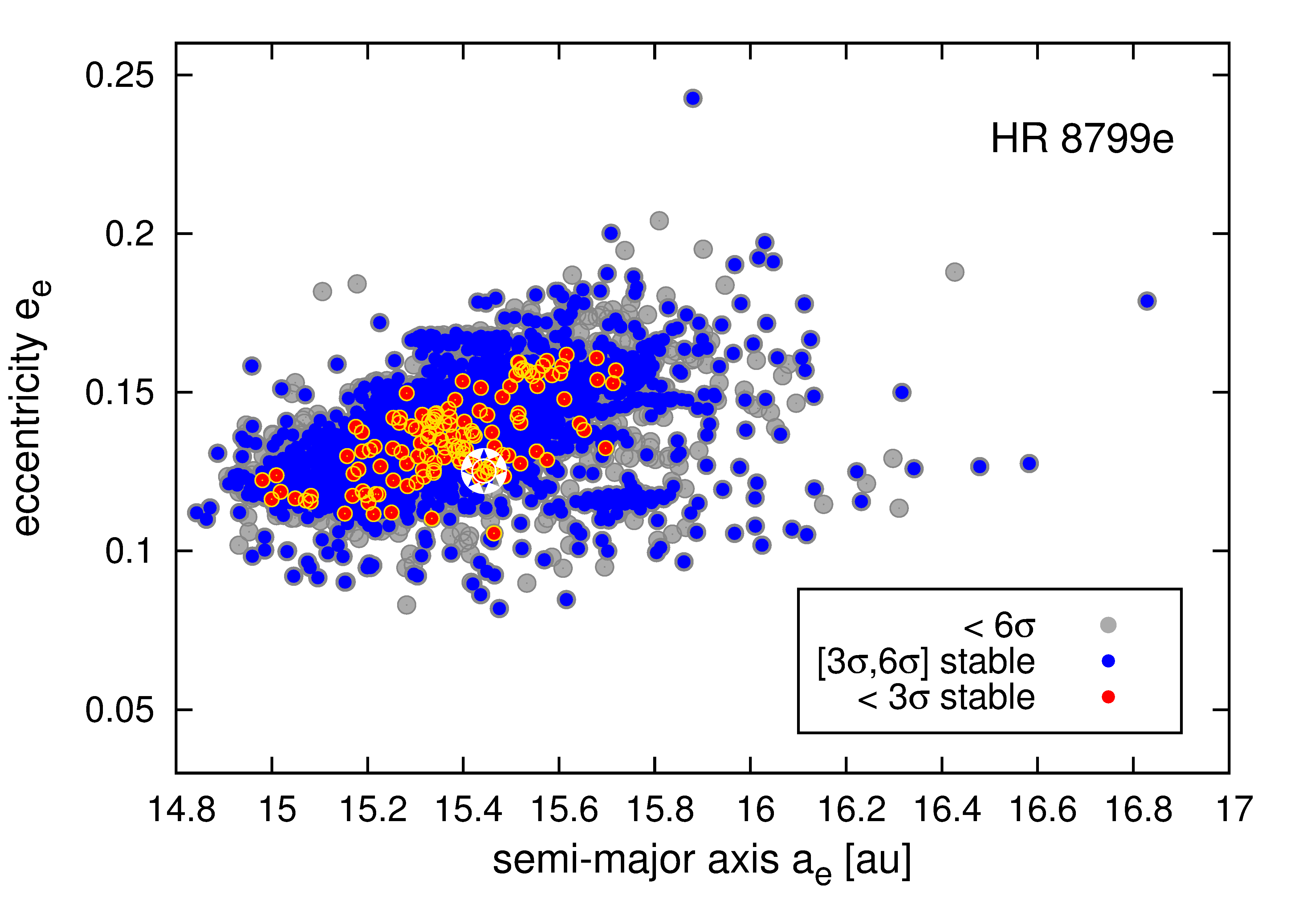}
\includegraphics[width=0.49\textwidth]{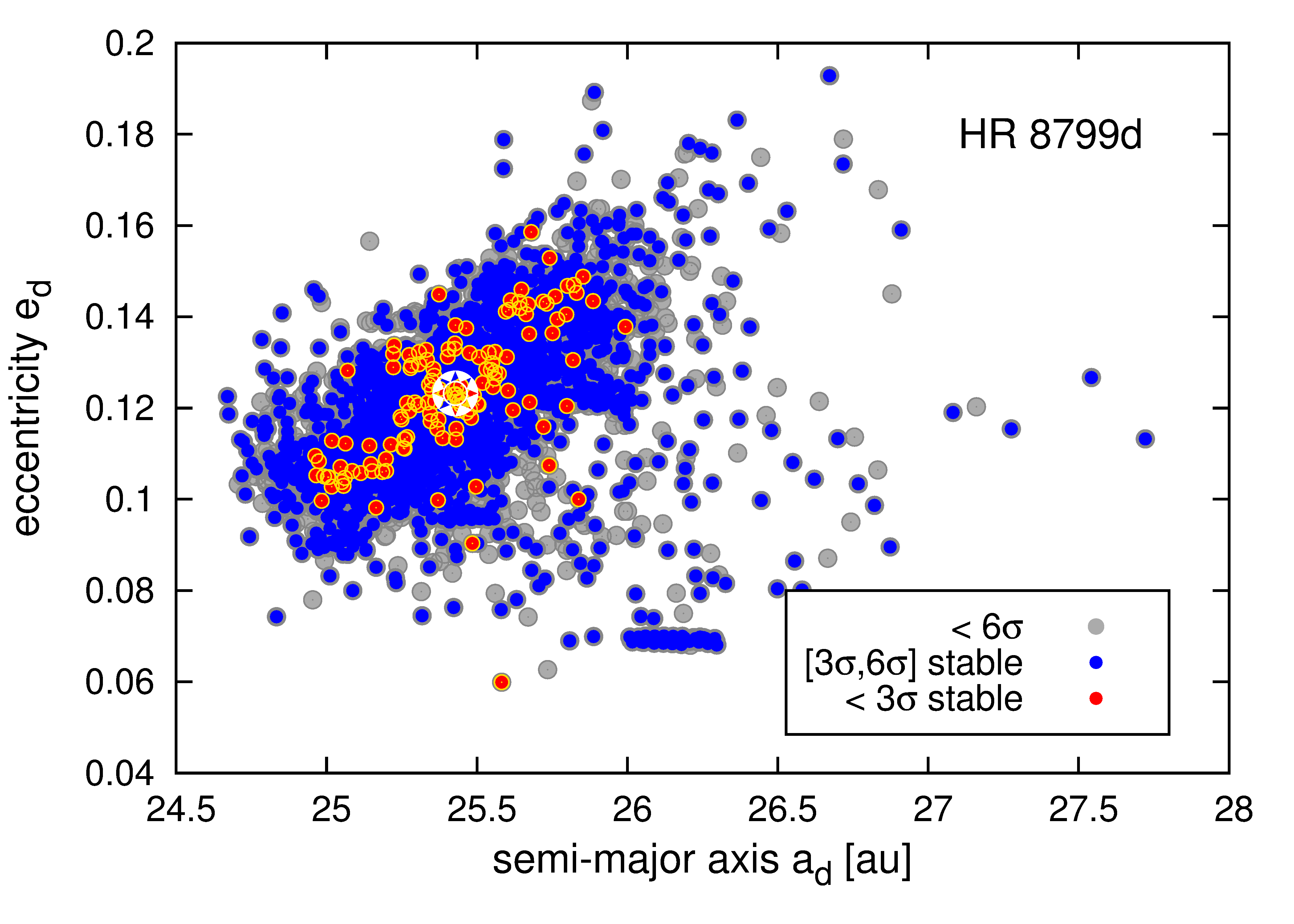}
}
\hbox{
\includegraphics[width=0.49\textwidth]{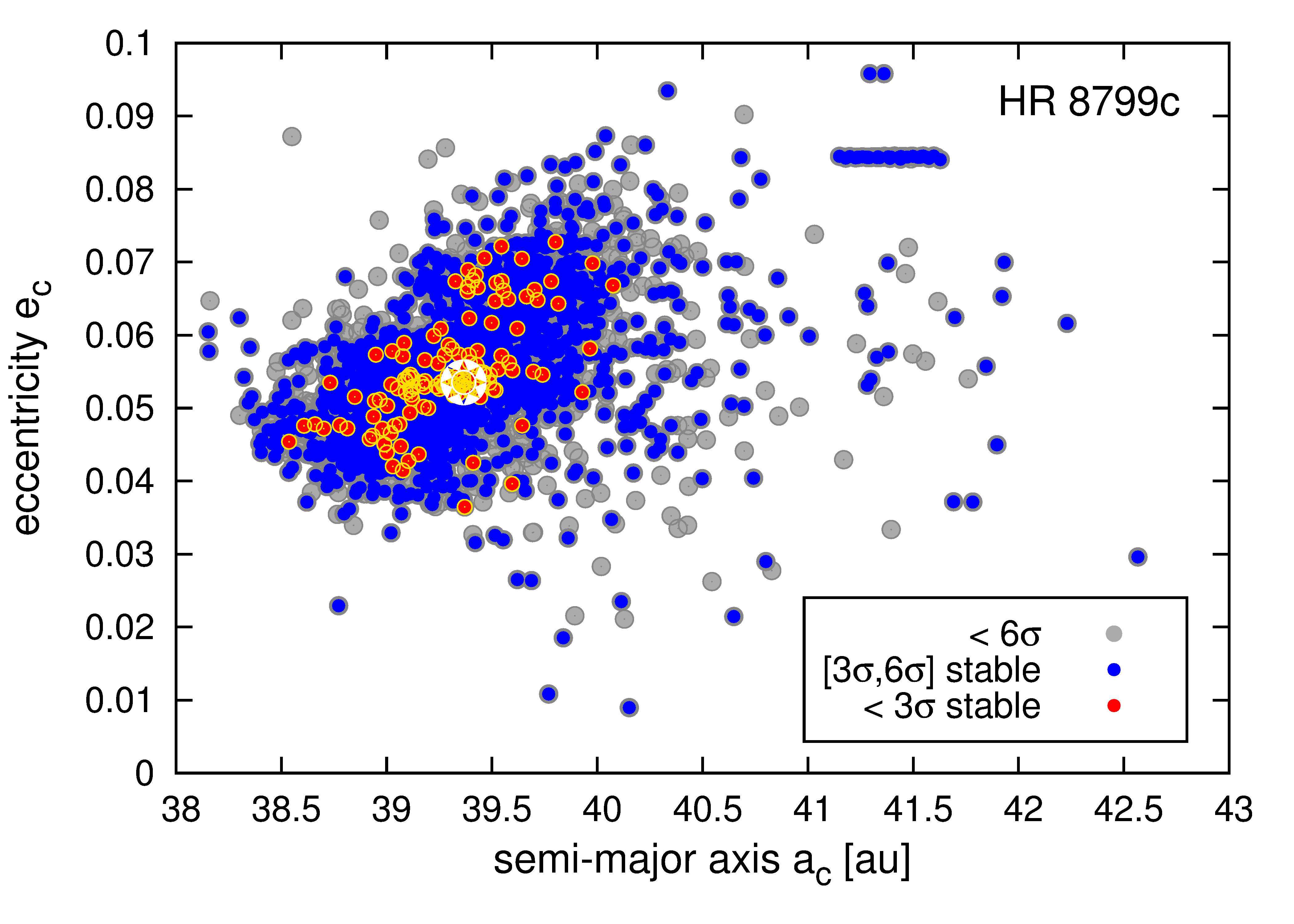}
\includegraphics[width=0.49\textwidth]{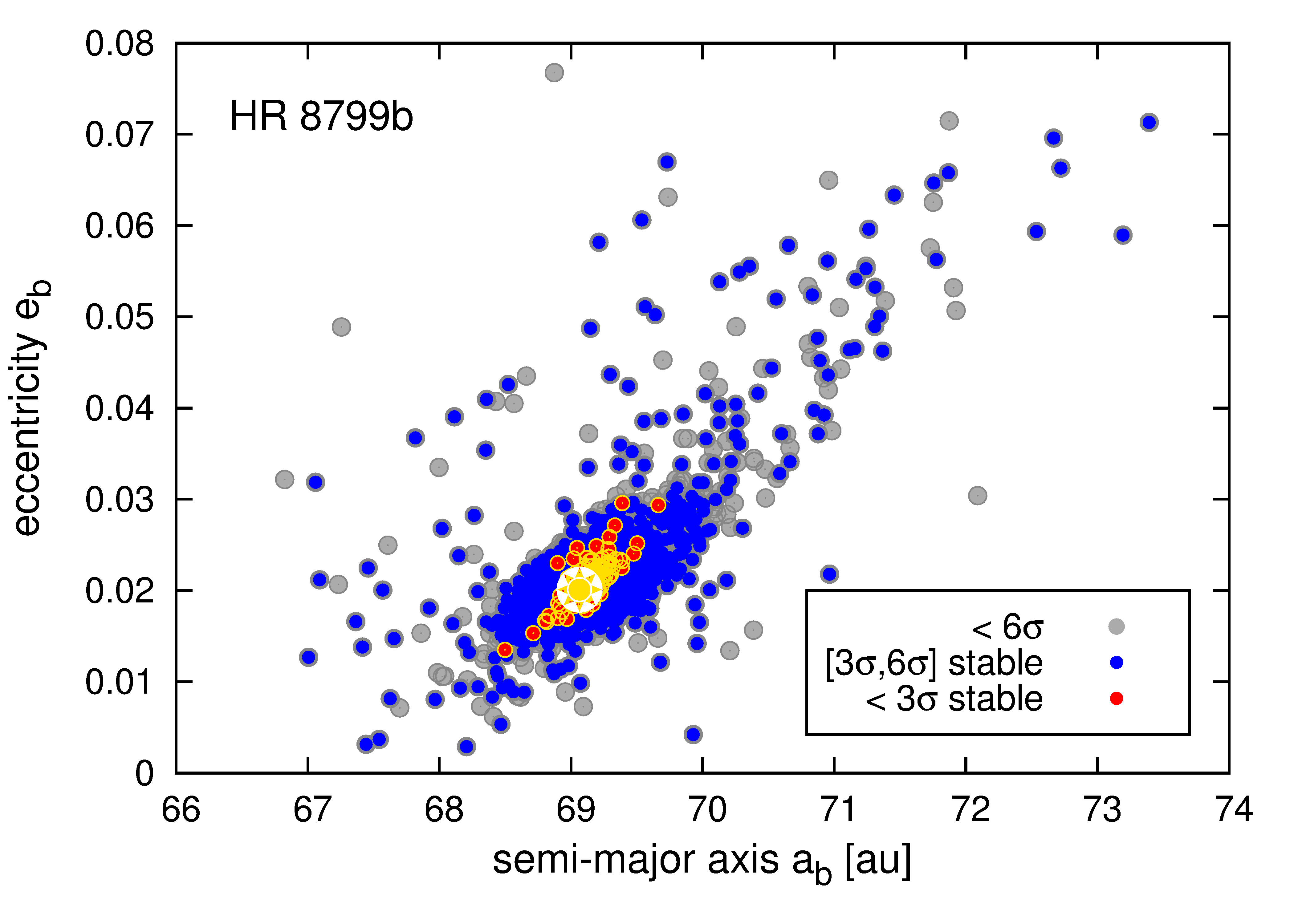}
}
}
}
\caption{
Best-fitting solutions projected onto semi-major axis -- eccentricity planes for subsequent planets. The star symbol marks the nominal, best-fitting solution in Tab.~\ref{tab:table1}. Grey circles are for all solutions within $6\sigma$ confidence interval ($\cchi < 1.5$). Blue and red circles are for {\em rigorously stable}  models in the range of $(3\sigma,6\sigma)$, ($1.2<\cchi <1.5$) and within $3\sigma$ ($\cchi\leq 1.2$), respectively. Their $|\left<Y\right>-2|<0.05$ for the integration time-span of 160~Myr, covering a few estimates of the HR~8799~lifetime in the literature.
}
\label{fig:fig7}
\end{figure*}

\begin{figure*}
\centerline{ 
\vbox{
\hbox{
\includegraphics[width=0.49\textwidth]{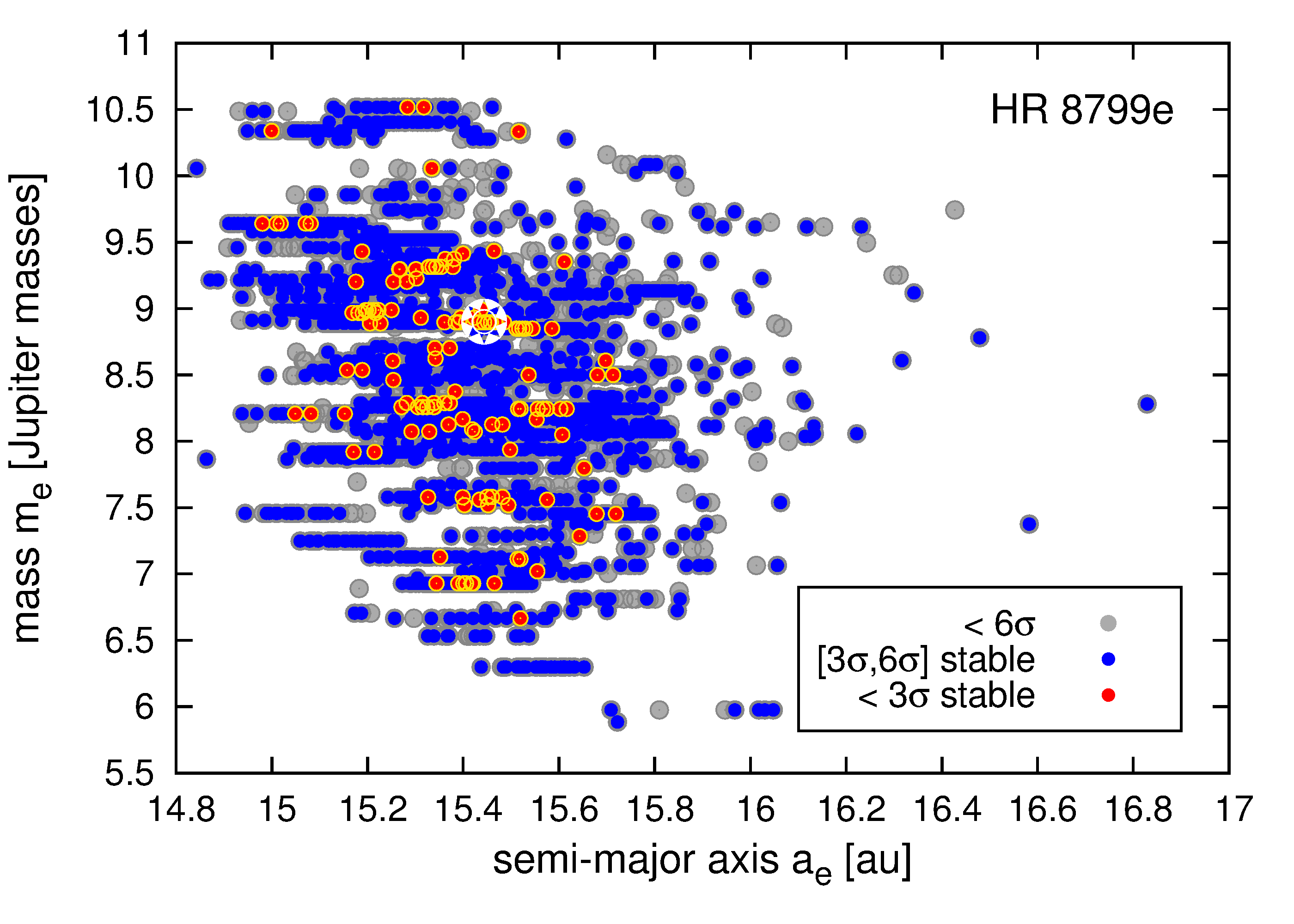}
\includegraphics[width=0.49\textwidth]{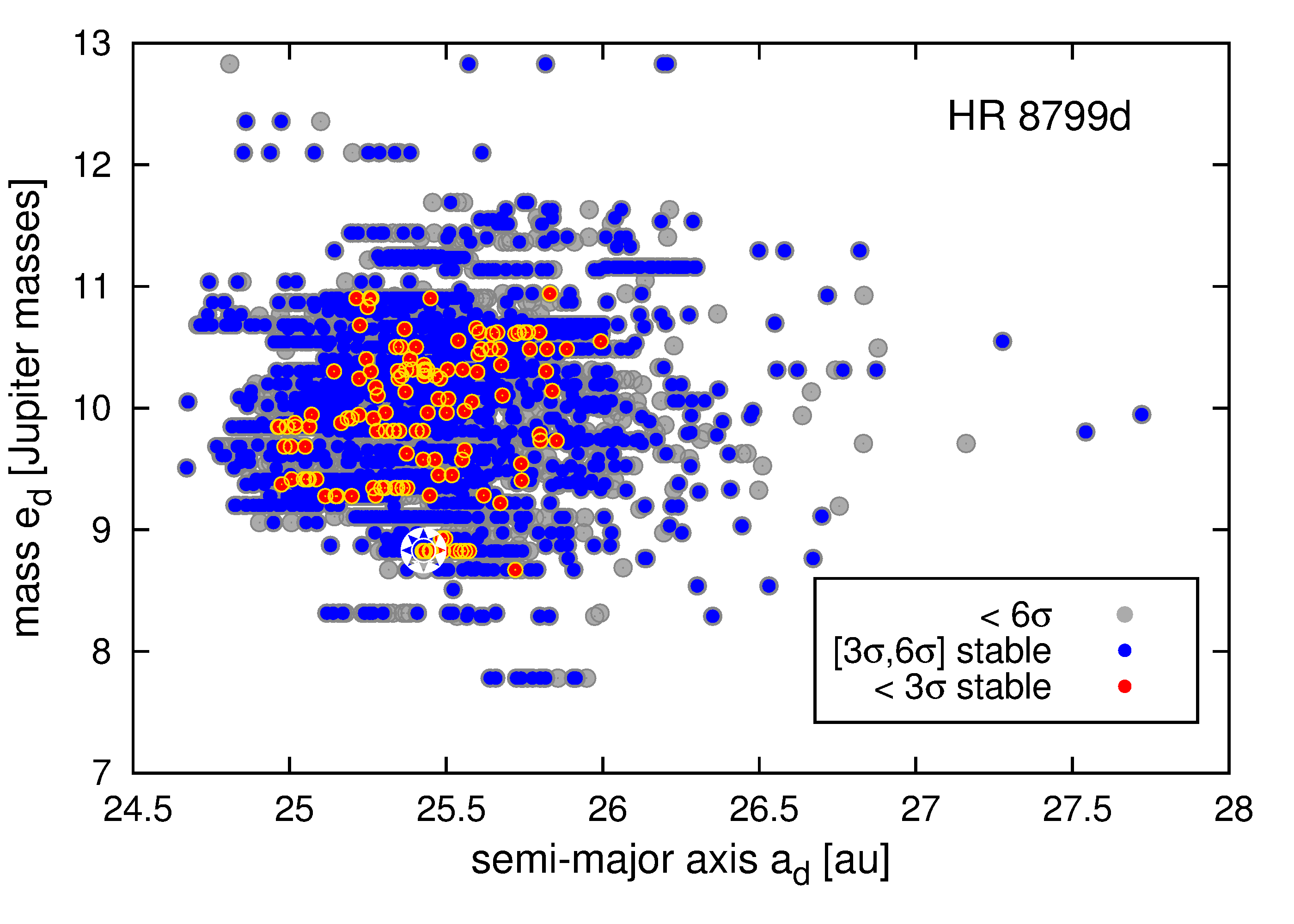}
}
\hbox{
\includegraphics[width=0.49\textwidth]{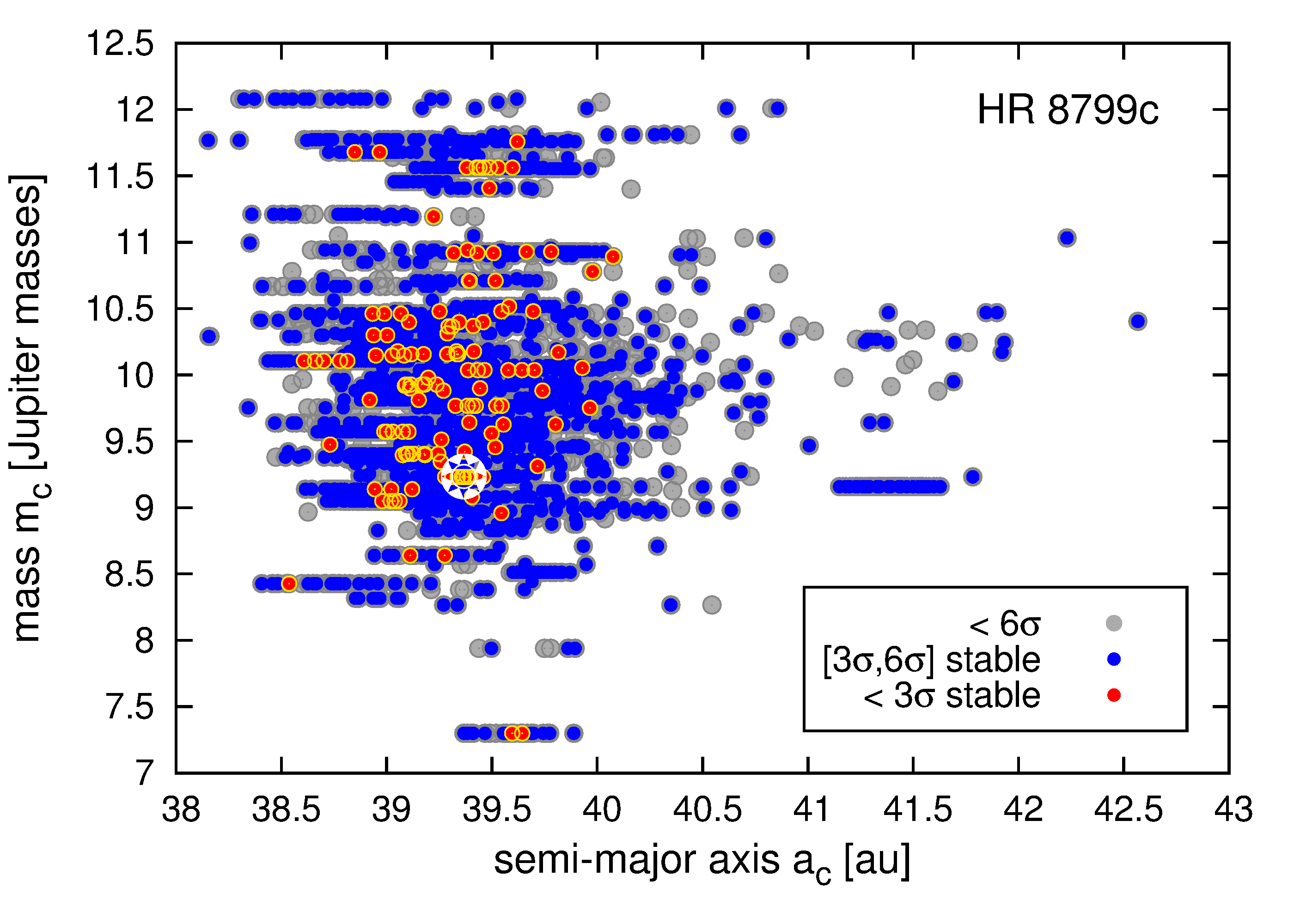}
\includegraphics[width=0.49\textwidth]{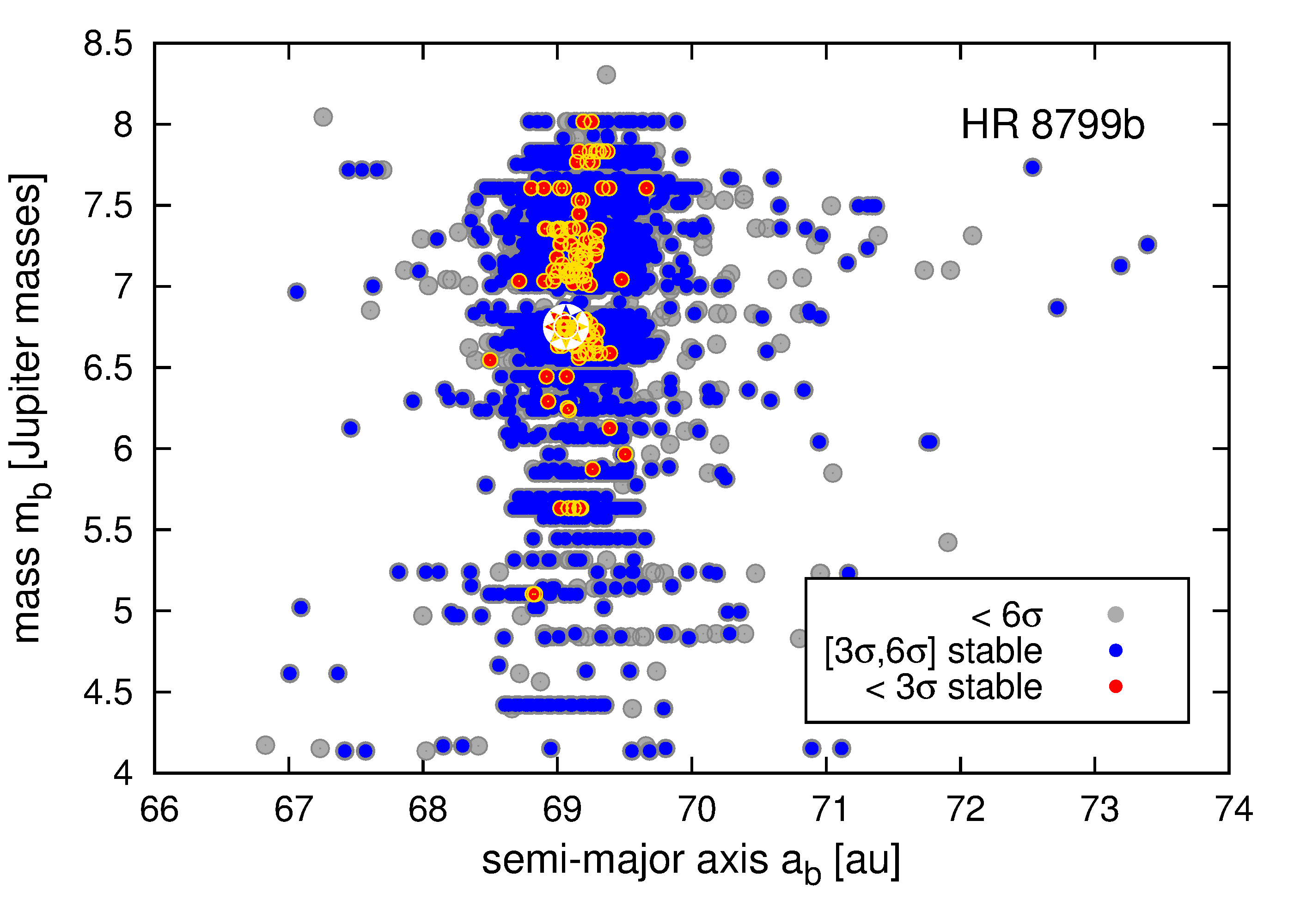}
}
}
}
\caption{
Statistics of the best-fitting solutions to four-planet model IVa and dataset~D1  illustrated as projections onto semi-major axis -- masses planes for subsequent planets. The star symbol marks the nominal, best-fitting solution~IVa in Tab.~\ref{tab:table1}. Grey circles are for all $6\sigma$ solutions with $\cchi < 1.5$. Blue and red circles are for {\em rigorously stable}  models in the range of
$(3\sigma,6\sigma)$, $1.2<\cchi <1.5$, and within $3\sigma$ ($\cchi\leq 1.2$), respectively. Their $|\left<Y\right>-2|<0.05$ for the integration time-span of 160~Myr, covering a few estimates of the HR~8799~lifetime. 
}
\label{fig:fig8}
\end{figure*}

Statistics of all models gathered in $\sim 10^5$ runs of \moa{} are presented on Figs.~\ref{fig:fig7} and~\ref{fig:fig8}. Fig.~\ref{fig:fig7} shows projections of the orbital parameters onto $(a_i, e_i)-$planes where $i=\mbox{e}, \mbox{d}, \mbox{c}, \mbox{b}$ and marked with grey filled circles.  We computed the maximal Lyapunov exponent expressed through the MEGNO indicator (Sect.~\ref{sec:stability}) for all these initial conditions for 160~Myr. Solutions providing $|\left<Y\right>-2|<0.05$ at the end of the integration time are considered as quasi-periodic and rigorously stable.  Blue and red filled circles in  Fig.~\ref{fig:fig7} mark such stable (quasi-periodic, regular) models within the  $(3\sigma,6\sigma)$ and $3\sigma$ joint confidence levels, respectively. The statistics demonstrate that orbital parameters in model IVa are well constrained. This is also illustrated on Fig.~\ref{fig:fig4}. Green curves illustrate geometrically  a dispersion of the orbital arcs of stable solutions within $3\sigma$ level around the best-fitting model. 

Figure~\ref{fig:fig8} illustrates best-fitting models projected onto $(a_i, m_i)-$planes in the same manner as shown in Fig.~\ref{fig:fig7}. Nominal masses are $(9, 10, 10, 7)\,\mJ$ for planets e, d, c, b, respectively. At the beginning of each optimization run, the actual masses were selected from the normal distribution within $2-3\,\mJ$ standard deviations around the nominal masses.  The statistics presented in Fig.~\ref{fig:fig8} reveal a relatively extended range of \corr{dynamical masses} in the $4$--$12$\,$\mJ$~rang e providing stable (quasi-periodic) solutions. This experiment shows that the astrophysical determination of the masses through cooling models are well consistent with dynamical constraints driven by the migration. This somehow contradicts previous studies concluding that the stability of HR~8799 system is possible only when the masses are at the lowest estimates \citep{Sudol2012,Esposito2013}. 

\begin{figure*}
\centerline{
\vbox{
\hbox{
{\includegraphics[width=0.49\textwidth]{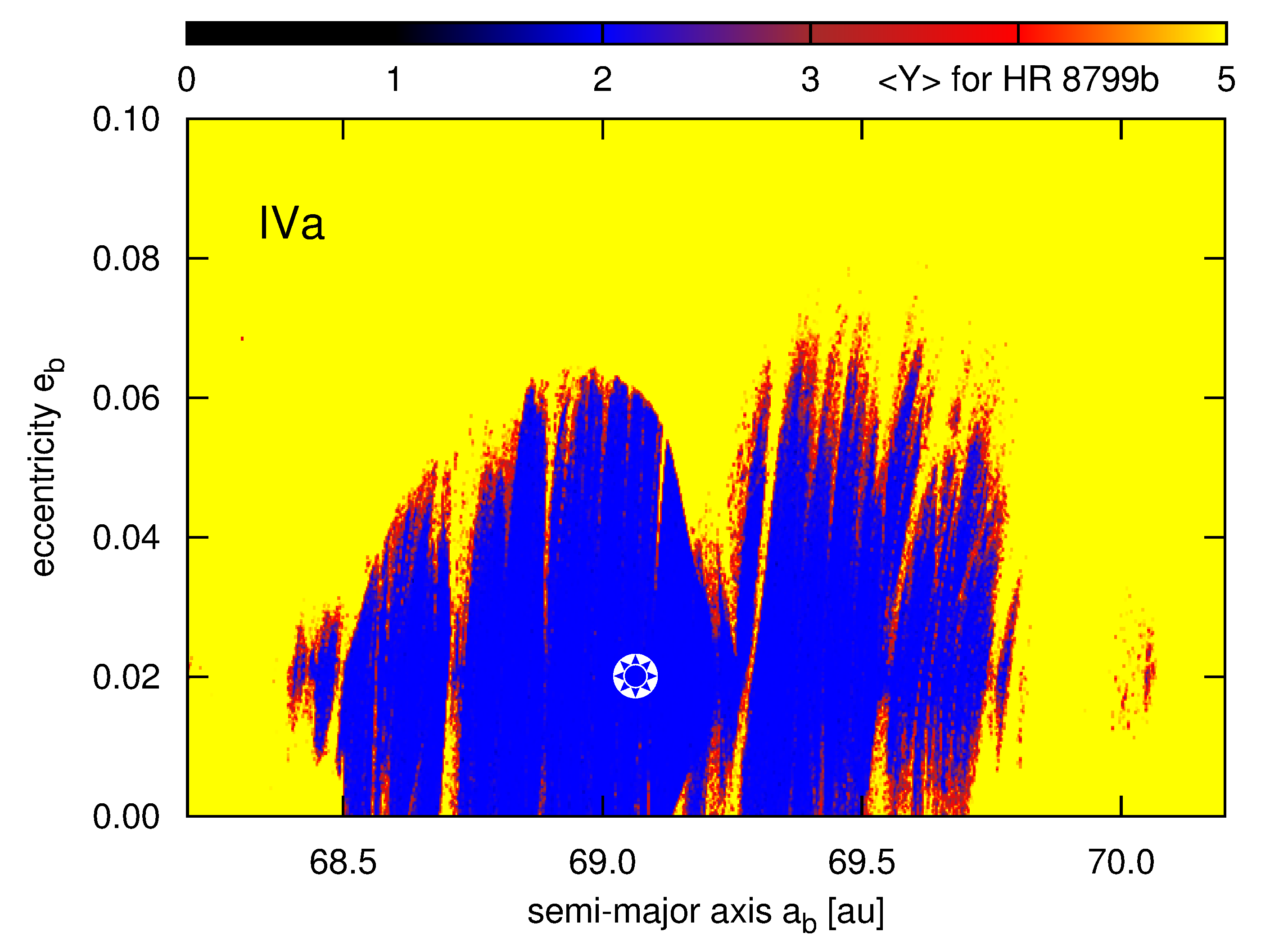}}
{\includegraphics[width=0.49\textwidth]{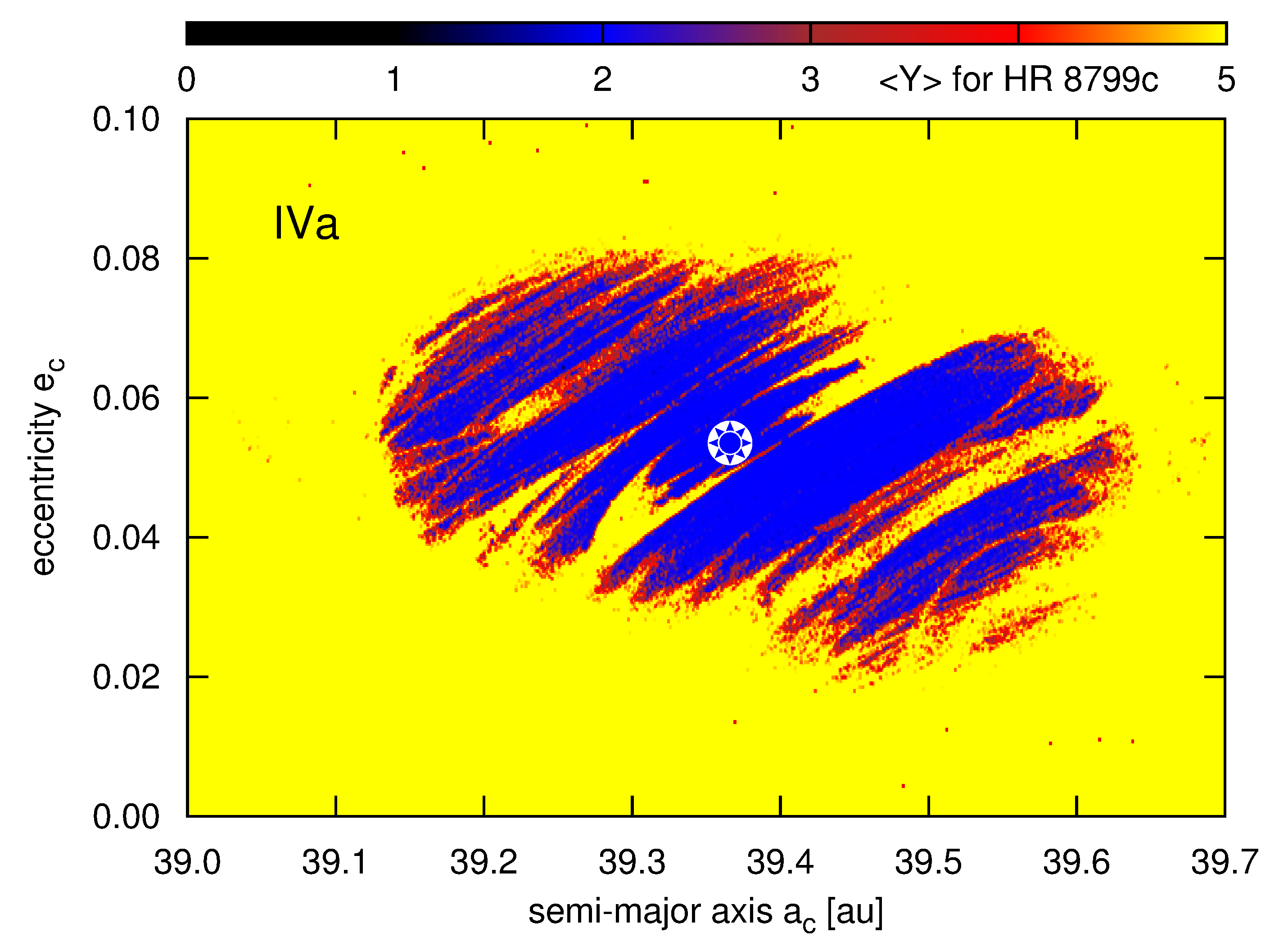}}
}
\hbox{
{\includegraphics[width=0.49\textwidth]{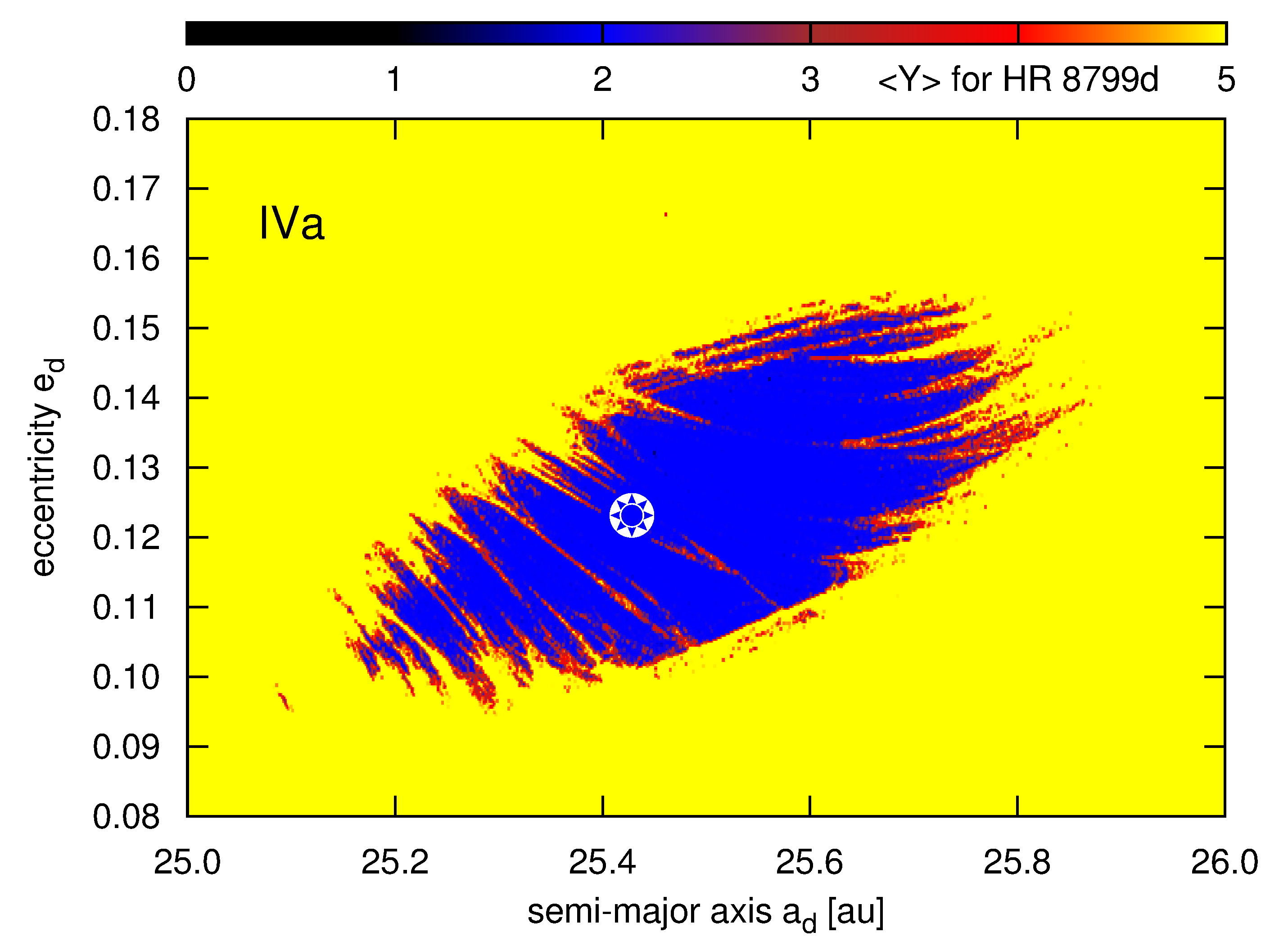}}
{\includegraphics[width=0.49\textwidth]{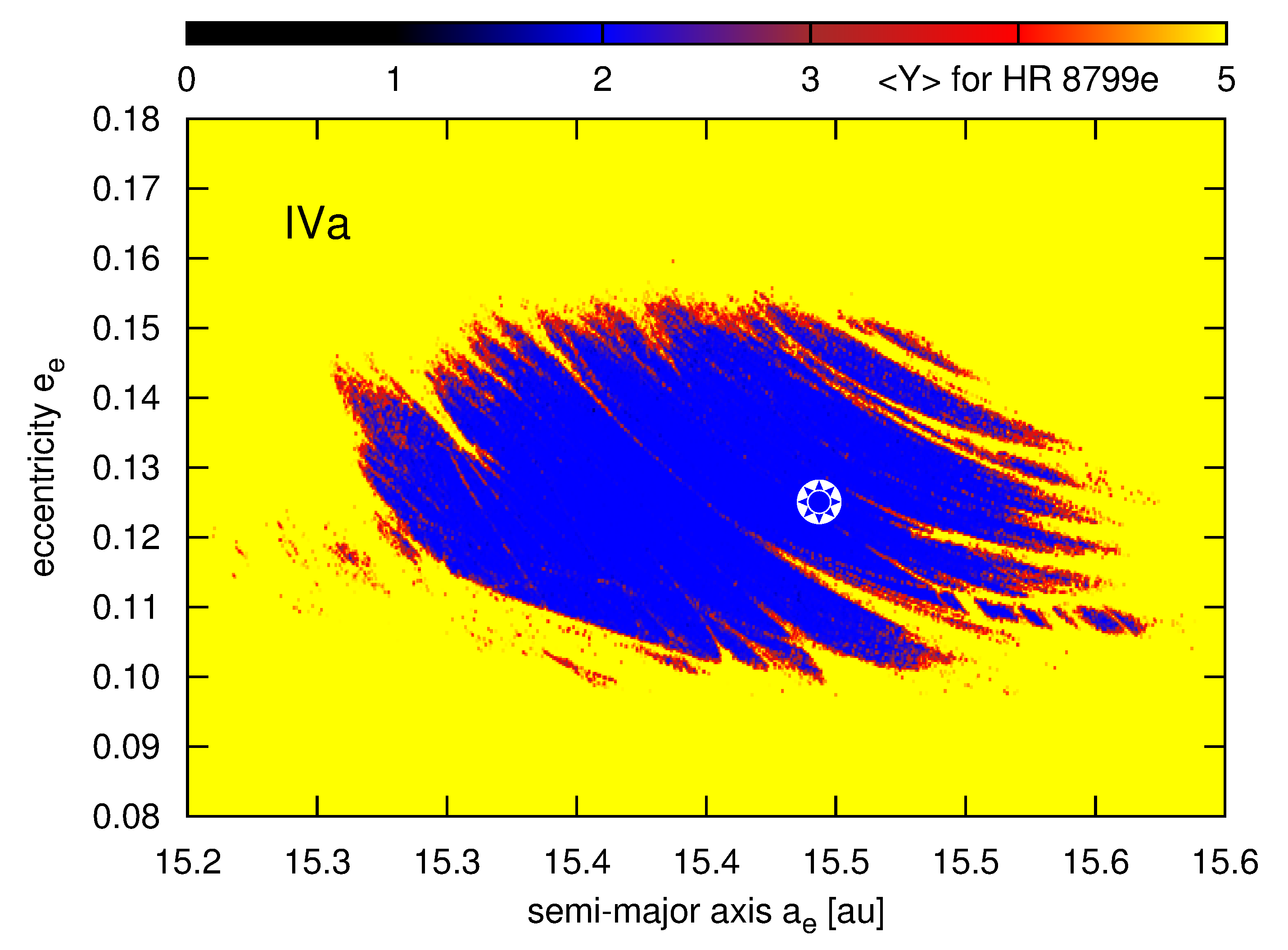}}
}
}
}
\caption{
Dynamical maps in terms of the maximal Lyapunov exponent, expressed by the MEGNO fast indicator $\left<Y\right>$, and shown in the semi-major--eccentricity planes for the four-planet model IVa.   The star symbol marks the best-fitting model (Tab.~\ref{tab:table1}), stable, quasi-periodic solutions are marked in blue.  The resolution of each map is $720\times360$ pixels. Integration time for each pixel is $\sim 20,000$ orbital periods of the outermost planet ($\simeq 10$~Myr).
}
\label{fig:fig9}
\end{figure*}

Finally, we computed dynamical MEGNO maps in the $(a_i,e_i$)-planes of osculating elements for all planets (see model IVa in Table~\ref{tab:table2}). The results are shown on Fig.~\ref{fig:fig9}. The nominal best-fitting model IVa, which is marked with a star symbol, is found in relatively extended zones of stable motions. These islands reveal complex structure of the 1e:2d:4c:8b~MMR, demonstrating that solution IVa is deeply locked in this multiple resonance. See also Sect.~5 for the event time $T_{\idm{E}}$ maps and the critical argument $\theta_{\idm{1:2:4:8}}$ in a close neighborhood of the best-fitting model~IVa. 

\corr{
%
\subsubsection{Independent refinement of Fit~IVa through {\sc GAMP}}
%
%
MCOA is CPU demanding and complex algorithm that requires many trials to derive the best-fit model. To verify that Fit~IVa is indeed optimal, and to better characterize this solution statistically, we applied other, independent method of constrained optimization which is called {\sc GAMP} (acronym of the Genetic Algorithm with the MEGNO Penalty) \citep{Gozdziewski2008a}. GAMP relies on penalizing unstable dynamical models by some large value of $\cchi$. In this experiment, the inclinations and masses are free  parameters of the model. Masses were searched  in the $[4,20]~\mJ$ range, and inclinations in the $[10^{\circ},33^{\circ}$] range, respectively. Other parameters are confined to a hyper-cube around Fit~IVa bounded by roughly $\pm 20$\% deviations from each orbital element in model IVa. At the optimization stage, the penalty term (system stability) is examined through the MEGNO indicator, equivalent to the Lyapunov exponent. This indicator must be evaluated for possibly short time-span 1,000--2,000~outermost periods of planet~b, due to significant CPU overhead. 
As the result, we gathered $\simeq 3000$ stable models with $1.13<\cchi<1.15$ computed with 4~degrees of freedom (DOF), in accord with Sect.~3.2. Each solution in this set formally better in terms of
$\cchi$ than Fit~IVa, and roughly within $3\sigma$ of the best fit model found in this independent search
with $\cchi \approx 1.5$ for 28~DOF (all orbital elements and masses). For each of these models, we calculated MEGNO for 32,000 periods of planet~b (16~Myr). We verified (Sect.~5) that MEGNO converged to $(2\pm 0.05)$ indicates a solution dynamically stable for at least 10 times longer motion times.

The results (stable solutions in the $1\sigma$ range of the best fit  GAMP model) are shown in panels of Fig.~\ref{fig:fig7add}, which are projections of the found models onto two parameter planes: the semi-major axis vs. eccentricity, and the inclination vs. mass. In all cases, the nominal fit is located roughly in the middle of the stable set. All inclinations remain within a few degrees around Fit~IVa, and masses are found systematically below $\sim 14\mJ$ and $\sim 18\mJ$ for two inner planets, and two outer planets, respectively. The results confirm that Fit~IVa is robust to relatively large masses, still well consistent with the estimates from cooling models. The spread of individual inclinations may be estimated as $\sim 5^{\circ}$ around IVa value of $\sim 25^{\circ}$.

\begin{figure*}
\centerline{ 
\vbox{
\hbox{
\includegraphics[width=0.43\textwidth]{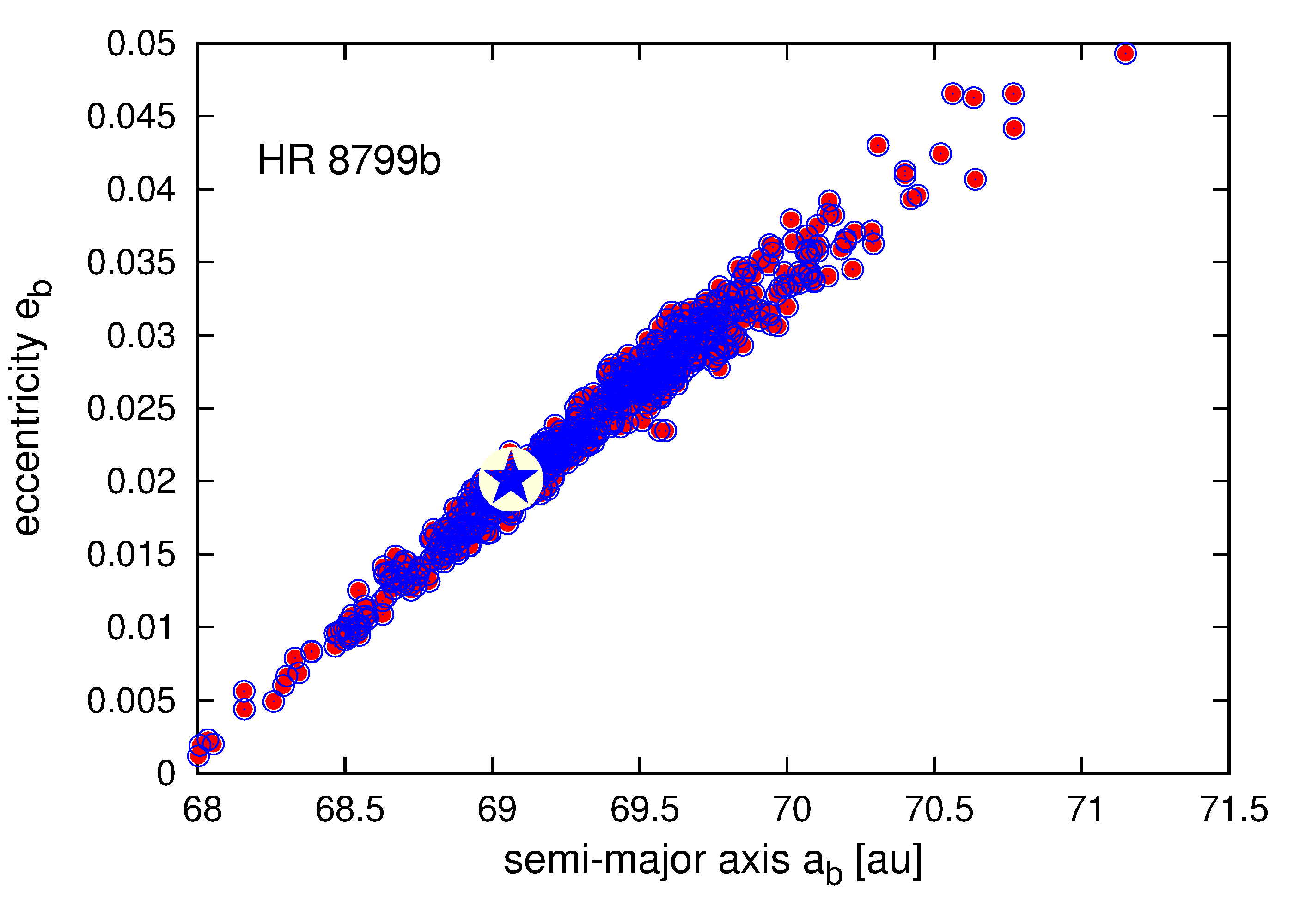}
\includegraphics[width=0.43\textwidth]{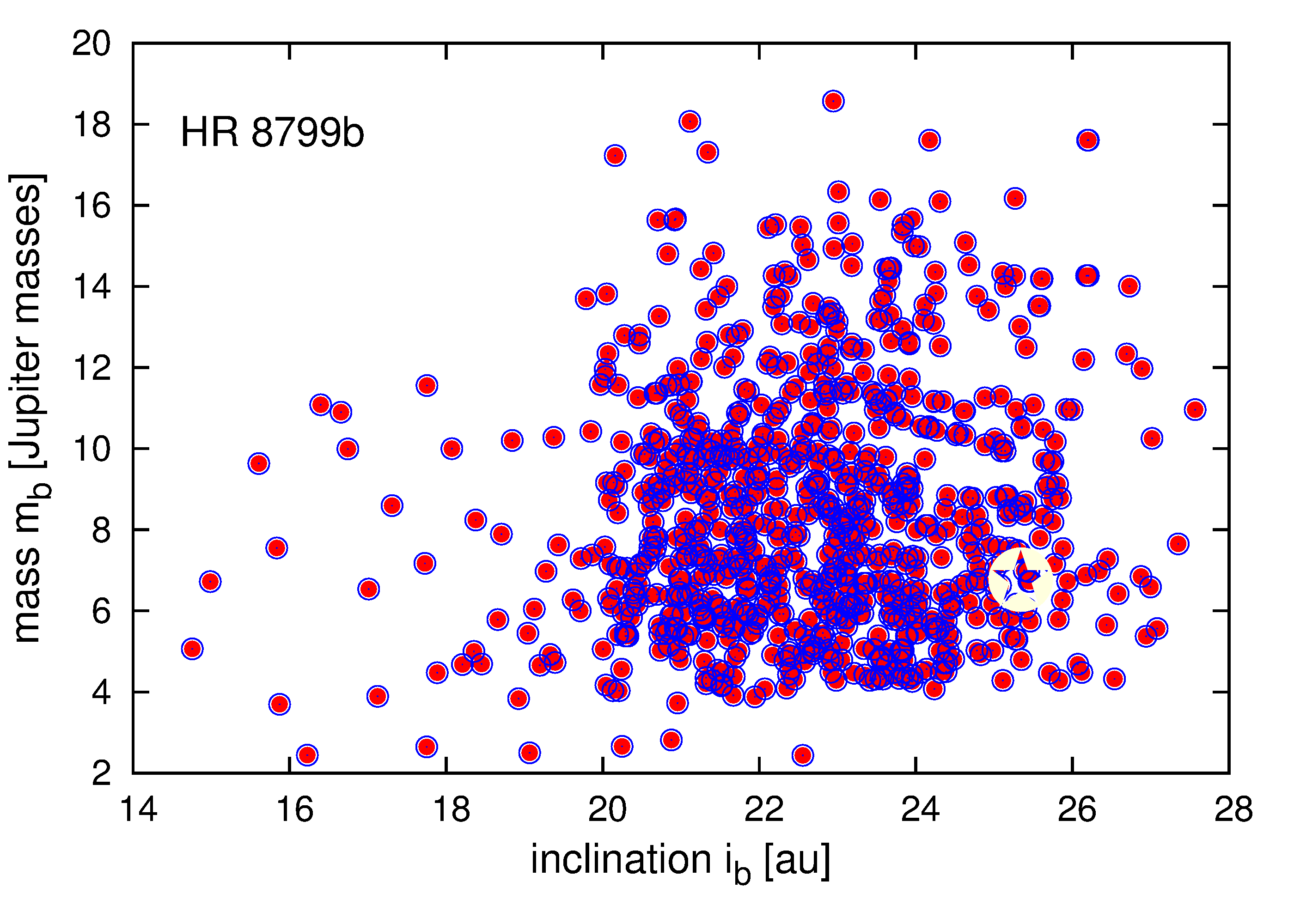}
}
\hbox{
\includegraphics[width=0.43\textwidth]{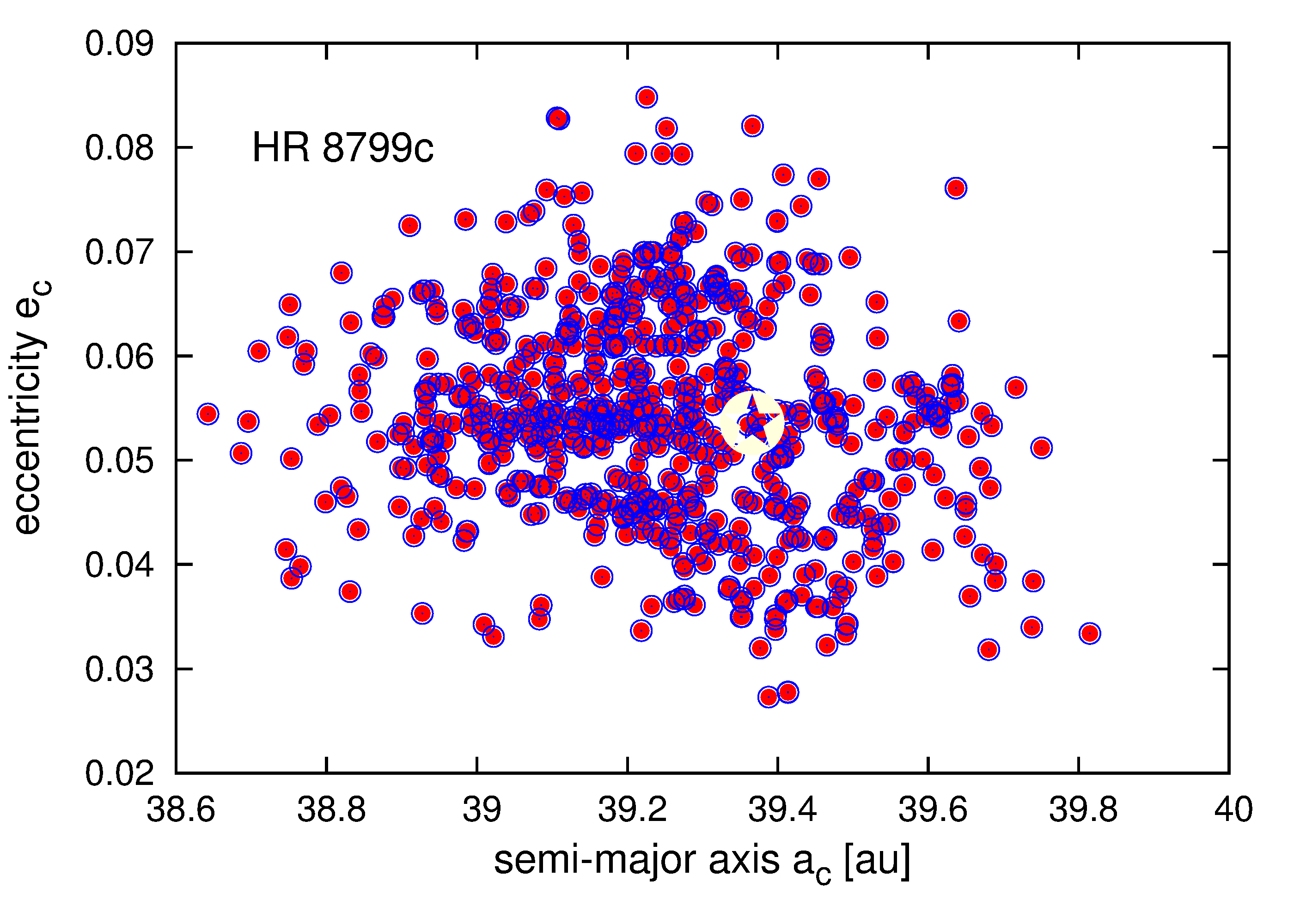}
\includegraphics[width=0.43\textwidth]{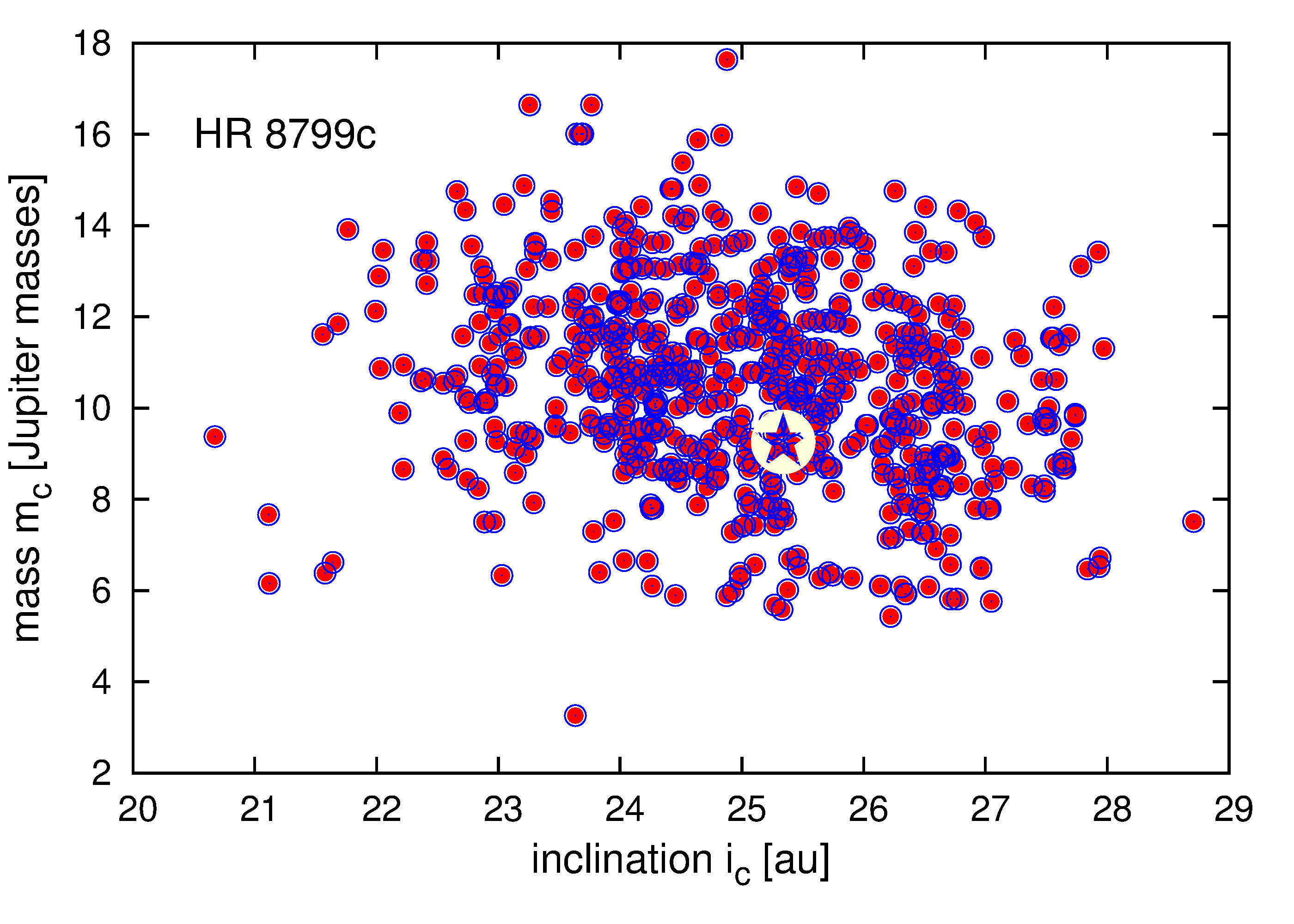}
}
\hbox{
\includegraphics[width=0.43\textwidth]{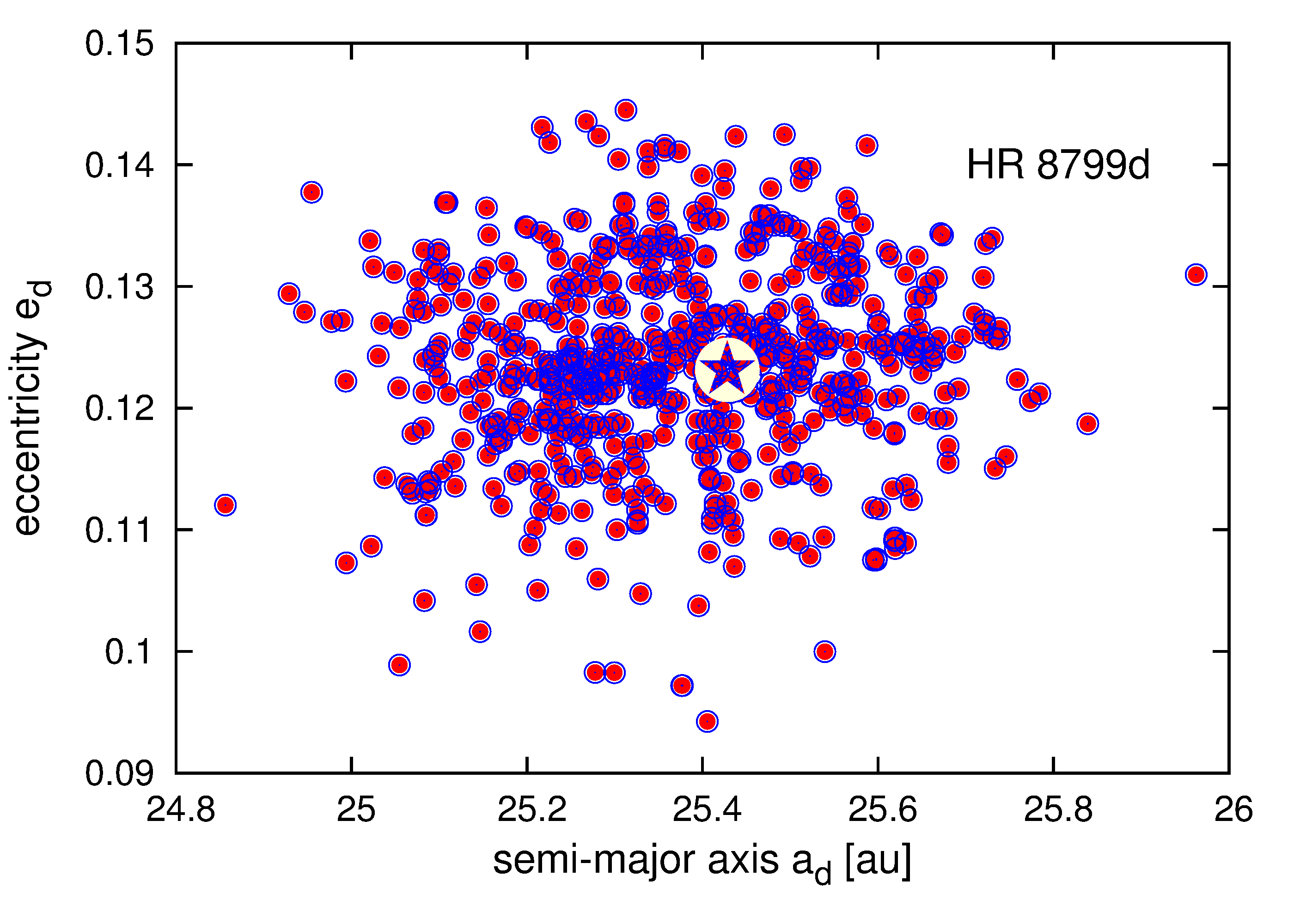}
\includegraphics[width=0.43\textwidth]{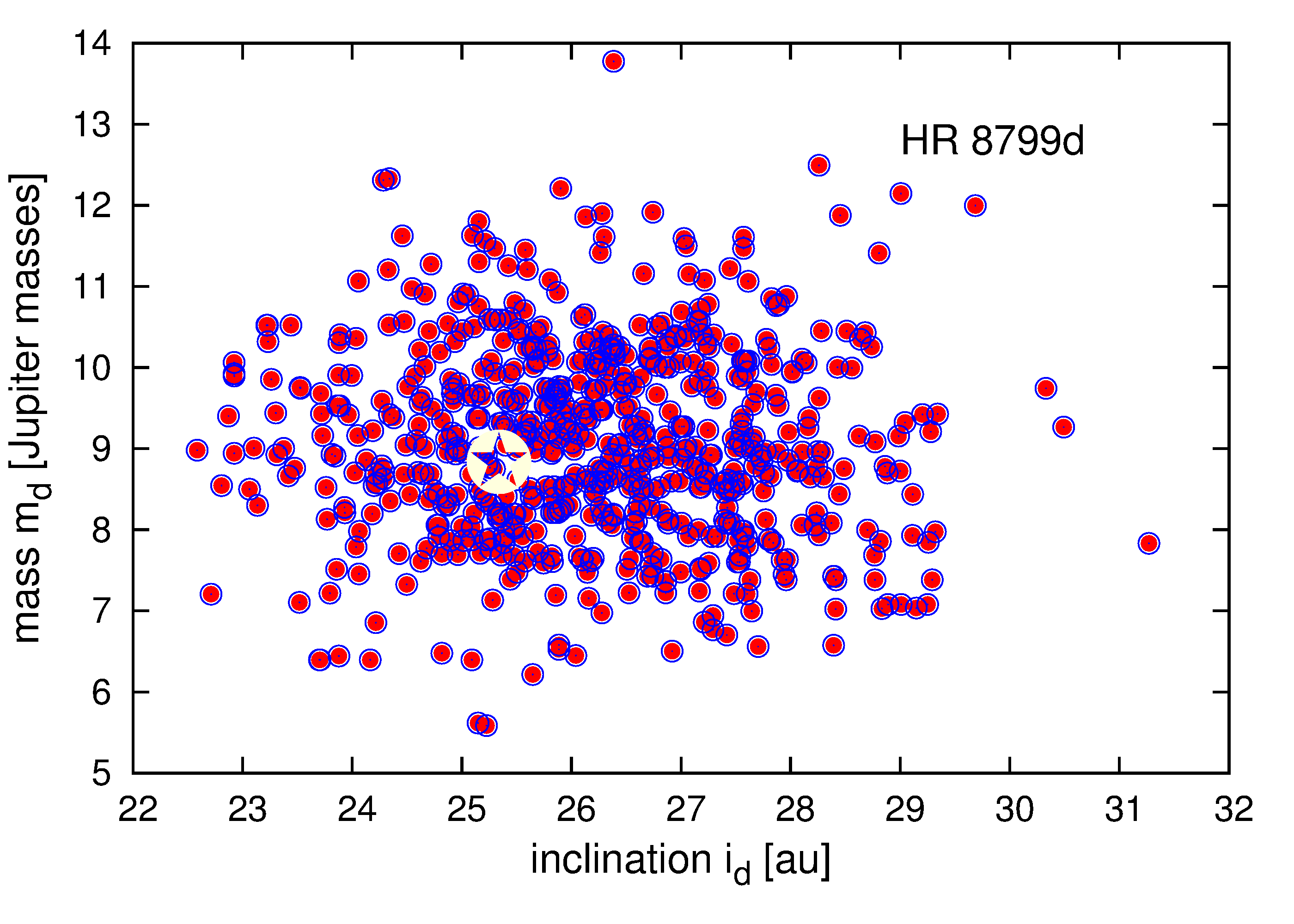}
}
\hbox{
\includegraphics[width=0.43\textwidth]{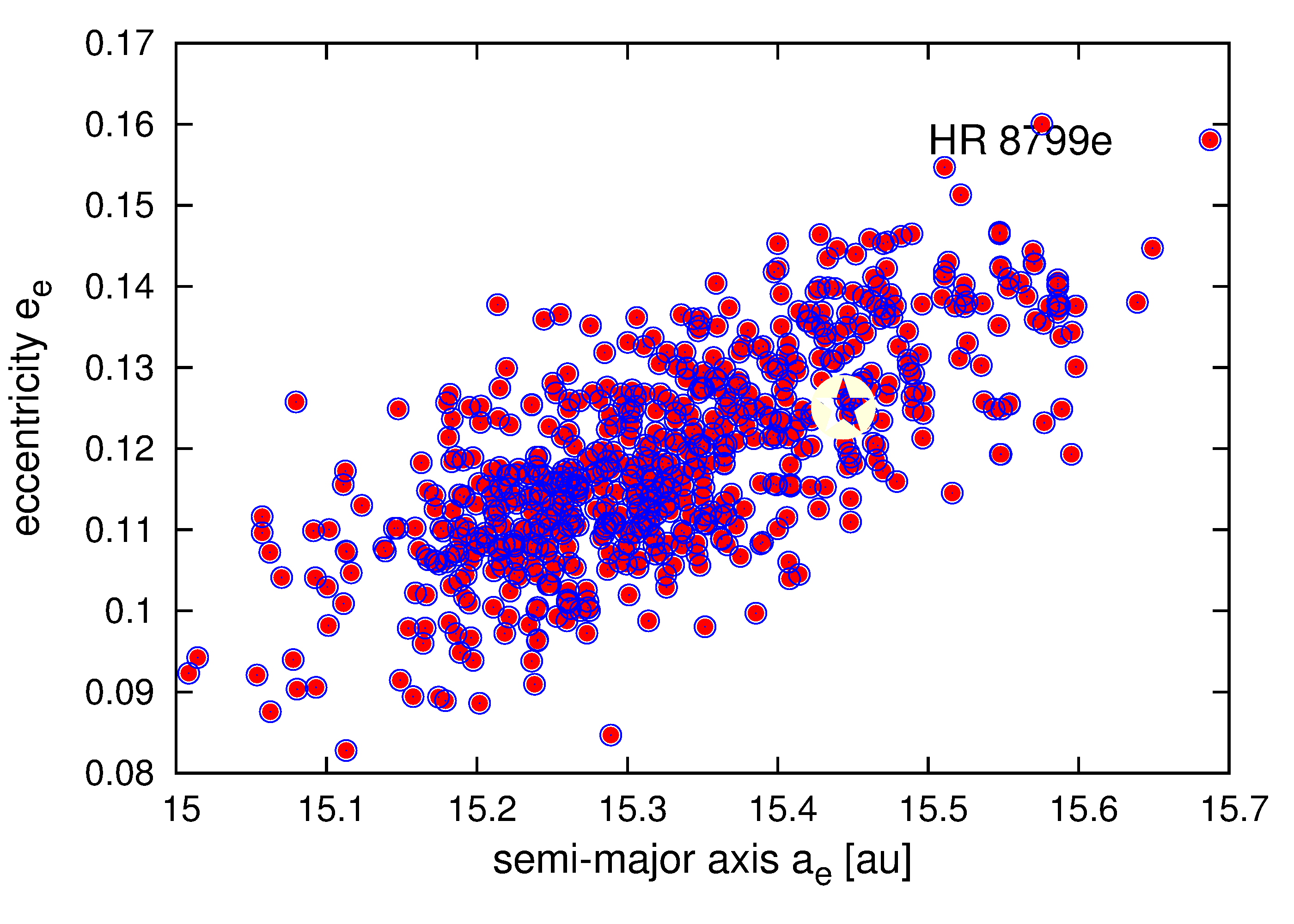}
\includegraphics[width=0.43\textwidth]{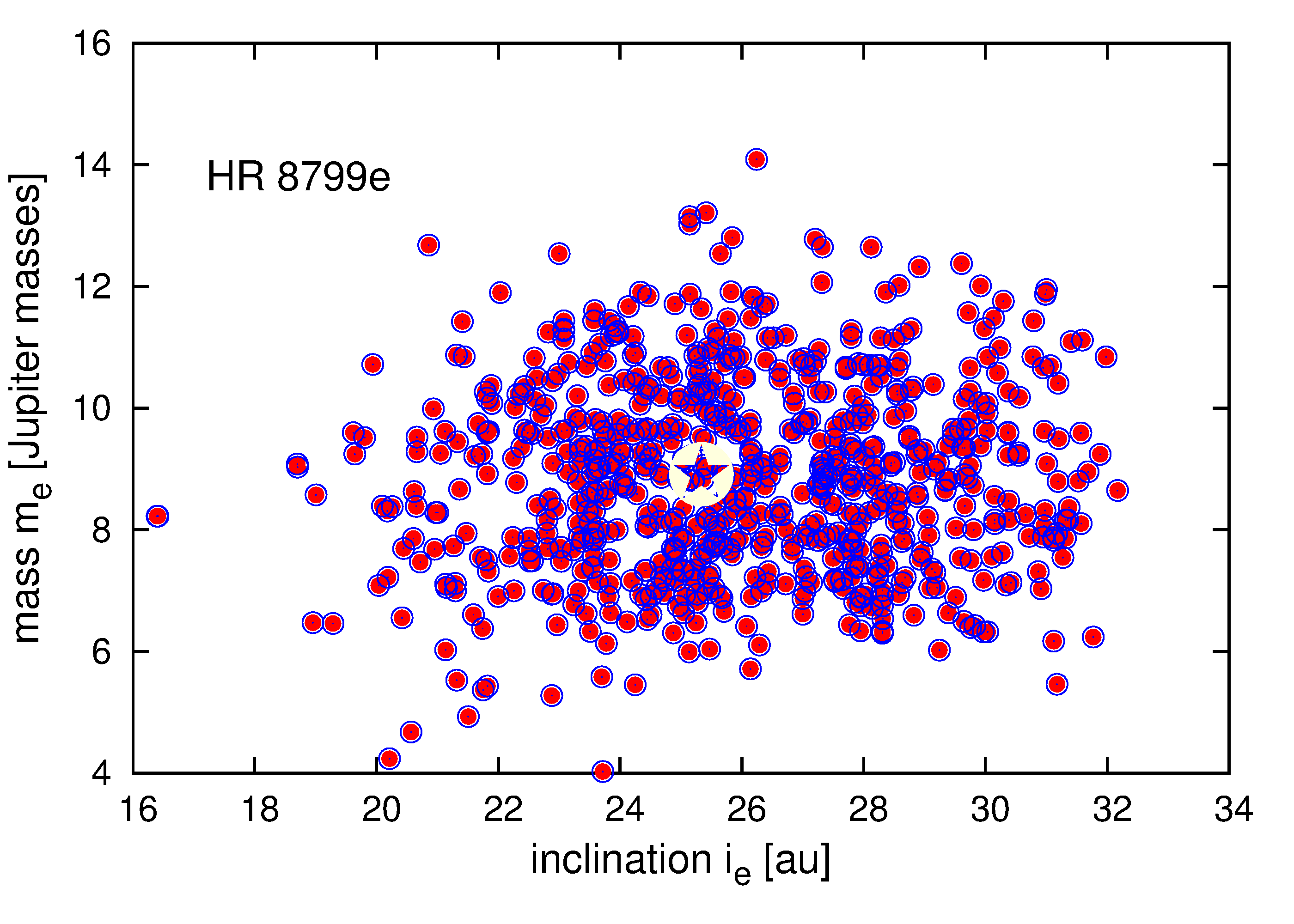}
}
}
}
\caption{
Best-fitting stable solutions derived with the GAMP algorithm (see text for details) projected onto semi-major axis -- eccentricity planes  ({\em left column)}), and onto inclination -- mass planes ({\em right column}). The star symbol marks the nominal, best-fitting solution in Tab.~\ref{tab:table1}.  Filled circles are for stable
solutions within $\cchi < 1.15$ that corresponds to $1\sigma$ of the best fit model found in the GAMP search with $\cchi \simeq 1.13$. Their $|\left<Y\right>-2|<0.05$ for the integration time-span of 16~Myr.
}
\label{fig:fig7add}
\end{figure*}

}

\corr{
%
\subsection{Statistical study of migration}
%
\begin{table}
\caption{Percentage of simulated MMRs for different masses of planets, i.e., the nominal masses of $9$, $10$, $10$ and $7\,\mJ$ from the innermost to the outermost planet, respectively ({\em middle column}), all masses increased by $50\,\%$ ({\em left-hand column}) and decreased by $50\,\%$ ({\em  right-hand column}), respectively. 
A notion of "$<0.1$" means that there was one or at most two solutions of a given type, while "$-$" means no solution of this type.}
\label{tab:res_freq}
\begin{tabular}{l | c c c}
\hline
\hline
MMR chain & \multicolumn{3}{c}{percentage of solutions}\\
e:d, d:c, c:b & $1.5 \times m$ & $1 \times m$ & $0.5 \times m$\\
\hline
$2:1, 2:1, 2:1$	&			$-$	&	$10.2$	&	$20.3$\\
$2:1, 2:1, 3:1$	&			$-$	&	$-$	&	$<0.1$\\
$2:1, 2:1, 3:2$	&			$-$	&	$-$	&	$12.9$\\
$2:1, 2:1, 4:1$	&			$-$	&	$0.2$	&	$-$\\
$2:1, 5:3, 2:1$	&			$-$	&	$-$	&	$2.3$\\
$2:1, 3:1, 2:1$	&			$-$	&	$-$	&	$0.3$\\
$2:1, 3:1, 3:1$	&			$-$	&	$-$	&	$<0.1$\\
$2:1, 3:1, 5:2$	&			$-$	&	$-$	&	$<0.1$\\
$2:1, 3:2, 2:1$	&			$-$	&	$-$	&	$<0.1$\\
\hline
$3:1, 2:1, 2:1$	&			$57.5$	&	$51.8$	&	$37.7$\\
$3:1, 2:1, 3:1$	&			$-$	&	$0.1$	&	$0.5$\\
$3:1, 2:1, 4:1$	&			$-$	&	$-$	&	$<0.1$\\
$3:1, 2:1, 3:2$	&			$-$	&	$0.1$	&	$1.8$\\
$3:1, 3:1, 2:1$	&			$13.5$	&	$17.0$	&	$13.1$\\
$3:1, 3:1, 3:1$	&			$-$	&	$0.2$	&	$0.2$\\
$3:1, 5:2, 2:1$	&			$-$	&	$0.1$	&	$0.3$\\
$3:1, 5:2, 3:1$	&			$-$	&	$-$	&	$<0.1$\\
\hline
$4:1, 2:1, 2:1$	&			$6.4$	&	$3.1$	&	$0.8$\\
$4:1, 2:1, 3:1$	&			$0.1$	&	$0.1$	&	$<0.1$\\
$4:1, 3:1, 2:1$	&			$20.9$	&	$14.0$	&	$3.6$\\
$4:1, 3:1, 3:1$	&			$-$	&	$0.2$	&	$0.5$\\
$4:1, 3:1, 4:1$	&			$-$	&	$0.1$	&	$-$\\
$4:1, 3:1, 5:2$	&			$-$	&	$0.1$	&	$-$\\
$4:1, 4:1, 2:1$	&			$0.1$	&	$0.2$	&	$0.2$\\
$4:1, 4:1, 4:1$	&			$0.1$	&	$0.1$	&	$<0.1$\\
$4:1, 4:1, 5:2$	&			$-$	&	$0.1$	&	$0.1$\\
$4:1, 5:2, 4:1$	&			$-$	&	$0.1$	&	$-$\\
$4:1, 5:2, 2:1$	&			$-$	&	$0.1$	&	$<0.1$\\
$4:1, 5:2, 5:2$	&			$-$	&	$-$	&	$<0.1$\\
\hline
$5:2, 2:1, 2:1$	&			$1.3$	&	$2.5$	&	$4.0$\\
$5:2, 2:1, 3:2$	&			$-$	&	$-$	&	$0.3$\\
$5:2, 3:1, 2:1$	&			$-$	&	$-$	&	$0.4$\\
$5:2, 3:1, 4:1$	&			$0.1$	&	$0.1$	&	$-$\\
$5:2, 5:2, 2:1$	&			$-$	&	$-$	&	$0.1$\\
$5:2, 5:2, 4:1$	&			$0.1$	&	$-$	&	$-$\\
$5:2, 5:2, 5:2$	&			$-$	&	$-$	&	$<0.1$\\
\hline
$3:2, 2:1, 2:1$	&			$-$	&	$-$	&	$0.2$\\
\hline
$5:3, 2:1, 3:2$	&			$-$	&	$-$	&	$<0.1$\\
\hline
total number of solutions	&	$1196$	&	$1739$	&	$2872$\\
\hline
\hline
\end{tabular}
\end{table}

We used the heuristic model of migration (Eq.~\ref{eq:moore}) to find how frequently a particular four-planet MMR chain forms, dependent on planetary masses and time-scales of migration. In a first experiment we fixed planetary masses at their nominal values in the HR~8799 system, i.e., $9, 10, 10, 7\,\mJ$ (counting from the innermost to the outermost planet, respectively). Parameters of the migration model were chosen randomly from wide ranges. Using notation in Sections~3.1 and~3.2, these ranges are the following: $\log_{10} \tau_{1,0}(\mbox{yr}) \in [6,10]$, $\log_{10} T_1(\mbox{yr}) \in [6, 10]$, $K \in [1, 100]$, $\alpha_1 \in [-2.0, -0.1]$. The second term $\tau_{2,0}$ was not taken into account. The initial orbits are distributed exponentially, with $a_1 \in [30, 160]\,\au$ and $\beta \in [0.3, 1.8]$. The orbits calculated from the exponential distribution are then shifted randomly with Gaussian distribution ($\sigma = 0.2$ of the nominal values). Initial eccentricities were random within $[0, 0.05]$~range. A given initial system was integrated for $100\,$Myr.

We found that slightly less than $20\,\%$ of simulations ($18\,\%$, $1739$ of $9488$) ended up as stable resonant configurations. The most frequent chain of MMRs is $3:1, 2:1, 2:1$, i.e., $1{\mbox{e}}:3{\mbox{d}}:6\mbox{c}:12\mbox{b}$~MMR. More than a half of stable final systems are of this type. Another solutions $3:1, 3:1, 2:1$ ($17\,\%$) and $4:1, 3:1, 2:1$ ($14\,\%$) are also relatively common in a sample of final configurations. The double Laplace MMR ($2:1, 2:1, 2:1$ or equivalently $1\mbox{e}:2\mbox{d}:4\mbox{c}:8\mbox{b}$~MMR) appeared in $\sim 10\,\%$ of the stable runs. By a resonant four-planet configuration we mean a configuration characterized by its critical argument $\phi$ librating around 
a particular libration center.
Table~\ref{tab:res_freq} summarizes the results obtained for the nominal masses of the planets, as well as masses $50\,\%$ higher and $50\,\%$ lower than these values, respectively.

When masses of the planets are smaller than the nominal values, i.e., $4.5, 5, 5, 3.5\,\mJ$, a fraction of systems that survived the integration time increases to $\sim 29\,\%$ ($2878$ of $9960$). The number of possible final configurations also increases. Some of MMR chains absent in the previous test appear for smaller masses, e.g., $2:1, 2:1, 3:2$ ($13\,\%$ of stable systems). The double Laplace MMR now appears more frequently ($\sim 20\,\%$), while $3:1, 2:1, 2:1$ appears in $\sim 38\,\%$ cases, i.e., less frequently when compared to the previous test.

When masses are increased ($13.5, 15, 15, 10.5\,\mJ$), only $\sim 12\,\%$ ($1196$ of $10235$) of systems end up as stable chains of MMRs. Moreover, there are fewer possible final configurations. Particularly, the double Laplace resonance does not appear at all. Three most common chains of MMRs constitute $\sim 92\,\%$ of stable systems. This test could provide us an upper limit on planetary masses in the HR~8799 system. For planets massive enough, the double Laplace resonance, which best matches the observations, may unlikely form through convergent migration of already formed planets. We would like to warn the reader, that the migration model may be too simplistic: for instance,  it does not take into account any (possibly substantial) mass increase during the migration. Any definite conclusions should be taken with much caution here. 

\begin{figure}
\vbox{
\includegraphics[width=0.47\textwidth]{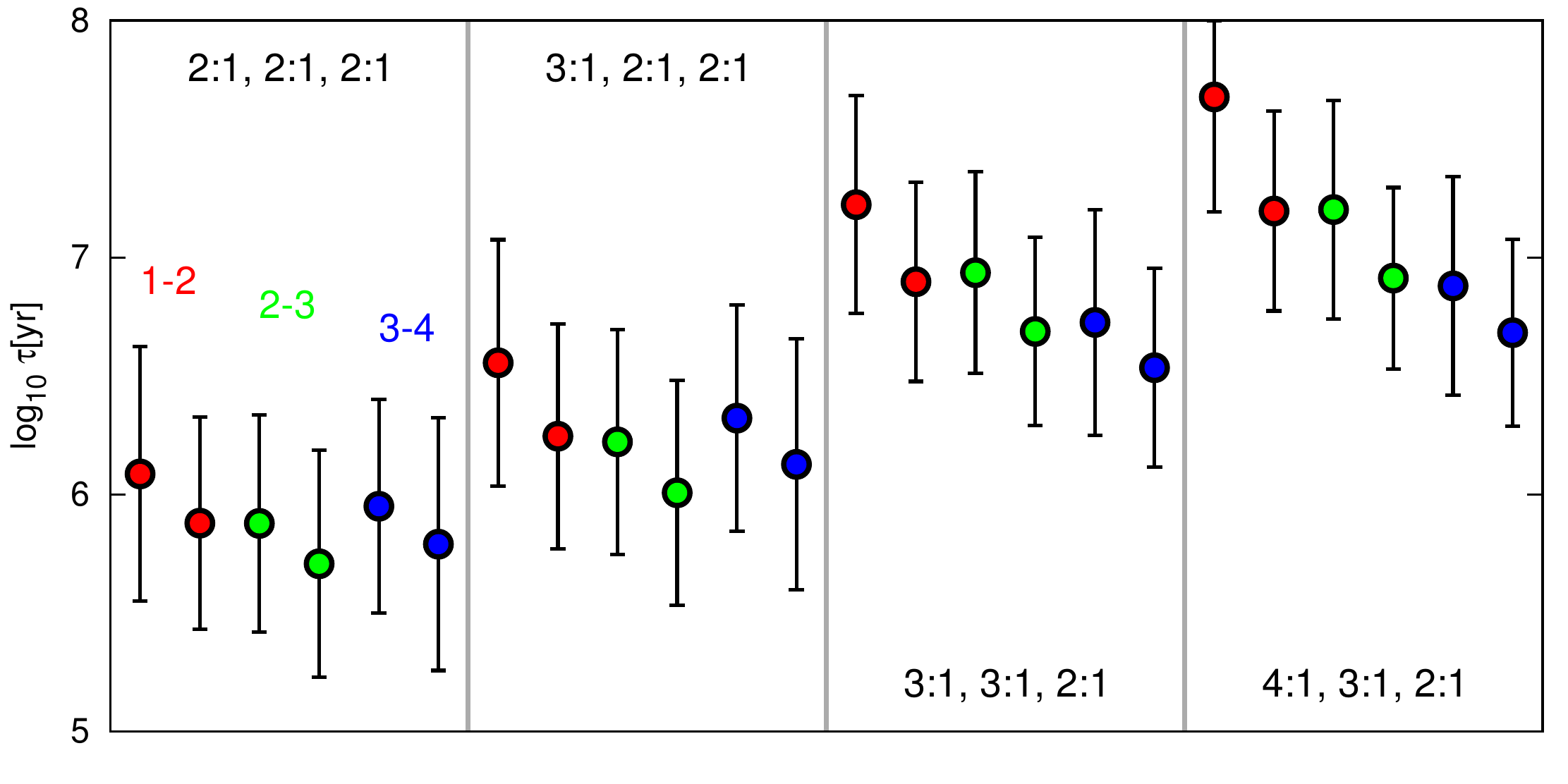}
}
\caption{
Time scales of migration $\tau \equiv a/\dot{a}$ for a given planet before MMR capture with a neighboring planet. Each column is for a different multiple MMR chain. Red color marks values of $\tau$ for the innermost pair (planets numbered as $1$ and $2$), green color is for the middle pair (planets $2$ and $3$), and blue color is for the outermost pair (planets $3$ and $4$), respectively.}
\label{fig:time_scales}
\end{figure}

Because parameters of the migration model were chosen from very wide ranges, planets migrate in different time-scales, from one simulation to another. This leads to different final MMR configurations. Figure~\ref{fig:time_scales} presents the results for four most common MMRs obtained during simulations with the nominal masses. The left-hand part of the diagram shows the time-scales for the double Laplace resonance. The red filled circles present average values of $L \equiv \log_{10} \tau(yr) \equiv \log_{10} a/\dot{a}$ for two innermost planets just before they are captured into the $2:1$~MMR. The standard deviation of $\sigma_L$ is depicted by the bars. The planets migrate in average time-scales of $\approx 1\,$Myr, i.e, $L \approx 6$ (note that the second planet migrates slightly faster), with $\sigma_L \approx 0.5$.
Green filled circles present $L$ for the middle pair of planets, while blue filled circles are for the outermost pair. The migration time-scale required to form the double Laplace MMR is of the order of $1\,$Myr.

The MMRs chain $3:1, 2:1, 2:1$ form when the migration occurs slightly slower ($2-3$~Myr on average). Another two chains $3:1, 3:1, 2:1$ and $4:1, 3:1, 2:1$ appear for even larger $L$. The latter MMR chain form typically for $L \approx 7$ for a middle pair of planets. The inner pair migrates slower, while the outer pair migrates typically faster. The general conclusion is that the characteristic time-scales of migration leading to a particular chain of MMRs, are different for different chains.
}

%
\subsection{Model IVb: could be planet~HR8799~e predicted?}

{Decades of observations are required to constrain orbital parameters of very long-period planets with standard methods.  Relying on discrete and in some sense {\em deterministic} outcomes of the migration algorithm, we may consider different architectures of the resonant systems with even very limited observations. At an extreme case, we may turn back to 2008, when measurements for three outer planets were published in the discovery paper \citep{Marois2008}. Two years later, the fourth planet was detected. Having in mind the multiple MMR model of the four-planet system, we may ask: could be the fourth planet predicted or found in the \corr{present, observed place if we did not have even a single} data point for this planet? 

\begin{figure*}
\centerline{
\vbox{
\hbox{
\includegraphics[width=0.49\textwidth]{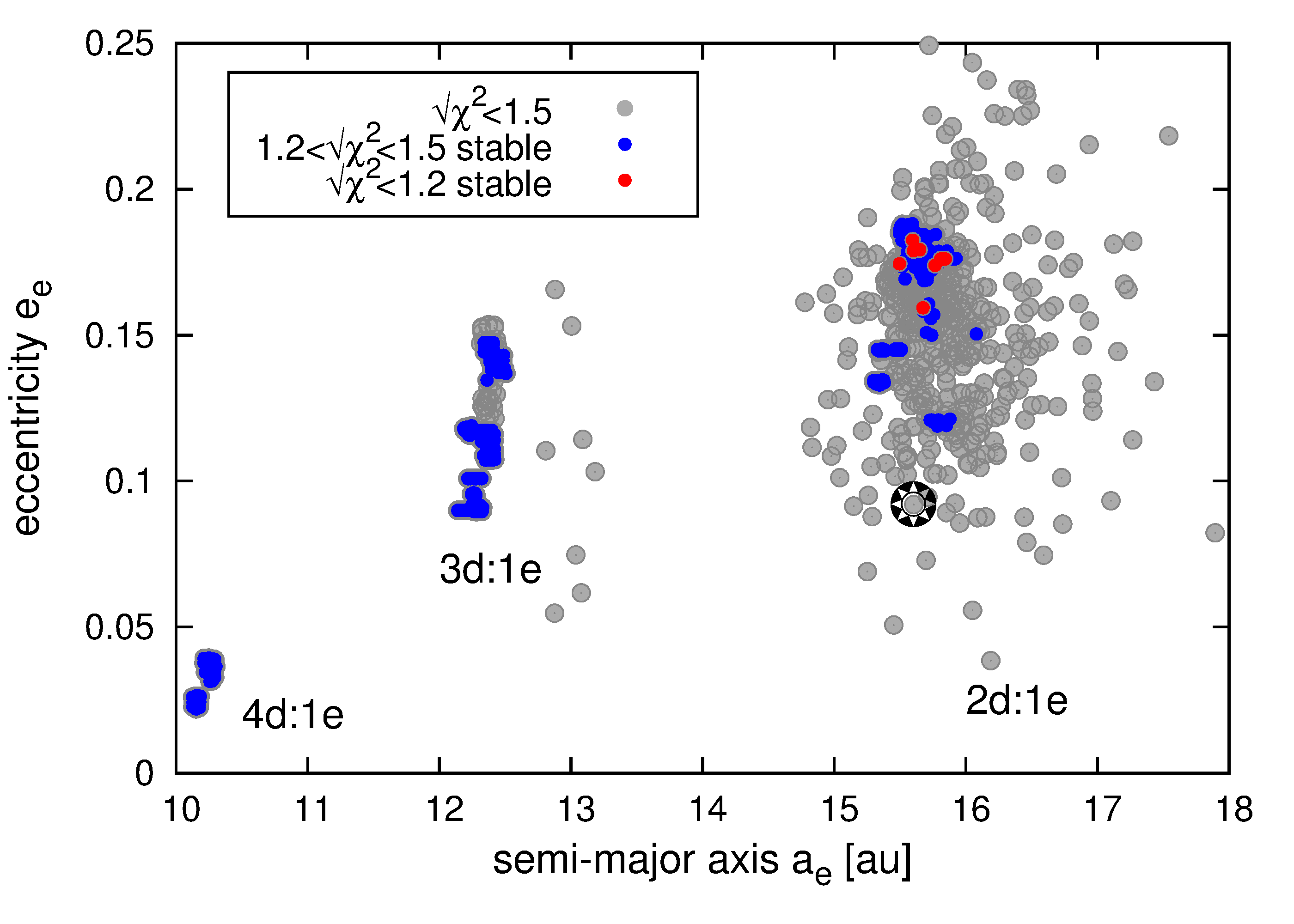}
\includegraphics[width=0.49\textwidth]{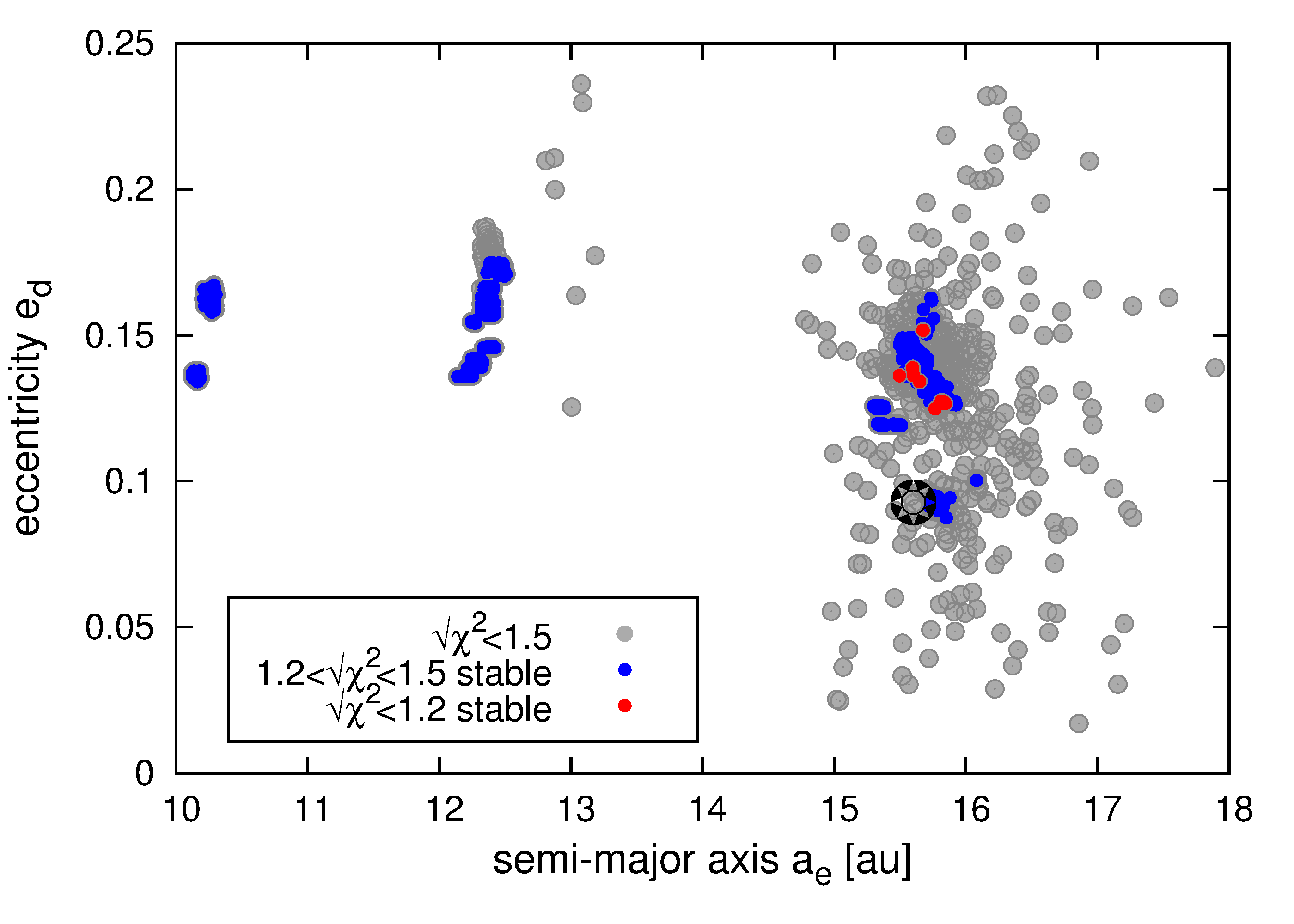}
}
\hbox{
\includegraphics[width=0.49\textwidth]{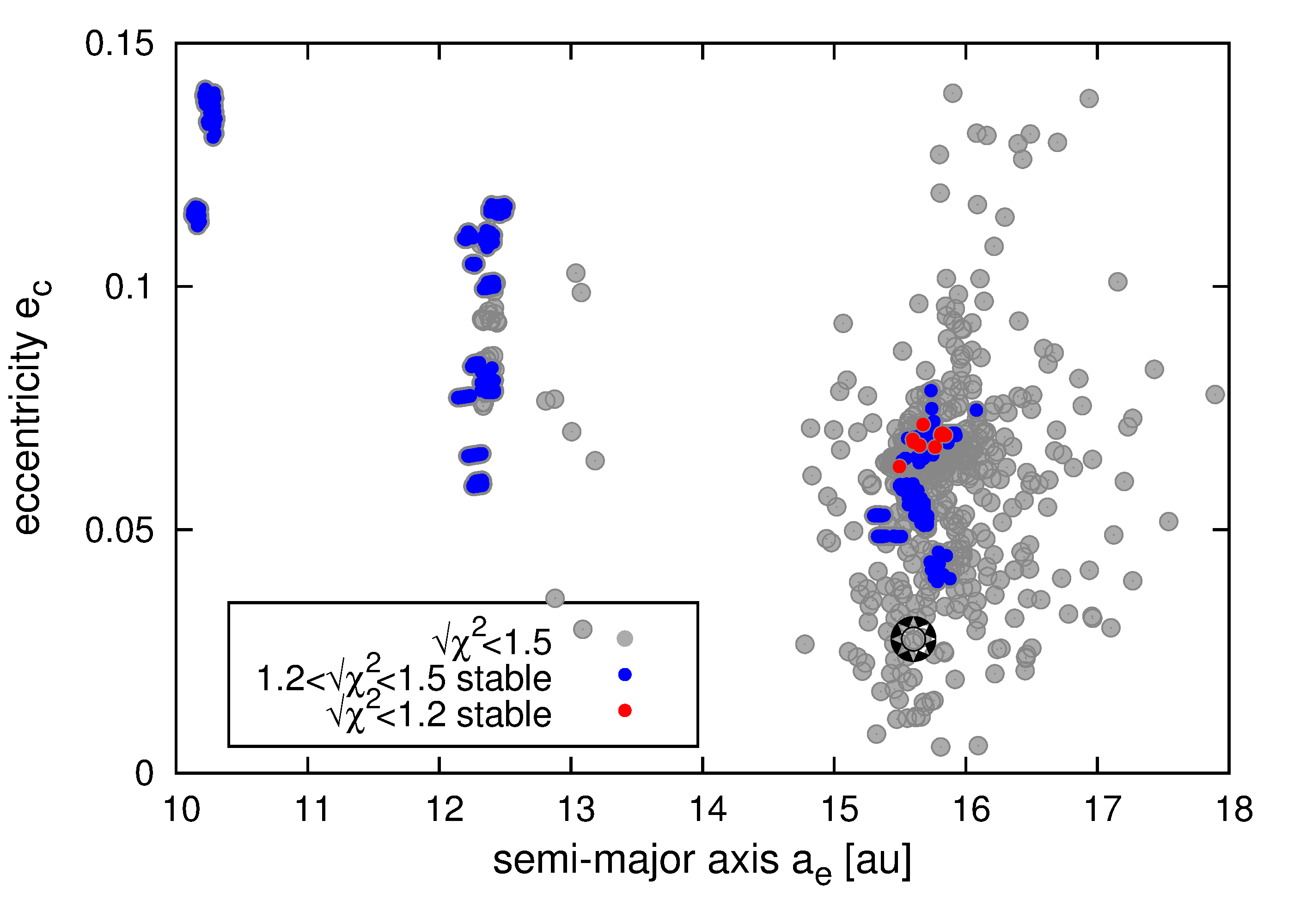}
\includegraphics[width=0.49\textwidth]{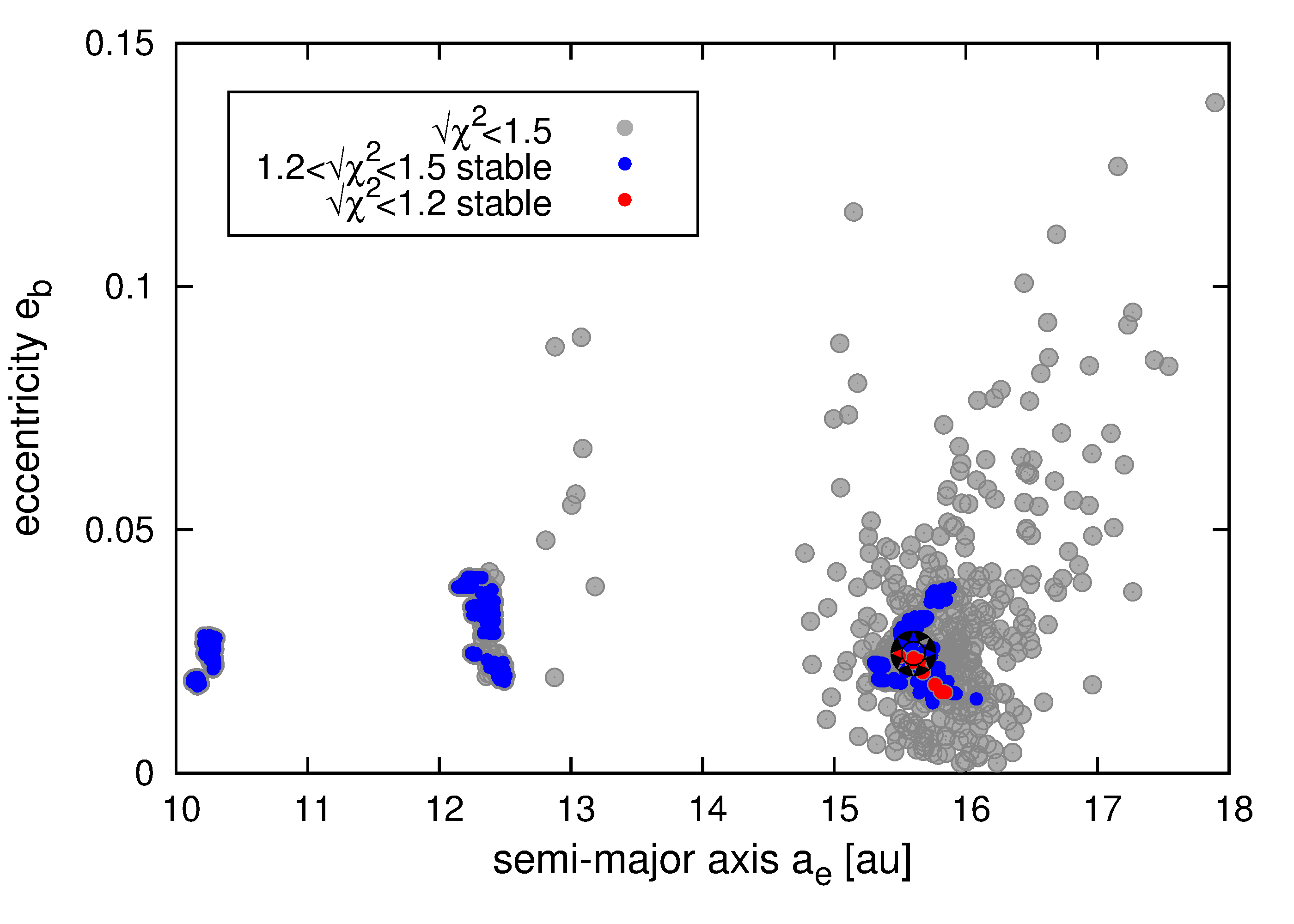}
}
}
}
\caption{
Best-fitting solutions to four-planet model IVb (with ``unseen'' planet~e) and dataset~D5  comprising of all observations of three planets b, c, and d, {\em without data for planet~e},  projected onto $(a_{\idm{e}}, x)-$plane, where $x$ is $e_\idm{e}$, $e_\idm{d}$, $e_\idm{c}$, $e_\idm{b}$. Grey symbols denote solutions within $6\sigma$ confidence interval ($\cchi < 1.5$), blue and red symbols are for stable solutions with  $(3\sigma,6\sigma)$, equivalent to $1.2 < \cchi < 1.5$, and for $3\sigma$ models ($\cchi < 1.2$), respectively. 
}
\label{fig:fig10}
\end{figure*}

\begin{figure*}
\centerline{
\hbox{
{\includegraphics[width=0.49\textwidth]{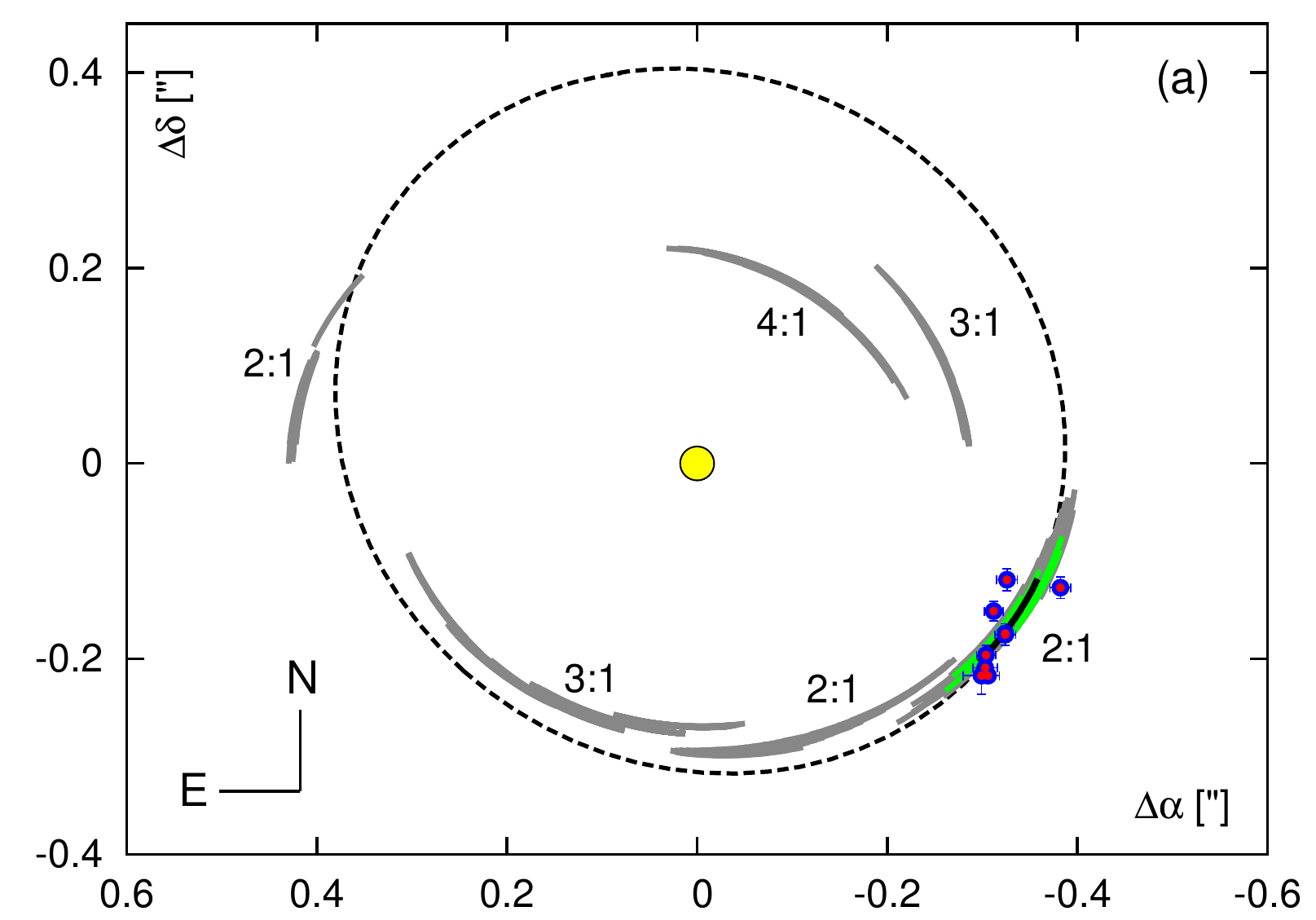}}
{\includegraphics[width=0.49\textwidth]{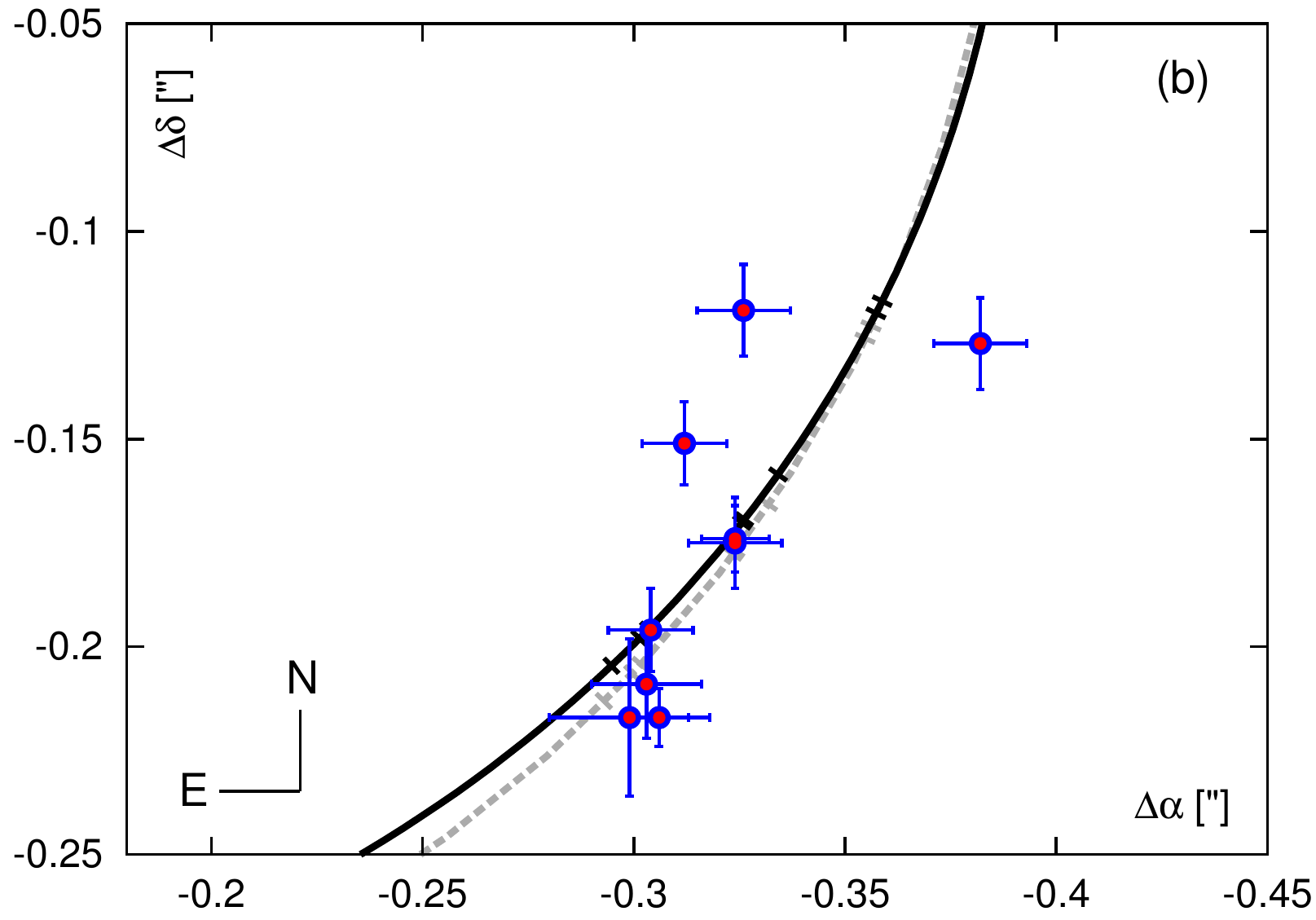}}
}
}
\caption{
{\em Left panel:} Relative astrometry of planet~e (red/blue symbols) with best-fitting four-planet model~IVa to all observations (dashed curve). Grey, green and black curves show arcs of different families of the four-planet model~IVb to the most recent observations of planets b, c and d (dataset D5). Black curve is for the best-fitting model, green curves are for $3\sigma$ solutions and grey curves are for $6\sigma$ models ($\cchi < 1.5$). {\em Right panel:} A close-up of the left panel. The black curve is for the best-fitting stable solution~IVb, grey dashed curve is for the best-fitting nominal, four-planet model~IVa.
}
\label{fig:fig11}
\end{figure*}

Following the general idea, we assume that orbits of planets~b, c and~d, as well as the ``unseen'' planet~e are outcomes of the migration scenario. The Laplace 1d:2c:4b~MMR of the outermost planets found in the early dynamical papers must be not necessarily preserved by the four-planet architecture. Actually, the literature is not consistent about this problem. For instance, due to mutual interactions and complex dissipative evolution, the 1c:2b~MMR between planets~c and~d might be  changed to 1c:3b~MMR. Many other two-body resonances are possible as well (see Sect.~4.2). The primary factor that makes it possible to distinguish between these cases are the observations. Assuming that the system as a whole evolves towards certain, discrete and small number of states (multiple MMRs), the observations ''decide''  which configuration is the right one. 

We performed a series of simulations concerning \corr{the four-planet model and D5 dataset} to test this idea. In this scenario the innermost planet~e is unseen, and the four-planet model combined with dataset~D5 is called model~IVb.  The results of the \moa{} search are illustrated in Figs.~\ref{fig:fig10},\ref{fig:fig11} and~\ref{fig:fig12}.  Similarly to the model IVa, Fig.~\ref{fig:fig10} illustrates  projections of the best-fitting osculating elements onto different planes. Grey filled circles are for solutions with $\cchi$ within the $6\sigma$ confidence level, while red and blue filled circles are for $3\sigma$ models. The statistics  reveals that configuration involving $1\mbox{e}:2\mbox{d}$~MMR  describes dataset D5 better than two other configurations with  $1\mbox{e}:3\mbox{d}$~MMR and $1\mbox{e}:4\mbox{d}$~MMR.  
 
We should realize here that multiple resonant configurations are determined not only by $\aee$, but may be also distinguished due to different $\ed, \ec, \eb$.     For instance,  the initial conditions for a system involved in the 1e:2d:4c:8b~MMR  and in the 1e:3d:6c:12b~MMR are significantly different. This is clearly seen in the $(\aee, \ec)$-panel in Fig.~\ref{fig:fig10}. Because $e_i$ are different in these two concurrent solutions, also the best-fitting inclinations are different, as shown in the top-left panel of Fig.~\ref{fig:fig12}}). While the $1\mbox{e}:2\mbox{d}$~MMR implies $I \approx 25^{\circ}$, the best-fitting models with  $1\mbox{e}:3\mbox{d}$~MMR and  $1\mbox{e}:4\mbox{e}$~MMR provide much larger inclination, i.e., $I \sim 35^{\circ}$. {The 1e:2d:4d:8b~MMR is favored due to lowest value of $\cchi$ and because its inclination matches closely the inclination of the stellar equator $\sim 25^{\circ}$}. 

\begin{figure*}
\centerline{
\vbox{
\hbox{
\includegraphics[width=0.49\textwidth]{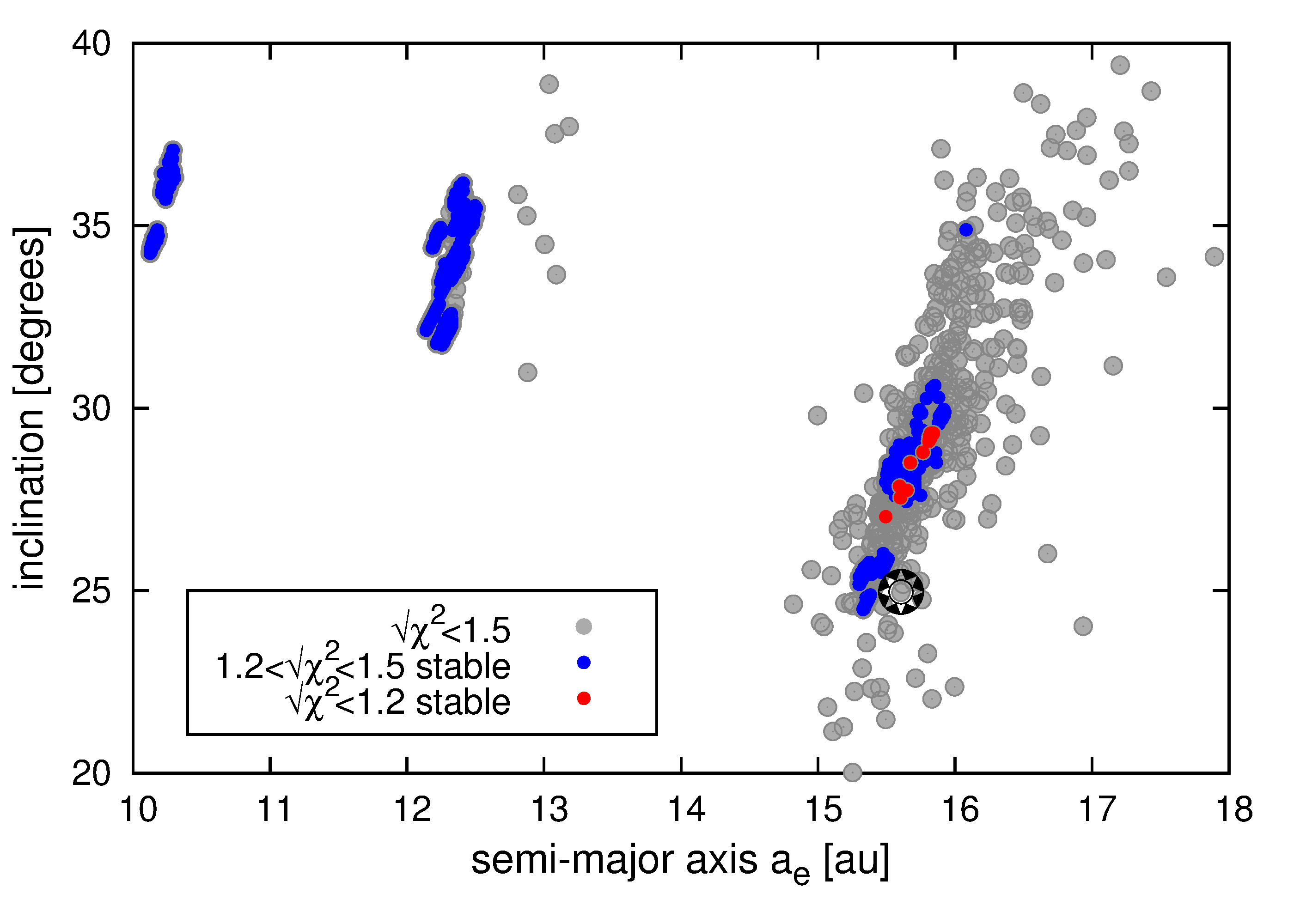}
\includegraphics[width=0.49\textwidth]{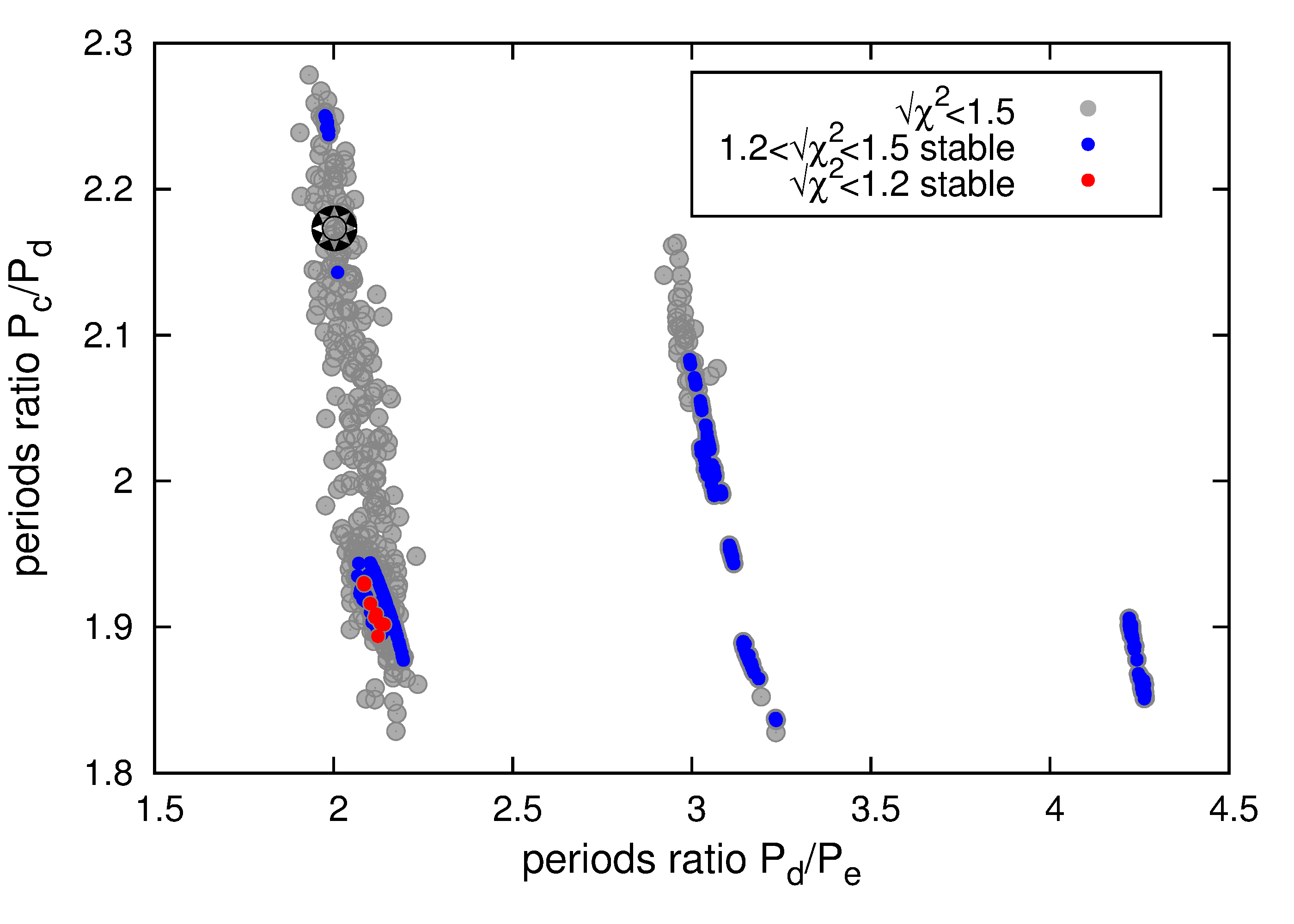}
}
\hbox{
\includegraphics[width=0.49\textwidth]{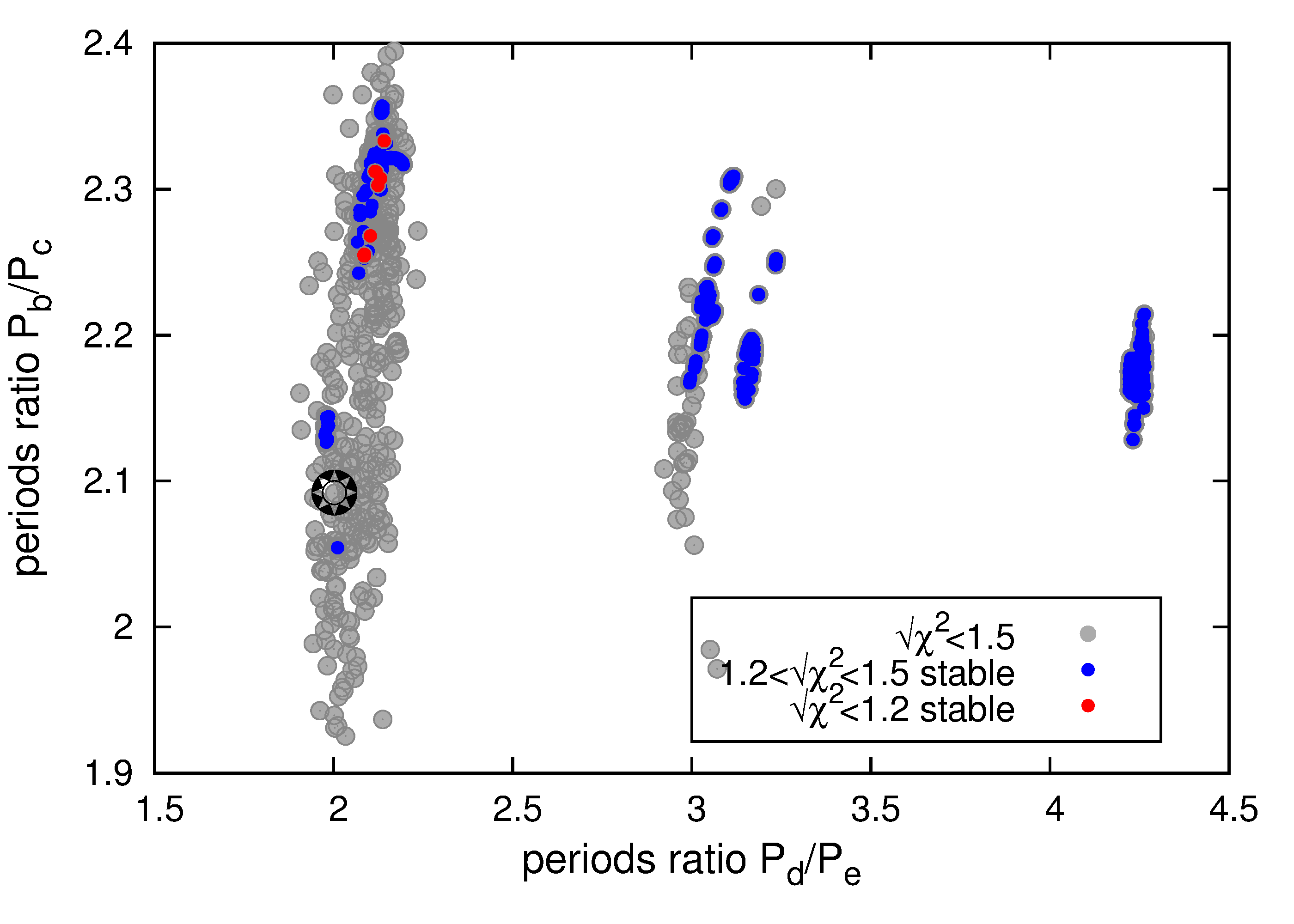}
\includegraphics[width=0.49\textwidth]{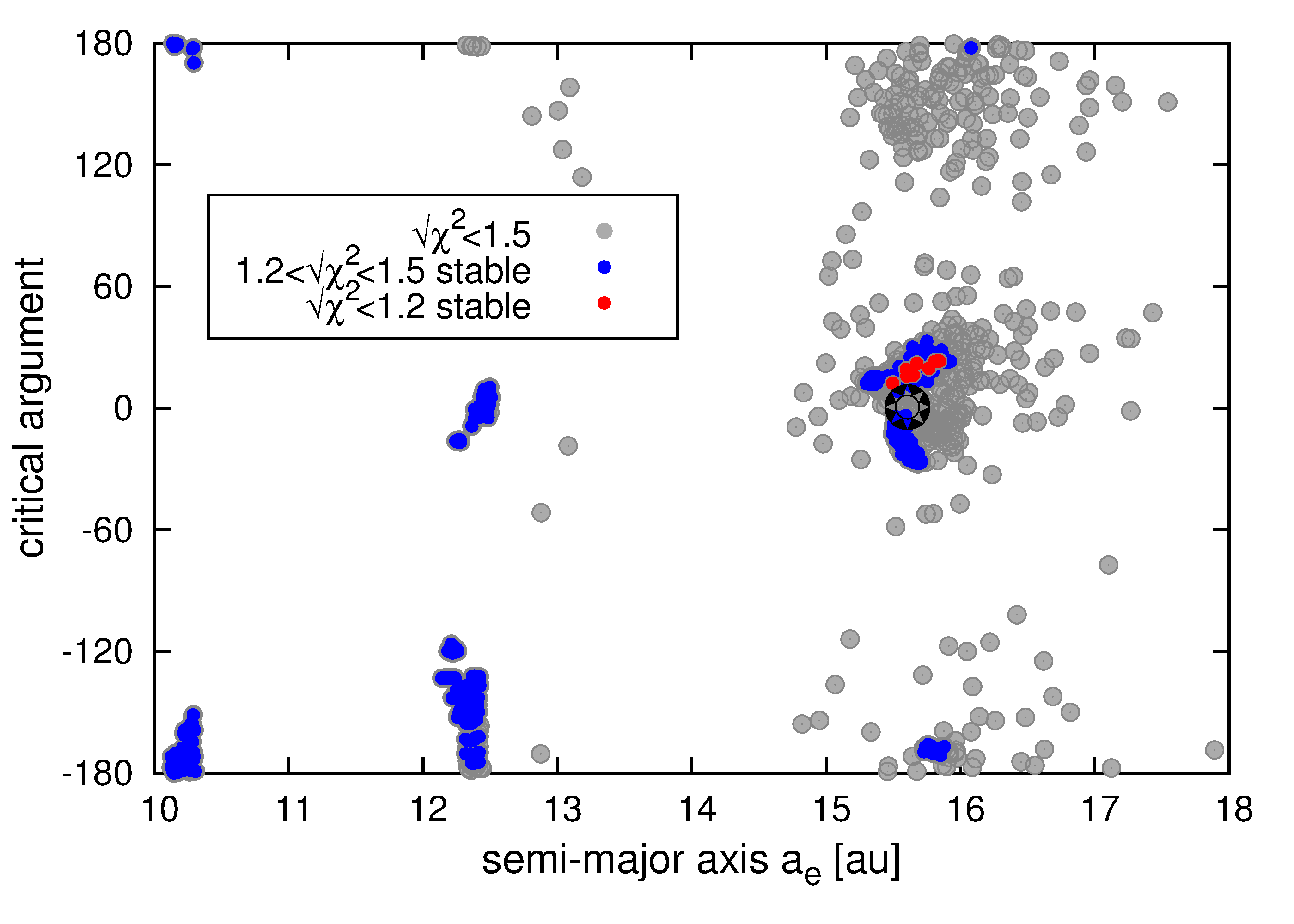}
}
}
}
\caption{
Parameters of the best-fitting model IVb to dataset D5 projected onto $(\aee, I)-$, $(P_{\idm{d}}/P_{\idm{e}}, P_{\idm{c}}/P_{\idm{d}})-$, $(P_{\idm{d}}/P_{\idm{e}}, P_{\idm{b}}/P_{\idm{c}})-$, and $(\aee, \mbox{critical angle})-$planes. 
The star symbol marks the nominal, best-fitting solution. Grey filled circles are for all solutions within $6\sigma$ confidence interval, $\cchi < 1.5$. Blue and red filled circles are for {\em rigorously stable}  models in the range of ($3\sigma,6\sigma)$, equivalent to $1.2<\cchi <1.5$, and within $3\sigma$ confidence level ($\cchi\leq 1.2$), respectively. Stable models are defined by $|\left<Y\right>-2|<0.05$ for the integration time-span of 160~Myr.
The critical arguments for 1e:2d:4c:8b~MMR, 1e:3d:6c:12b~MMR, and 1e:4d:8c:16b~MMR are 
$\theta_{\idm{1:4:8:16}} =\lambda_{\idm{e}} - 2\,\lambda_{\idm{d}} - \lambda_{\idm{c}} + 2\,\lambda_{\idm{b}}$, 
$\theta_{\idm{1:3:6:12}} = \lambda_{\idm{e}} - 4\,\lambda_{\idm{d}} + \lambda_{\idm{c}} + 2\,\lambda_{\idm{b}}$ and 
$\theta_{\idm{1:4:8:16}} = \lambda_{\idm{e}} - 3\,\lambda_{\idm{d}} - 6\,\lambda_{\idm{c}} + 8\,\lambda_{\idm{b}}$, respectively.
}
\label{fig:fig12}
\end{figure*}

These results suggest that the optimization of the four-planet model with yet undetected  innermost planet might at least help to narrow the search areas for such a putative object and to interpret the speckle images.  Figure~\ref{fig:fig11} shows the actual observations of planet~e in dataset D1 (let us recall: these data {\em were not used} in the IVb~experiment) with the best-fitting orbits of model IVb over-plotted. {For a reference, the left-hand panel of Fig.~\ref{fig:fig11} illustrates the best-fitting orbit IVa (dashed curve) and arcs of  rigorously stable configurations (solid curves) in model~IVb, respectively. Black curve is for the best-fitting model IVb, grey curves are for configurations IVb within $6\sigma$, while green curves illustrate solutions within $3\sigma$ confidence levels}. Remarkably, the best-fitting stable solutions {IVb are consistent with the actual observations as well as with the full model~IVa}. 
The right-hand panel of Fig.~\ref{fig:fig11} shows a close-up of the previous plot. The best-fitting orbits IVb (black curve) are plotted together with the best-fitting nominal four-planet model IVa (grey dashed curve). Positions of the planets at the observational epochs are marked with small tics.  Clearly, these two orbits almost overlap.

Finally, we carried out a number of simulations to examine whether the four-planet model with unseen planet~e might provide  better fits than the three-planet model. Both these models appear to match observations in dataset D5 equally well.  

\begin{table*}
\caption{
Osculating parameters of the best-fitting four-planet solutions.  The stellar mass $m_0 = 1.56\,\msun$. Osculating epoch is~$1998.83$.
}
\label{tab:table2}
\begin{center}
\begin{tabular}{l c c c c c c c c c}
\hline
model & planet & $\cchi$ & $m\,[\mJ]$ & $a\,[\au]$ & e & $I$[deg] & $\Omega$[deg] & $\omega\,$[deg] & $\Mmean_0\,$[deg]\\
\hline
& e & & $8.895706$ & $15.443557$ & $0.124958$ & & & $112.198950$ & $325.667983$\\
IVa & d & $1.147$ & $8.825311$ & $25.428138$ & $0.123029$ & $25.337113$ & $64.180486$ & $26.598297$ & $57.901471$\\
& c & & $9.231718$ & $39.366093$ & $0.053442$ & & & $87.154893$ & $147.870426$\\
& b & & $6.748302$ & $69.063963$ & $0.020022$ & & & $30.350808$ & $321.261401$\\
\hline
& e & & $7.957365$ & $15.675018$ & $0.159392$ & & & $110.340454$ & $325.747637$\\
IVb & d & $1.172$ & $9.961193$ & $25.896792$ & $0.151664$ & $28.504826$ & $65.493989$ & $24.572894$ & $55.420303$\\
& c & & $10.417128$ & $39.638907$ & $0.071622$ & & & $84.322520$ & $149.892256$\\
& b & & $7.920258$ & $69.121225$ & $0.020655$ & & & $27.328258$ & $322.721614$\\
\hline
& e & & $7.511610$ & $15.600838$ & $0.130436$ & & & $110.787818$ & $334.739478$\\
IVc & d & $0.474$ & $8.871059$ & $25.451257$ & $0.108087$ & $27.616454$ & $56.264572$ & $29.153001$ & $63.162502$\\
& c & & $8.479021$ & $39.679862$ & $0.065732$ & & & $103.182135$ & $137.630090$\\
& b & & $8.722503$ & $68.747691$ & $0.018124$ & & & $42.802760$ & $317.361404$\\
\hline
\end{tabular}
\end{center}
\end{table*}

%
\subsection{Models IIIa, IIIb and IVc: single-image characterisation}
%

The orbital periods of directly imaged planets usually count in tens or hundreds of years. The common optimization techniques require long period of observations to establish the true orbital architecture. The literature devoted to the HR~8799 system is a good evidence of this apparently unavoidable problem. However, if a multiple, resonant configuration is observed, the migration algorithm may be helpful to constrain its orbits even by {\em one single-epoch} observation. This means basically single-image orbital characterisation of the system, although in the real world much longer observational time span is required, for instance, to confirm common proper motion and the same parallax
\corr{at least two, well separated epochs are required.}

\subsubsection{Model IIIa: the data in (Marois et. al 2008) revisited}

Dataset D2 in \citep{Marois2008} serves as a good example of time-limited observations, which were examined by many groups.  We performed the migration constrained optimization of these data with three-planet model (IIIa). This example might be also thought as an excellent test-bed of the
\moa{}. 

\begin{figure}
\centerline{
\hbox{
{\includegraphics[width=0.49\textwidth]{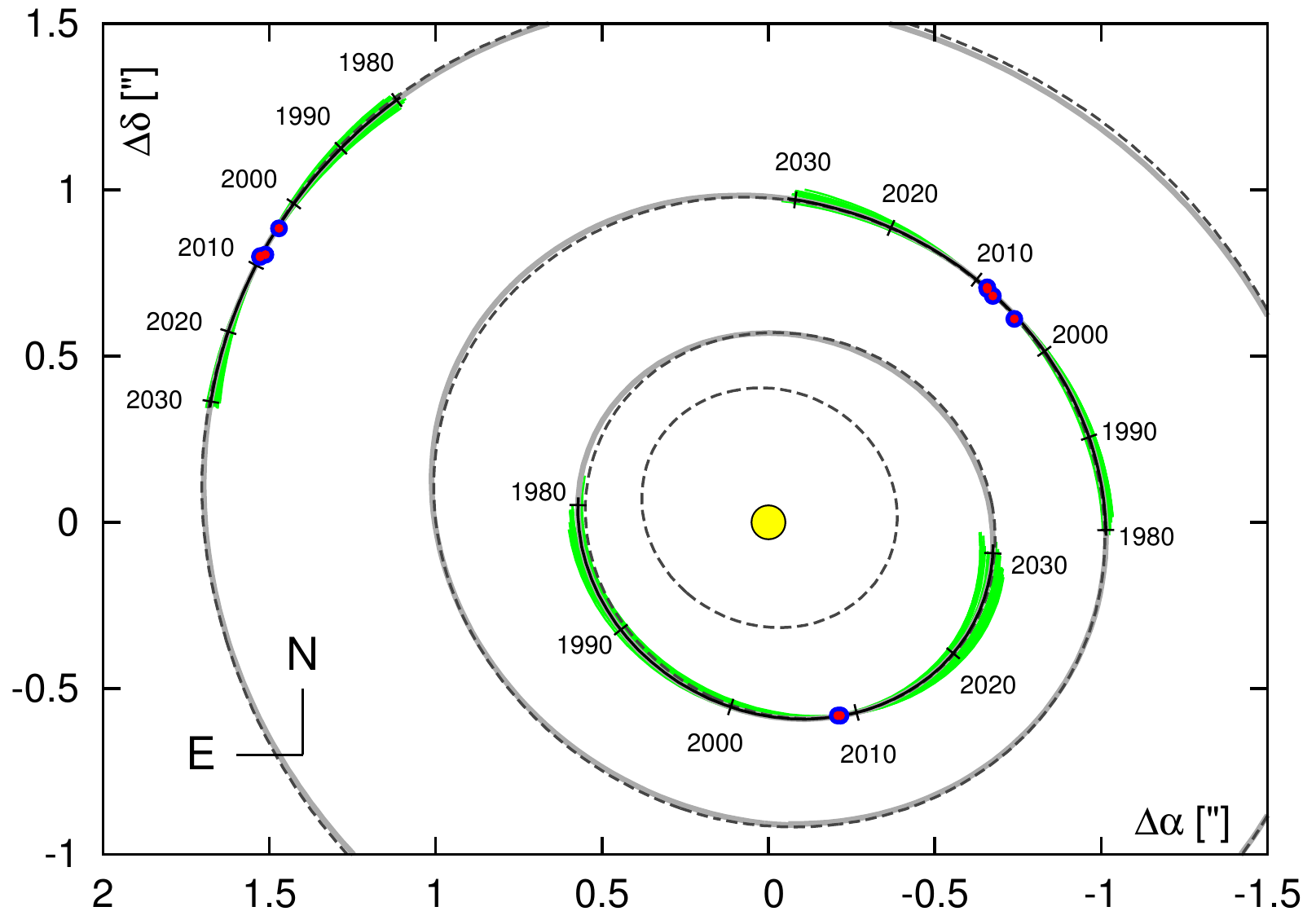}}
}
}
\caption{
Relative astrometric positions of the HR~8799 planets (red filled circles) and orbital arcs for the best-fitting model~IIIa combined with dataset~D2 in the discovery paper \protect\citep{Marois2008}. Best-fitting orbits of this model are plotted with solid grey curves. For a reference, black, dashed curves are for the best-fitting nominal four-planet model IVa related to the whole available data in the literature (dataset~D1).
}
\label{fig:fig13}
\end{figure}

Osculating elements of best-fitting model~IIIa for the epoch $t_0=1998.71$ are given in Tab.~\ref{tab:table3}. Figure~\ref{fig:fig13} shows the orbital geometry of model~IIIa (solid grey curves) compared to model~IVa (dashed curves), and overplotted on orbital arcs of stable solutions within $3\sigma$-confidence level (green curves). The three- and four-planet models  can be hardly distinguished, suggesting that the Laplace resonance of three outer planets is particularly robust for a perturbation, as large as the inner planet~e may introduce. 

\begin{figure*}
\centerline{
\vbox{
\hbox{
\includegraphics[width=0.49\textwidth]{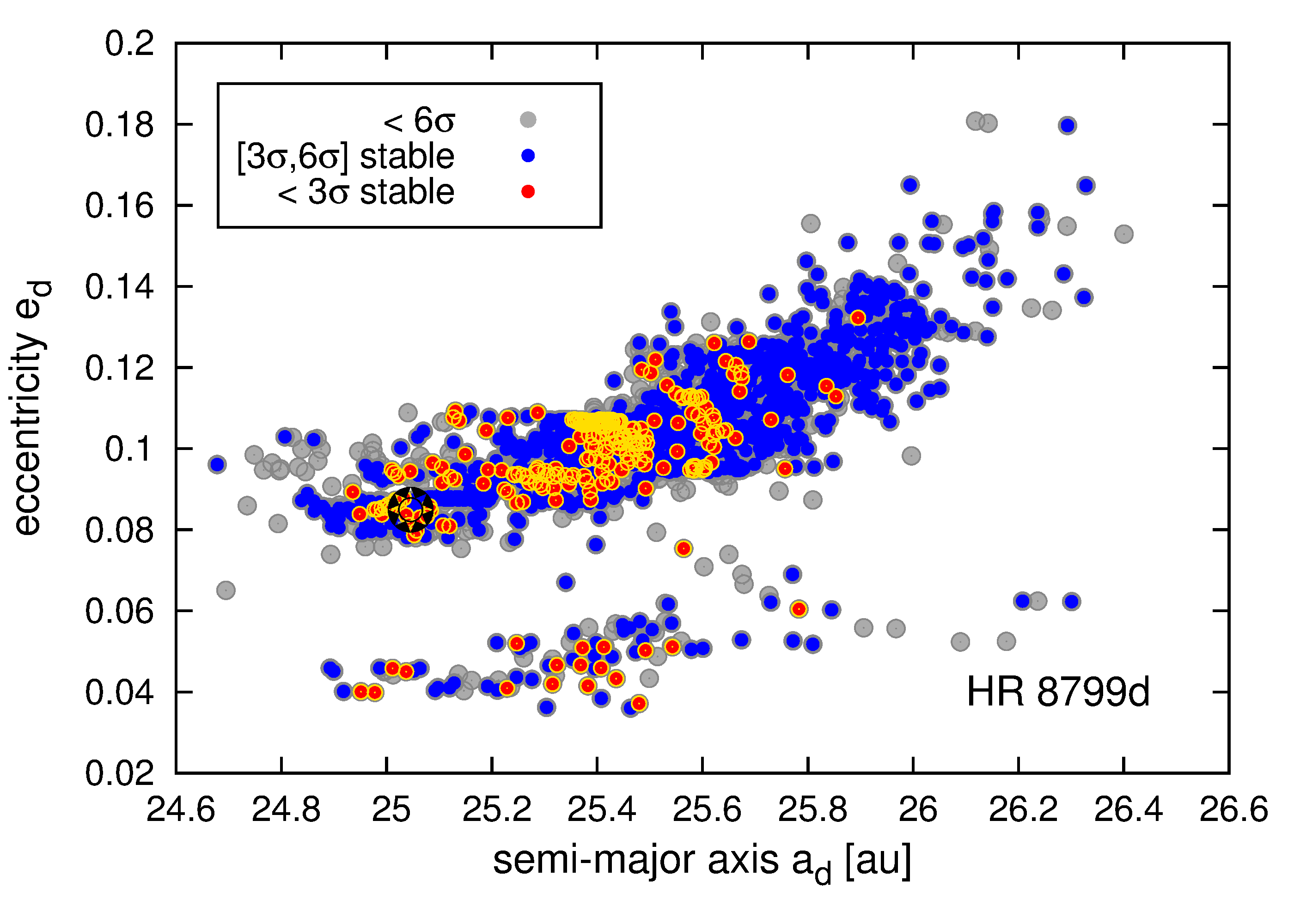}
\includegraphics[width=0.49\textwidth]{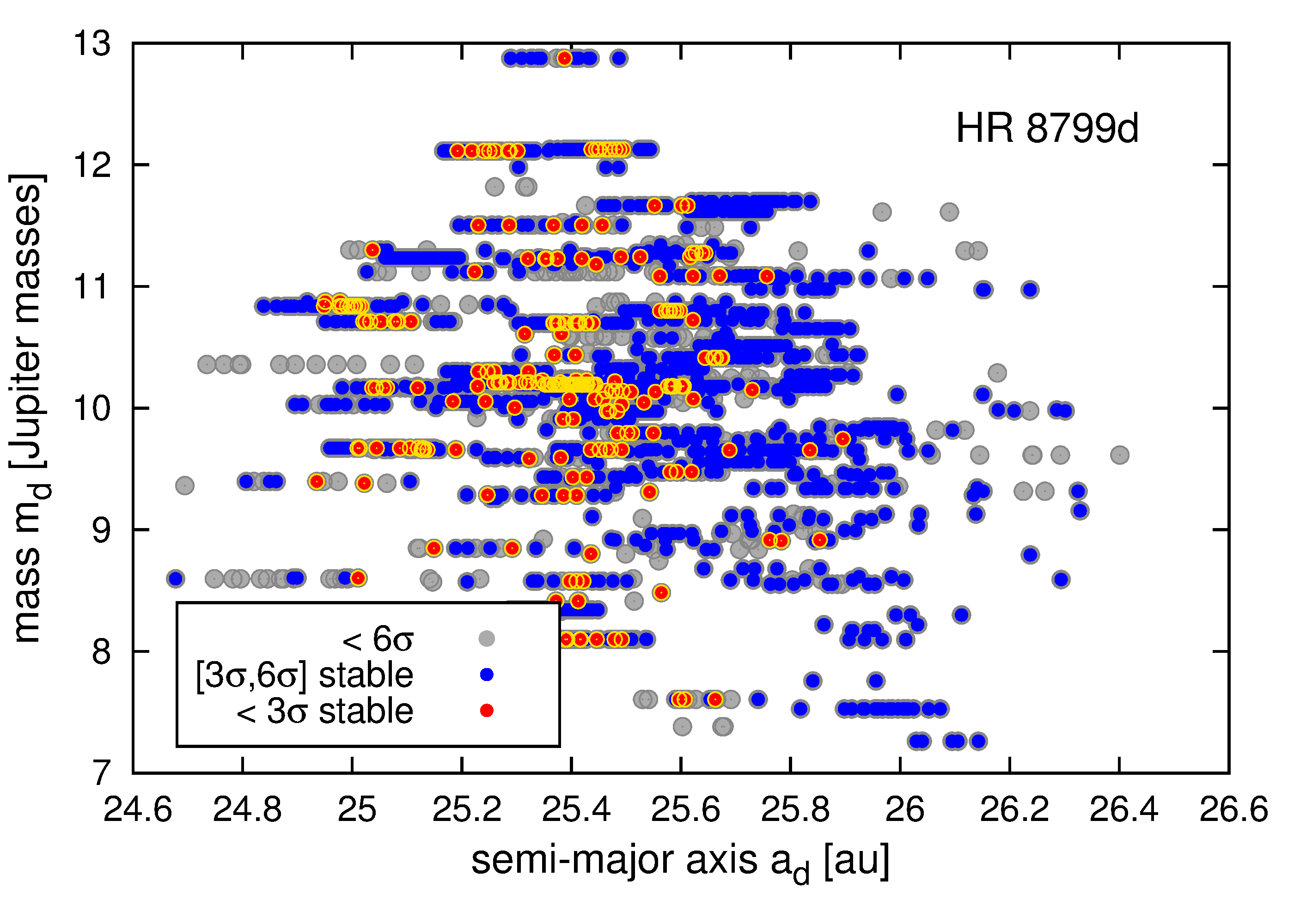}
}
\hbox{
\includegraphics[width=0.49\textwidth]{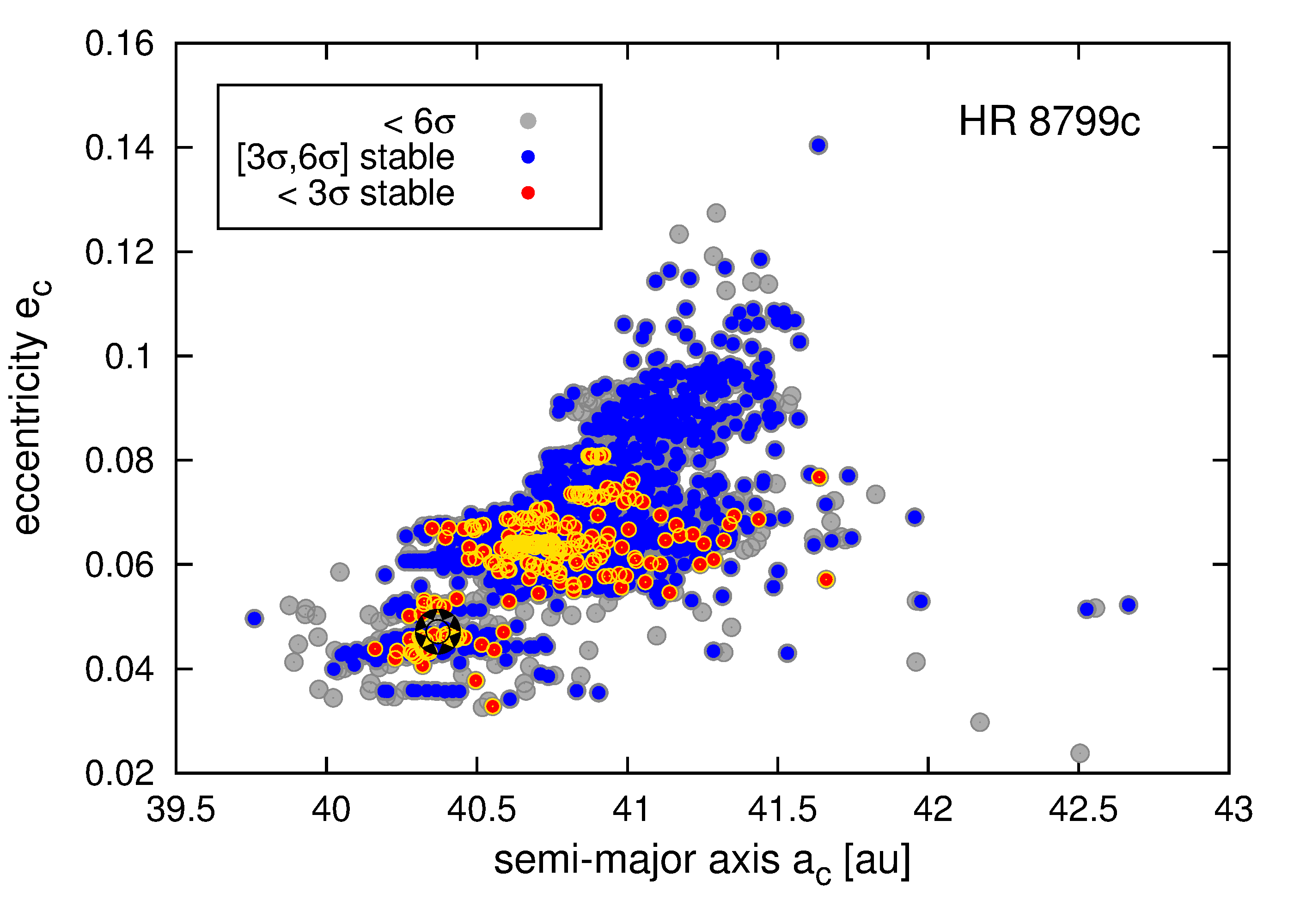}
\includegraphics[width=0.49\textwidth]{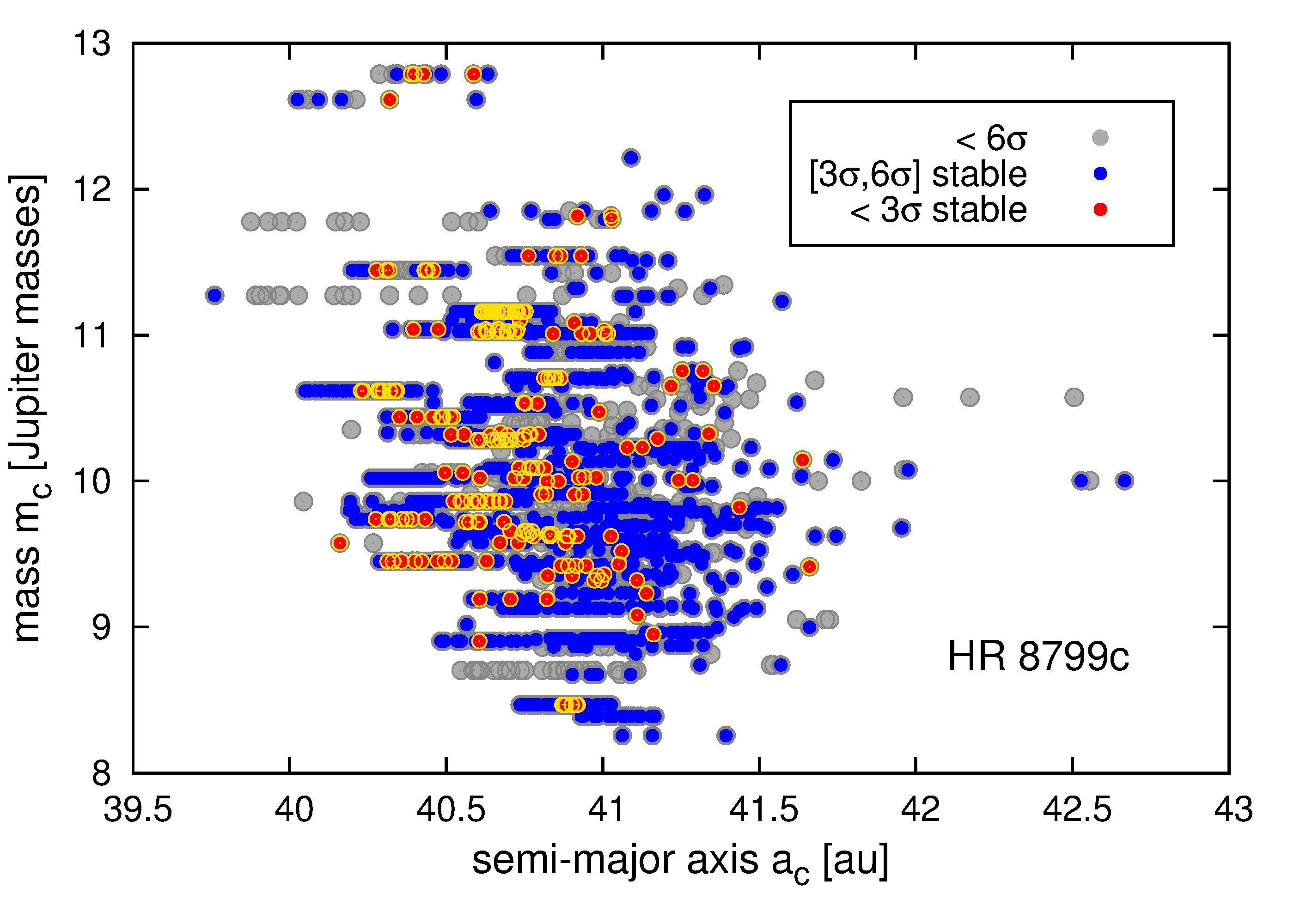}
}
\hbox{
\includegraphics[width=0.49\textwidth]{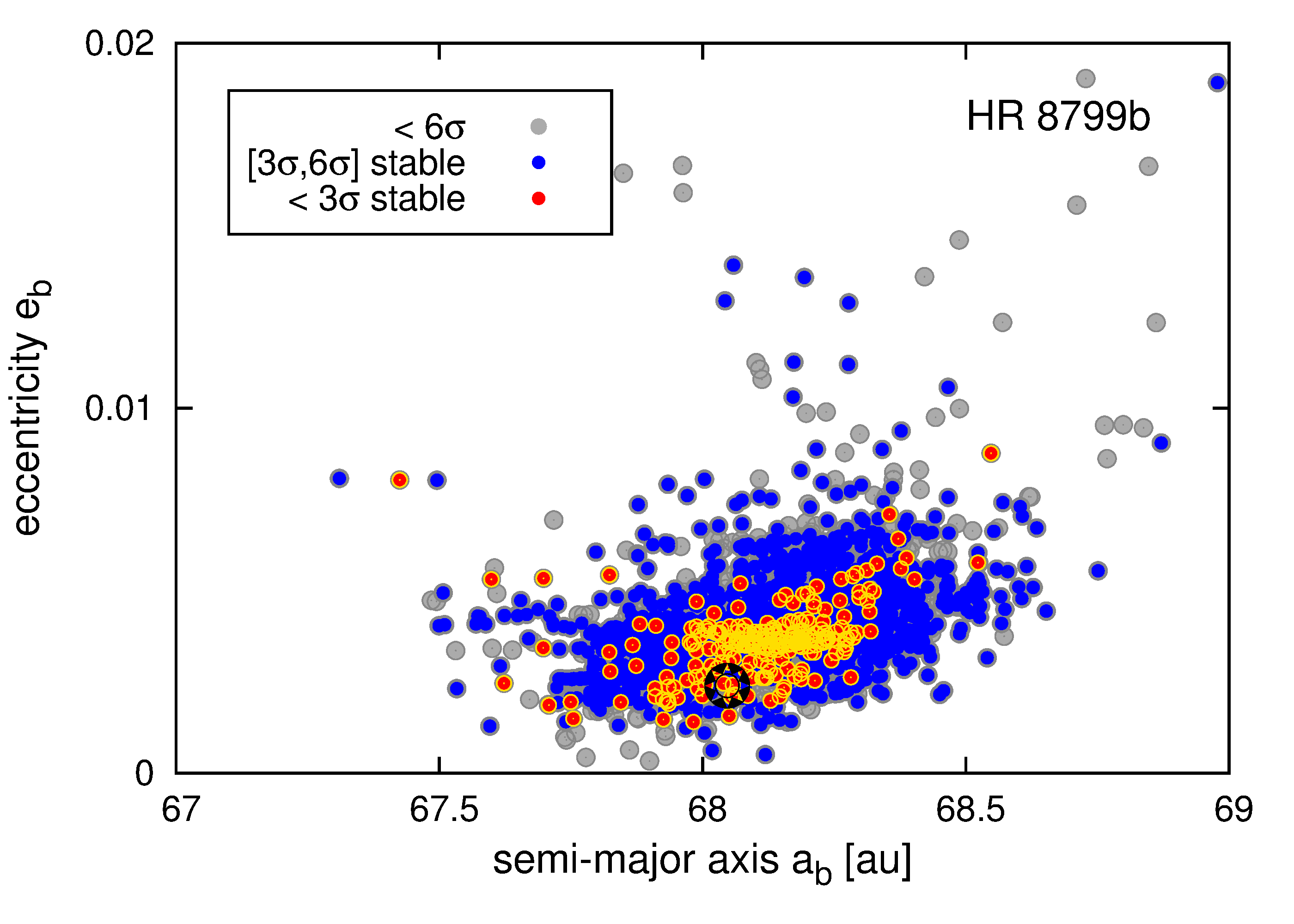}
\includegraphics[width=0.49\textwidth]{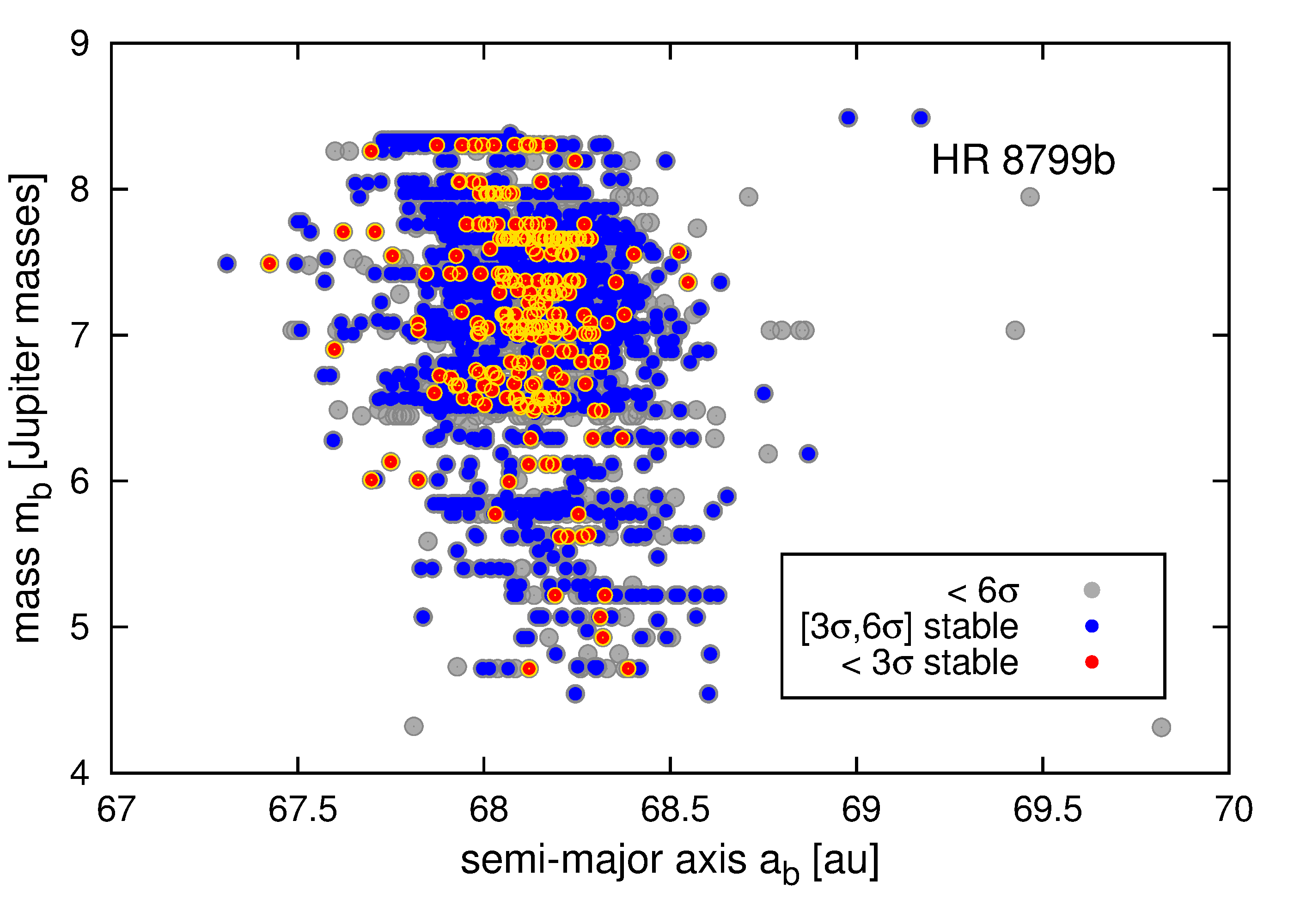}
}
}
}
\caption{
Best-fitting three-planet models~IIIa to the data published in the discovery paper \protect\citep{Marois2008}, {dataset~D2}, projected onto planes of the semi-major axis -- eccentricity ({\em left column}) and the semi-major axis -- masses ({\em right column}), respectively, for subsequent planets.  The star symbol marks the nominal, best-fitting solution.  {Grey filled circles are for all solutions within $\cchi < 1.5$ (roughly $6\sigma$ confidence level).  Blue and red circles are for {\em rigorously stable} models in the range of $(3\sigma,6\sigma)$  ($<1.2 \cchi < 1.5$), and for the $3\sigma$ ($\cchi\leq 1.2$) confidence levels, respectively}.  Their $|\left<Y\right>-2|<0.05$ for  the integration time-span of 160~Myr covering assumed lifetime of HR~8799.
}
\label{fig:fig14}
\end{figure*}

The statistics of best-fitting models with $\cchi$ within the $6\sigma$ confidence interval is shown in Fig.~\ref{fig:fig14}.  This result confirms and complements the earlier literature data. Similarly to model IVa, we found relatively wide mass ranges of {\em rigorously stable} solutions (the right-hand column in Fig.~\ref{fig:fig14}) which are fully consistent with astrophysical estimates, independent on the geometric model of orbits. Our stable models cover smoothly both the 10-10-7~$\mJ$ range as well as the 7-7-5~$\mJ$. The statistics of stable models suggests the upper limit of masses $\sim 13$\,$\mJ$. In spite of many trials \corr{with masses extended to $\sim 20~\mJ$}, we could not find any stable configurations with masses above this limit which are {\em consistent with observations} at least at the $6\sigma$ confidence level. Such a limitation \corr{indicates} that the HR~8799 companions are really planets with masses below the deuterium burning limit. 

\begin{figure*}
\centerline{
\vbox{
\hbox{
{\includegraphics[width=0.49\textwidth]{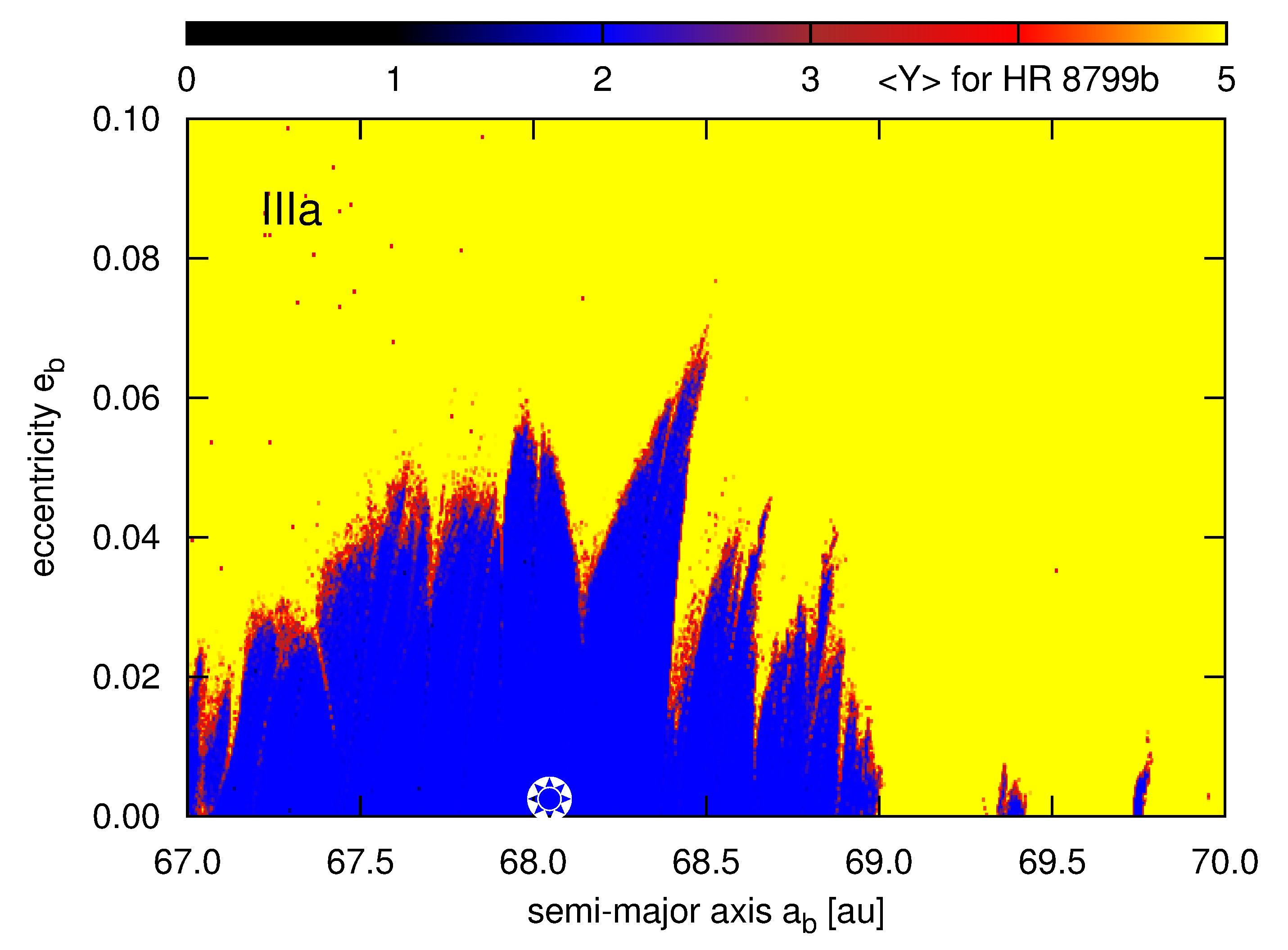}}
{\includegraphics[width=0.49\textwidth]{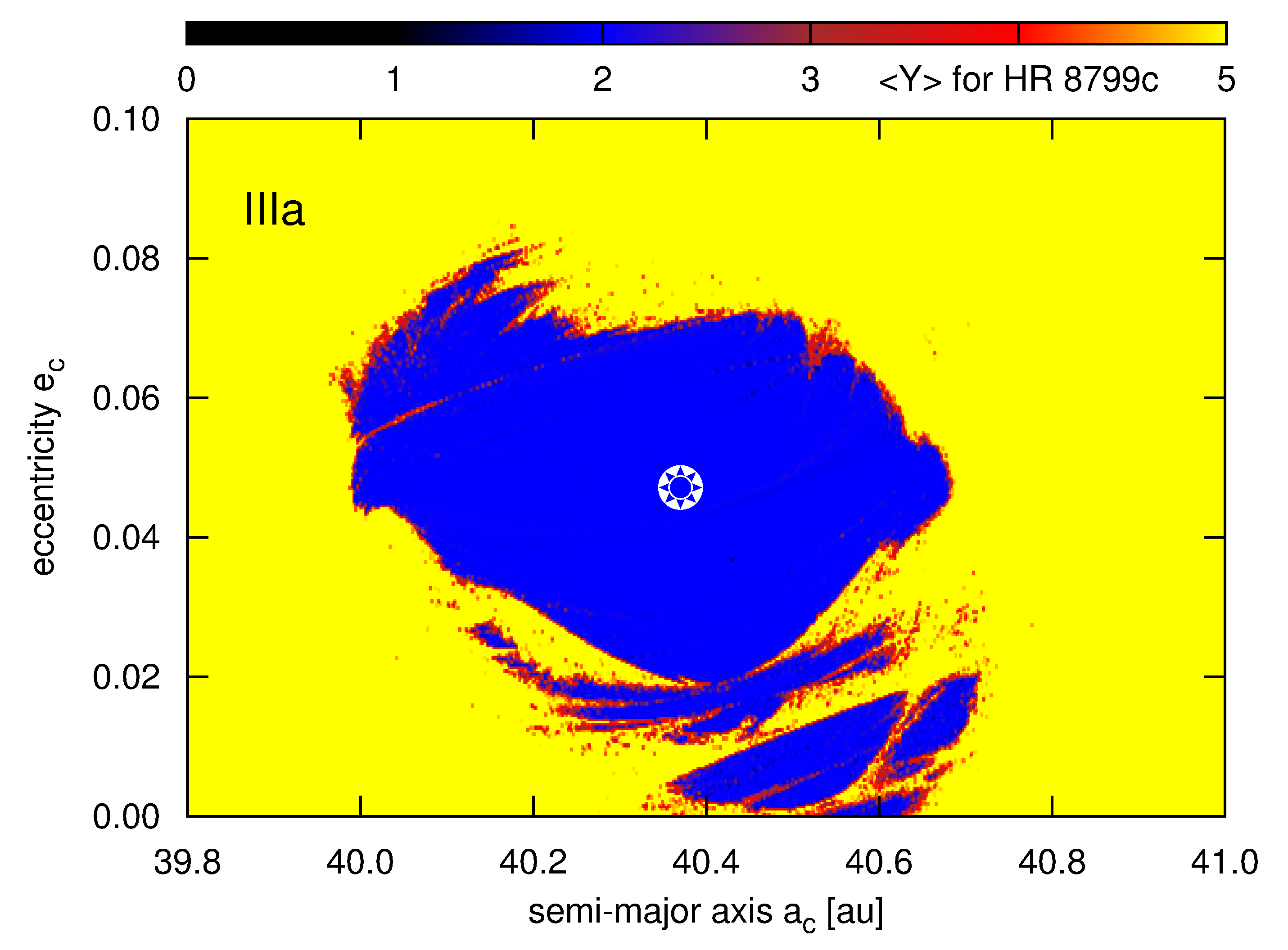}}
}
\hbox{
{\includegraphics[width=0.49\textwidth]{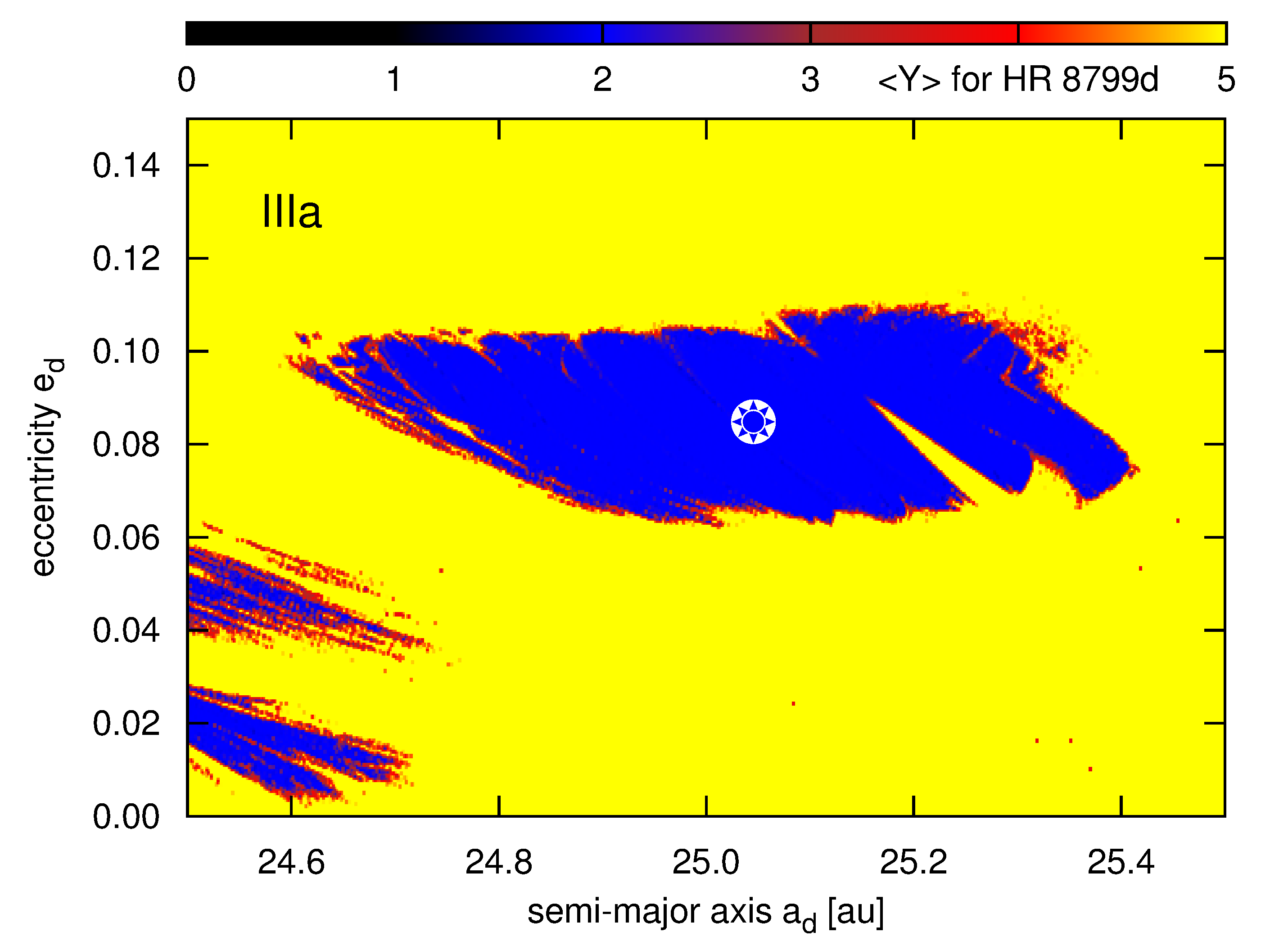}}
{\includegraphics[width=0.49\textwidth]{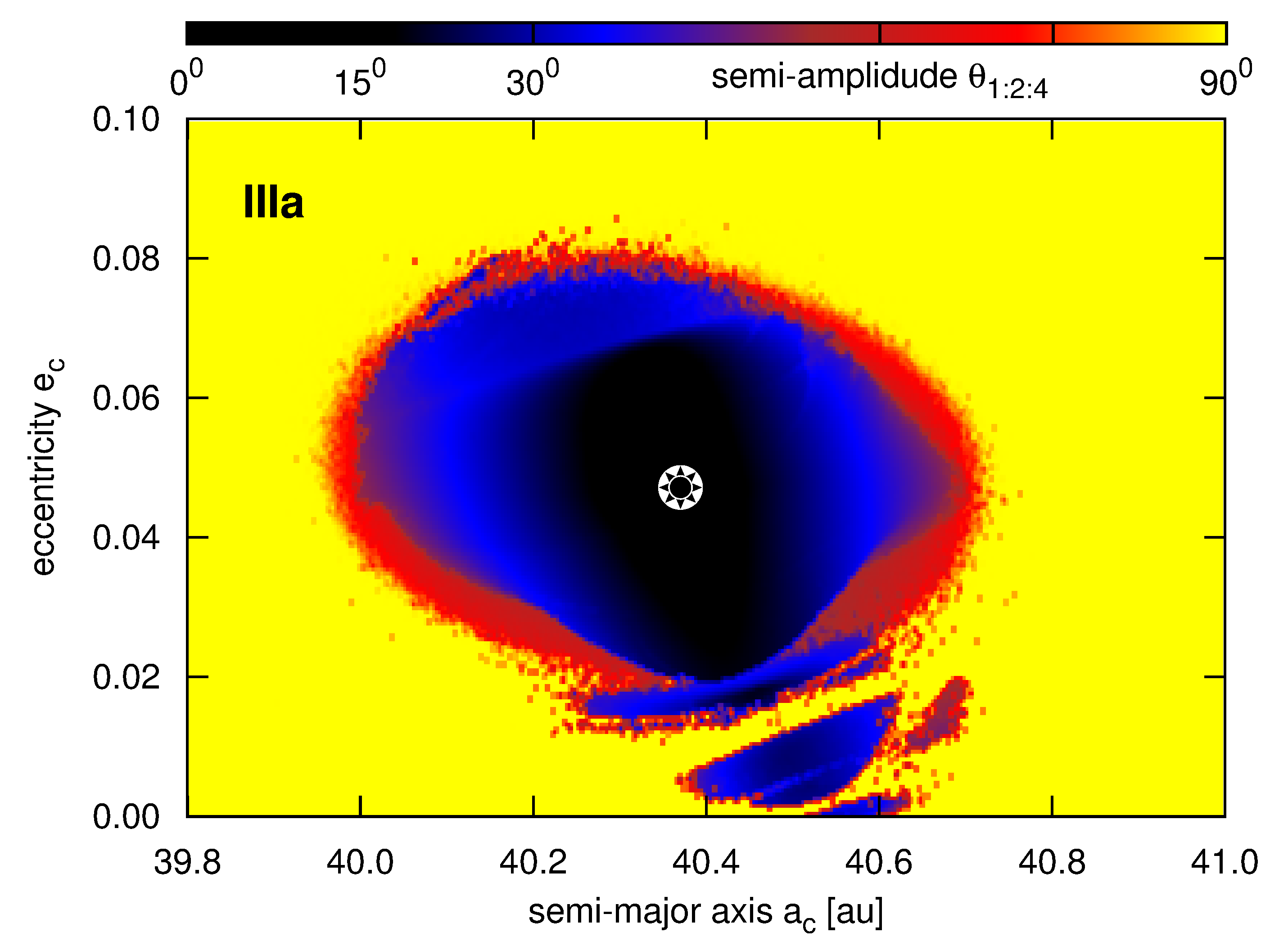}}
}
}
}
\caption{
Dynamical maps in terms of the maximal Lyapunov exponent, expressed by the MEGNO fast indicator $\left<Y\right>$, in the semi-major--eccentricity planes for subsequent planets in the HR~8799 system with three planets. This is the best-fitting model IIIa to astrometric data in the discovery paper \protect\citep{Marois2008}. The bet-fitting configuration is deeply locked in the classic Laplace resonance. The semi-amplitude of this MMR around $90^{\circ}$ is shown in the bottom-right panel.  
}
\label{fig:fig15}
\end{figure*}

We also computed dynamical maps for the best-fitting model (Fig.~\ref{fig:fig15}), which corresponds to three-planet configuration deeply involved in the classic Laplace 1d:2c:4b~MMR.   The bottom-right panel illustrates the critical argument $\theta_{\idm{1d:2c:4b}}=\lambda_{\idm{d}} - 3\lambda_{\idm{c}} + 2\lambda_{\idm{b}}$ that librates around $90^{\circ}$ .
The best-fitting configuration is found in the central part  which perfectly overlaps with the
minimum of the critical angle $\theta{\idm{1d:2c:4b}}$ librating with a semi-amplitude $\sim 20^{\circ}$. Remarkably, the system is found long term-stable for at least  160~Myr in quite an extended area which is much wider than the MEGNO zone of quasi-periodic, stable motions ({not shown here}).

%
\subsubsection{Single-image characterization of a three-planet model}
\begin{figure}
\centerline{
\vbox{
\hbox{\includegraphics[width=0.48\textwidth]{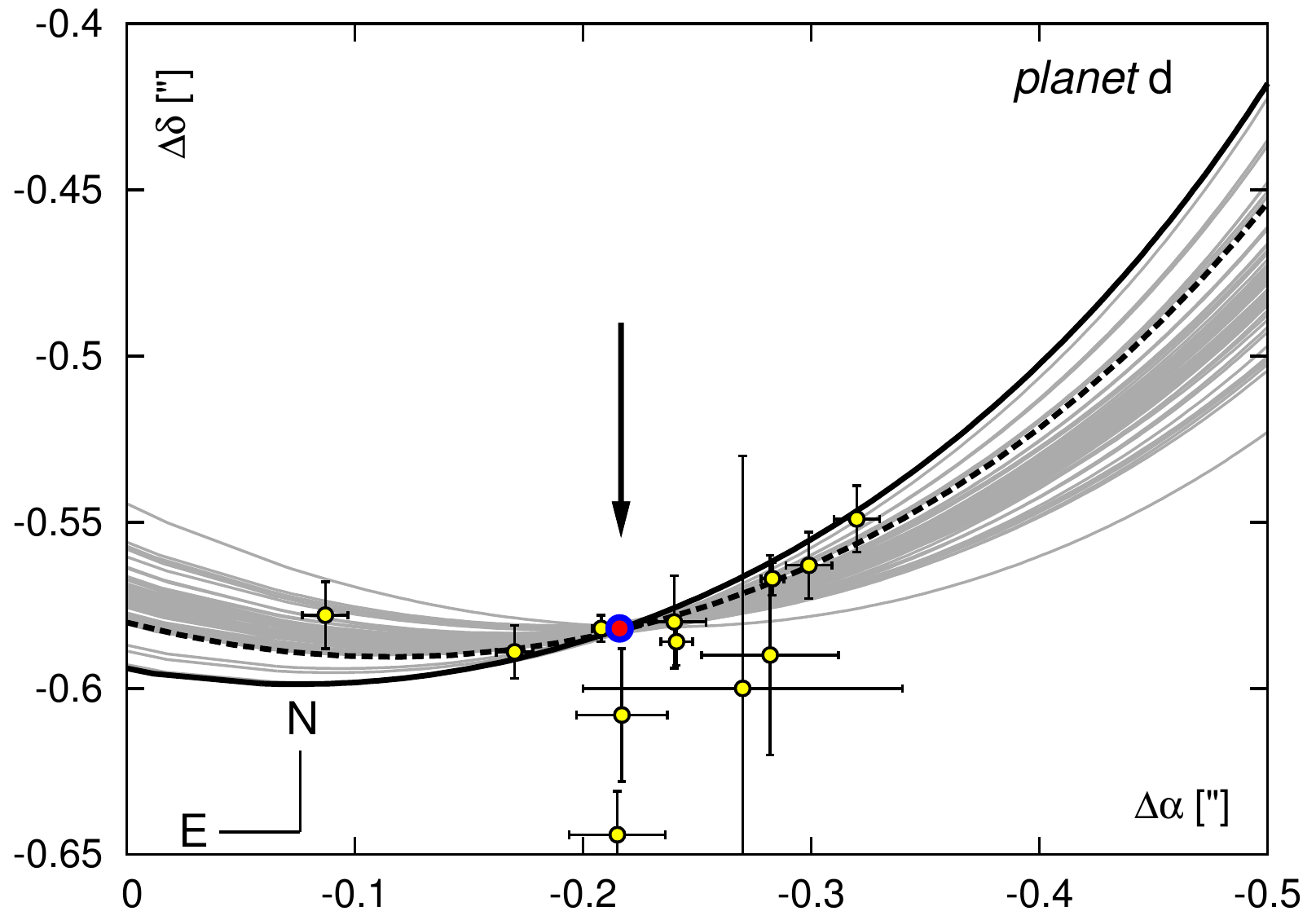}}
\hbox{\includegraphics[width=0.48\textwidth]{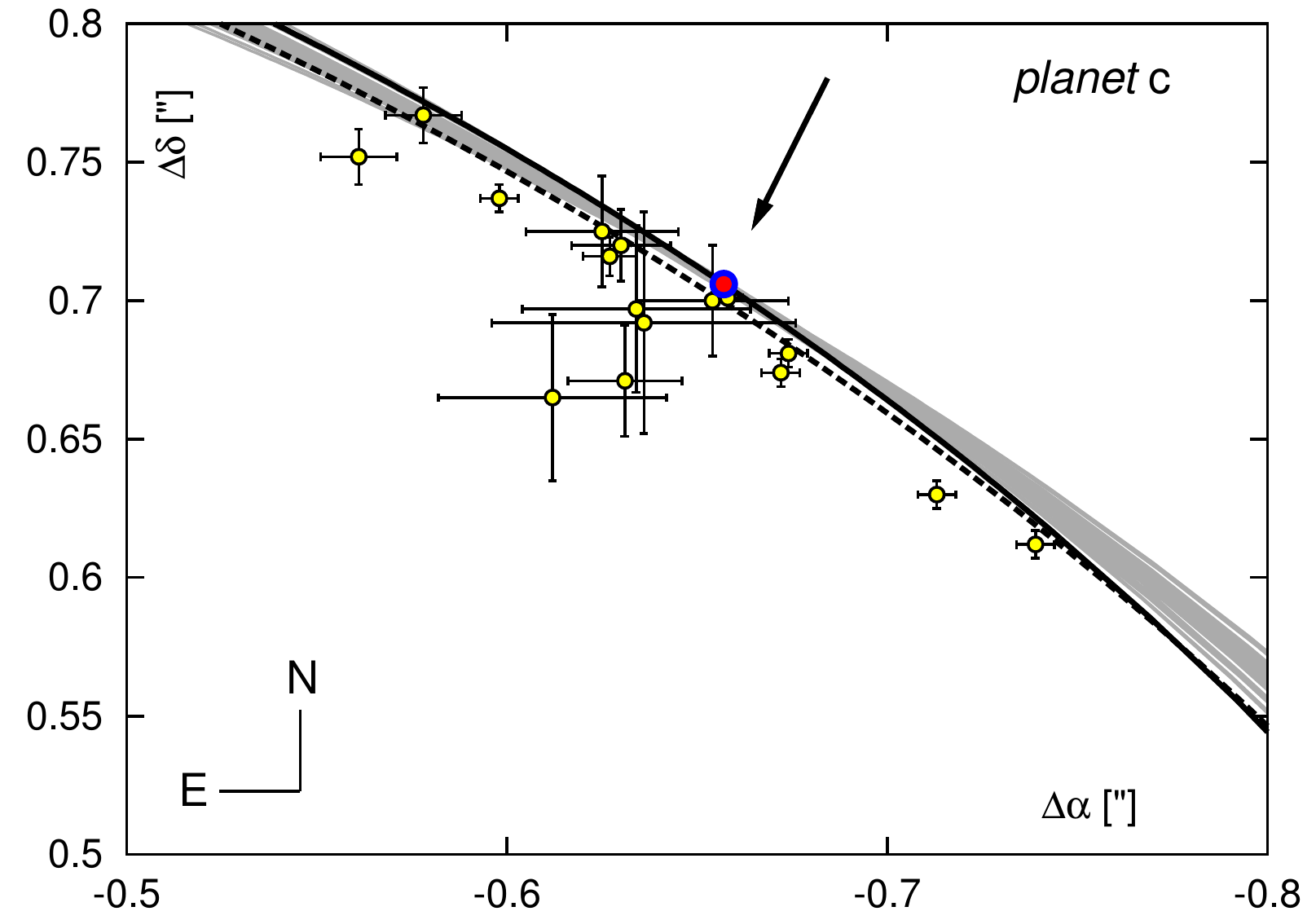}}
\hbox{\includegraphics[width=0.48\textwidth]{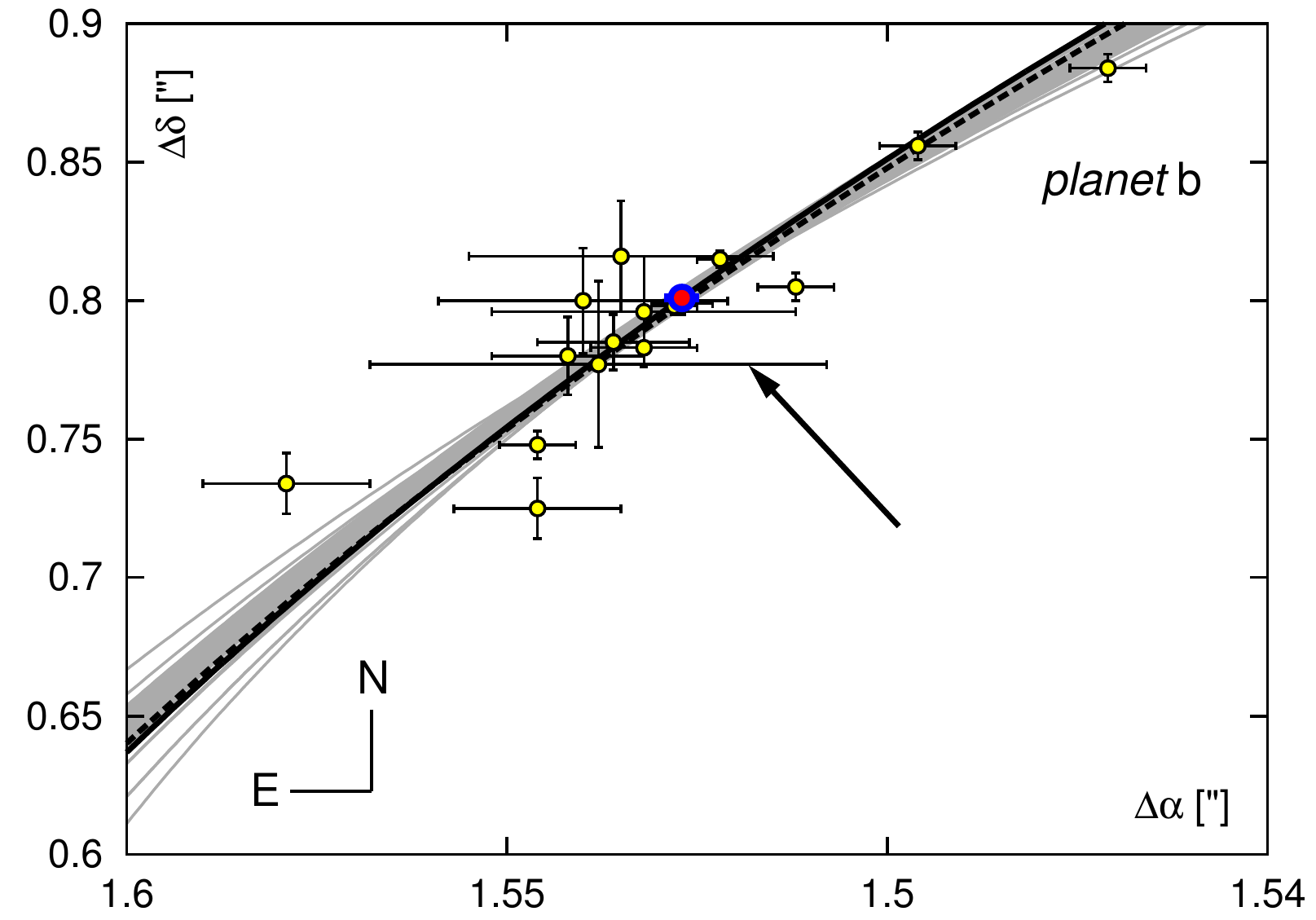}}
}
}
\caption{
Relative astrometric positions $\Delta\alpha$~[arcsec] vs $\Delta\delta$~[arcsec] of the HR~8799 planets and orbital arcs for the best-fitting four-planet model IVa (black dashed curves), and for the best-fitting, single-epoch model IIIb  (black solid curves). Grey curves are for model IIIb and solutions with $\cchi<1$.
}
\label{fig:fig16}
\end{figure}

%
Going further, to mimic a single-image detection of three outermost planets, we chose only one, single epoch $2008.61$ and three measurements for outer planets from the discovery paper \citep{Marois2008}. Epoch $2008.61$ refers to the most precise measurement. We optimized three-planet model (IIIb) with this dataset~D3. The results are presented on Fig.~\ref{fig:fig16}. Red filled circles pointed out with an arrow are for the D3 observations, while yelow/black circles mark all remaining observations {(dataset D5)}, as the reference data. The best-fitting orbits of model~IIIb are plotted with black solid curves, while the dashed curves are for the best-fitting solution~IVa (four-planets, full dataset~D1). Orbits plotted in grey have $\cchi < 1.0$. Surprisingly, even the single-epoch measurements are sufficient to constrain the orbits. Clearly, both orbital geometries closely overlap.

\begin{table*}
\caption{
Osculating elements and masses of the best-fitting three-planet models~IIIa and IIIb.  The stellar mass $m_0 = 1.56\,\msun$. The osculating is epoch~$1998.83$.
}
\label{tab:table3}
\begin{center}
\begin{tabular}{l c c c c c c c c c}
\hline
model & planet & $\cchi$ & $m\,[\mJ]$ & $a\,[\au]$ & e & $I$[deg] & $\Omega$[deg] & $\omega\,$[deg] & $\Mmean_0\,$[deg]\\
\hline
& d & & $10.135136$ & $25.045527$ & $0.084720$ & & & $12.364179$ & $96.623071$\\
IIIa & c & $1.060$ & $10.018128$ & $40.370405$ & $0.047036$ & $26.566822$ & $64.680038$ & $103.509878$ & $140.603388$\\
& b & & $6.953918$ & $68.047272$ & $0.002379$ & & & $1.142328$ & $352.986933$\\
\hline
& d & & $7.923625$ & $24.234342$ & $0.072091$ & & & $203.586910$ & $95.496437$\\
IIIb & c & $0.053$ & $10.656124$ & $39.482602$ & $0.019226$ & $18.867507$ & $236.887633$ & $296.672146$ & $135.283261$\\
& b & & $4.587625$ & $68.108515$ & $0.002603$ & & & $146.618350$ & $35.666508$\\
\hline
\end{tabular}
\end{center}
\end{table*}

The stable Laplace~MMR island of best-fitting model IIIb is illustrated in dynamical MEGNO map for planet HR~8799c shown in the left-hand panel of Fig.~{\ref{fig:fig17}. A quasi-periodic character of this solution assures us that this configuration survives for more than 160~Myr without any sign of instability. In fact, this particular solution IIIb has the MEGNO signature $\sim 2$ after 160~Myr that guarantees its stability for times 1-2 orders of magnitude longer, for 1~Gyr or longer.

\begin{figure*}
\centerline{
\hbox{
{\includegraphics[width=0.49\textwidth]{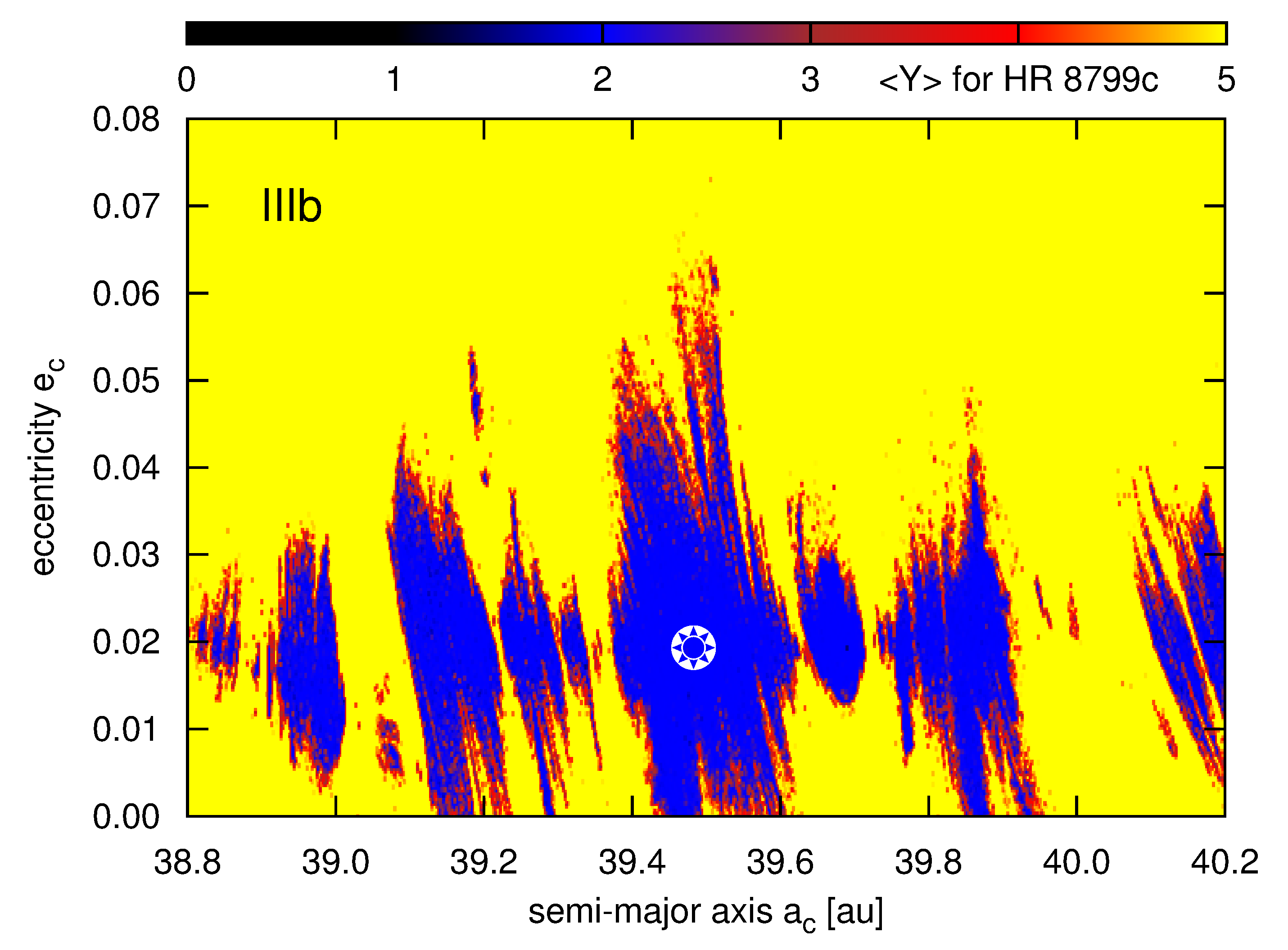}}
{\includegraphics[width=0.49\textwidth]{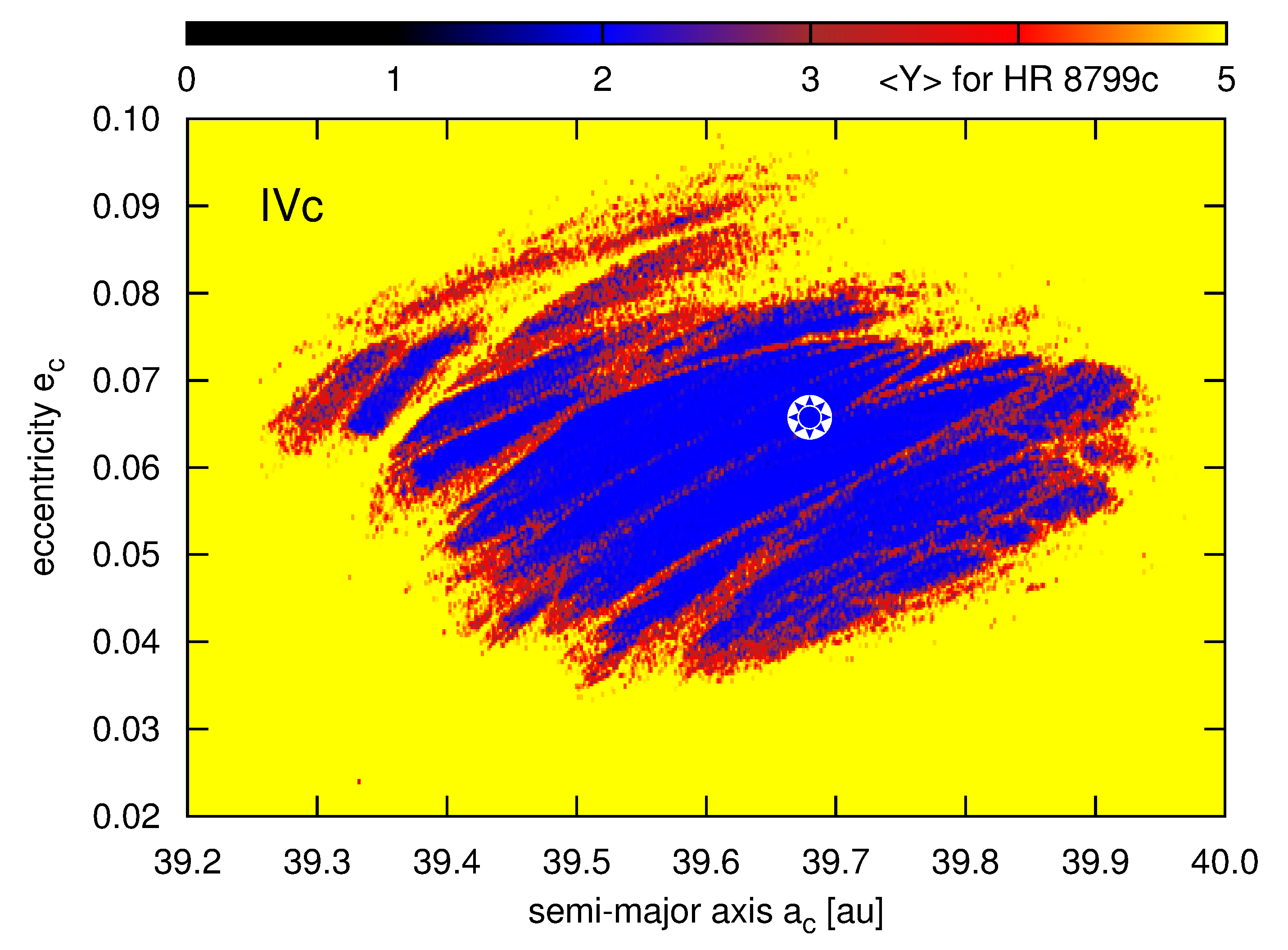}}
}
}
\caption{
Dynamical maps in terms of the MEGNO $\left<Y\right>$ in the semi-major axis --eccentricity planes for planet HR~8799~c.  {\em Left-hand} panel illustrates the best-fitting solution IIIc to only {\em three} data points in the discovery paper \protect\citep{Marois2008} mimicking a single--image detection of three outer planets. Similarly, {\em right-hand} panel is for the best-fit model IVc to {\em four} data points in \protect\citep{Currie2011} that might represent a detection of four planets at single image.  The star symbol marks the best-fitting model.  Resolution of each map is $720\times360$ initial conditions.  The integration time for each pixel is $\sim 10,000$ orbital periods of the outermost planet ($\simeq 5$~Myr).
}
\label{fig:fig17}
\end{figure*}

%
\subsubsection{Single-image characterization of a four-planet model}
%
We did a similar experiment simulating single-epoch detection of four planets (model IVc). Dataset D4 which is optimized consists of most accurate four data points at epoch $2009.77$ in \citep{Currie2011}. The results are shown on Fig.~\ref{fig:fig18} in the same manner as  Fig.~\ref{fig:fig16}). Clearly, the best-fitting orbits derived from the full dataset D1 and from a single-epoch image D4 agree amazingly well. The best-fitting solution is found in the center of stable MMR island (the right hand panel of Fig.\ref{fig:fig17}). The double Laplace MMR lock is so robust and  bounded in the orbital parameter-space that even the minimal data are sufficient to constrain its orbital configuration in space.

\begin{figure*}
\centerline{
\vbox{
\hbox{\includegraphics[width=0.48\textwidth]{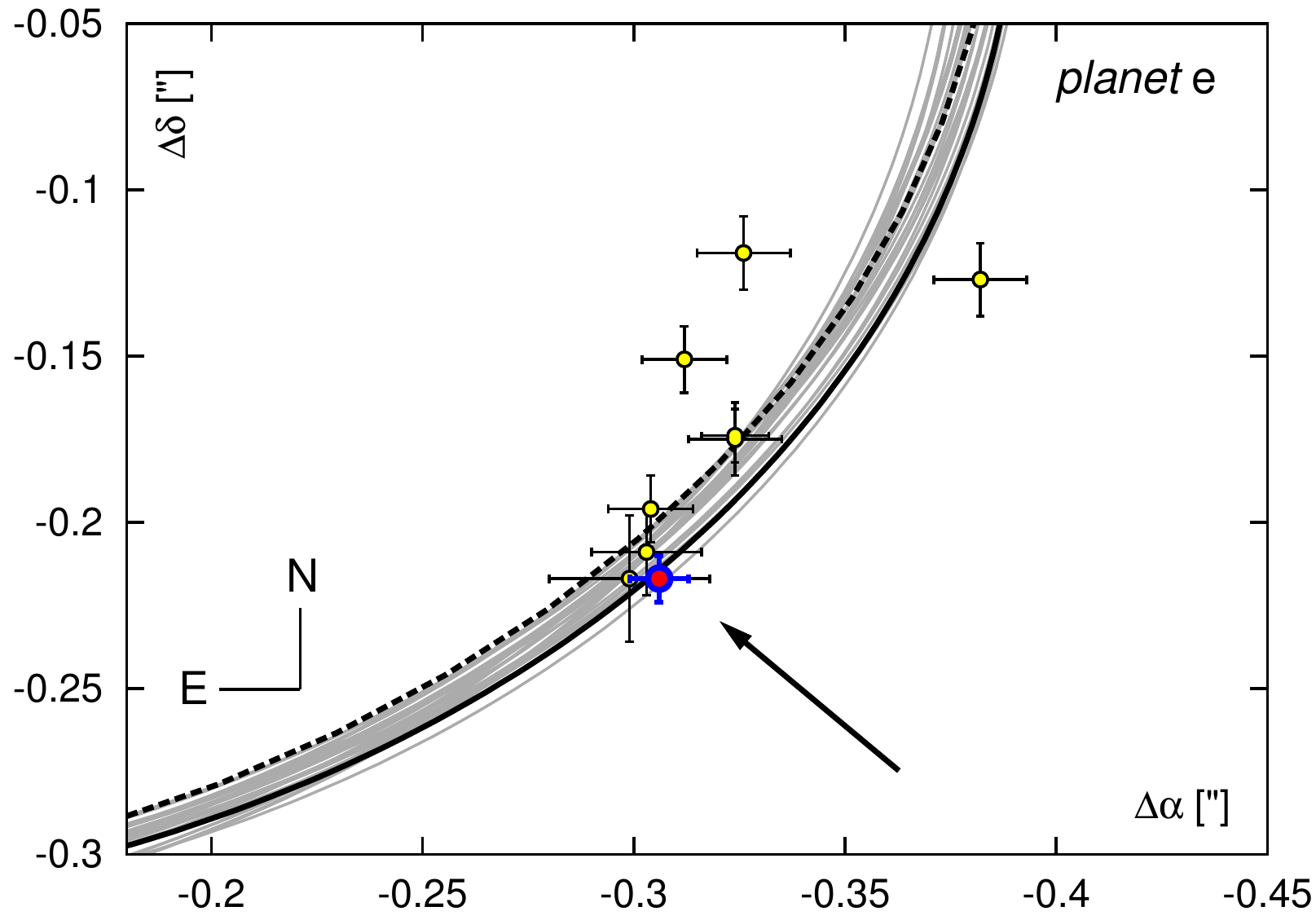}}
\hbox{\includegraphics[width=0.48\textwidth]{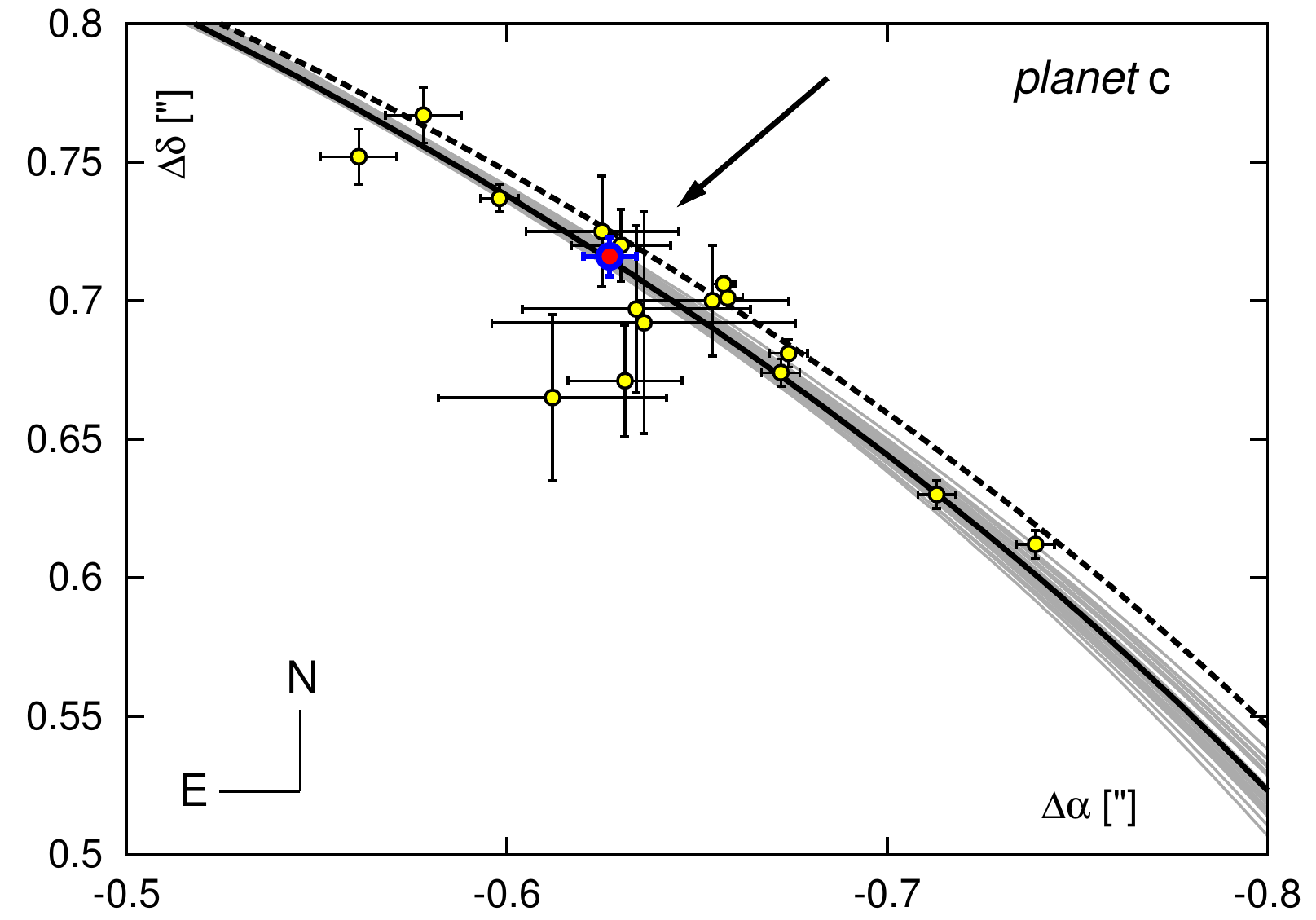}}
}
\vbox{
\hbox{\includegraphics[width=0.48\textwidth]{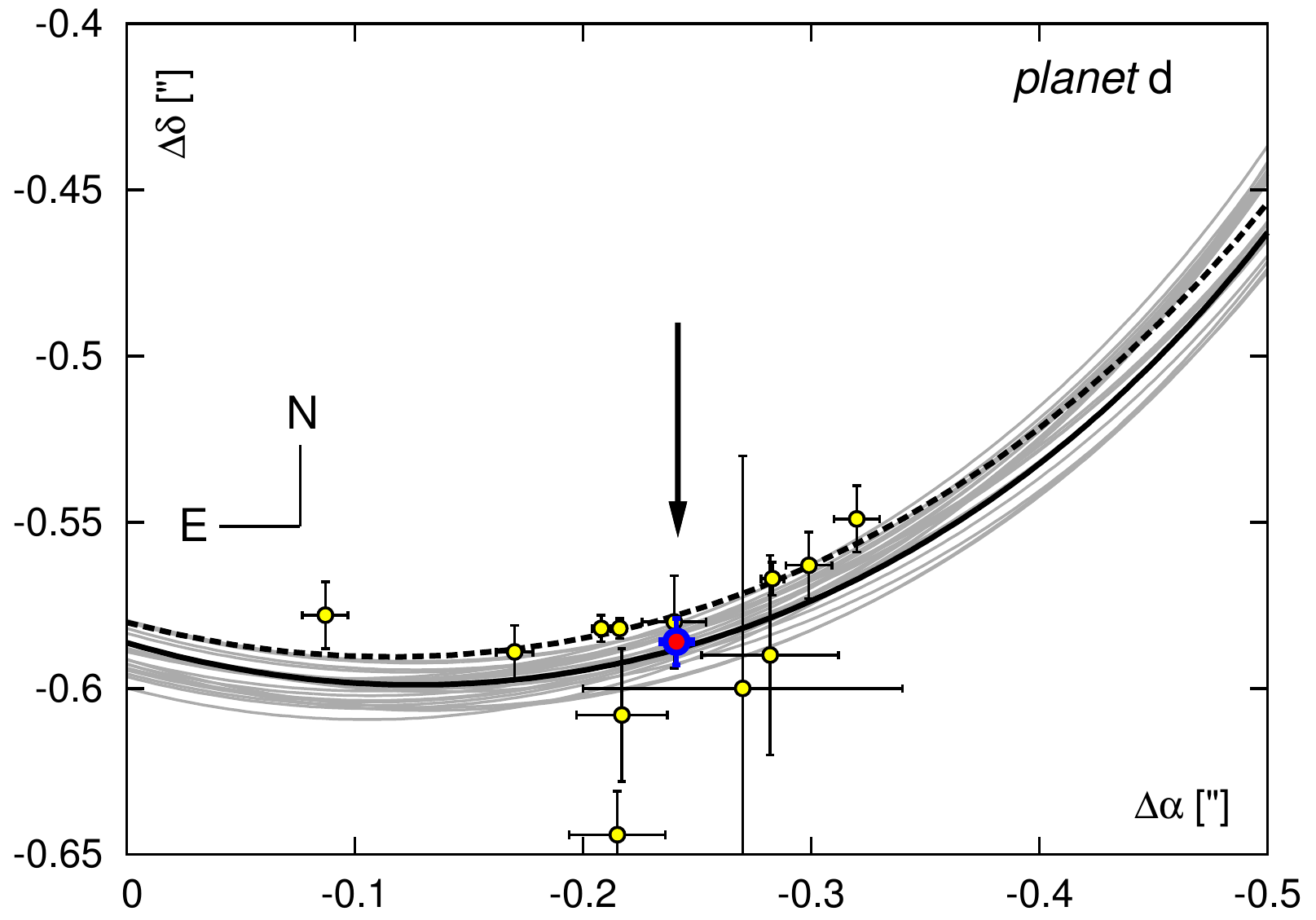}}
\hbox{\includegraphics[width=0.48\textwidth]{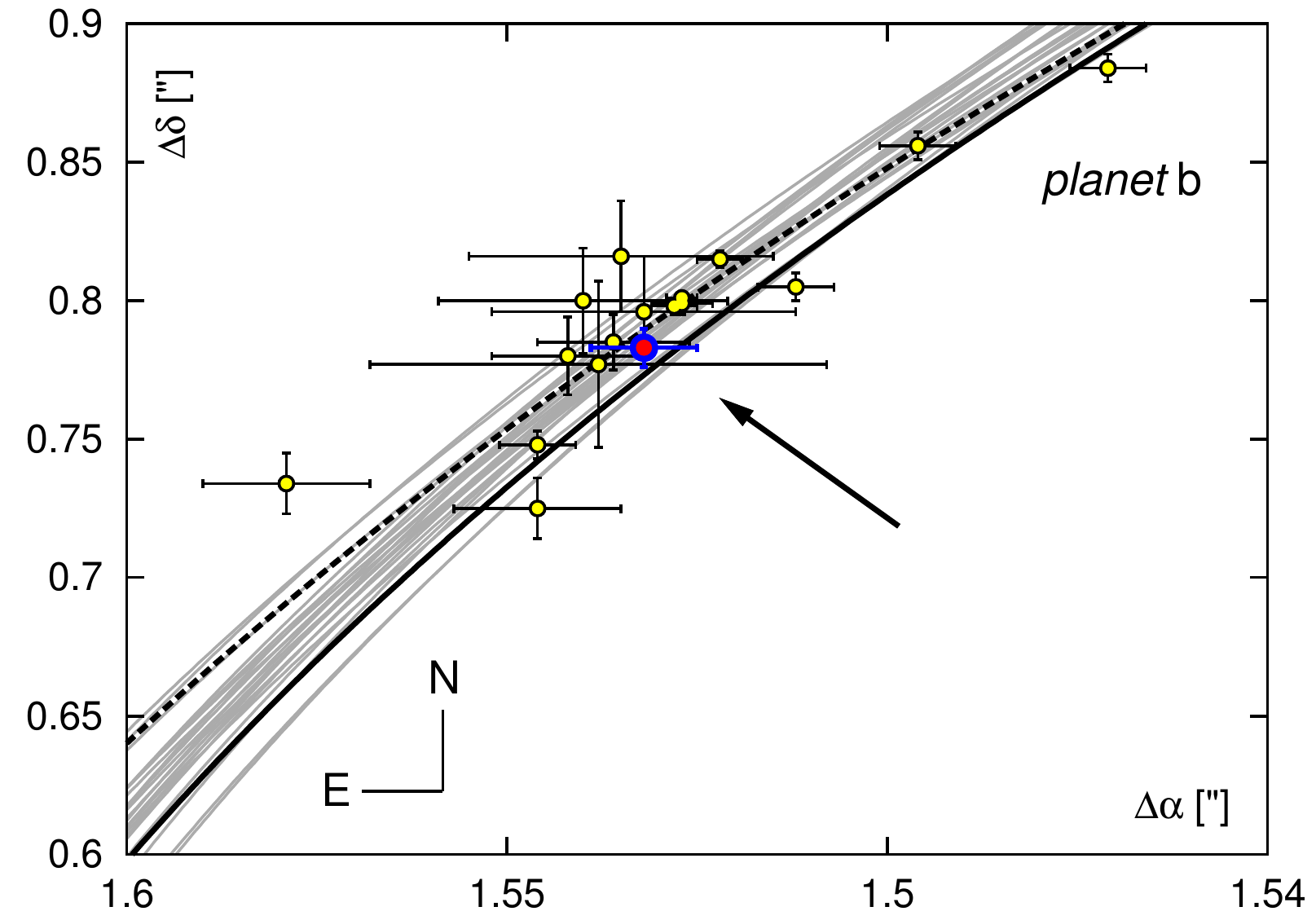}}
}
}
\caption{
Relative astrometric positions $\Delta\alpha$~[arcsec] vs $\Delta\delta$~[arcsec] of  planets  and orbital arcs for the best-fitting four-planet models~IVa (black dashed curves), and for the best-fitting single-epoch model~IVc (black solid curves).  Grey curves are for solutions IVc with formal $\cchi<1$.
}
\label{fig:fig18}
\end{figure*}

%
\subsection{Model V: the fith planet in the HR 8799 system?}
%
The migrating HR~8799 planetary system stabilized by the MMRs might involve more planets orbiting interior to the orbit of planet~e. There is a free space up to the distance $\sim 15$~au comprising of a few low order MMRs with planet~e, like 1f:2e ($\sim 10$~au), 1f:3e ($\sim 7.5$~au), and 2f:5e~($\sim 8.5$~au). A detection of so close objects overshined by the star is certainly very difficult. The contrast requirements and angular resolution are extreme in such a case, at the technical limit of the direct imaging. The negative results of the search for the fifth planet by \cite{Skemer2012} put the upper orbit limit for the hypothetical object to 1f:2e~MMR with the innermost planet, provided that this new planet has similar mass to planet~e.  We may note that the zone inner to $\sim 8$~au orbit suffers from exponential degrading of the contrast (see Fig.~3 in \cite{Skemer2012}), and, actually, less massive and luminous object might be still present. 

\begin{figure}
\centerline{
\hbox{
{\includegraphics[width=0.50\textwidth]{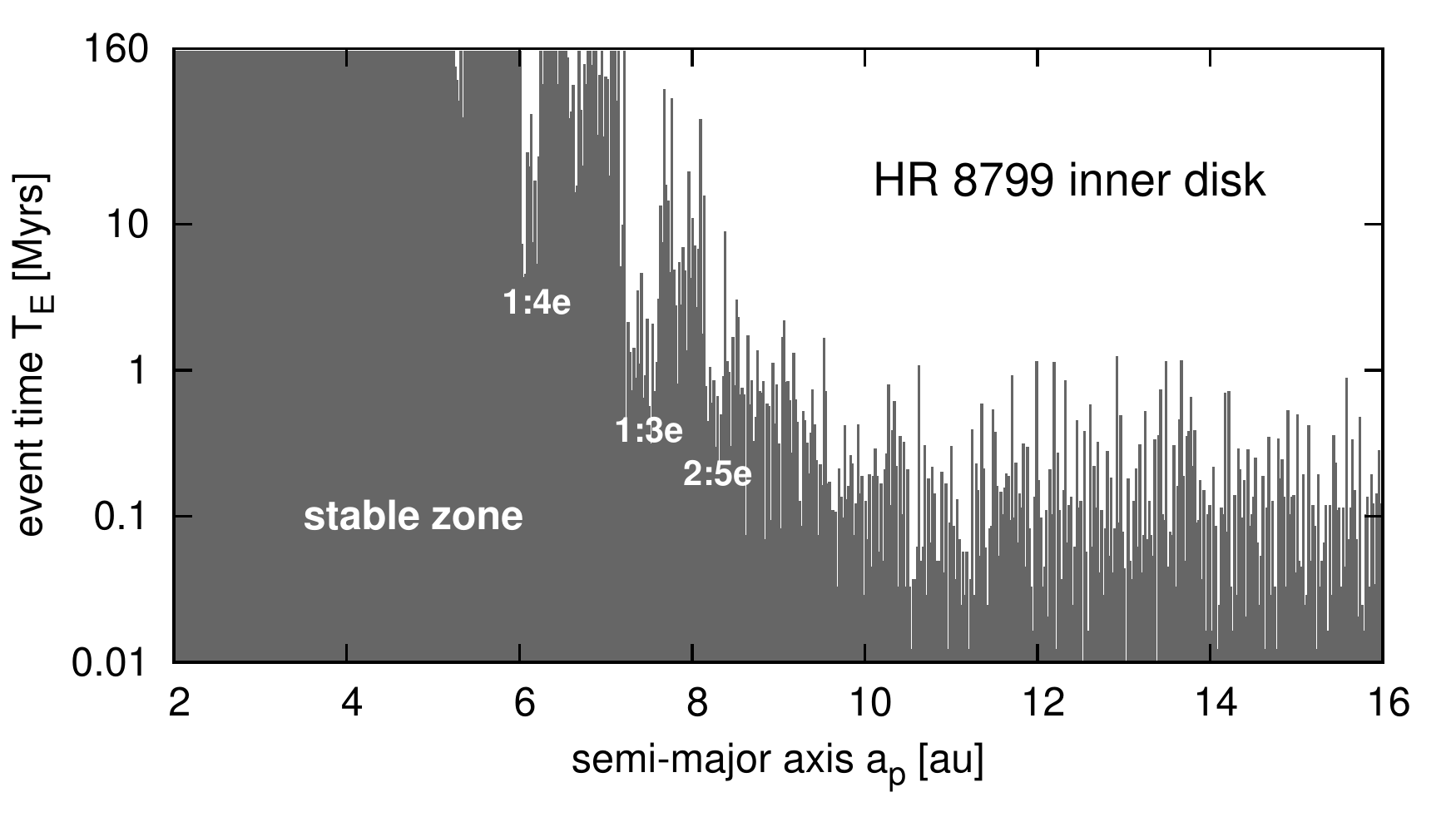}}
}
}
\caption{
\corr{Statistical} event times $T_{\idm{E}}$ for mass-less particles in the inner zone of the HR~8799 four planet model~IVa. Particles placed interior to $\sim 7$~au with random eccentricities $e_p \in [0,0.5]$ and random orbital phases survive for at least 160~Myr. 
}
\label{fig:fig19}
\end{figure}

The results \cite{Hinkley2011} also do not exclude a low-mass object with the mass below 11\,$\mJ$ between 0.8~au and 10~au. Moreover, the warm disk gap interior to $\sim 6$~au \citep{Su2009,Hinkley2011,Oppenheimer2013} is incompatible with the results of simulating the lifetimes of massless particles in the four-planet system (our model IVa). We integrated $\sim1000$ probe particles placed at $a_p \in [2,16]$~au with initially random eccentricities $e_p \in [0,0.3]$ and with random orbital phases $\sim [0,360^{\circ}]$. The event time $T_{\idm{E}}$ graph in Fig.~\ref{fig:fig19} reveals that low-mass objects (\corr{large dust particles, colliding asteroids producing that dust}) could survive for at least 160~Myr in the innermost zone of the system that ends at $\sim 6$--$7$~au. 

\begin{figure}
\centerline{
\vbox{
\vbox{
{\includegraphics[width=0.47\textwidth]{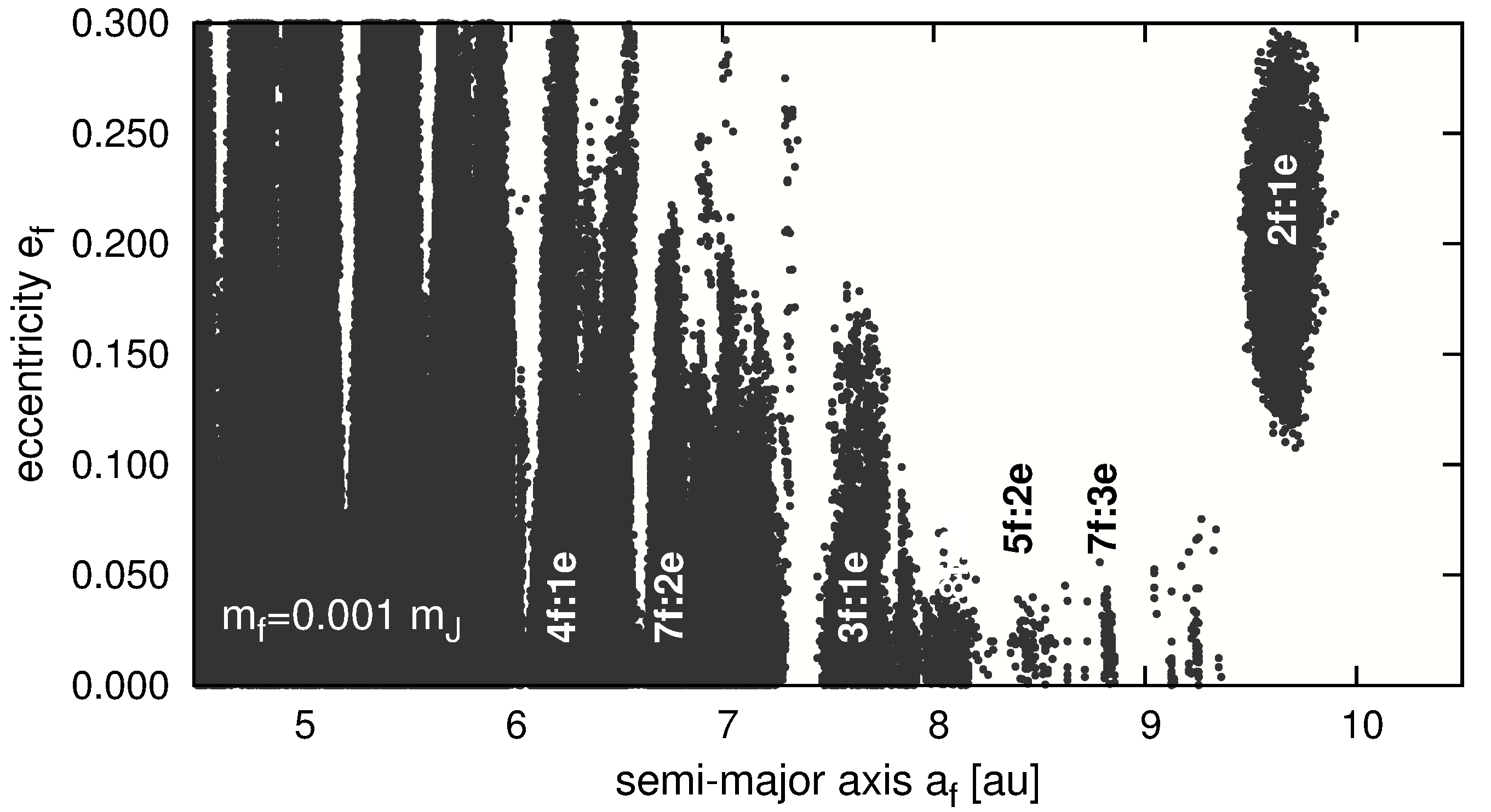}}
{\includegraphics[width=0.47\textwidth]{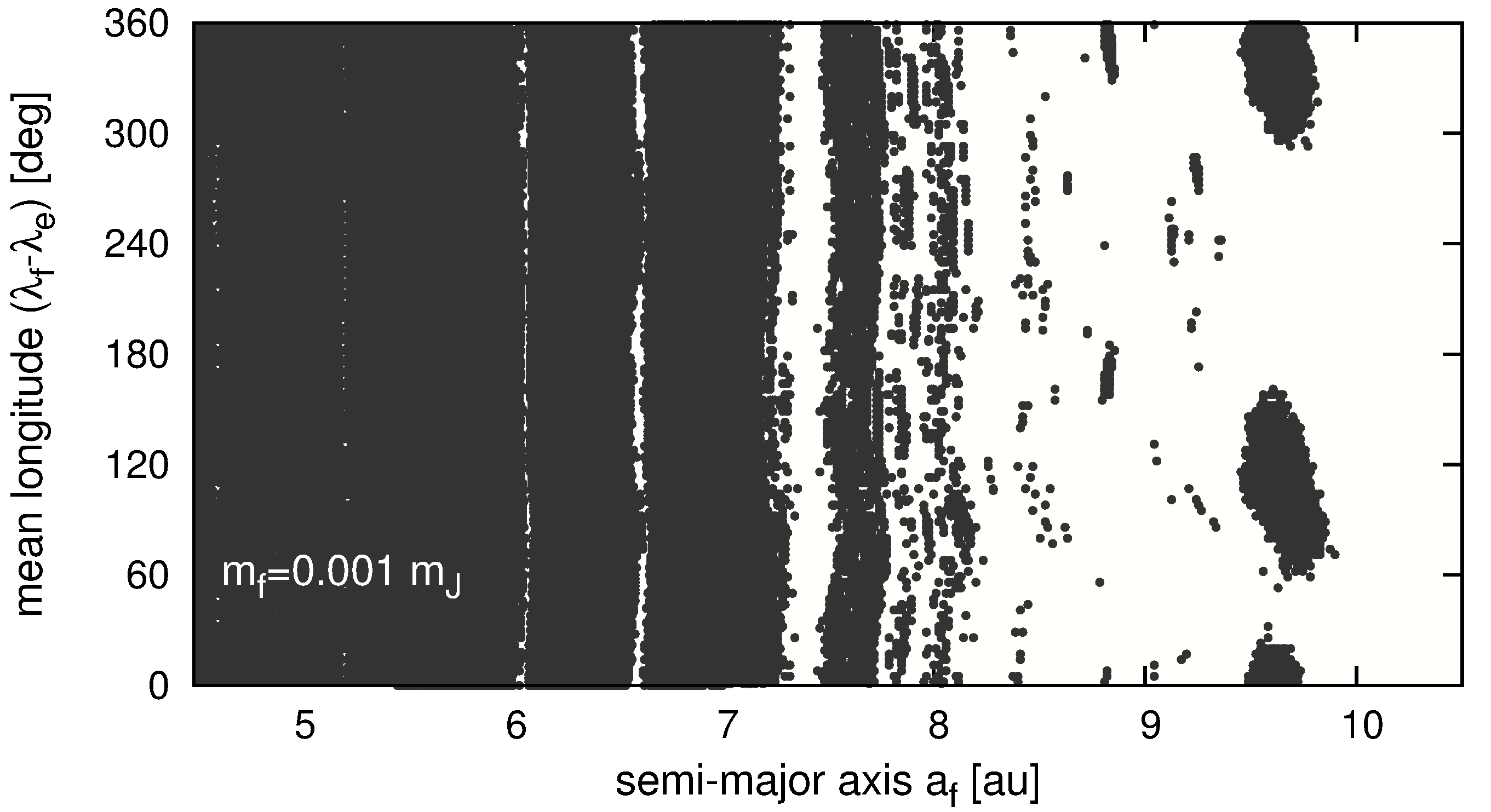}}
}
}
}
\caption{
\corr{
A simulation of stable regions in the semi-major axis v.s eccentricity plane ({\em top}) and in the semi-major vs. relative orbital phase plane ({\em bottom}) for hypothetical  massive, Mars-like asteroids $\sim 0.001~\mJ$. Low order two--body MMRs with planet HR~8799e are labelled. Each dot represent a quasi-periodic, stable orbit for orbital phase illustrated in the bottom panel. The MEGNO integration time of $1$~Myr is equal to $\sim20,000$ orbital periods of planet~HR~8799~e. Black filled circles are for stable
orbits. See the text for details.
}
}
\label{fig:fig19A}
\end{figure}

\corr{
Note that due to random distribution of initial eccentricity,  $T_{\idm{E}}$ cannot be uniquely determined, hence the plot illustrates a~border of stable orbits in statistical sense. To confirm this result for non-zero mass objects, we carried out much more extensive multi-dimensional scan in the orbital elements space. The probe mass of $\sim 0.001~\mJ$ was placed in orbit interior to planet HR~8799e, with the semi-major axis $a_f \in [4.5,10.5]$~au. Next, for each point of the pericenter argument and the mean anomaly at the grid with $3^{\circ} \times 3^{\circ}$ resolution and spanning full angle,  the initial eccentricity was sampled from uniform distribution $e_f \in [0,0.3]$. Each orbit was then MEGNO-integrated for $1$~Myr. The results are shown in Fig.~\ref{fig:fig19A}. Each point in this Figure marks quasi-periodic (stable) orbit. The tested region of innermost orbits is filled with multiple MMRs. The bottom plot shows that stable MMRs are possible 
only for particular initial relative phases of the probe mass. Overall, interior
to the $\sim 6$--$7$~au his zone is basically stable, in accord with two orders of magnitude longer, direct
integrations illustrated in Fig.~\ref{fig:fig19}. The results may be also helpful to determine
possible locations of small planets in the system, below the current detection limit.
}

This experiment also suggest that the presence of a relatively massive object interior to planet~e could further clear or sculpt dynamically the innermost debris disk.

Before examining such a hypothesis of less-massive fifth planetary object, we carefully verified that the migration algorithm ``predicts'' correctly planet~e. We would like to recall that the observational circumstances were easily simulated by optimizing four-planet model to all observations of only three outer planets d,c, and~b (dataset D5). Indeed, the best-fitting simulated configuration found the ``missing'' planet~e perfectly in place of the actual detection. Moreover, stability constraints even narrow possible outcomes of \moa{} to agree with the best-fitting, nominal four-planet model~IVa  (see Figs.~\ref{fig:fig10}--\ref{fig:fig12}). 

We carried out a similar experiment regarding yet unseen, hypothetical innermost planet~f. An extension of model~IVb to five planets is quite straightforward. This model~V combines the full dataset~D1 with a five-planet system.  The simulated mass range of planet~f is assumed similar to planet~e, around $6\,\mJ$ with a significant dispersion $\pm 4\,\mJ$. 

The results of more than $10^5$ single runs of \moa{} are illustrated in Figs.~\ref{fig:fig20}, \ref{fig:fig21} and~\ref{fig:fig22}. Figure~\ref{fig:fig20} illustrates the sky-plane geometry of the statistics of solutions within $6\sigma$ confidence level. Figure~\ref{fig:fig21} shows the osculating elements and masses of the hypothetical planet projected onto the planes of the semi-major axis -- eccentricity (the left-hand panel) and onto the semi-major axis -- mass (the right-hand panel).  The measurements are consistent with two MMRs chains resulting in similar $\cchi\sim 1.18$ --- the five-planet {\em triple} Laplace resonance, 1f:2e:4d:8c:16b~MMR, and the five-planet 1f:3e:6d:12e:24b~MMR, which we refer to as the 1f:2e~MMR and the 1f:3e~MMR for short, respectively. A multiple MMR comprising of the 2f:5e~MMR combined with double Laplace MMR of four outer planets is also possible but this solution has significantly larger $\cchi \sim 1.25$.    

Each of the two dominant MMRs appears as two well bounded families of orbits which are differently phase-spaced. These families are distinguished by particular critical argument of the MMR. {We call them Va (1f:3e~MMR with $\cchi \sim 1.176$),  Vb (1f:3e~MMR with $\cchi \sim 1.177$), Vc (1f:2e~MMR with $\cchi \sim 1.169$), and Vd (1f:2e~MMR with $\cchi \sim 1.25$)} from hereafter (see also Tab.~\ref{tab:table4} and  ephemeris tables,
Tabs.~\ref{tab:ephemeris2}--\ref{tab:ephemeris5} for all models).  We note that the outer planets are always involved in a double Laplace resonance, 1e:2d:4c:8b~MMR.

Overall, the best-fitting mass ranges of the putative fifth planet seem well correlated with the MMR type.  The triple Laplace MMR favours planets with $\sim 1$--$4$\,$\mJ$, while stable 1f:3e~MMR solutions permits the masses much larger, $\sim 2$--$8$\,$\mJ$ (see the right-hand panel in Fig.~\ref{fig:fig21}).

\begin{figure}
\centerline{
\hbox{
{\includegraphics[width=0.48\textwidth]{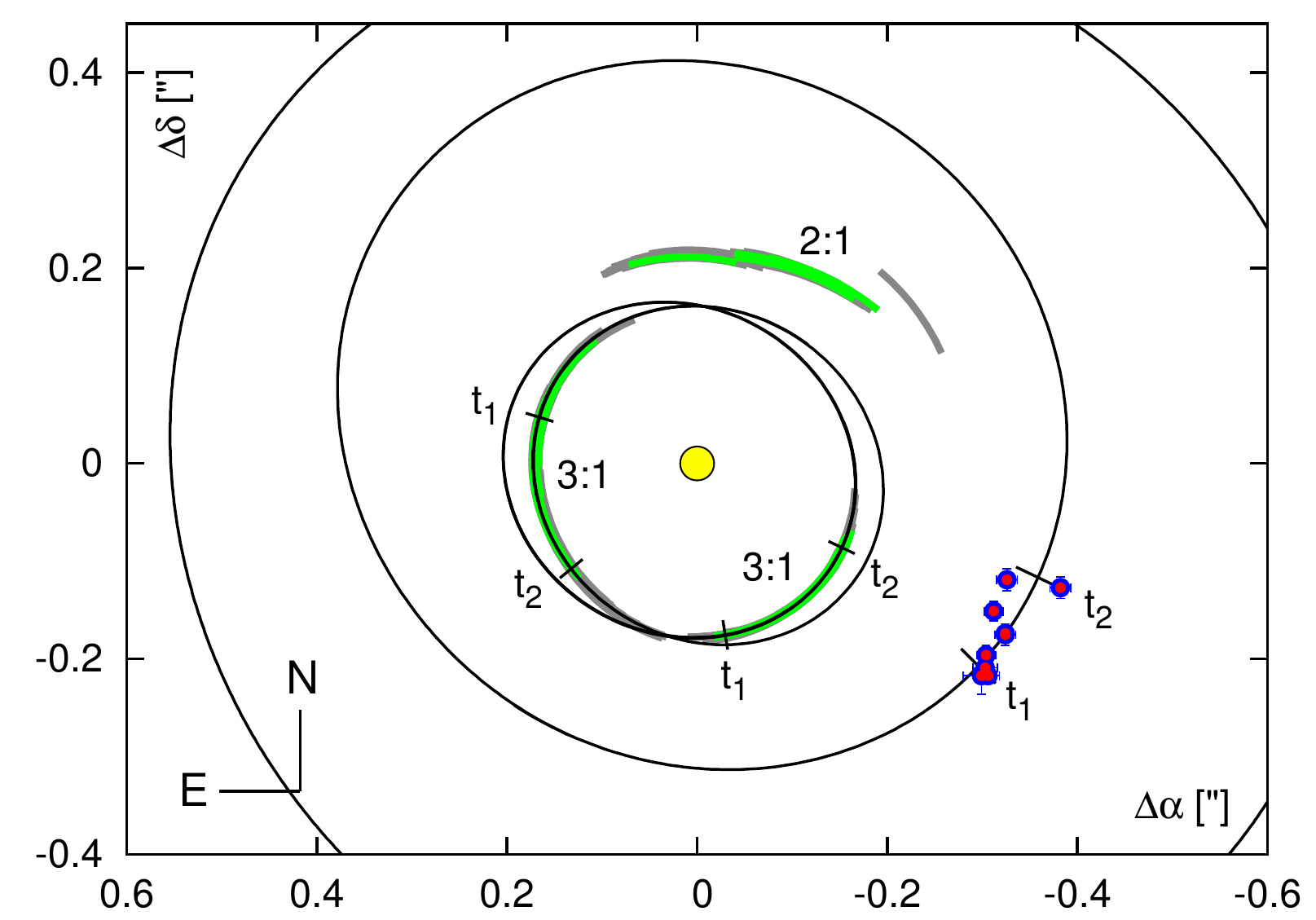}}
}
}
\caption{
Observations of planet~e (red/blue filled circles) over-plotted on best-fitting orbits of planets~e and~f in model~Va (black solid curves). Orbital arcs of planet~f  between epochs~2009.58 and epoch~2011.86 are plotted with solid grey curves ($\cchi < 1.5$, equivalent to 6$\sigma$ solutions) and with green curves ($\cchi < 1.3$). {Arcs of stable orbits within $3\sigma$ confidence interval are marked with tics}.  
}
\label{fig:fig20}
\end{figure}

\begin{figure*}
\centerline{
\hbox{
{\includegraphics[width=0.49\textwidth]{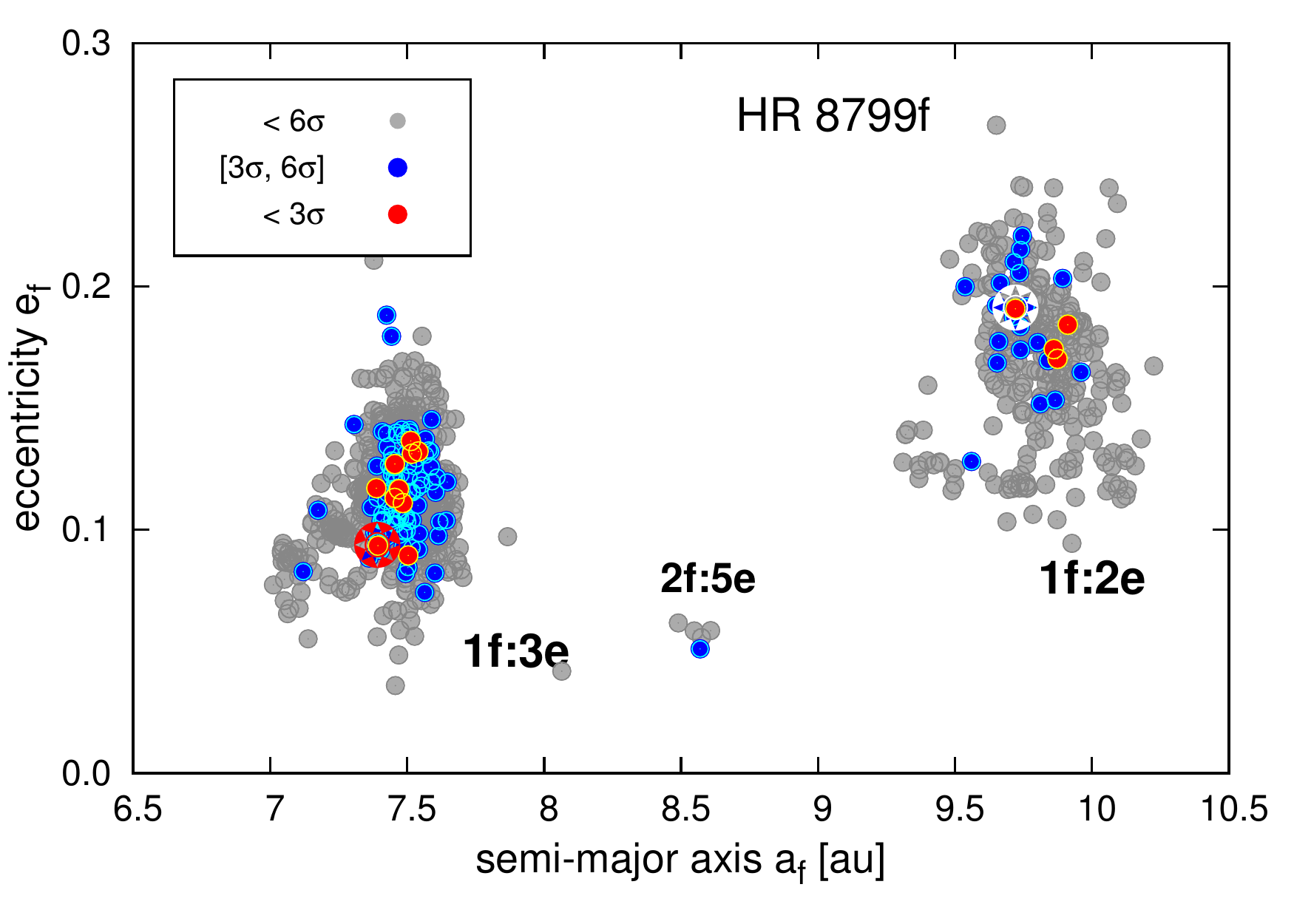}}
{\includegraphics[width=0.49\textwidth]{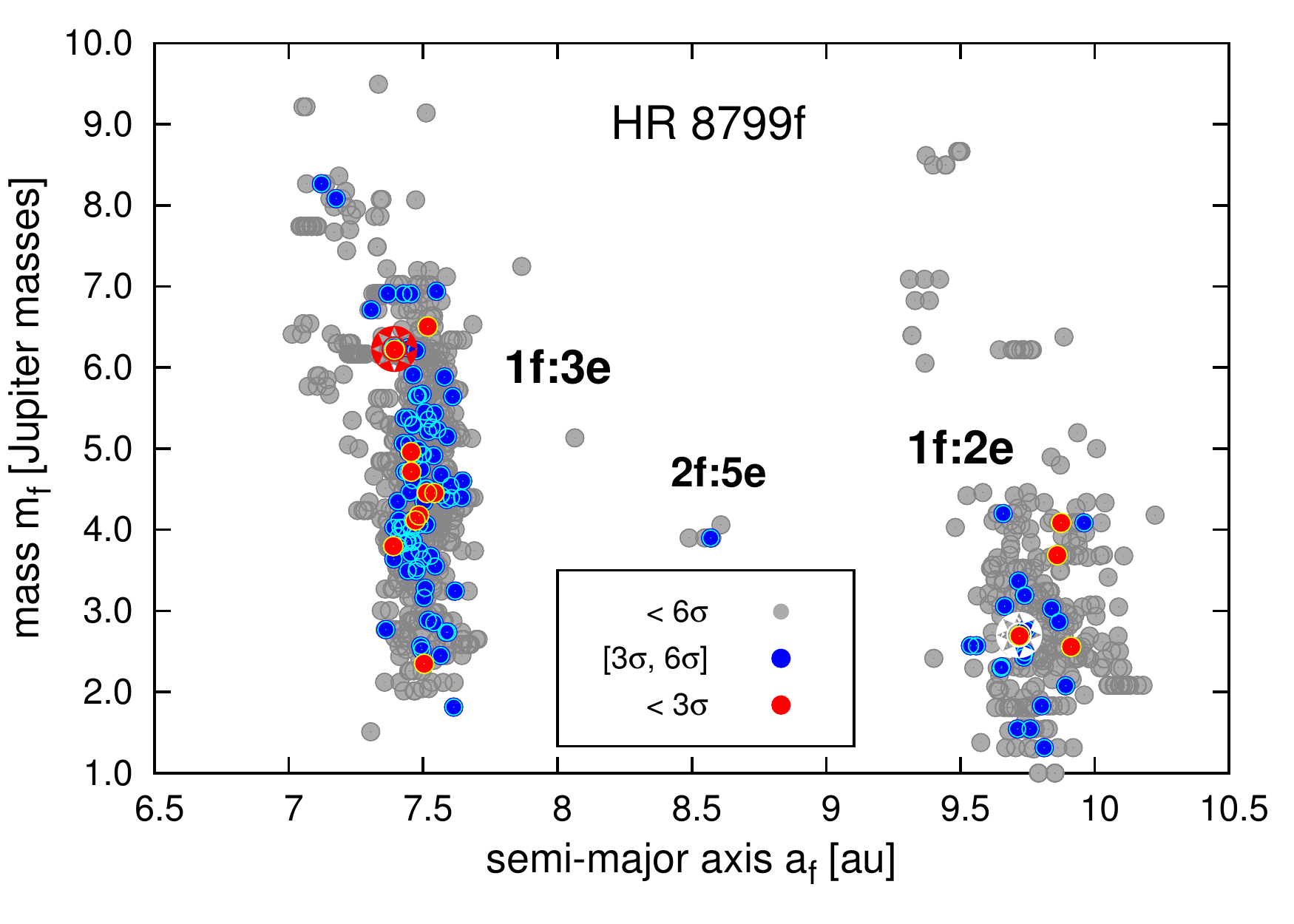}}
}
}
\caption{
Orbital parameters of the five-planet model~V regarding yet unseen, hypothetical planet HR~8799f. These parameters are projected onto semi-major axis -- eccentricity  ({\em left}) and the semi-major axis -- mass ({\em right})  planes of this planet, respectively. The red and white star symbols mark the nominal, best-fitting solutions that correspond to distinct families of these models:  the 3f:1e~MMR ($\cchi=1.176$) and 2f:1e~MMR ($\cchi=1.169$), respectively. Grey circles are for all (also \corr{unstable}) solutions with $\cchi < 1.5$. Compare these results
with simulations of the stability of less-massive objects interior to planet~e, shown in Fig.~\ref{fig:fig19A}.
}
\label{fig:fig21}
\end{figure*}

\begin{figure*}
\centerline{
\hbox{
{\includegraphics[width=0.72\textwidth]{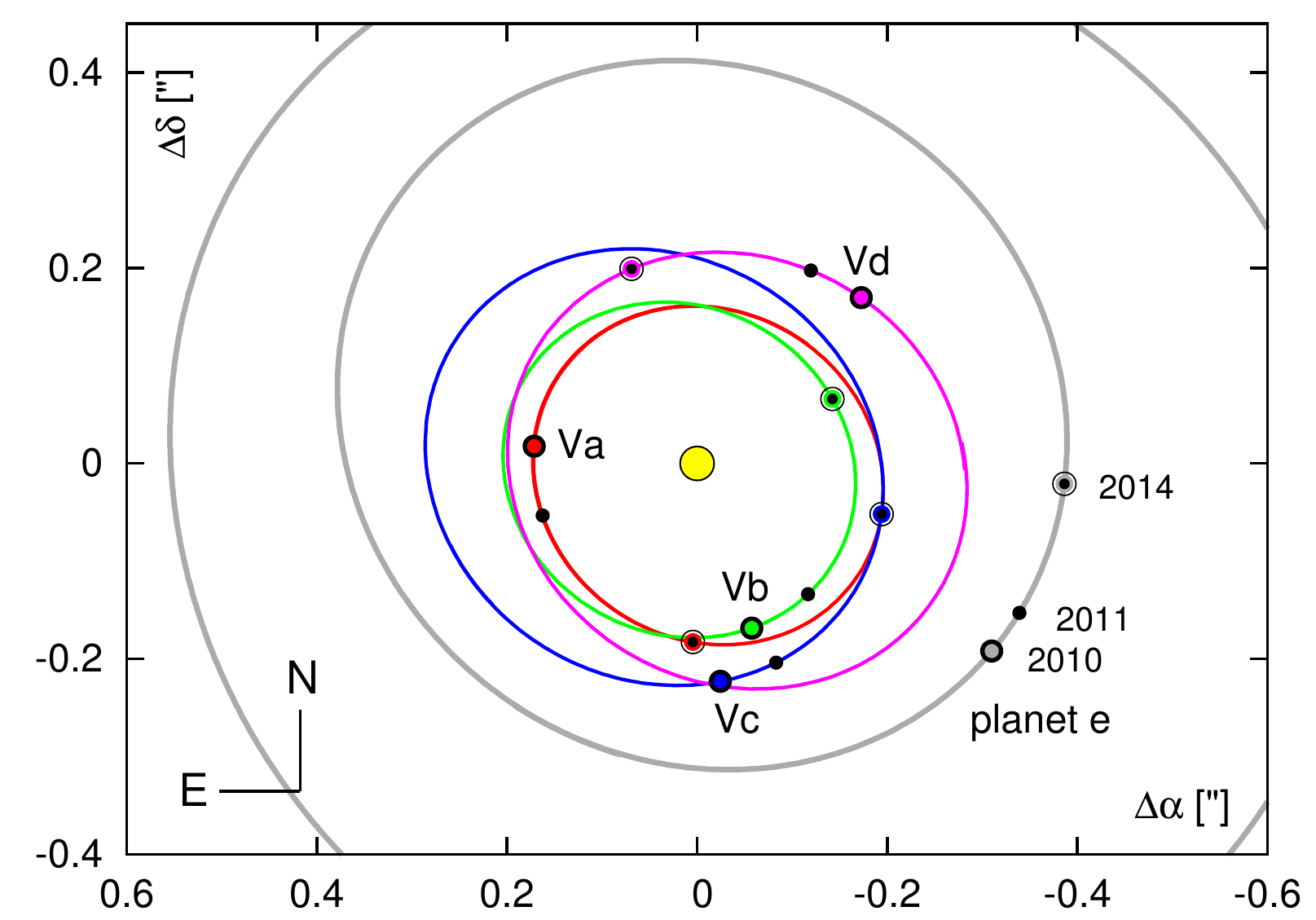}}
}
}
\caption{
{Best-fitting orbits Va (1f:3e~MMR), Vb (1f:3e~MMR), Vc (1f:2e~MMR) and Vd (1f:2e~MMR) of planets~e and hypothetical planet~f. Orbital positions of both planets at epochs 2010.0, 2011.0 and 2014.0 are marked with filled circles and labeled, accordingly. See also ephemeris Tabs.~\ref{tab:ephemeris2}--\ref{tab:ephemeris5}.
} 
}
\label{fig:fig22}
\end{figure*}

We tested the long-term stability of $\sim 2000$ solutions within $6\sigma$ confidence interval of the best solutions with $\cchi \sim 1.18$,  by integrating MEGNO for 160~Myr. The results of detailed dynamical analysis combined with $\cchi$  seems to narrow the likely positions of planet~f to basically four well bounded locations. The dynamical maps computed for lowest $\cchi$ solutions in each family are illustrated in Fig.~\ref{fig:fig23}. Except for the 1f:2e~MMR model Vc, these solutions are found in the centres of islands of long-term stable motions. Models Va, Vb, and Vd are long term stable. For a reference,  the orbital architecture of model Va (a family of the 1f:3e~MMR) for 160~Myr is illustrated in
Fig.~\ref{fig:fig24}. This plot marks also the inner, thin disc of the habitable zone extending roughly to $\sim 4$~au, interior to the orbit of planet~f. The region inner to orbit of planet~e may contain a warm debris disk detected by \spitzer{} \citep{Su2009,Reidemeister2009}. The dynamical structure of this region is certainly very complex. We postpone its analysis to a future paper.   

The Vc solution (1f:2e~MMR) is found as marginally stable. This is confirmed by Fig.~\ref{fig:fig25} which illustrates the MEGNO behaviour for this model, as compared to the rigorously stable model Va. The best-fitting configuration disrupts after $\sim 120$~Myr, although is found at the edge of quite extended stability zone (see also the $T_{\idm{E}}$ dynamical map in Fig.~\ref{fig:fig27} and a discussion in Sect.~\ref{sec:stability}).

Our predictions do not contradict a negative result of the search for the fifth planet in \citep{Skemer2012}. For instance, the 1f:3e~MMR orbits are systematically closer to the star than the 1f:2e~MMR orbits (see Fig.~\ref{fig:fig22}). At distances $\sim 8$~au the contrast of LBT~images drops exponentially, see Figs.~3 in \citep{Skemer2012} and in \citep{Esposito2013}. We also note that the 1f:2e~MMR models systematically tend to low masses of $\sim 2$--3\,$\mJ$, hence the object would have much lower luminosity than      \citep{Skemer2012} tested on fake images.

In all examined models the resonance islands span up to $\sim 0.1$~au, hence are extremely narrow when compared with the overall dimension of the system ($\sim 100$~au). This is a clear warning that a search for the fifth planet with the common, ``traditional'' optimization methods would be difficult and unlikely possible due to large number of free parameters and narrow observational window. With the help of \moa{}, planet~f may be ``pre-detected'' without even seen it first at the images.  The predictions and ephemeris data (Tabs.\ref{tab:ephemeris2}--\ref{tab:ephemeris5}) might be helpful to confirm or withdraw this intriguing hypothesis. 

\begin{figure*}
\centerline{
\vbox{
\hbox{
{\includegraphics[width=0.49\textwidth]{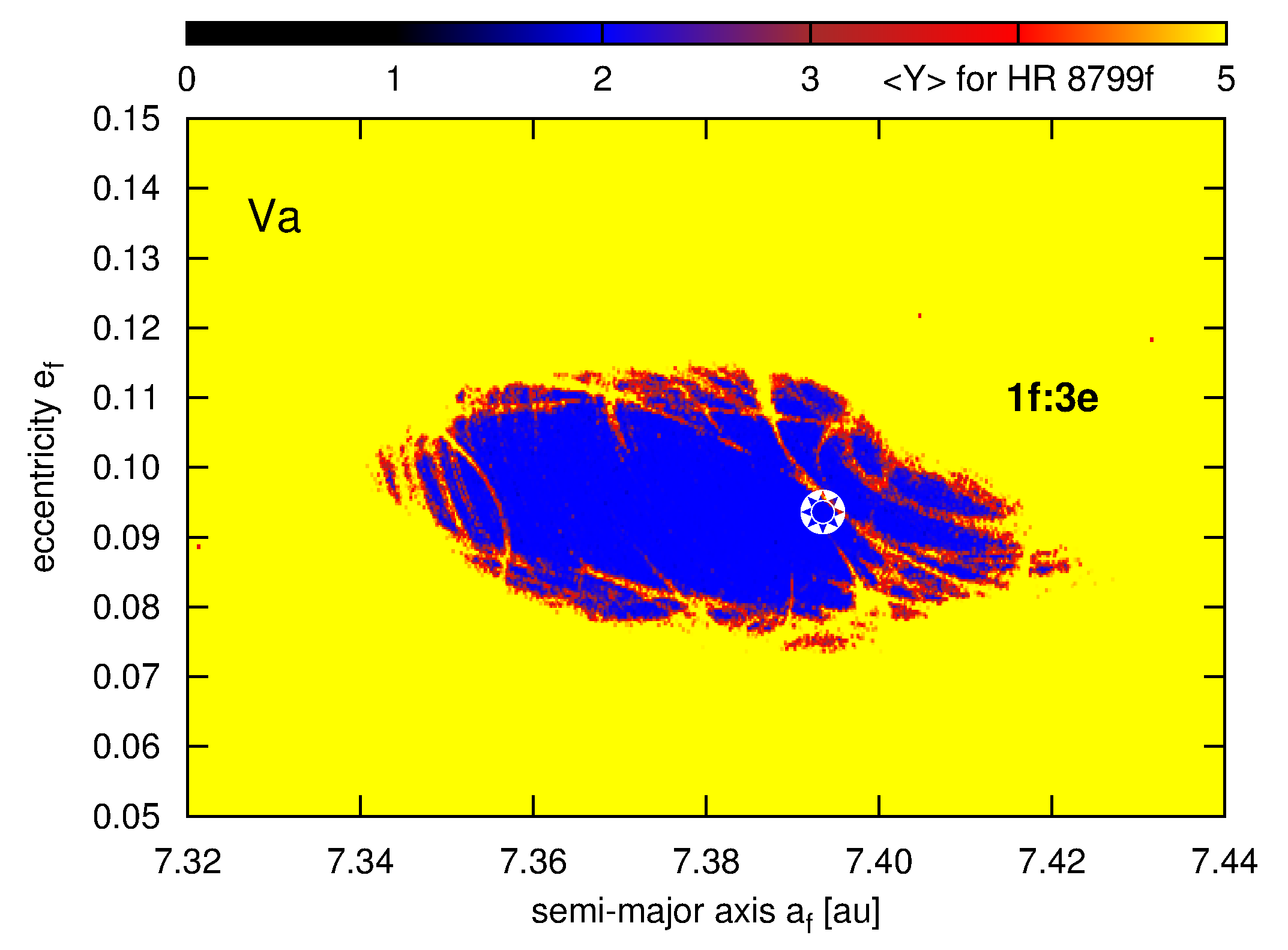}}
{\includegraphics[width=0.49\textwidth]{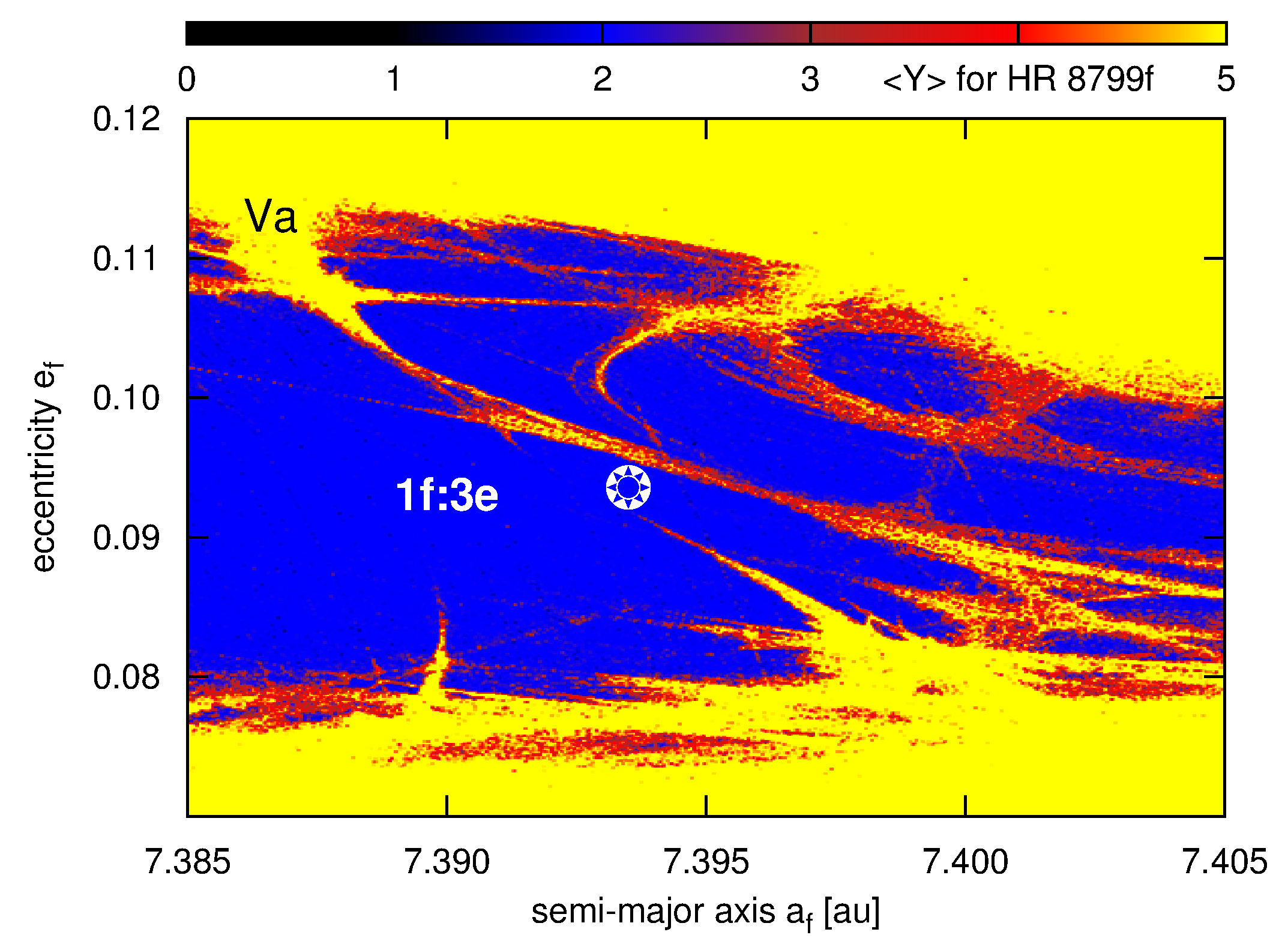}}
}
\hbox{
{\includegraphics[width=0.49\textwidth]{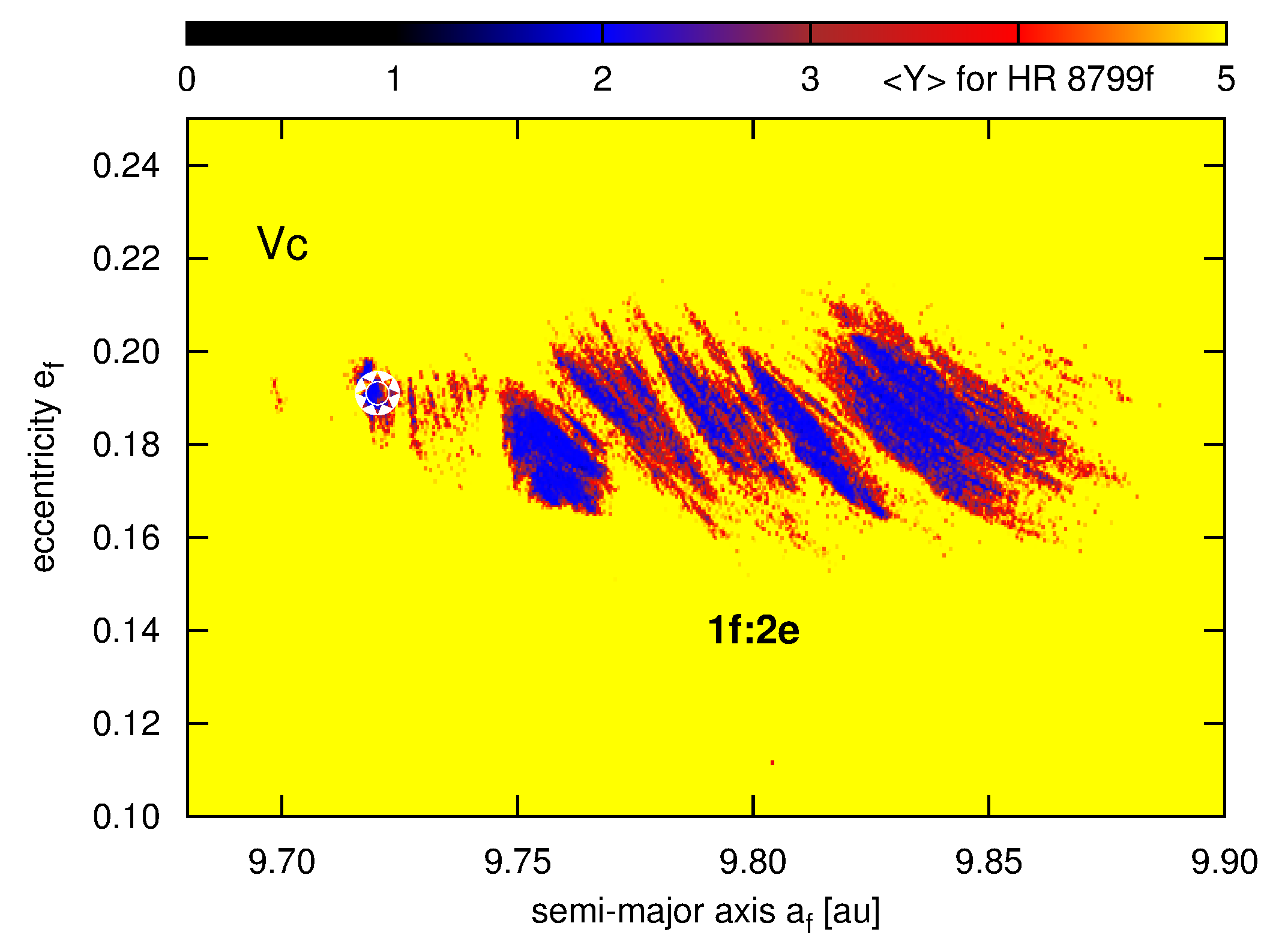}}
{\includegraphics[width=0.49\textwidth]{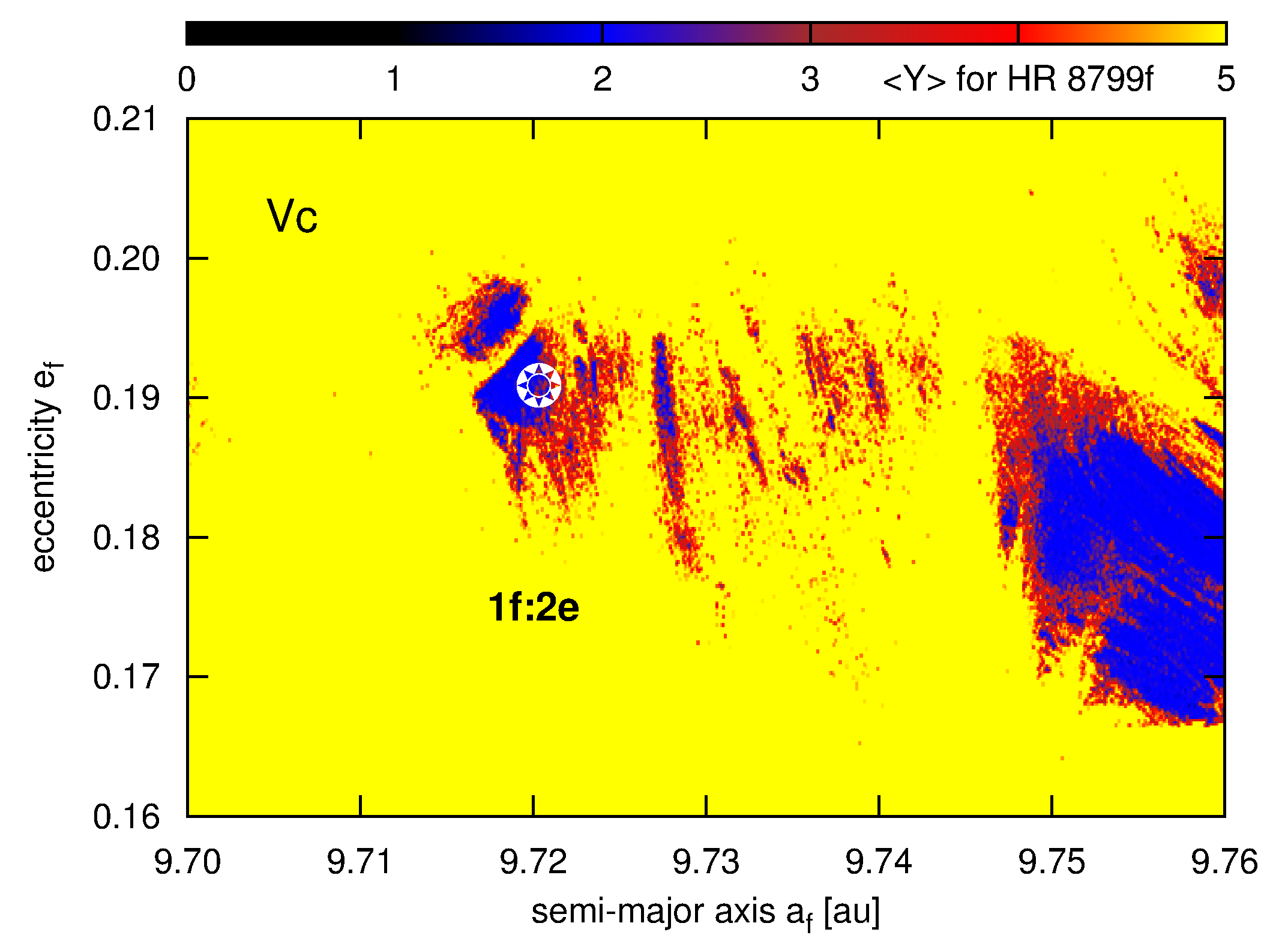}}
}
\hbox{
{\includegraphics[width=0.49\textwidth]{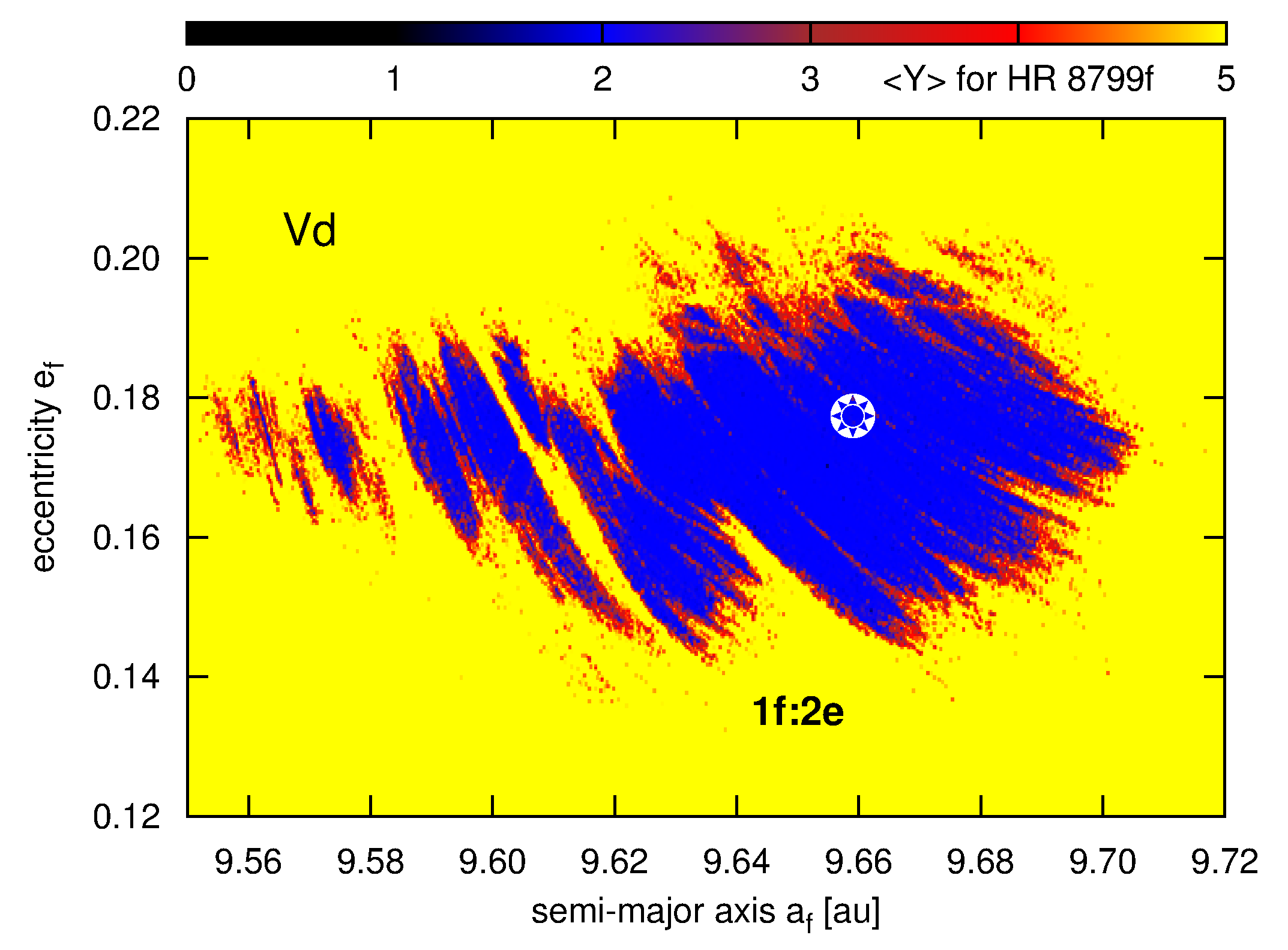}}
{\includegraphics[width=0.49\textwidth]{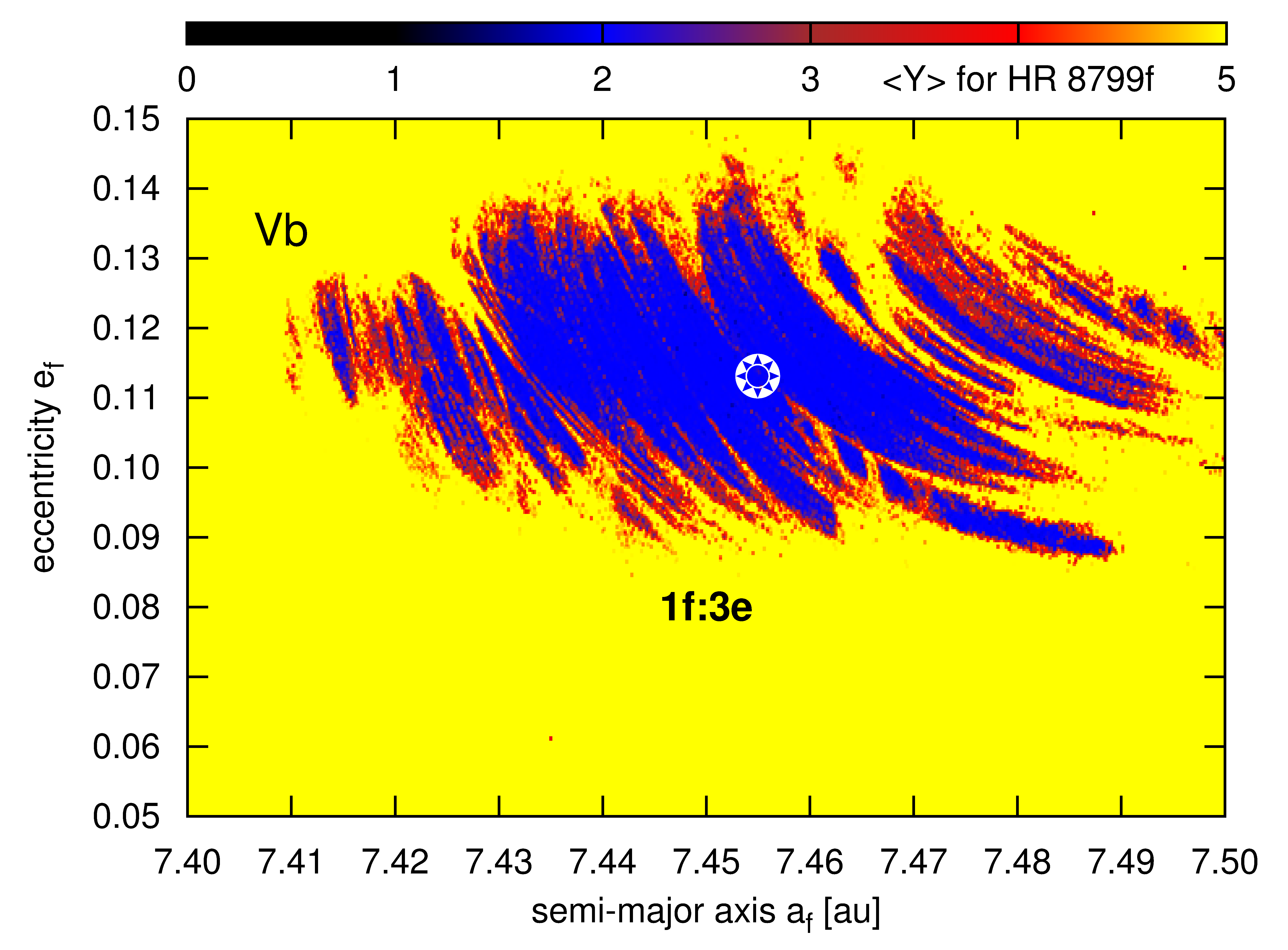}}
}
}
}
\caption{
Dynamical maps  $\left<Y\right>$,  in the semi-major--eccentricity plane for the best-fitting five-planet models including a hypothetical planet HR~8799~f. This planet might be involved in the 1e:3f~MMR, or in the 1f:2e~MMR with planet HR~8799~e. The star symbol marks the best-fitting models in Tab.~\ref{tab:table4} (subsequent plots are labelled accordingly), see also Figure~\ref{fig:fig22}. Two upper right-hand panels are for close-ups of solutions Va and Vc with smallest $\cchi$. The resolution of all maps is $640\times320$ pixels.  The integration time for each pixel is  $\sim 20,000$--$30,000$  orbital periods of the outermost planet ($\simeq 10$--$15$~Myr). }
\label{fig:fig23}
\end{figure*}

\begin{figure}
\centerline{
\hbox{
\includegraphics[width=0.48\textwidth]{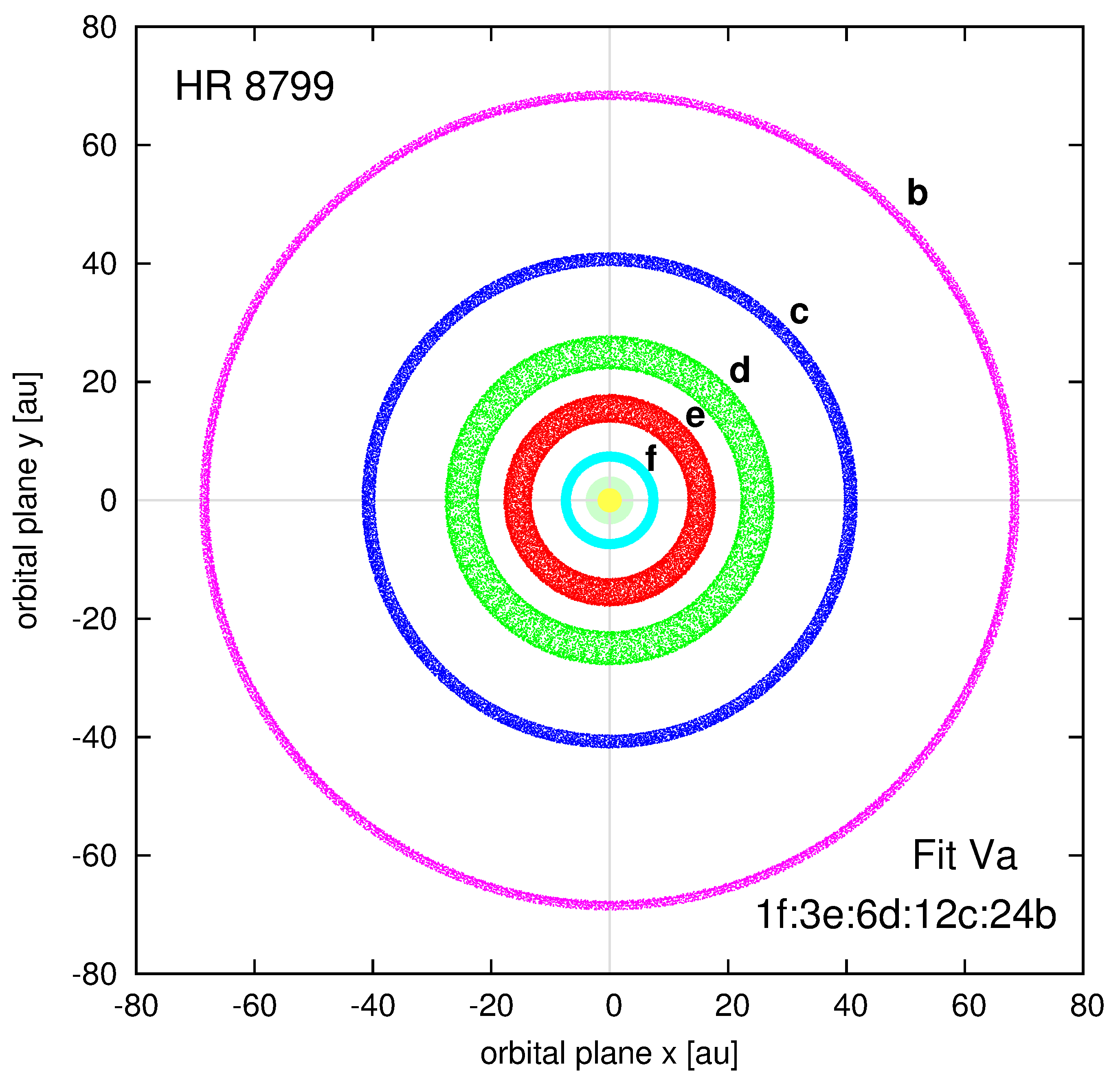}}
}
\caption{
The orbital architecture of the HR 8799 system, in accord with the
best-fitting, coplanar and long-term stable model~Va of {\em five} planets
involved in the 1f:3e:6d:12e:24b~MMR.  This solution results in $\cchi =
1.176$ and its osculating elements are given in Table~\ref{tab:table4}.  The
new planet would be in the 1e:3f~MMR with HR~8799e.  Orbital positions of
HR~8799 planets are shown for 160~Myr in the astrocentric coordinate frame,
coplanar with the orbits.  The innermost green circle in the center has the
radius of $\sim 4$~au and corresponds to the outer edge of the habitable
zone in the HR~8799 system.
}
\label{fig:fig24}
\end{figure}


\begin{table*}
\caption{
Osculating elements and masses for the best-fitting five-planet models. The stellar mass $m_0 = 1.56\,\msun$.  The osculating epoch is~$1998.83$.
}
\label{tab:table4}
\begin{center}
\begin{tabular}{l c c c c c c c c c}
\hline
model & planet & $\cchi$ & $m\,[\mJ]$ & $a\,[\au]$ & e & $I$[deg] & $\Omega$[deg] & $\omega\,$[deg] & $\Mmean_0\,$[deg]\\
\hline
& f & & $6.218270$ & $7.393497$ & $0.093498$ & & & $162.267947$ & $143.405566$\\
& e & & $7.620773$ & $15.809728$ & $0.153788$ & & & $290.348791$ & $328.757318$\\
Va & d & $1.176$ & $7.818042$ & $26.006470$ & $0.124417$ & $25.542107$ & $243.667749$ & $215.954602$ & $49.813246$\\
& c & & $8.749394$ & $39.680175$ & $0.049666$ & & & $272.353696$ & $142.869706$\\
& b & & $4.585532$ & $68.187050$ & $0.016927$ & & & $253.022770$ & $279.634792$\\
\hline
& f & & $4.959466$ & $7.454967$ & $0.113033$ & & & $51.797812$ & $23.180967$\\
& e & & $7.785460$ & $15.466841$ & $0.170185$ & & & $290.256931$ & $322.820611$\\
Vb & d & $1.177$   & $9.496197$ & $25.529862$ & $0.125842$ & $26.869698$ & $245.474714$ & $205.284684$ & $57.312742$\\
& c & & $8.852039$ & $40.465447$ & $0.037989$ & & & $274.813041$ & $139.910933$\\
& b & & $5.169680$ & $69.434385$ & $0.020676$ & & & $174.060193$ & $354.774655$\\
\hline
& f & & $2.691369$ & $9.720335$ & $0.190823$ & & & $221.621749$ & $116.829209$\\
& e & & $6.887347$ & $15.804942$ & $0.183061$ & & & $113.782165$ & $326.274813$\\
Vc & d & $1.169$ & $8.425855$ & $25.747574$ & $0.144288$ & $27.627645$ & $59.952718$ & $35.857501$ & $50.193242$\\
& c & & $9.448184$ & $39.984360$ & $0.061401$ & & & $93.794336$ & $145.271020$\\
& b & & $7.707015$ & $69.811606$ & $0.027531$ & & & $17.258200$ & $338.762294$\\
\hline
& f & & $4.197927$ & $9.659082$ & $0.177303$ & & & $201.043842$ & $80.245495$\\
& e & & $6.698260$ & $15.660910$ & $0.168239$ & & & $290.074917$ & $324.85032$\\
Vd & d & $1.252$ & $7.178005$ & $25.705213$ & $0.137796$ & $25.568815$ & $243.953507$ & $213.186128$ & $50.262837$\\
& c & & $11.868008$ & $39.486207$ & $0.054102$ & & & $269.946387$ & $145.097451$\\
& b & & $6.527020$ & $68.597249$ & $0.017425$ & & & $217.631852$ & $314.164304$\\
\hline
\end{tabular}
\end{center}
\end{table*}

%
\section{Dynamical stability and numerical setup}
\label{sec:stability}
%
A marginal dynamical stability of HR~8799 system is a common problem  highlighted in the literature. Similarly to most papers published so far,  we solved numerically the Newtonian, $N$-body equations of motion to track the dynamical  evolution of particular best-fitting solutions. The results of  stability analysis in the same dynamical framework may be then easily compared with the previous studies. 

Planets and the host star in our numerical experiments are  approximated by point masses. The estimated system age in the range of $[30,160]$~Myr \citep{Marois2010} is  equivalent to $[6,32] \times 10^4$ orbital periods of the outermost planet. In such time-scales, the short-term dynamics is governed by low-order MMRs. Our dynamical model neglects the general relativity,  as well as conservative and dissipative tidal body--body interactions. Such perturbations are scaled with large negative powers of semi-major axes \cite[e.g.,][and references therein]{Migaszewski2012}. We also do not include tidal interactions of the planets with two remnant disks. In the recent paper, \cite{Moore2013} show that the outer disk might influence the system stability. However, their $N$-body four-planet model is marginally stable only for a few Myr. We postpone a similar study making use of updated initial conditions of the HR~8799 system to another work.  

The conservative, Newtonian $N$-body model  permits to introduce at least two notions of the dynamical stability. The direct, long term numerical integrations make it possible to investigate the Lagrange or astronomical stability. The astronomical stability may be expressed by the event time $T_{\idm{E}}$ of a collision between planets (orbits crossing) or an ejection of a body from the system.  The direct 160~Myr integrations of the equations of motion for one initial condition of five-planet system require CPU time counted in hours.  Such a significant CPU overhead is not suitable to illustrate the global dynamics of the system.

Stability of planetary configurations may be also expressed through the maximal Lyapunov characteristic exponent $\lambda$ (MLCE). The MLCE is a fundamental measure of the divergence of initially close trajectories in the phase space. A non-zero value of MLCE indicates a chaotic (unstable) system.  Chaotic motions in a regime of strong, low-order MMRs may lead to short event times
\cite[e.g.][]{Gozdziewski2008a}.  To compute MLCE, we use an effective algorithm of the Mean Exponential Growth factor of Nearby Orbits $\left<Y\right>$
\cite[MEGNO][]{Cincotta2000,Cincotta2003}. An uniform definition of this  fast indicator describes two basic classes of motions in the phase space:
\[
 \lim_{t\rightarrow \infty} \left<Y(t)\right> = \frac{1}{2}\lambda t + d,  
\]
where for the regular, stable quasi-periodic solutions $\lambda=0$, $d \simeq 2$, and for chaotic (unstable solutions) $\lambda>0$ and $d \simeq 0$.  MLCE measures the slope of linear function for the chaotic solutions, hence an approximation of MLCE after time~$t$:
\[
 \lambda(t) \sim 2 \frac{\left<Y(t)\right>}{t}.
\]
This technique requires relatively short arcs of the phase space trajectories, equivalent to a few of $10^4$ orbital periods of the outermost planet (characteristic periods; let us recall that $10^4$ outermost orbital periods of HR~8799b  translates to roughly 5~Myr). This makes it possible to construct high-resolution dynamical maps in selected planes of orbital parameters with much smaller CPU overhead than required by the direct $N$-body integrations.

There is no simple nor uniform relation between the MLCE and the event time. We performed a number of experiments to calibrate a link between the MEGNO interval and the direct integration time. 

To examine the dynamical stability of isolated best-fitting models gathered in
Figs.~\ref{fig:fig7},\ref{fig:fig8},\ref{fig:fig10}, \ref{fig:fig12}
and~\ref{fig:fig14}, we computed MEGNO for the upper limit of the system age (160~Myr).  To conserve the total energy and angular momentum with the relative accuracy $\sim 10^{-9}$, we  apply the tangent map algorithm
\citep{Gozdziewski2008} that makes use of the fourth-order  ${\cal SABA}_4$ symplectic integration scheme \citep{Laskar2001}. Step sizes of the symplectic tangent mapping are usually 256 or 384~days for the four-planet configurations and 128 or 256~days for the five-planet configurations. We verified these settings with an accurate but 2-3 times less CPU-effective Bulirsh-Stoer-Gragg scheme [the ODEX code \cite{Wanner2001}]. 

A typical convergence of MEGNO indicating strictly quasi-periodic best-fitting model IVa is shown in the left panel  of Fig~\ref{fig:fig25}. The right panel compares two other solutions corresponding to five-planet models~Va and Vc, labelled with 1f:3e and 1f:2e, accordingly (see their orbital geometry in Fig.~\ref{fig:fig22} and elements in Table~\ref{tab:table4}).   The MEGNO convergence of the 1f:3e solution Va is similarly perfect for $100$~Myr.  This is not the case for the 1f:2e~MMR solution Vc. After $\sim 15-20$~Myr, the indicator starts to grow roughly in linear rate indicating weakly chaotic configuration. The direct $N$-body integration shows that this system disrupts due to crossing orbits after $\sim 120$~Myr.  This illustrates a well known instability due to secular interactions in multiple MMRs
\corr{\citep[e.g.][]{Murray2001,Guzzo2005,Gozdziewski2008,Quillen2011}}. A proper choice of the integration time to compute MLCE is a delicate matter
\citep{Sussman1992}.
  
\begin{figure*}
\centerline{
\hbox{
{\includegraphics[width=0.49\textwidth]{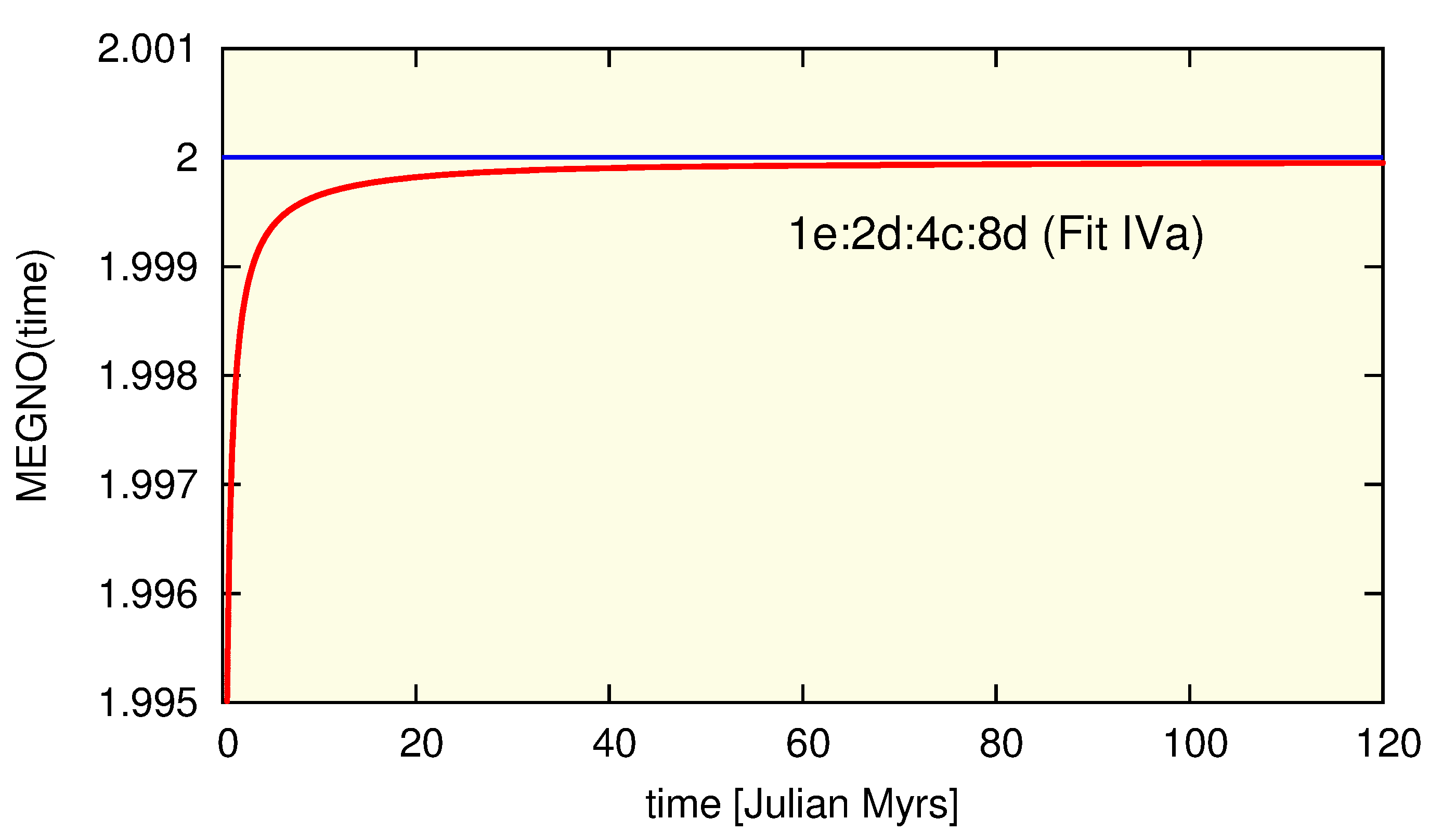}}
{\includegraphics[width=0.49\textwidth]{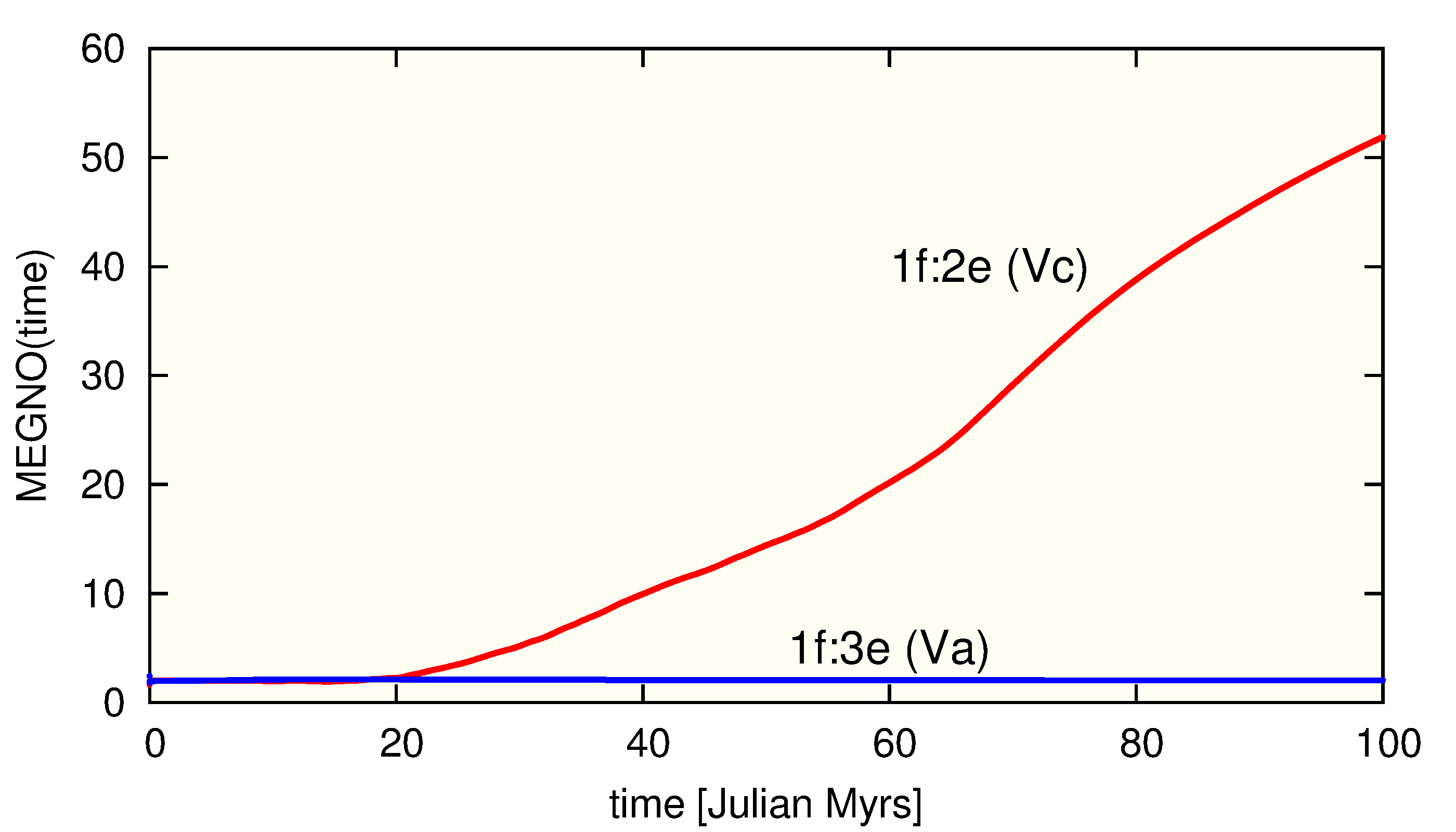}}
}
}
\caption{
{\em Left panel}: A typical convergence of the MEGNO fast indicator to 2 for a regular, quasi-periodic configuration. The initial condition corresponds to the best-fitting model IVa with four planets (Tab.~\ref{tab:table1}). {\em Right panel}: Temporal evolution of MEGNO for the best-fitting  five-planet models Va and Vb. See the dynamical maps in Fig.~\ref{fig:fig25} for a reference.
}
\label{fig:fig25}
\end{figure*}

To resolve the structure of the phase space, and to measure the width of MMRs, we computed a number of dynamical MEGNO maps. The MEGNO integration time was set to $\sim 5$--$12$~Myr, which is equivalent to $\sim 10,000$--$25,000$ characteristic periods. We carefully validated this choice. The MEGNO maps in the ($a_{\idm{c}},e_{\idm{c}}$)-plane for the best-fitting four-planets model IVa (Fig.~\ref{fig:fig9}), and the five-planet models Va  and Vc (Fig.~\ref{fig:fig23}) were compared with the event time maps in Fig.~\ref{fig:fig26} and Fig.~\ref{fig:fig27}.  In the MEGNO maps, similar to all such maps in the text, stable, quasi-periodic solutions with $\left<Y\right> \simeq 2$ are always marked in blue.  Yellow color encodes strongly unstable (chaotic) models with $\left<Y\right> \geq 5$, as usually the integrations were stopped if $\left<Y\right> \geq 5$.   Intermediate values of MEGNO between 2 and 5 correspond to chaotic solutions too.  

For the event time maps in Fig.~\ref{fig:fig26} (top-left panel) and Fig.~\ref{fig:fig27}, the equations of motions were integrated  for the maximum time span of 160~Myr. Maps in the top-right and bottom-left panels of Fig.~\ref{fig:fig26}) were integrated for longer intervals.  In this way, three subsequent panels in Fig.~\ref{fig:fig26} are for gradually increased limit of the integration time from 160~Myr (top-left panel), to 500~Myr (top-right panel) and 1~Gyr (bottom-left panel). We may observe that the maximal $T_{\idm{E}}$ zone shrinks with the longer integration times. However, the boundary of $T_{\idm{E}}$ map for 1~Gyr perfectly matches the boundary of quasi-periodic motions in the MEGNO map (see top-rght panel in Fig.\ref{fig:fig9}). This experiment assures us that initial conditions of four-planet models identified with MEGNO as quasi-periodic for time $T \sim 10$~Myr (typical integration period of MEGNO) may be safely considered as astronomically stable for $\sim 10^2~T$ that translates to $\sim 1$~Gyr interval encompassing all current estimates of the HR~8799 lifetime. This estimate for five-planet systems might be too optimistic, however the MEGNO maps match at least the $T_{\idm{E}}$ maps computed for the maximal time of 160~Myr.

Remarkably, a single integration of symplectic MEGNO for $\sim 10$--$15$~Myr takes roughly up to 10~minutes of CPU per initial condition, depending on the $N$-body model, integrator and the mapping step size.  Even with such reasonable CPU overhead, a quasi-global analysis of the dynamical stability still require significant CPU power which is inaccessible on a single workstation. For instance, 
a single symplectic MEGNO map in the $720\times360$ resolution occupied up to 512~CPU cores for up to 1-2~days supercomputers {\tt chimera} and {\tt cane} installed in the Pozna\'n Supercomputer Centre PCSS, Poland.
To perform massive numerical experiments in this work, we applied our new CPU cluster environment \mechanic{} \citep{Slonina2014} and simple codes written with the standard Message Passing Interface (MPI).  

\begin{figure*}
\centerline{
\vbox{
\hbox{
{\includegraphics[width=0.49\textwidth]{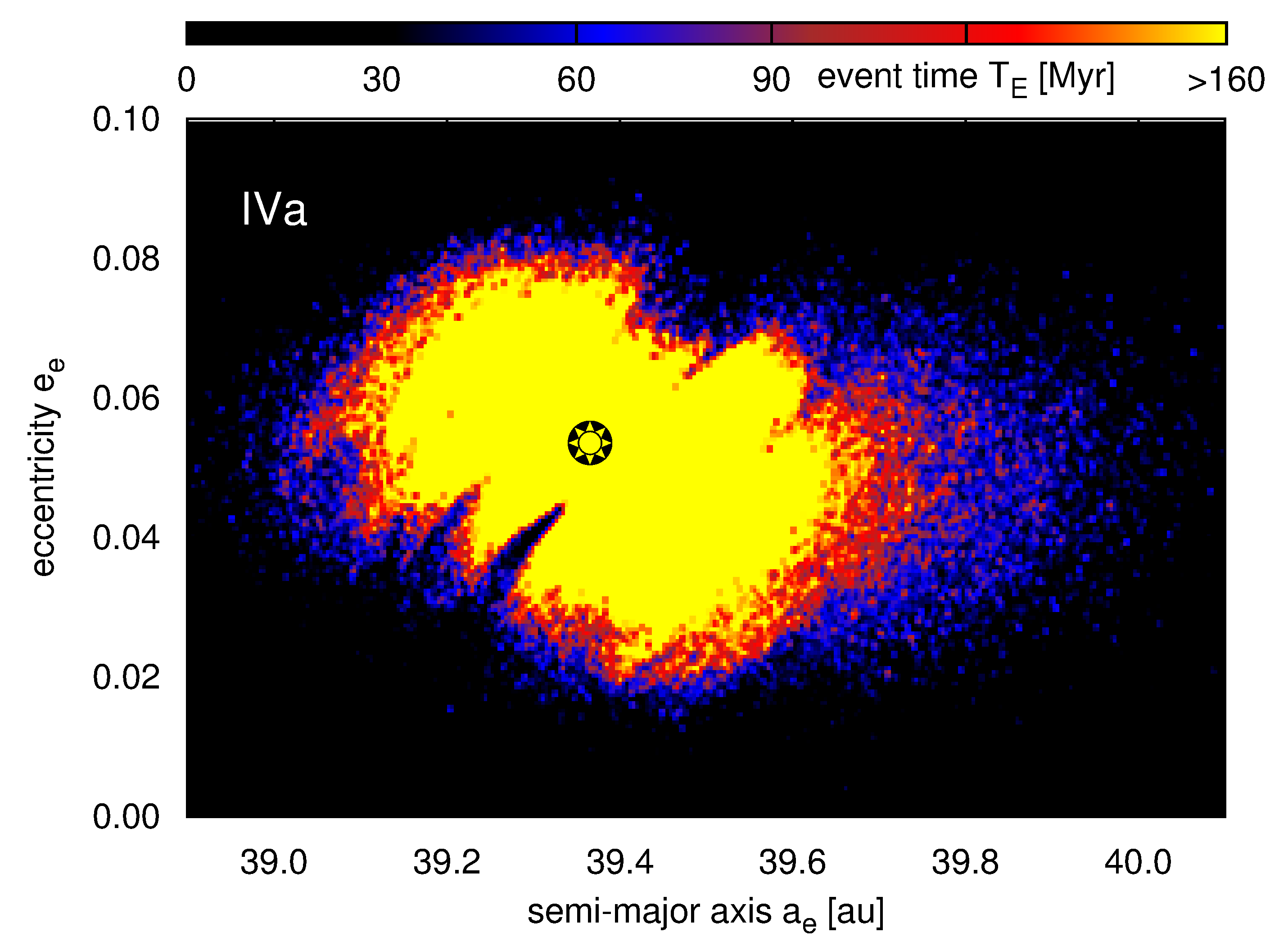}} 
{\includegraphics[width=0.49\textwidth]{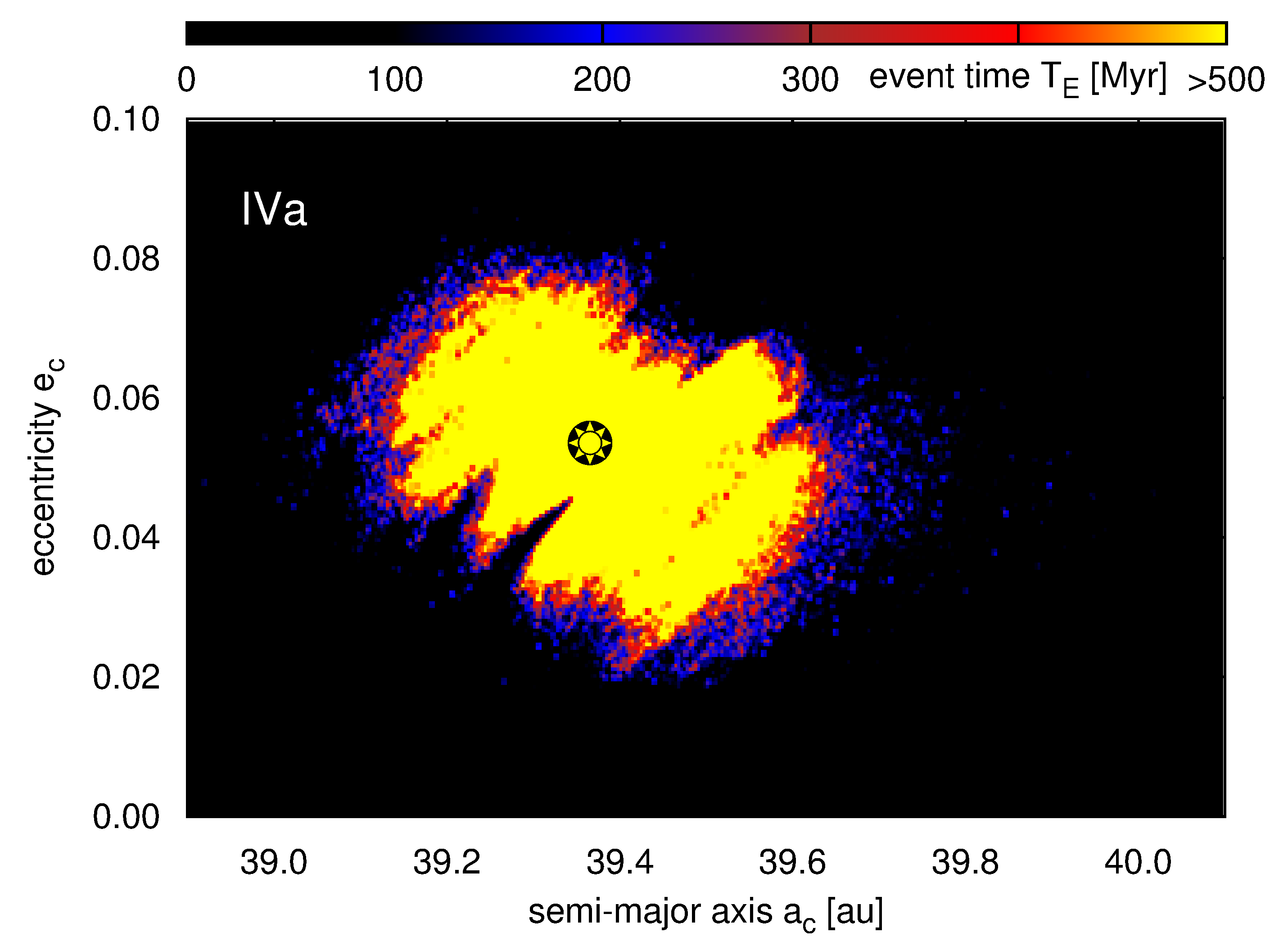}} 
}
\hbox{
{\includegraphics[width=0.49\textwidth]{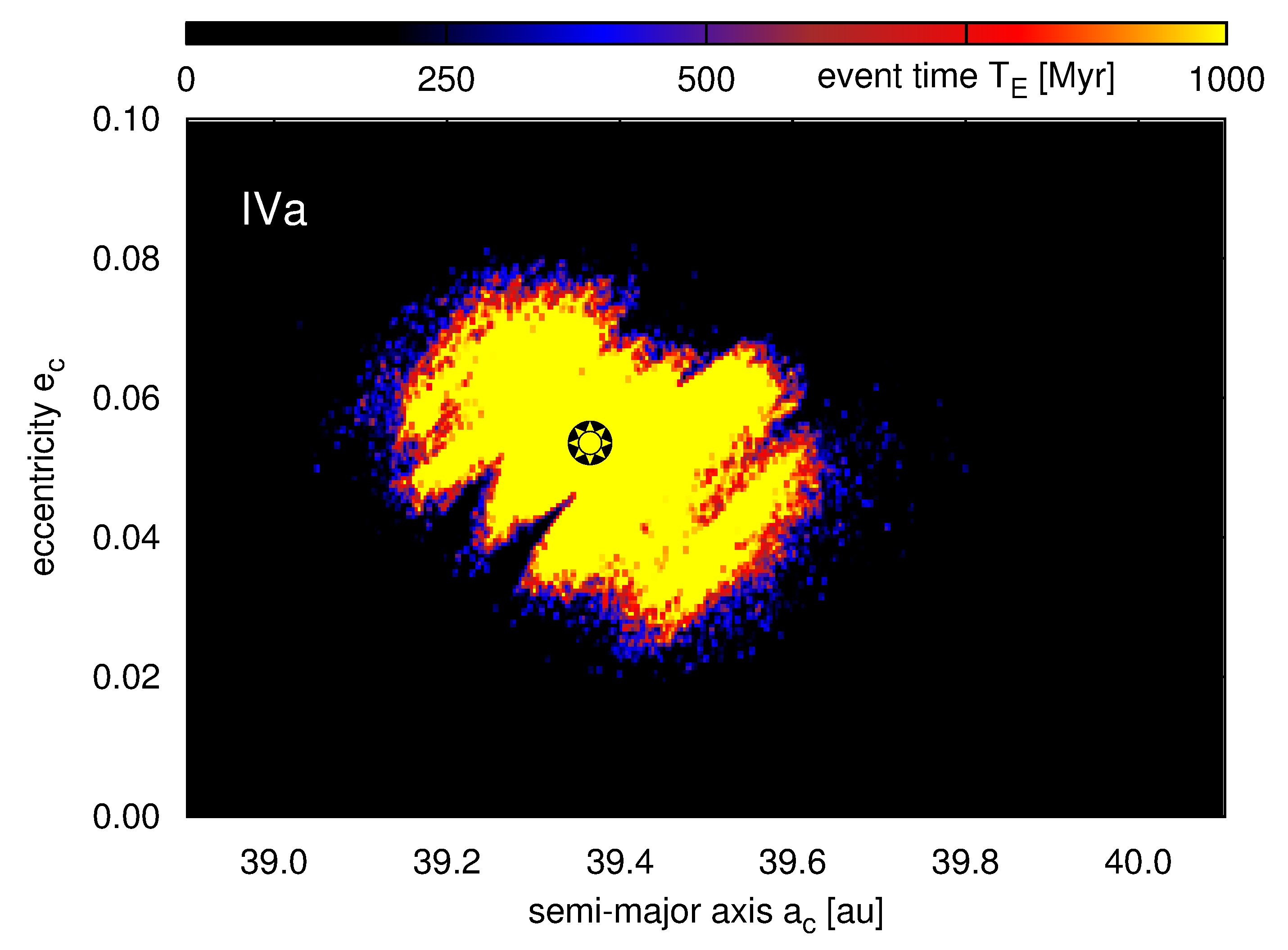}} 
{\includegraphics[width=0.49\textwidth]{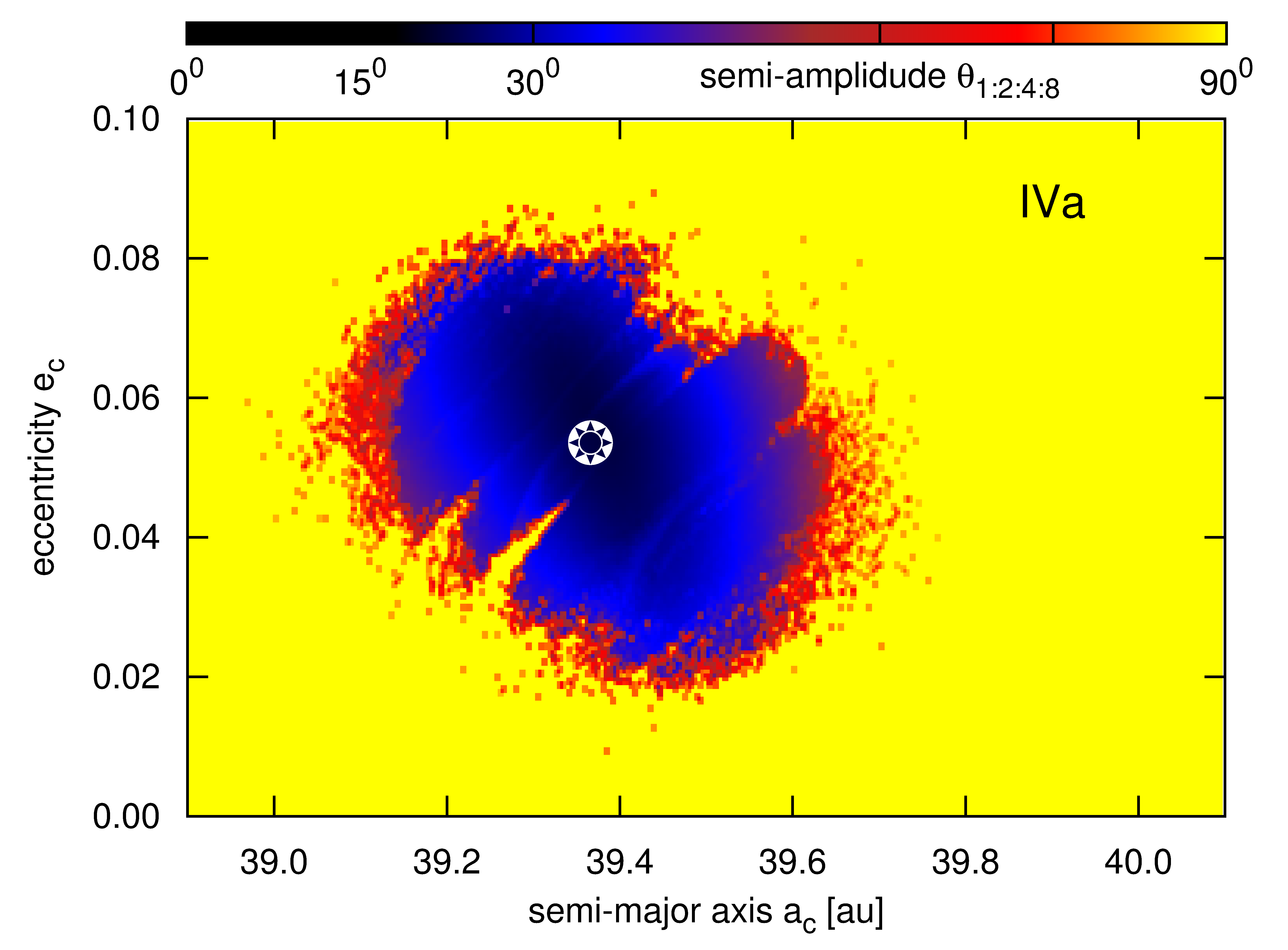}} 
}
}
}
\caption{
Event time $T_{\idm{E}}$ maps in the semi-major--eccentricity plane of planet HR8799~c in the four-planet model IVa (Tab.~\ref{tab:table1}). The maps are computed for gradually increased integration time: {\em top-left}: 160~Myr ($\sim 3\times10^5$~outermost periods), {\em top-right}: 500~Myr ($\sim 10^6$~outermost periods), {\em bottom-left}: 1~Gyr ($\sim 2\times 10^6$~outermost periods).  The event time corresponds to orbit crossing or ejection a planet from the system and is colour-coded. Yellow colour encodes configurations lifetime is longer than the integration interval. The best-fitting system IVa marked with the star symbol is locked deeply in the 1e:2d:4c:8b~MMR. The {\em bottom-right} panels shows the semi-amplitude of the critical angle librating around $0^{\circ}$ in the exact resonance. The resolution of each map is $360\times180$ pixels.  
}
\label{fig:fig26}
\end{figure*}

\begin{figure*}
\centerline{
\vbox{
\hbox{
{\includegraphics[width=0.49\textwidth]{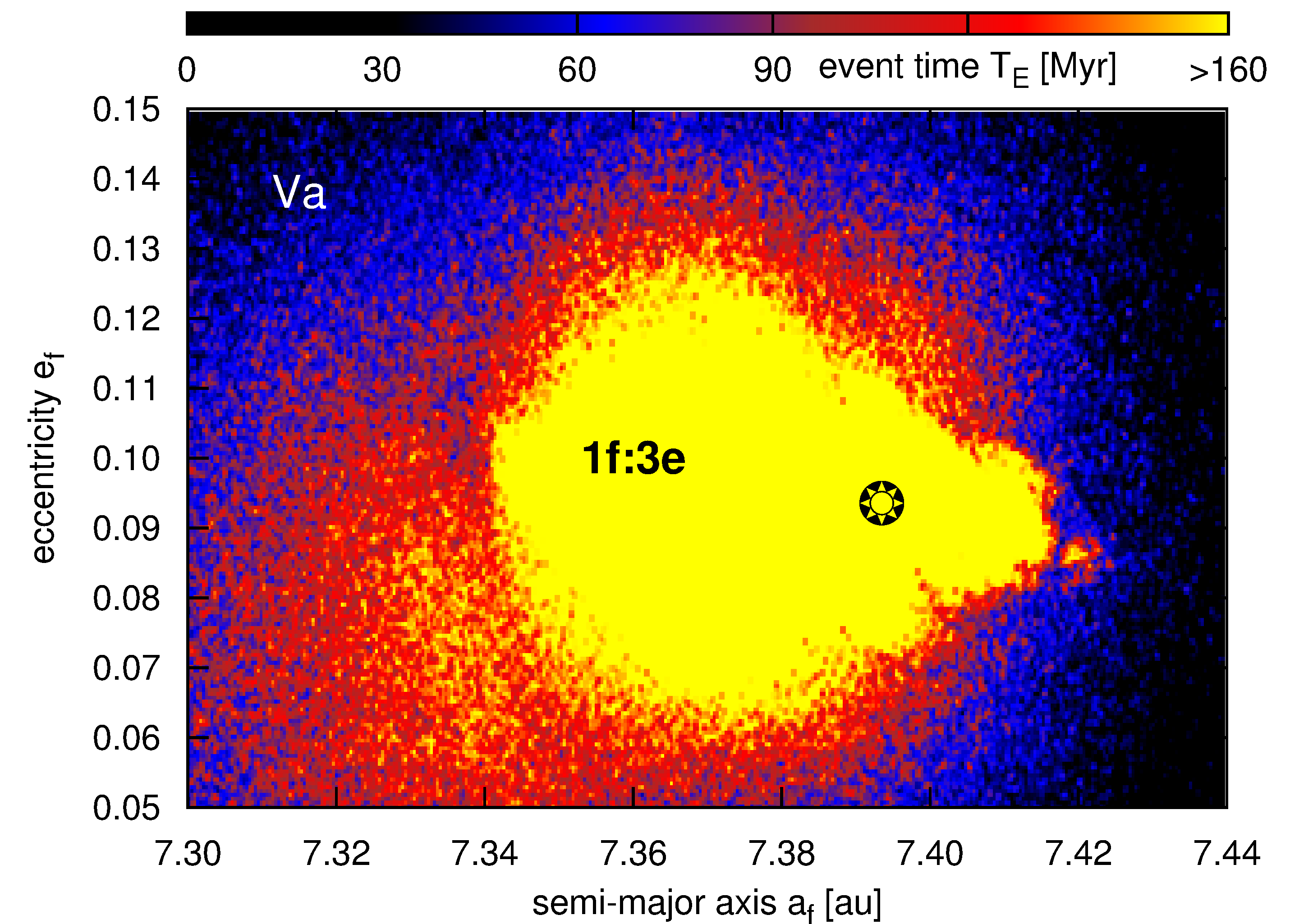}}
{\includegraphics[width=0.49\textwidth]{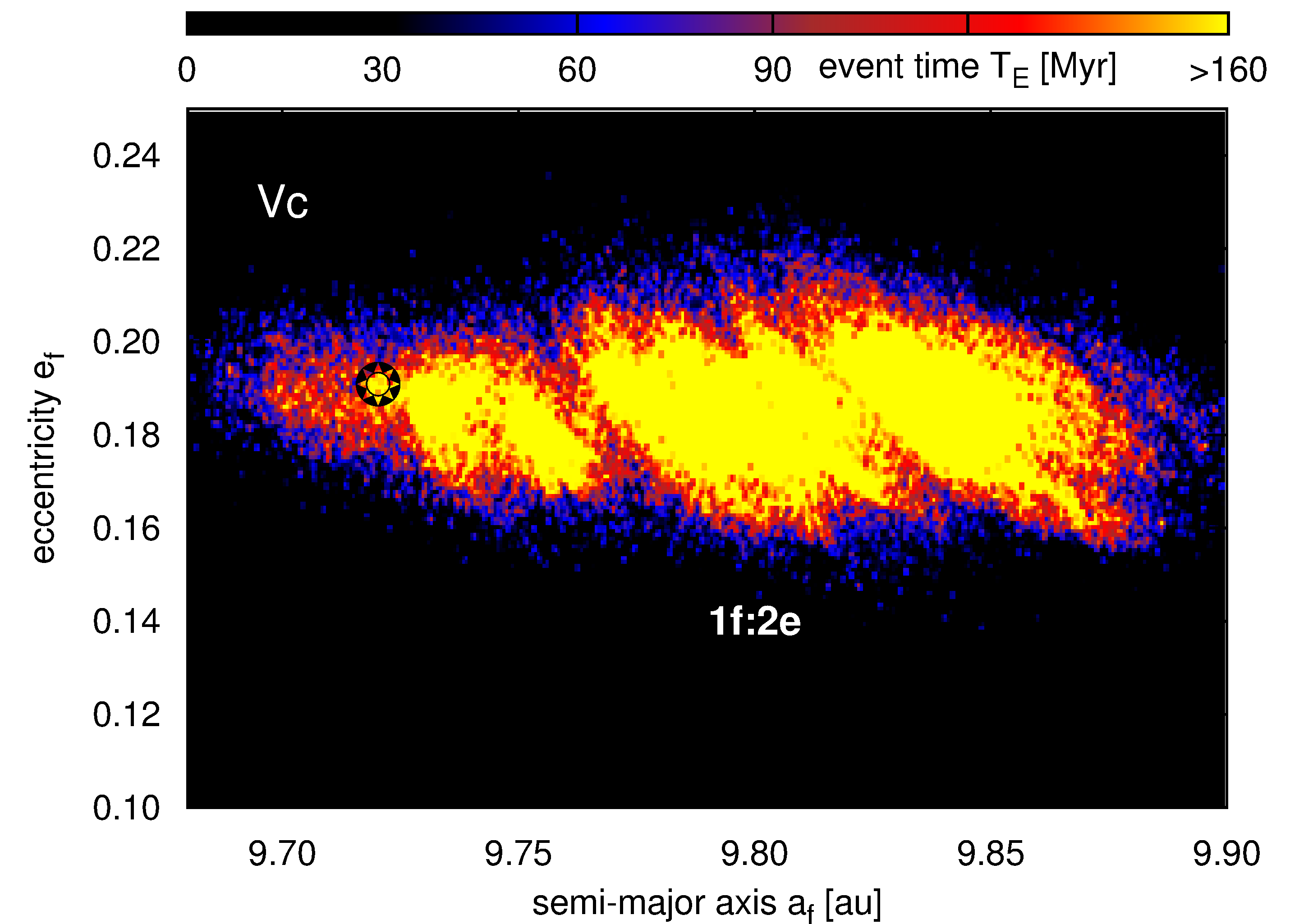}}
}
\hbox{
{\includegraphics[width=0.49\textwidth]{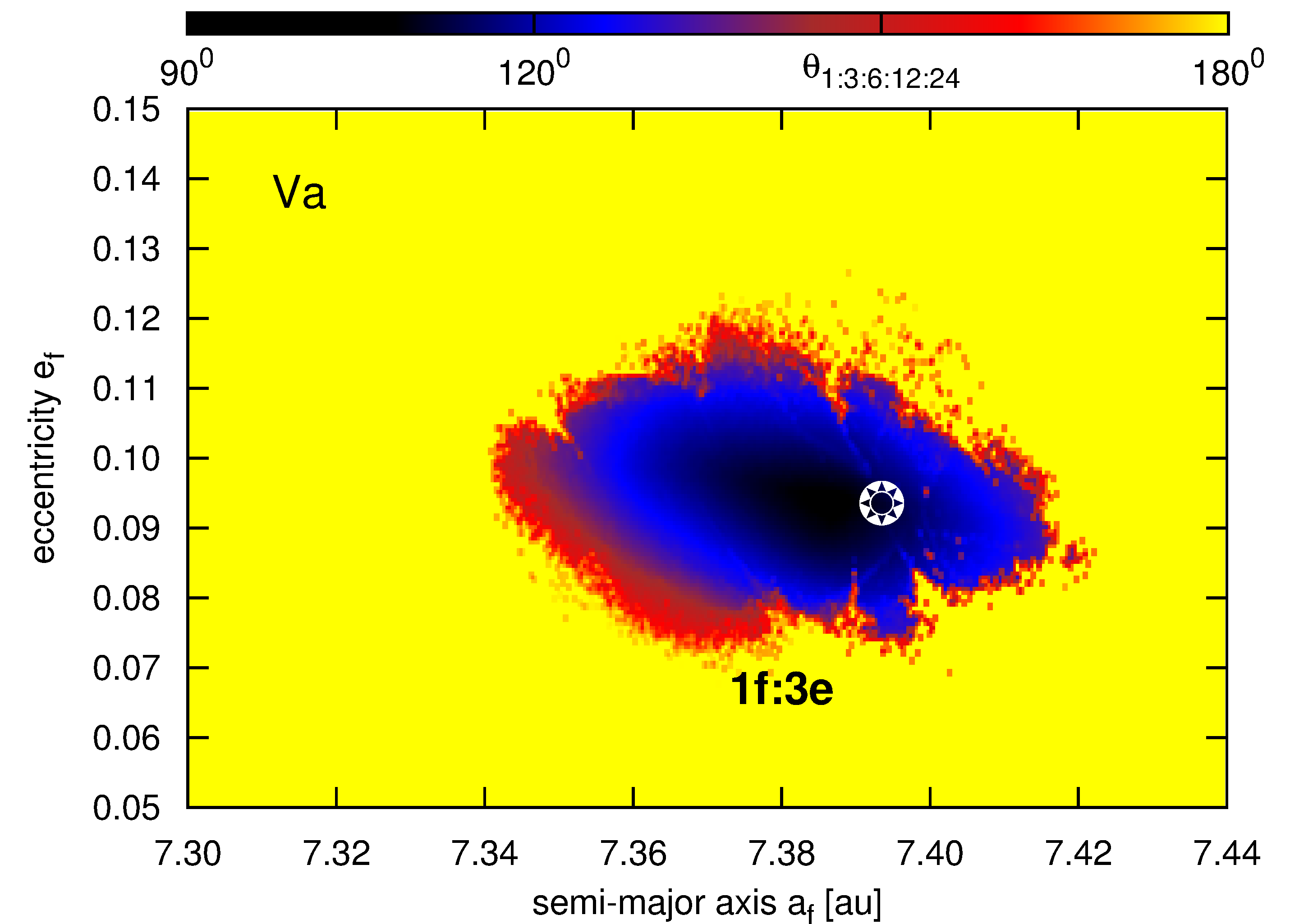}}
{\includegraphics[width=0.49\textwidth]{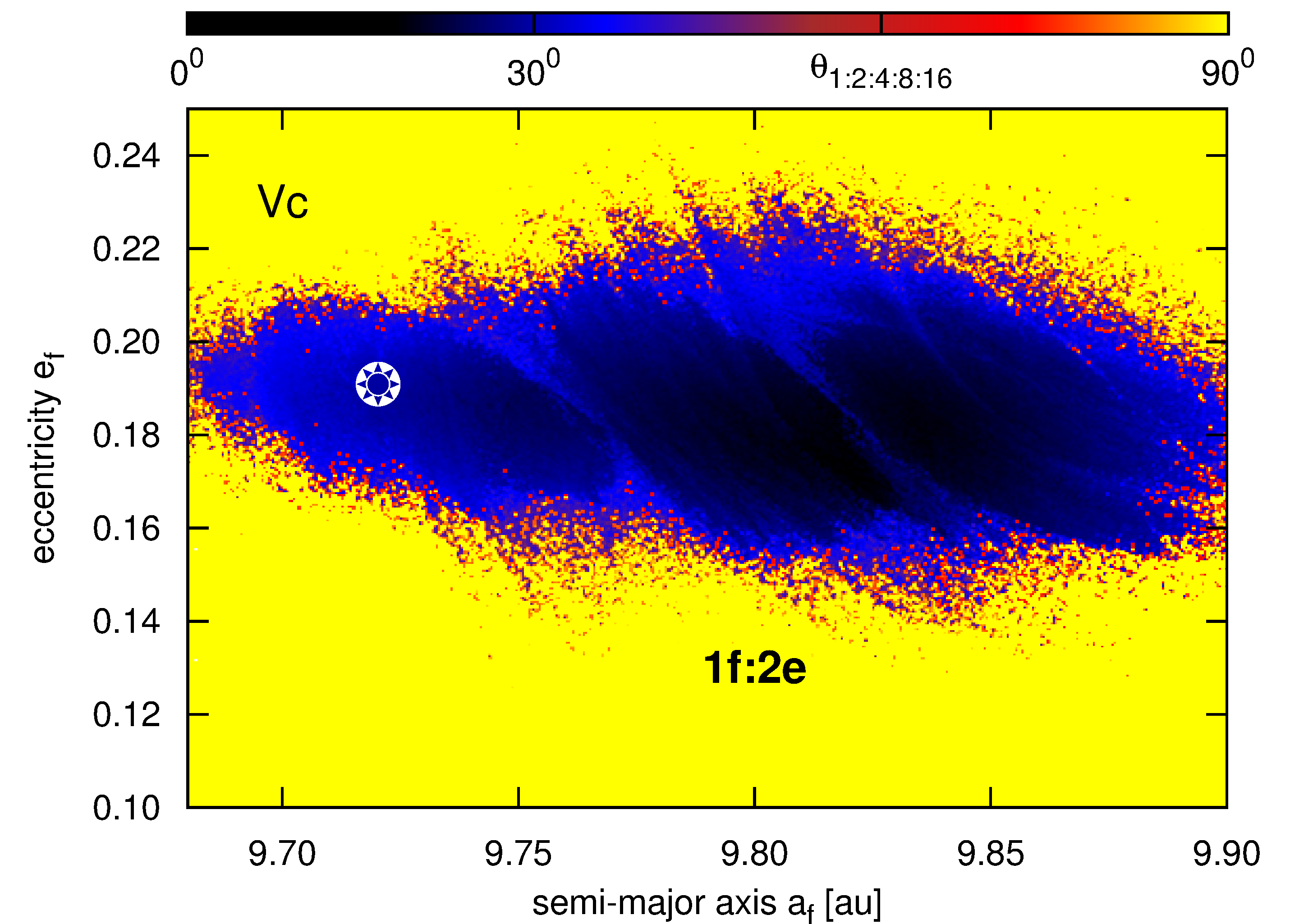}}
}
}
}
\caption{
\corr{
Event time $T_{\idm{E}}$ and critical angle $\theta$ maps in the semi-major
axis -- eccentricity plane for planet HR8799~f in five-planet models Va ({\em left column}) and Vc ({\em right column}). {\em Top row}: the event time corresponds to orbit crossing or ejection a planet from the system and is colour-coded. The resolution of each map is $360\times180$ pixels.   The maximal integration time for each pixel is $160$~Myr ($3.2 \times 10^5$ characteristic periods).  {\em Bottom row}: semi-amplitude of the critical angles  $\theta_{\idm{1:3:6:12:24}}= \lambda_{\idm{f}} - 4\lambda_{\idm{e}} +2\lambda_{\idm{d}}-\lambda_{\idm{c}} + 2\lambda_{\idm{b}}$ (which librates around $90^{\circ}$, {\em left}) and  $\theta_{\idm{1:3:6:12:24}}= \lambda_{\idm{f}} - 3\lambda_{\idm{e}} +3\lambda_{\idm{d}}-3\lambda_{\idm{c}} + 2\lambda_{\idm{b}}$ (which librates around $\approx15^{\circ}$, {\em right}), respectively. The best-fitting systems Va and Vc are marked with the star symbol \corr{(see their elements in Tab.~\ref{tab:table4})}.
}
}
\label{fig:fig27}
\end{figure*}

%
\section{Conclusions}
\label{sec:conclusions}
%
The HR~8799 is one of most exotic and intriguing extrasolar planetary systems detected so far. Any definite conclusion on its  current state suffers from ambiguities concerning its formation history, age and companion masses.  The astrometric data still cover only tiny arcs of the orbits and are relatively inaccurate. To overcome this problem,
we invented new optimization algorithm constrained through planetary migration (\moa{}).  This algorithm makes it possible to derive a self-consistent dynamical model of the HR~8799 system perfectly matching all observations and independent mass estimates. This model is rigorously stable for any estimate of the star age, between 30~Myr and 1~Gyr. Our results are suggestive for the formation of all HR~8799 planets in wider orbits that migrated shortly to their observed orbits. Indeed, the signatures of an extremely massive and dense protoplanetary disk revealed by  \spitzer{} observations, indicate both a rapid formation of massive planets and their common migration. Moreover, the orbits are locked into an amazingly ordered chain of double Laplace MMR 1e:2d:4c:8de. The mass ranges remain below the brown dwarf limit and confirm dynamically the planetary nature of the HR~8799 companions.

False assumptions may yield correct conclusions. However, we did not find any outcome of the \moa{}  that contradicts the results of independent astrophysical theory. Moreover, our model of the HR~8799 system predicts basically unique ephemeris of the four- and five-planet configurations that might be verified shortly on the observational basis. A reanalysis of earlier data in the discovery paper \citep{Marois2008} confirms the stability analysis in the literature. We found that the Laplace MMR fully consistent with the astrophysical mass estimates $\sim 10$\,$\mJ$. \corr{The dynamically determined inclination and nodal line of the system orbits closely matches independent determination of the inclination and node of the outer debris disk \citep{Matthews2014}.} 

We also demonstrate that \moa{} is able to derive meaningful orbital characterization of a resonant system on the basis of short-interval data, essentially using a single-image detection. \corr{Our method may be useful to characterize long-period, resonant systems with massive planets detected by the direct imaging.} 

The migration mechanism and resonance trapping is likely responsible, although in quite smaller scale when concerning the orbits and planetary masses, for creating multiple systems of super-Earth planets discovered by the \kepler{} mission \citep{Borucki2011,Batalha2013}. A significant sample of \kepler{} systems involving four and more planets are found very close to multiple MMRs which might be also explained by a common, inward migration \citep[][and references therein]{Migaszewski2013}.

The early dynamical models of the three-planet HR~8799 system configurations indicated marginally stable, chaotic system. After the discovery of the fourth planet, these models become apparently even more unstable. \corr{In contrast, our new models} derived with the migration constraints, and comprising of even five giant planets are suggestive for a completely ordered configuration which could be stable forever, if no significant perturbations are introduced. 

\section*{Acknowledgments}
Many thanks to Daniel Fabrycky for a review and comments that improved this manuscript.
This work was supported by the Polish Ministry of Science and Higher  Education, Grant N/N203/402739.  C.M.  is a recipient of the stipend of the Foundation for Polish Science (programme START, editions 2010 and 2011). This research was supported by computational resources provided through the POWIEW project, co-financed by the European Regional Development Fund under the Innovative Economy Operational Programme (POIG.02.03.00-00-018/08).

\begin{table*}
\caption{
Astrometric data in the literature for planet HR~8799~b.
}
\label{tab:data1}
\begin{tabular}{c c c c c c c c c c c c c c c}
\hline
epoch & \multicolumn{2}{c}{planet b} & reference &  \multicolumn{5}{c}{dataset}\\
& N [sec] & E [sec]  &  & D1 & D2 & D3 & D4 & D5\\
\hline
$1998.83$ & $1.411 \pm 0.009$ & $0.986 \pm 0.009$  & \cite{Lafreniere2009} & x & - & - & - & x\\
$1998.83$ & $1.418 \pm 0.022$ & $1.004 \pm 0.020$  & \cite{Soummer2011} & x & - & - & - & x\\
$2002.54$ & $1.481 \pm 0.023$ & $0.919 \pm 0.017$  & \cite{Fukagawa2009} & x & - & - & - & x\\
$2004.53$ & $1.471 \pm 0.005$ & $0.884 \pm 0.005$  & \cite{Marois2008} & x & x & - & - & x\\
$2005.54$ & $1.496 \pm 0.005$ & $0.856 \pm 0.005$  & \cite{Currie2012} & x & - & - & - & x\\
$2007.58$ & $1.522 \pm 0.003$ & $0.815 \pm 0.003$  & \cite{Metchev2009} & x & - & - & - & x\\
$2007.81$ & $1.512 \pm 0.005$ & $0.805 \pm 0.005$  & \cite{Marois2008} & x & x & - & - & x\\
$2008.52$ & $1.527 \pm 0.004$ & $0.799 \pm 0.004$  & \cite{Marois2008} & x & x & - & - & x\\
$2008.61$ & $1.527 \pm 0.002$ & $0.801 \pm 0.002$  & \cite{Marois2008} & x & x & x & - & x\\
$2008.71$ & $1.528 \pm 0.003$ & $0.798 \pm 0.003$  & \cite{Marois2008} & x & x & - & - & x\\
$2008.89$ & $1.532 \pm 0.020$ & $0.796 \pm 0.020$  & \cite{Currie2011} & x & - & - & - & x\\
$2008.89$ & $1.542 \pm 0.010$ & $0.780 \pm 0.014$  & \cite{Hinz2010} & x & - & - & - & x\\
$2009.62$ & $1.536 \pm 0.010$ & $0.785 \pm 0.010$  & \cite{Currie2011} & x & - & - & - & x\\
$2009.70$ & $1.538 \pm 0.030$ & $0.777 \pm 0.030$  & \cite{Currie2011} & x & - & - & - & x\\
$2009.76$ & $1.535 \pm 0.020$ & $0.816 \pm 0.020$  & \cite{Bergfors2011} & x & - & - & - & x\\
$2009.77$ & $1.532 \pm 0.007$ & $0.783 \pm 0.007$  & \cite{Currie2011} & x & - & - & \corr{x} & x\\
$2009.84$ & $1.540 \pm 0.019$ & $0.800 \pm 0.019$  & \cite{Galicher2011} & x & - & - & - & x\\
$2010.83$ & $1.546 \pm 0.005$ & $0.748 \pm 0.005$  & \cite{Currie2012} & x & - & - & - & x\\
$2011.79$ & $1.579 \pm 0.011$ & $0.734 \pm 0.011$  & \cite{Esposito2013} & x & - & - & - & x\\
$2011.86$ & $1.546 \pm 0.011$ & $0.725 \pm 0.011$  & \cite{Esposito2013} & x & - & - & - & x\\
\hline
\end{tabular}
\end{table*}

\begin{table*}
\caption{
Astrometric data in the literature for planet HR~8799~c.
}
\label{tab:data2}
\begin{tabular}{c c c c c c c c c c c c c c c}
\hline
epoch & \multicolumn{2}{c}{planet c} & reference &  \multicolumn{5}{c}{dataset}\\
& N [sec] & E [sec]  &  & D1 & D2 & D3 & D4 & D5\\
\hline
$1998.83$ & $-0.837 \pm 0.026$ & $0.483 \pm 0.023$ & \cite{Soummer2011} & x & - & - & - & x\\
$2004.53$ & $-0.739 \pm 0.005$ & $0.612 \pm 0.005$ & \cite{Marois2008} & x & x & - & - & x\\
$2005.54$ & $-0.713 \pm 0.005$ & $0.630 \pm 0.005$ & \cite{Currie2012} & x & - & - & - & x\\
$2007.58$ & $-0.672 \pm 0.005$ & $0.674 \pm 0.005$ & \cite{Metchev2009} & x & - & - & - & x\\
$2007.81$ & $-0.674 \pm 0.005$ & $0.681 \pm 0.005$ & \cite{Marois2008} & x & x & - & - & x\\
$2008.52$ & $-0.658 \pm 0.004$ & $0.701 \pm 0.004$ & \cite{Marois2008} & x & x & - & - & x\\
$2008.61$ & $-0.657 \pm 0.002$ & $0.706 \pm 0.002$ & \cite{Marois2008} & x & x & x & - & x\\
$2008.71$ & $-0.657 \pm 0.003$ & $0.706 \pm 0.003$ & \cite{Marois2008} & x & x & - & - & x\\
$2008.89$ & $-0.654 \pm 0.020$ & $0.700 \pm 0.020$ & \cite{Currie2011} & x & - & - & - & x\\
$2008.89$ & $-0.631 \pm 0.015$ & $0.671 \pm 0.020$ & \cite{Hinz2010} & x & - & - & - & x\\
$2009.02$ & $-0.612 \pm 0.030$ & $0.665 \pm 0.030$ & \cite{Hinz2010} & x & - & - & - & x\\
$2009.70$ & $-0.625 \pm 0.020$ & $0.725 \pm 0.020$ & \cite{Hinz2010} & x & - & - & - & x\\
$2009.70$ & $-0.634 \pm 0.030$ & $0.697 \pm 0.030$ & \cite{Currie2011} & x & - & - & - & x\\
$2009.76$ & $-0.636 \pm 0.040$ & $0.692 \pm 0.040$ & \cite{Bergfors2011} & x & - & - & - & x\\
$2009.77$ & $-0.627 \pm 0.007$ & $0.716 \pm 0.007$ & \cite{Currie2011} & x & - & - & x & x\\
$2009.84$ & $-0.630 \pm 0.013$ & $0.720 \pm 0.013$ & \cite{Galicher2011} & x & - & - & - & x\\
$2010.83$ & $-0.598 \pm 0.005$ & $0.737 \pm 0.005$ & \cite{Currie2012} & x & - & - & - & x\\
$2011.79$ & $-0.561 \pm 0.010$ & $0.752 \pm 0.010$ & \cite{Esposito2013} & x & - & - & - & x\\
$2011.86$ & $-0.578 \pm 0.010$ & $0.767 \pm 0.010$ & \cite{Esposito2013} & x & - & - & - & x\\
\hline
\end{tabular}
\end{table*}

\begin{table*}
\caption{
Astrometric data in the literature for planet HR~8799~d.
}
\label{tab:data3}
\begin{tabular}{c c c c c c c c c c c c c c c}
\hline
epoch & \multicolumn{2}{c}{planet d} & reference &  \multicolumn{5}{c}{dataset}\\
& N [sec] & E [sec]  &  & D1 & D2 & D3 & D4 & D5\\
\hline
$1998.83$ & $ 0.133 \pm 0.035$ & $-0.533 \pm 0.034$ & \cite{Soummer2011} & x & - & - & - & x\\
$2005.54$ & $-0.087 \pm 0.010$ & $-0.578 \pm 0.010$ & \cite{Currie2012} & x & - & - & - & x\\
$2007.58$ & $-0.170 \pm 0.008$ & $-0.589 \pm 0.008$ & \cite{Metchev2009} & x & - & - & - & x\\
$2008.52$ & $-0.208 \pm 0.004$ & $-0.582 \pm 0.004$ & \cite{Marois2008} & x & x & - & - & x\\
$2008.61$ & $-0.216 \pm 0.002$ & $-0.582 \pm 0.002$ & \cite{Marois2008} & x & x & x & - & x\\
$2008.71$ & $-0.216 \pm 0.003$ & $-0.582 \pm 0.003$ & \cite{Marois2008} & x & x & - & - & x\\
$2008.89$ & $-0.217 \pm 0.020$ & $-0.608 \pm 0.020$ & \cite{Currie2011} & x & - & - & - & x\\
$2008.89$ & $-0.215 \pm 0.021$ & $-0.644 \pm 0.013$ & \cite{Hinz2010} & x & - & - & - & x\\
$2009.70$ & $-0.282 \pm 0.030$ & $-0.590 \pm 0.030$ & \cite{Hinz2010} & x & - & - & - & x\\
$2009.76$ & $-0.270 \pm 0.070$ & $-0.600 \pm 0.070$ & \cite{Bergfors2011} & x & - & - & - & x\\
$2009.77$ & $-0.241 \pm 0.007$ & $-0.586 \pm 0.007$ & \cite{Currie2011} & x & - & - & x & x\\
$2009.84$ & $-0.240 \pm 0.014$ & $-0.580 \pm 0.014$ & \cite{Galicher2011} & x & - & - & - & x\\
$2010.83$ & $-0.283 \pm 0.005$ & $-0.567 \pm 0.005$ & \cite{Currie2012} & x & - & - & - & x\\
$2011.79$ & $-0.299 \pm 0.010$ & $-0.563 \pm 0.010$ & \cite{Esposito2013} & x & - & - & - & x\\
$2011.86$ & $-0.320 \pm 0.010$ & $-0.549 \pm 0.010$ & \cite{Esposito2013} & x & - & - & - & x\\
\hline
\end{tabular}
\end{table*}

\begin{table*}
\caption{
Astrometric data in the literature for planet HR~8799~e.
}
\label{tab:data4}
\begin{tabular}{c c c c c c c c c c c c c c c}
\hline
epoch & \multicolumn{2}{c}{planet e} & reference &  \multicolumn{5}{c}{dataset}\\
& N [sec] & E [sec]  &  & D1 & D2 & D3 & D4 & D5\\
\hline
$2009.58$ & $-0.299 \pm 0.019$ & $-0.217 \pm 0.019$ & \cite{Marois2010} & x & - & - & - & -\\
$2009.58$ & $-0.303 \pm 0.013$ & $-0.209 \pm 0.013$ & \cite{Marois2010} & x & - & - & - & -\\
$2009.77$ & $-0.306 \pm 0.007$ & $-0.217 \pm 0.007$ & \cite{Currie2011} & x & - & - & x & -\\
$2009.83$ & $-0.304 \pm 0.010$ & $-0.196 \pm 0.010$ & \cite{Marois2010} & x & - & - & - & -\\
$2010.53$ & $-0.324 \pm 0.008$ & $-0.174 \pm 0.008$ & \cite{Marois2010} & x & - & - & - & -\\
$2010.55$ & $-0.324 \pm 0.011$ & $-0.175 \pm 0.011$ & \cite{Marois2010} & x & - & - & - & -\\
$2010.83$ & $-0.312 \pm 0.010$ & $-0.151 \pm 0.010$ & \cite{Marois2010} & x & - & - & - & -\\
$2011.79$ & $-0.326 \pm 0.011$ & $-0.119 \pm 0.011$ & \cite{Esposito2013} & x & - & - & - & -\\
$2011.86$ & $-0.382 \pm 0.011$ & $-0.127 \pm 0.011$ & \cite{Esposito2013} & x & - & - & - & -\\
\hline
\end{tabular}
\end{table*}

\setcounter{table}{0}
\makeatletter 
\renewcommand{\thetable}{A\@arabic\c@table}

\begin{table*}
\caption{
Ephemeris of the best-fitting four-planet model IVa (1e:2d:4c:8b~MMR) between epochs 
{$1995.0$ and $2020.0$}.  
Astrometric data in the $[E, N]$-plane.
}
\label{tab:ephemeris1}
\begin{tabular}{l c c c c c c c c}
\hline
epoch & \multicolumn{2}{c}{planet e} & \multicolumn{2}{c}{planet d} & \multicolumn{2}{c}{planet c} & \multicolumn{2}{c}{planet b} \\
& $\Delta \alpha$ [sec] & $\Delta \delta$ [sec] & $\Delta \alpha$ [arc
sec] & $\Delta \delta$ [sec] & $\Delta \alpha$ [sec] & $\Delta \delta$
 [sec] & $\Delta \alpha$ [sec] & $\Delta \delta$ [sec]\\
\hline
$1995$ & $0.3598$ & $-0.0464$ & $0.3055$ & $-0.4425$ & $-0.9031$ & $0.3842$ & $1.3624$ & $1.0481$ \\
$1996$ & $0.3395$ & $-0.0931$ & $0.2712$ & $-0.4679$ & $-0.8894$ & $0.4101$ & $1.3764$ & $1.0308$ \\
$1997$ & $0.3123$ & $-0.1380$ & $0.2356$ & $-0.4909$ & $-0.8748$ & $0.4356$ & $1.3901$ & $1.0133$ \\
$1998$ & $0.2785$ & $-0.1801$ & $0.1989$ & $-0.5117$ & $-0.8596$ & $0.4608$ & $1.4035$ & $0.9956$ \\
$1999$ & $0.2385$ & $-0.2181$ & $0.1613$ & $-0.5301$ & $-0.8436$ & $0.4856$ & $1.4166$ & $0.9778$ \\
$2000$ & $0.1930$ & $-0.2512$ & $0.1229$ & $-0.5460$ & $-0.8269$ & $0.5100$ & $1.4293$ & $0.9597$ \\
$2001$ & $0.1430$ & $-0.2783$ & $0.0840$ & $-0.5596$ & $-0.8094$ & $0.5340$ & $1.4418$ & $0.9416$ \\
$2002$ & $0.0895$ & $-0.2988$ & $0.0448$ & $-0.5706$ & $-0.7913$ & $0.5576$ & $1.4539$ & $0.9233$ \\
$2003$ & $0.0338$ & $-0.3118$ & $0.0053$ & $-0.5793$ & $-0.7725$ & $0.5807$ & $1.4657$ & $0.9048$ \\
$2004$ & $-0.0228$ & $-0.3172$ & $-0.0341$ & $-0.5854$ & $-0.7530$ & $0.6034$ & $1.4773$ & $0.8863$ \\
$2005$ & $-0.0788$ & $-0.3147$ & $-0.0735$ & $-0.5892$ & $-0.7329$ & $0.6256$ & $1.4886$ & $0.8676$ \\
$2006$ & $-0.1329$ & $-0.3045$ & $-0.1125$ & $-0.5906$ & $-0.7121$ & $0.6473$ & $1.4995$ & $0.8488$ \\
$2007$ & $-0.1838$ & $-0.2871$ & $-0.1511$ & $-0.5896$ & $-0.6906$ & $0.6685$ & $1.5102$ & $0.8299$ \\
$2008$ & $-0.2304$ & $-0.2630$ & $-0.1891$ & $-0.5864$ & $-0.6686$ & $0.6892$ & $1.5207$ & $0.8108$ \\
$2009$ & $-0.2718$ & $-0.2330$ & $-0.2264$ & $-0.5809$ & $-0.6459$ & $0.7092$ & $1.5308$ & $0.7916$ \\
$2010$ & $-0.3073$ & $-0.1980$ & $-0.2628$ & $-0.5733$ & $-0.6226$ & $0.7287$ & $1.5407$ & $0.7723$ \\
$2011$ & $-0.3366$ & $-0.1589$ & $-0.2983$ & $-0.5636$ & $-0.5987$ & $0.7476$ & $1.5504$ & $0.7529$ \\
$2012$ & $-0.3592$ & $-0.1168$ & $-0.3327$ & $-0.5519$ & $-0.5743$ & $0.7659$ & $1.5597$ & $0.7334$ \\
$2013$ & $-0.3751$ & $-0.0724$ & $-0.3659$ & $-0.5383$ & $-0.5494$ & $0.7835$ & $1.5688$ & $0.7137$ \\
$2014$ & $-0.3844$ & $-0.0268$ & $-0.3979$ & $-0.5228$ & $-0.5239$ & $0.8005$ & $1.5777$ & $0.6939$ \\
$2015$ & $-0.3871$ & $0.0193$ & $-0.4284$ & $-0.5056$ & $-0.4979$ & $0.8168$ & $1.5862$ & $0.6740$ \\
$2016$ & $-0.3837$ & $0.0651$ & $-0.4576$ & $-0.4866$ & $-0.4715$ & $0.8324$ & $1.5945$ & $0.6539$ \\
$2017$ & $-0.3743$ & $0.1098$ & $-0.4852$ & $-0.4661$ & $-0.4446$ & $0.8472$ & $1.6026$ & $0.6337$ \\
$2018$ & $-0.3594$ & $0.1530$ & $-0.5112$ & $-0.4441$ & $-0.4173$ & $0.8614$ & $1.6103$ & $0.6134$ \\
$2019$ & $-0.3393$ & $0.1940$ & $-0.5356$ & $-0.4207$ & $-0.3896$ & $0.8747$ & $1.6178$ & $0.5929$ \\
$2020$ & $-0.3146$ & $0.2323$ & $-0.5582$ & $-0.3960$ & $-0.3616$ & $0.8874$ & $1.6249$ & $0.5723$ \\
\hline
\end{tabular}
\end{table*}

\begin{table*}
\caption{
Ephemeris of the best-fitting five-planet model 
Va (1f:3e~MMR). Epochs between $1995.0$ and $2020.0$.
Astrometric coordinates in the $[E, N]$-plane.
}
\label{tab:ephemeris2}
\begin{tabular}{l c c c c c c c c c c}
\hline
epoch & \multicolumn{2}{c}{planet f} & \multicolumn{2}{c}{planet e} & \multicolumn{2}{c}{planet d} & \multicolumn{2}{c}{planet c} & \multicolumn{2}{c}{planet b} \\
& $\Delta \alpha$ [sec] & $\Delta \delta$ [sec] & 
$\Delta \alpha$ [sec] & $\Delta \delta$ [sec] & $\Delta \alpha$ [arc
sec] & $\Delta \delta$ [sec] & $\Delta \alpha$ [sec] & $\Delta \delta$ [sec] & $\Delta \alpha$ [sec] & $\Delta \delta$ [sec]\\
\hline
$1995$ & $0.1595$ & $-0.0585$ & $0.3625$ & $-0.0384$ & $0.3076$ & $-0.4439$ & $-0.9050$ & $0.3826$ & $1.3646$ & $1.0480$ \\
$1996$ & $0.1203$ & $-0.1197$ & $0.3421$ & $-0.0857$ & $0.2732$ & $-0.4693$ & $-0.8912$ & $0.4083$ & $1.3785$ & $1.0305$ \\
$1997$ & $0.0625$ & $-0.1626$ & $0.3148$ & $-0.1312$ & $0.2373$ & $-0.4925$ & $-0.8767$ & $0.4337$ & $1.3920$ & $1.0129$ \\
$1998$ & $-0.0042$ & $-0.1829$ & $0.2806$ & $-0.1739$ & $0.2003$ & $-0.5132$ & $-0.8615$ & $0.4589$ & $1.4051$ & $0.9951$ \\
$1999$ & $-0.0703$ & $-0.1798$ & $0.2402$ & $-0.2126$ & $0.1624$ & $-0.5316$ & $-0.8456$ & $0.4837$ & $1.4179$ & $0.9773$ \\
$2000$ & $-0.1281$ & $-0.1551$ & $0.1941$ & $-0.2463$ & $0.1238$ & $-0.5474$ & $-0.8289$ & $0.5083$ & $1.4304$ & $0.9595$ \\
$2001$ & $-0.1709$ & $-0.1122$ & $0.1434$ & $-0.2739$ & $0.0847$ & $-0.5608$ & $-0.8114$ & $0.5325$ & $1.4427$ & $0.9415$ \\
$2002$ & $-0.1937$ & $-0.0563$ & $0.0891$ & $-0.2946$ & $0.0454$ & $-0.5717$ & $-0.7932$ & $0.5564$ & $1.4548$ & $0.9234$ \\
$2003$ & $-0.1931$ & $0.0065$ & $0.0325$ & $-0.3078$ & $0.0059$ & $-0.5801$ & $-0.7742$ & $0.5798$ & $1.4667$ & $0.9053$ \\
$2004$ & $-0.1676$ & $0.0683$ & $-0.0247$ & $-0.3131$ & $-0.0335$ & $-0.5862$ & $-0.7544$ & $0.6027$ & $1.4783$ & $0.8869$ \\
$2005$ & $-0.1185$ & $0.1203$ & $-0.0813$ & $-0.3104$ & $-0.0727$ & $-0.5898$ & $-0.7339$ & $0.6251$ & $1.4898$ & $0.8685$ \\
$2006$ & $-0.0510$ & $0.1534$ & $-0.1357$ & $-0.2998$ & $-0.1116$ & $-0.5912$ & $-0.7127$ & $0.6470$ & $1.5011$ & $0.8498$ \\
$2007$ & $0.0251$ & $0.1599$ & $-0.1867$ & $-0.2820$ & $-0.1499$ & $-0.5903$ & $-0.6908$ & $0.6682$ & $1.5121$ & $0.8309$ \\
$2008$ & $0.0963$ & $0.1361$ & $-0.2333$ & $-0.2575$ & $-0.1877$ & $-0.5873$ & $-0.6682$ & $0.6888$ & $1.5228$ & $0.8117$ \\
$2009$ & $0.1485$ & $0.0854$ & $-0.2746$ & $-0.2272$ & $-0.2249$ & $-0.5822$ & $-0.6452$ & $0.7087$ & $1.5331$ & $0.7923$ \\
$2010$ & $0.1714$ & $0.0179$ & $-0.3099$ & $-0.1919$ & $-0.2613$ & $-0.5750$ & $-0.6216$ & $0.7280$ & $1.5431$ & $0.7727$ \\
$2011$ & $0.1626$ & $-0.0531$ & $-0.3389$ & $-0.1527$ & $-0.2969$ & $-0.5658$ & $-0.5976$ & $0.7466$ & $1.5527$ & $0.7530$ \\
$2012$ & $0.1260$ & $-0.1153$ & $-0.3613$ & $-0.1105$ & $-0.3316$ & $-0.5545$ & $-0.5732$ & $0.7647$ & $1.5619$ & $0.7332$ \\
$2013$ & $0.0701$ & $-0.1600$ & $-0.3771$ & $-0.0662$ & $-0.3651$ & $-0.5413$ & $-0.5483$ & $0.7821$ & $1.5708$ & $0.7133$ \\
$2014$ & $0.0044$ & $-0.1827$ & $-0.3863$ & $-0.0206$ & $-0.3975$ & $-0.5261$ & $-0.5229$ & $0.7990$ & $1.5794$ & $0.6934$ \\
$2015$ & $-0.0619$ & $-0.1823$ & $-0.3890$ & $0.0254$ & $-0.4285$ & $-0.5091$ & $-0.4971$ & $0.8153$ & $1.5877$ & $0.6734$ \\
$2016$ & $-0.1210$ & $-0.1600$ & $-0.3856$ & $0.0711$ & $-0.4580$ & $-0.4903$ & $-0.4707$ & $0.8309$ & $1.5957$ & $0.6534$ \\
$2017$ & $-0.1659$ & $-0.1191$ & $-0.3764$ & $0.1158$ & $-0.4859$ & $-0.4698$ & $-0.4439$ & $0.8460$ & $1.6035$ & $0.6334$ \\
$2018$ & $-0.1915$ & $-0.0643$ & $-0.3617$ & $0.1589$ & $-0.5122$ & $-0.4478$ & $-0.4166$ & $0.8603$ & $1.6111$ & $0.6132$ \\
$2019$ & $-0.1939$ & $-0.0017$ & $-0.3420$ & $0.1998$ & $-0.5368$ & $-0.4244$ & $-0.3888$ & $0.8739$ & $1.6185$ & $0.5929$ \\
$2020$ & $-0.1713$ & $0.0610$ & $-0.3178$ & $0.2381$ & $-0.5596$ & $-0.3996$ & $-0.3605$ & $0.8868$ & $1.6257$ & $0.5726$ \\
\hline
\end{tabular}
\end{table*}

\begin{table*}
\caption{
Ephemeris of best-fitting five-planet model Vb (1f:3e~MMR) 
between epochs $1995.0$ and $2020.0$.
Astrocentric coordinates in the $[E, N]$-plane.
}
\label{tab:ephemeris3}
\begin{tabular}{l c c c c c c c c c c}
\hline
epoch & \multicolumn{2}{c}{planet f} & \multicolumn{2}{c}{planet e} & \multicolumn{2}{c}{planet d} & \multicolumn{2}{c}{planet c} & \multicolumn{2}{c}{planet b} \\
& $\Delta \alpha$ [sec] & $\Delta \delta$ [sec] & $\Delta \alpha$ [sec] & $\Delta \delta$ [sec] & $\Delta \alpha$ [sec] & $\Delta \delta$ [sec] & $\Delta \alpha$ [sec] & $\Delta \delta$ [sec] & $\Delta \alpha$ [sec] & $\Delta \delta$ [sec]\\
\hline
$1995$ & $-0.1281$ & $-0.1244$ & $0.3684$ & $0.0086$ & $0.3078$ & $-0.4348$ & $-0.9060$ & $0.3909$ & $1.3609$ & $1.0443$ \\
$1996$ & $-0.1616$ & $-0.0642$ & $0.3529$ & $-0.0392$ & $0.2733$ & $-0.4601$ & $-0.8920$ & $0.4164$ & $1.3749$ & $1.0275$ \\
$1997$ & $-0.1645$ & $0.0080$ & $0.3303$ & $-0.0861$ & $0.2376$ & $-0.4832$ & $-0.8772$ & $0.4417$ & $1.3886$ & $1.0105$ \\
$1998$ & $-0.1339$ & $0.0784$ & $0.3006$ & $-0.1311$ & $0.2009$ & $-0.5040$ & $-0.8615$ & $0.4666$ & $1.4021$ & $0.9933$ \\
$1999$ & $-0.0754$ & $0.1325$ & $0.2638$ & $-0.1731$ & $0.1633$ & $-0.5225$ & $-0.8451$ & $0.4910$ & $1.4154$ & $0.9759$ \\
$2000$ & $-0.0017$ & $0.1604$ & $0.2205$ & $-0.2109$ & $0.1251$ & $-0.5388$ & $-0.8279$ & $0.5150$ & $1.4285$ & $0.9583$ \\
$2001$ & $0.0725$ & $0.1595$ & $0.1714$ & $-0.2432$ & $0.0863$ & $-0.5528$ & $-0.8100$ & $0.5385$ & $1.4412$ & $0.9404$ \\
$2002$ & $0.1354$ & $0.1333$ & $0.1176$ & $-0.2689$ & $0.0470$ & $-0.5644$ & $-0.7915$ & $0.5615$ & $1.4536$ & $0.9223$ \\
$2003$ & $0.1796$ & $0.0884$ & $0.0603$ & $-0.2871$ & $0.0076$ & $-0.5738$ & $-0.7724$ & $0.5840$ & $1.4656$ & $0.9041$ \\
$2004$ & $0.2014$ & $0.0323$ & $0.0013$ & $-0.2971$ & $-0.0320$ & $-0.5807$ & $-0.7527$ & $0.6060$ & $1.4773$ & $0.8856$ \\
$2005$ & $0.2000$ & $-0.0275$ & $-0.0579$ & $-0.2986$ & $-0.0715$ & $-0.5854$ & $-0.7324$ & $0.6276$ & $1.4886$ & $0.8670$ \\
$2006$ & $0.1766$ & $-0.0843$ & $-0.1154$ & $-0.2916$ & $-0.1108$ & $-0.5877$ & $-0.7115$ & $0.6487$ & $1.4995$ & $0.8483$ \\
$2007$ & $0.1340$ & $-0.1320$ & $-0.1699$ & $-0.2767$ & $-0.1497$ & $-0.5877$ & $-0.6901$ & $0.6693$ & $1.5101$ & $0.8295$ \\
$2008$ & $0.0766$ & $-0.1649$ & $-0.2199$ & $-0.2544$ & $-0.1881$ & $-0.5854$ & $-0.6682$ & $0.6894$ & $1.5204$ & $0.8106$ \\
$2009$ & $0.0101$ & $-0.1782$ & $-0.2644$ & $-0.2258$ & $-0.2258$ & $-0.5809$ & $-0.6456$ & $0.7090$ & $1.5304$ & $0.7917$ \\
$2010$ & $-0.0575$ & $-0.1683$ & $-0.3028$ & $-0.1919$ & $-0.2627$ & $-0.5742$ & $-0.6225$ & $0.7281$ & $1.5401$ & $0.7727$ \\
$2011$ & $-0.1166$ & $-0.1337$ & $-0.3345$ & $-0.1537$ & $-0.2986$ & $-0.5655$ & $-0.5988$ & $0.7468$ & $1.5496$ & $0.7537$ \\
$2012$ & $-0.1560$ & $-0.0769$ & $-0.3592$ & $-0.1123$ & $-0.3334$ & $-0.5547$ & $-0.5745$ & $0.7649$ & $1.5589$ & $0.7346$ \\
$2013$ & $-0.1660$ & $-0.0058$ & $-0.3771$ & $-0.0687$ & $-0.3670$ & $-0.5420$ & $-0.5496$ & $0.7824$ & $1.5680$ & $0.7154$ \\
$2014$ & $-0.1422$ & $0.0662$ & $-0.3881$ & $-0.0239$ & $-0.3993$ & $-0.5274$ & $-0.5241$ & $0.7992$ & $1.5769$ & $0.6961$ \\
$2015$ & $-0.0889$ & $0.1246$ & $-0.3927$ & $0.0212$ & $-0.4301$ & $-0.5113$ & $-0.4980$ & $0.8154$ & $1.5856$ & $0.6766$ \\
$2016$ & $-0.0178$ & $0.1583$ & $-0.3912$ & $0.0660$ & $-0.4595$ & $-0.4935$ & $-0.4715$ & $0.8308$ & $1.5942$ & $0.6569$ \\
$2017$ & $0.0567$ & $0.1637$ & $-0.3840$ & $0.1097$ & $-0.4874$ & $-0.4743$ & $-0.4445$ & $0.8454$ & $1.6024$ & $0.6370$ \\
$2018$ & $0.1225$ & $0.1431$ & $-0.3715$ & $0.1519$ & $-0.5138$ & $-0.4537$ & $-0.4171$ & $0.8593$ & $1.6103$ & $0.6169$ \\
$2019$ & $0.1713$ & $0.1023$ & $-0.3543$ & $0.1920$ & $-0.5387$ & $-0.4318$ & $-0.3894$ & $0.8724$ & $1.6179$ & $0.5965$ \\
$2020$ & $0.1986$ & $0.0486$ & $-0.3328$ & $0.2298$ & $-0.5620$ & $-0.4086$ & $-0.3614$ & $0.8846$ & $1.6251$ & $0.5761$ \\
\hline
\end{tabular}
\end{table*}

\begin{table*}
\caption{Ephemeris for the best-fitting five-planet model Vc 
(1e:2f~MMR) between epochs $1995.0$ and $2020.0$. 
Astrocentric coordinates in the $[E, N]$-plane.
}
\label{tab:ephemeris4}
\begin{tabular}{l c c c c c c c c c c}
\hline
epoch & \multicolumn{2}{c}{planet f} & \multicolumn{2}{c}{planet e} & \multicolumn{2}{c}{planet d} & \multicolumn{2}{c}{planet c} & \multicolumn{2}{c}{planet b} \\
& $\Delta \alpha$ [sec] & $\Delta \delta$ [sec] & $\Delta \alpha$ [sec] & $\Delta \delta$ [sec] & $\Delta \alpha$ [sec] & $\Delta \delta$ [sec] & $\Delta \alpha$ [sec] & $\Delta \delta$ [sec] & $\Delta \alpha$ [sec] & $\Delta \delta$ [sec]\\
\hline
$1995$ & $0.0211$ & $0.2113$ & $0.3653$ & $0.0044$ & $0.3072$ & $-0.4169$ & $-0.9002$ & $0.3883$ & $1.3692$ & $1.0467$ \\
$1996$ & $0.0856$ & $0.2149$ & $0.3483$ & $-0.0436$ & $0.2732$ & $-0.4439$ & $-0.8862$ & $0.4139$ & $1.3828$ & $1.0296$ \\
$1997$ & $0.1442$ & $0.2034$ & $0.3242$ & $-0.0907$ & $0.2379$ & $-0.4688$ & $-0.8716$ & $0.4390$ & $1.3961$ & $1.0123$ \\
$1998$ & $0.1942$ & $0.1795$ & $0.2931$ & $-0.1358$ & $0.2013$ & $-0.4913$ & $-0.8562$ & $0.4639$ & $1.4091$ & $0.9948$ \\
$1999$ & $0.2337$ & $0.1461$ & $0.2552$ & $-0.1778$ & $0.1637$ & $-0.5115$ & $-0.8402$ & $0.4883$ & $1.4217$ & $0.9771$ \\
$2000$ & $0.2621$ & $0.1056$ & $0.2109$ & $-0.2154$ & $0.1253$ & $-0.5293$ & $-0.8236$ & $0.5123$ & $1.4340$ & $0.9592$ \\
$2001$ & $0.2788$ & $0.0603$ & $0.1610$ & $-0.2474$ & $0.0864$ & $-0.5447$ & $-0.8063$ & $0.5360$ & $1.4460$ & $0.9412$ \\
$2002$ & $0.2838$ & $0.0126$ & $0.1068$ & $-0.2727$ & $0.0470$ & $-0.5576$ & $-0.7883$ & $0.5592$ & $1.4577$ & $0.9230$ \\
$2003$ & $0.2774$ & $-0.0357$ & $0.0496$ & $-0.2904$ & $0.0074$ & $-0.5681$ & $-0.7698$ & $0.5820$ & $1.4690$ & $0.9046$ \\
$2004$ & $0.2599$ & $-0.0825$ & $-0.0091$ & $-0.2997$ & $-0.0322$ & $-0.5761$ & $-0.7506$ & $0.6044$ & $1.4801$ & $0.8862$ \\
$2005$ & $0.2319$ & $-0.1260$ & $-0.0675$ & $-0.3006$ & $-0.0717$ & $-0.5818$ & $-0.7308$ & $0.6263$ & $1.4908$ & $0.8676$ \\
$2006$ & $0.1941$ & $-0.1642$ & $-0.1241$ & $-0.2931$ & $-0.1109$ & $-0.5850$ & $-0.7104$ & $0.6478$ & $1.5012$ & $0.8488$ \\
$2007$ & $0.1478$ & $-0.1952$ & $-0.1773$ & $-0.2777$ & $-0.1497$ & $-0.5859$ & $-0.6895$ & $0.6688$ & $1.5113$ & $0.8300$ \\
$2008$ & $0.0945$ & $-0.2167$ & $-0.2258$ & $-0.2550$ & $-0.1879$ & $-0.5845$ & $-0.6680$ & $0.6893$ & $1.5212$ & $0.8111$ \\
$2009$ & $0.0361$ & $-0.2266$ & $-0.2689$ & $-0.2261$ & $-0.2254$ & $-0.5809$ & $-0.6459$ & $0.7093$ & $1.5307$ & $0.7920$ \\
$2010$ & $-0.0244$ & $-0.2229$ & $-0.3058$ & $-0.1920$ & $-0.2621$ & $-0.5752$ & $-0.6232$ & $0.7287$ & $1.5400$ & $0.7728$ \\
$2011$ & $-0.0831$ & $-0.2036$ & $-0.3360$ & $-0.1538$ & $-0.2979$ & $-0.5673$ & $-0.6001$ & $0.7477$ & $1.5490$ & $0.7536$ \\
$2012$ & $-0.1350$ & $-0.1677$ & $-0.3596$ & $-0.1124$ & $-0.3325$ & $-0.5574$ & $-0.5763$ & $0.7660$ & $1.5578$ & $0.7342$ \\
$2013$ & $-0.1739$ & $-0.1160$ & $-0.3764$ & $-0.0690$ & $-0.3660$ & $-0.5456$ & $-0.5521$ & $0.7838$ & $1.5663$ & $0.7147$ \\
$2014$ & $-0.1939$ & $-0.0518$ & $-0.3867$ & $-0.0244$ & $-0.3982$ & $-0.5320$ & $-0.5274$ & $0.8010$ & $1.5745$ & $0.6951$ \\
$2015$ & $-0.1907$ & $0.0186$ & $-0.3907$ & $0.0206$ & $-0.4291$ & $-0.5166$ & $-0.5022$ & $0.8176$ & $1.5826$ & $0.6754$ \\
$2016$ & $-0.1641$ & $0.0866$ & $-0.3888$ & $0.0654$ & $-0.4585$ & $-0.4995$ & $-0.4765$ & $0.8335$ & $1.5903$ & $0.6556$ \\
$2017$ & $-0.1180$ & $0.1445$ & $-0.3813$ & $0.1092$ & $-0.4864$ & $-0.4809$ & $-0.4503$ & $0.8487$ & $1.5978$ & $0.6356$ \\
$2018$ & $-0.0591$ & $0.1869$ & $-0.3688$ & $0.1515$ & $-0.5128$ & $-0.4609$ & $-0.4238$ & $0.8632$ & $1.6051$ & $0.6155$ \\
$2019$ & $0.0055$ & $0.2118$ & $-0.3515$ & $0.1919$ & $-0.5376$ & $-0.4394$ & $-0.3968$ & $0.8770$ & $1.6121$ & $0.5952$ \\
$2020$ & $0.0698$ & $0.2197$ & $-0.3300$ & $0.2300$ & $-0.5608$ & $-0.4167$ & $-0.3696$ & $0.8900$ & $1.6187$ & $0.5747$ \\
\hline
\end{tabular}
\end{table*}

\begin{table*}
\caption{
Ephemeris of the best-fitting five-planet model~Vd (1f:2e~MMR) 
between epochs $1995.0$ and $2020.0$. 
Astrometric coordinates in the $[E, N]$-plane. 
}
\label{tab:ephemeris5}
\begin{tabular}{l c c c c c c c c c c}
\hline
epoch & \multicolumn{2}{c}{planet f} & \multicolumn{2}{c}{planet e} & \multicolumn{2}{c}{planet d} & \multicolumn{2}{c}{planet c} & \multicolumn{2}{c}{planet b} \\
& $\Delta \alpha$ [sec] & $\Delta \delta$ [sec] & $\Delta \alpha$ [sec] & $\Delta \delta$ [sec] & $\Delta \alpha$ [sec] & $\Delta \delta$ [sec] & $\Delta \alpha$ [sec] & $\Delta \delta$ [sec] & $\Delta \alpha$ [sec] & $\Delta \delta$ [sec]\\
\hline
$1995$ & $0.1761$ & $-0.0811$ & $0.3686$ & $0.0002$ & $0.3100$ & $-0.4290$ & $-0.9033$ & $0.3851$ & $1.3574$ & $1.0459$ \\
$1996$ & $0.1356$ & $-0.1404$ & $0.3528$ & $-0.0480$ & $0.2757$ & $-0.4554$ & $-0.8895$ & $0.4109$ & $1.3714$ & $1.0285$ \\
$1997$ & $0.0812$ & $-0.1856$ & $0.3297$ & $-0.0952$ & $0.2399$ & $-0.4795$ & $-0.8750$ & $0.4364$ & $1.3851$ & $1.0110$ \\
$1998$ & $0.0196$ & $-0.2145$ & $0.2996$ & $-0.1404$ & $0.2030$ & $-0.5013$ & $-0.8598$ & $0.4615$ & $1.3984$ & $0.9933$ \\
$1999$ & $-0.0436$ & $-0.2269$ & $0.2625$ & $-0.1823$ & $0.1650$ & $-0.5205$ & $-0.8438$ & $0.4863$ & $1.4115$ & $0.9755$ \\
$2000$ & $-0.1040$ & $-0.2243$ & $0.2190$ & $-0.2197$ & $0.1263$ & $-0.5373$ & $-0.8271$ & $0.5108$ & $1.4242$ & $0.9575$ \\
$2001$ & $-0.1583$ & $-0.2085$ & $0.1700$ & $-0.2515$ & $0.0870$ & $-0.5516$ & $-0.8096$ & $0.5348$ & $1.4367$ & $0.9395$ \\
$2002$ & $-0.2044$ & $-0.1817$ & $0.1164$ & $-0.2767$ & $0.0474$ & $-0.5634$ & $-0.7915$ & $0.5584$ & $1.4488$ & $0.9212$ \\
$2003$ & $-0.2406$ & $-0.1461$ & $0.0596$ & $-0.2942$ & $0.0075$ & $-0.5727$ & $-0.7727$ & $0.5817$ & $1.4607$ & $0.9029$ \\
$2004$ & $-0.2660$ & $-0.1039$ & $0.0012$ & $-0.3036$ & $-0.0323$ & $-0.5795$ & $-0.7531$ & $0.6044$ & $1.4724$ & $0.8845$ \\
$2005$ & $-0.2798$ & $-0.0573$ & $-0.0572$ & $-0.3045$ & $-0.0720$ & $-0.5838$ & $-0.7329$ & $0.6267$ & $1.4837$ & $0.8659$ \\
$2006$ & $-0.2818$ & $-0.0081$ & $-0.1141$ & $-0.2971$ & $-0.1114$ & $-0.5858$ & $-0.7120$ & $0.6485$ & $1.4948$ & $0.8472$ \\
$2007$ & $-0.2718$ & $0.0414$ & $-0.1678$ & $-0.2816$ & $-0.1503$ & $-0.5854$ & $-0.6904$ & $0.6697$ & $1.5057$ & $0.8283$ \\
$2008$ & $-0.2499$ & $0.0891$ & $-0.2172$ & $-0.2589$ & $-0.1885$ & $-0.5827$ & $-0.6682$ & $0.6905$ & $1.5163$ & $0.8094$ \\
$2009$ & $-0.2166$ & $0.1327$ & $-0.2612$ & $-0.2298$ & $-0.2260$ & $-0.5778$ & $-0.6453$ & $0.7106$ & $1.5267$ & $0.7903$ \\
$2010$ & $-0.1727$ & $0.1697$ & $-0.2991$ & $-0.1953$ & $-0.2627$ & $-0.5708$ & $-0.6218$ & $0.7301$ & $1.5368$ & $0.7710$ \\
$2011$ & $-0.1195$ & $0.1976$ & $-0.3303$ & $-0.1566$ & $-0.2983$ & $-0.5617$ & $-0.5977$ & $0.7490$ & $1.5466$ & $0.7516$ \\
$2012$ & $-0.0591$ & $0.2136$ & $-0.3547$ & $-0.1146$ & $-0.3329$ & $-0.5506$ & $-0.5731$ & $0.7672$ & $1.5562$ & $0.7320$ \\
$2013$ & $0.0052$ & $0.2150$ & $-0.3722$ & $-0.0705$ & $-0.3663$ & $-0.5377$ & $-0.5479$ & $0.7848$ & $1.5656$ & $0.7122$ \\
$2014$ & $0.0689$ & $0.1994$ & $-0.3830$ & $-0.0251$ & $-0.3984$ & $-0.5229$ & $-0.5223$ & $0.8016$ & $1.5746$ & $0.6923$ \\
$2015$ & $0.1263$ & $0.1658$ & $-0.3872$ & $0.0207$ & $-0.4291$ & $-0.5064$ & $-0.4961$ & $0.8177$ & $1.5834$ & $0.6722$ \\
$2016$ & $0.1706$ & $0.1152$ & $-0.3853$ & $0.0662$ & $-0.4585$ & $-0.4884$ & $-0.4696$ & $0.8331$ & $1.5918$ & $0.6519$ \\
$2017$ & $0.1955$ & $0.0516$ & $-0.3775$ & $0.1106$ & $-0.4864$ & $-0.4687$ & $-0.4426$ & $0.8477$ & $1.5999$ & $0.6314$ \\
$2018$ & $0.1972$ & $-0.0181$ & $-0.3644$ & $0.1534$ & $-0.5128$ & $-0.4476$ & $-0.4153$ & $0.8616$ & $1.6077$ & $0.6108$ \\
$2019$ & $0.1759$ & $-0.0858$ & $-0.3464$ & $0.1941$ & $-0.5376$ & $-0.4251$ & $-0.3877$ & $0.8747$ & $1.6151$ & $0.5901$ \\
$2020$ & $0.1353$ & $-0.1442$ & $-0.3239$ & $0.2323$ & $-0.5608$ & $-0.4012$ & $-0.3598$ & $0.8871$ & $1.6222$ & $0.5693$ \\
\hline
\end{tabular}
\end{table*}

\bibliographystyle{mn2e}
\bibliography{ms}
\label{lastpage}
\end{document}